\documentclass[a4paper]{article}

\usepackage[utf8]{inputenc}
\usepackage{amsmath}
\usepackage{jheppub}
\usepackage{amsfonts}
\usepackage{amssymb}
\usepackage{mathtools}
\usepackage{graphicx}
\usepackage{relsize}
\usepackage{array}
\usepackage{pdfpages}
\usepackage{subfig}
\usepackage{multicol}
\usepackage{subfig}
\usepackage{pdfpages}
\usepackage{float}
\usepackage{epstopdf}

\newcommand{\D}[1]{(\mathbf{1}_{#1},0)}
\newcommand{\ta}[1]{\mathbf{T}^{a}_{#1}} 
\newcommand{\tb}[1]{\mathbf{T}^{b}_{#1}}
\newcommand{\tc}[1]{\mathbf{T}^{c}_{#1}}
\newcommand{\td}[1]{\mathbf{T}^{d}_{#1}}
\newcommand{\tf}[1]{\mathbf{T}^{f}_{#1}} 

\newcommand{\te}[1]{\mathbf{T}^{e}_{#1}}
\newcommand{\tj}[1]{\mathbf{T}^{j}_{#1}}
\newcommand{\thh}[1]{\mathbf{T}^{h}_{#1}}
\newcommand{\tkk}[1]{\mathbf{T}^{k}_{#1}}
\newcommand{\tm}[1]{\mathbf{T}^{m}_{#1}}

\newcommand{\tu}[1]{\mathbf{T}^{u}_{#1}}

\newcommand{\fabc}{f^{abc}}

\newcommand{\faeh}{f^{aeh}}

\newcommand{\fafk}{f^{afk}}
\newcommand{\fbcg}{f^{bcg}}
\newcommand{\fefg}{f^{efg}}
\newcommand{\fabm}{f^{abm}}
\newcommand{\nn}{\nonumber}
\newcommand{\secn}[1]{Section~\ref{#1}}
\newcommand{\be}{\begin{equation}}
\newcommand{\ee}{\end{equation}}
\newcommand{\beq}{\begin{eqnarray}}
\newcommand{\eeq}{\end{eqnarray}}
\newcommand{\bea}{\begin{eqnarray}}
\newcommand{\eea}{\end{eqnarray}}
\newcommand{\nnn}{\nonumber}
\newcommand{\e}{\epsilon}

\newcommand{\as}{\alpha_s}
\providecommand{\tabularnewline}{\\}
\def\bra#1{%
  \left\langle\smash{#1}{\vphantom1}\right|}
\def\ket#1{%
  \left|\smash{#1}{\vphantom1}\right\rangle}
\def\eq#1{Eq.~(\ref{#1})}

\preprint{\\ \rightline{CERN-TH-2021-015}}
                 
\title{Cwebs beyond three loops in multiparton amplitudes} 

\author[a]{Neelima Agarwal,}
\author[b,c]{Lorenzo Magnea,}
\author[d]{Sourav Pal,}
\author[d]{Anurag Tripathi}

\affiliation[a]{Department of Physics, Chaitanya Bharathi Institute of 
  Technology, \\ Gandipet, Hyderabad, Telangana State 500075, India}
\affiliation[b]{Theoretical Physics Department, CERN, CH-1211 Geneva 23, 
  Switzerland}
\affiliation[c]{Dipartimento di Fisica and Arnold-Regge Center, Universit\`a 
  di Torino, \\ and INFN, Sezione di Torino, Via Pietro Giuria 1, I-10125 
  Torino, Italy}
\affiliation[d]{Department of Physics, Indian Institute of Technology 
  Hyderabad, \\ Kandi, Sangareddy, Telangana State 502285, India}

\emailAdd{neelimaagarwal$\_$physics@cbit.ac.in}
\emailAdd{lorenzo.magnea@unito.it}
\emailAdd{spalexam@gmail.com}
\emailAdd{tripathi@phy.iith.ac.in}

\abstract{Correlators of Wilson-line operators in non-abelian gauge theories are 
known to exponentiate, and their logarithms can be organised in terms of collections 
of Feynman diagrams called {\it webs}. In \cite{Agarwal:2020nyc} we introduced 
the concept of {\it Cweb}, or {\it correlator web}, which is a set of skeleton diagrams
built with connected gluon correlators, and we computed the mixing matrices for 
all Cwebs connecting four or five Wilson lines at four loops. Here we complete
the evaluation of four-loop mixing matrices, presenting the results for all Cwebs
connecting two and three Wilson lines. We observe that the conjuctured column 
sum rule is obeyed by all the mixing matrices that appear at four-loops. We also
show how low-dimensional mixing matrices can be uniquely determined from
their known combinatorial properties, and provide some all-order results for
selected classes of mixing matrices. Our results complete the required colour 
building blocks for the calculation of the soft anomalous dimension matrix at 
four-loop order.}

\begin{document}

%%%%%%%%%%%%%%%%%%%%%%%%%%%%%%%%%%%%%%%%%

\maketitle

%%%%%%%%%%%%%%%%%%%%%%%%%%%%%%%%%%%%%%%%%

\section{Introduction}
\label{Intro}

The infrared (IR) structure of scattering amplitudes in gauge field theories is 
universal. It is this universality -- the independence from the specific nature of 
the hard process under consideration, that makes it an important object of study. 
Remarkable insights on how IR singularities organise themselves in perturbation 
theory have been accumulated over a long history spanning almost a century 
~\cite{Bloch:1937pw,Sudakov:1954sw,Yennie:1961ad,Kinoshita:1962ur,
        Lee:1964is,Grammer:1973db,Mueller:1979ih,Collins:1980ih,Sen:1981sd,
        Sen:1982bt,Korchemsky:1987wg,Korchemsky:1988hd,Magnea:1990zb,
        Dixon:2008gr,Gardi:2009qi,Becher:2009qa,Feige:2014wja}. 
While these studies are of course interesting in their own right, uncovering 
subtle properties of gauge field theories, it is important to note that they also 
have practical applications to high-energy scattering. First of all, even though 
IR singularities cancel when one constructs an experimentally measurable 
IR-safe observable, they often leave behind potentially large logarithms
of kinematic variables that jeopardise the applicability of fixed-order perturbation 
theory. The universal IR structure of scattering amplitudes is one of the key 
ingredients allowing for the all-order summation of these large logarithms, 
leading to more precise and controlled physical predictions~\cite{Sterman:1995fz,
Laenen:2004pm,Luisoni:2015xha}. Furthermore, even at fixed orders, implementing 
the cancellation of IR singularities for complex collider observables at high 
perturbative orders is a difficult and important undertaking: indeed, the 
construction of general and efficient subtraction procedures beyond 
next-to-leading order (NLO) is a broad ongoing effort (see, for example, 
~\cite{GehrmannDeRidder:2005cm,Somogyi:2005xz,Catani:2007vq,
        Czakon:2010td,Boughezal:2015dva,Sborlini:2016hat,Caola:2017dug,
        Herzog:2018ily,Magnea:2018hab,Magnea:2018ebr,Capatti:2020xjc}).

To be somewhat more precise, gauge theory scattering amplitudes in the IR limit 
factorise into universal soft and collinear functions~\cite{Sterman:1995fz,Dixon:2008gr,
Gardi:2009qi,Becher:2009qa,Gardi:2009zv,Feige:2014wja}, multiplying 
finite matching coefficients. These soft and collinear functions, in turn, can 
be expressed as matrix elements of field operators and Wilson lines, which 
are the object of the present study. We note in passing that such matrix 
elements play a ubiquitous role, not only for the factorisation of scattering 
amplitudes, but also in many effective theories based on QCD~\cite{Manohar:2000dt,
Brambilla:2004jw,Becher:2014oda}. The methods developed in~\cite{Agarwal:2020nyc} 
and the objects of interest -- Cwebs -- concern the evaluation of Wilson-line
correlators of the general form
\beq
  {\cal S}_n \left( \gamma_i \right) \, \equiv \, \bra{0} \prod_{k = 1}^n
  \Phi \left(  \gamma_k \right) \ket{0} \, ,
\label{genWLC}
\eeq
where $\Phi ( \gamma )$ are Wilson-line operators evaluated on smooth space-time 
contours $\gamma$,
\beq
  \Phi \left(  \gamma \right) \, \equiv \, P \exp \left[ {\rm i} g \!
  \int_\gamma d x \cdot {\bf A} (x) \right] \, ,
\label{genWL}
\eeq
where ${\bf A}^\mu (x) = A^\mu_a (x) \, {\bf T}^a$ is a non-abelian gauge field, 
and ${\bf T}^a$ is a generator of the gauge algebra, which can be taken to belong
to any desired representation. If the smooth contours $\gamma$ are closed, then 
the correlator is gauge-invariant, while, if they are open, the correlator is a colour 
tensor with open colour indices in the chosen representations, attached to the 
endpoints of each Wilson line. In the present paper, we will restrict ourselves to
soft colour operators associated with multi-particle scattering amplitudes in 
gauge theories, which encode all their soft singularities. These operators are
of the form of \eq{genWLC}, with the contours $\gamma_k$ given by rays 
extending from the origin along directions $\beta_k$, corresponding to the 
four-velocities of the particles participating in the scattering. In this case, we 
write the soft operator as
\beq
  {\cal S}_n \Big( \beta_i \cdot \beta_j, \as (\mu^2), \e \Big) \, \equiv \, 
  \bra{0} \prod_{k = 1}^n \Phi_{\beta_k} \left( \infty, 0 \right) \ket{0} , \quad \!
  \Phi_\beta \left( \infty, 0 \right) \, \equiv \, P \exp \left[ {\rm i} g \!
  \int_0^\infty d \lambda \, \beta \cdot {\bf A} (\lambda \beta) \right] \! ,
\label{softWLC}
\eeq
where each Wilson line is taken in the representation of the gauge algebra
corresponding to the high-energy particle it represents.

We note that soft operators of the form of \eq{softWLC} are affected by ultraviolet, 
soft, and, for $\beta_i^2 = 0$, collinear singularities: as a consequence, 
special care is required to evaluate them~\cite{Mitov:2010xw,Henn:2013wfa,
Gardi:2013saa,Falcioni:2014pka}. In what follows, we will generally assume
$\beta_i^2 \neq 0$, so that collinear divergences will not arise; for soft divergences,
we will make use of the smooth exponential suppression of gluon interactions
at large distances, discussed in Refs.~\cite{Gardi:2013saa,Falcioni:2014pka}; 
finally, we will retain dimensional regularisation for ultraviolet singularities.
With this scheme, the bare correlator is finite, and can be evaluated and 
renormalised. The physically relevant quantity to be extracted is then the
set of ultraviolet counterterms for the correlator: indeed, in the absence of
an IR regulator, all radiative corrections to \eq{softWLC} vanish, since all
relevant Feynman diagrams are scale-less; the renormalised correlator
is therefore given precisely (in a minimal scheme) by the set of its UV 
counterterms.

Renormalised correlators of the form  of \eq{softWLC}  obey renormalisation 
group equations which lead to exact exponentiation. This means that they can 
be written as
\beq
  \mathcal{S}_n \Big( \beta_i \cdot \beta_j, \as (\mu^2), \e \Big) \, = \, 
  \mathcal{P} \exp \left[ - \frac{1}{2} \int_{0}^{\mu^2} \frac{d \lambda^2}  
  {\lambda^2} \, {\bf \Gamma}_n \Big( \beta_i \cdot \beta_j, \alpha_s (\lambda^2), 
  \e \Big) \right]  \, ,
\label{softmatr}
\eeq
where ${\bf \Gamma}_n$ is the soft anomalous dimension matrix, which is the focus 
of our studies. ${\bf \Gamma}_n$ was computed at one loop in~\cite{Kidonakis-1998} 
(see also~\cite{Korchemskaya:1994qp}); at two loops in the massless case 
in~\cite{Aybat:2006wq,Aybat:2006mz}, and in the massive case in~\cite{Mitov:2009sv,
Ferroglia:2009ep,Ferroglia:2009ii,Kidonakis:2009ev,Chien:2011wz}; finally, at three 
loops in the massless case in~\cite{Almelid:2015jia,Almelid:2017qju}. Pushing these 
calculations to higher perturbative orders is of great interest, since the resulting 
structures are highly constrained by symmetries, and display non-trivial mathematical
properties, tied to interesting physical configurations: preliminary results on the
four-loop structure of ${\bf \Gamma_n}$ have been given in~\cite{Becher:2019avh,
Falcioni:2020lvv}.

In practice, an approach alternative to the above renormalisation-group based 
methods is often useful for studying general Wilson-line correlators of the form 
of \eq{genWLC}. All such correlators are known to obey a non-trivial form of 
{\it diagrammatic exponentiation}, so that one can write
\beq
  {\cal S}_n \left( \gamma_i \right) \, = \, \exp \Big[ {\cal W}_n \left( \gamma_i \right) 
  \Big]  \, ,
\label{diaxp}
\eeq
where the exponent, ${\cal W}_n \left( \gamma_i \right)$, can be directly computed 
in terms of a subset of the Feynman diagrams contributing to ${\cal S}_n (\gamma_i)$. 
For non-abelian gauge theories, this was first pointed out in Refs.~\cite{Sterman-1981,
Gatheral,Frenkel-1984}, in the case of two straight, semi-infinite Wilson lines. For 
general configurations, it was proven in Refs.~\cite{Mitov:2010rp,Gardi:2010rn}. 
These studies show that Feynman diagrams contributing to ${\cal W}_n \left( 
\gamma_i \right)$ are organised in sets that are called {\it webs}: in particular, 
\begin{itemize}
\item For an abelian theory, webs are given by connected photon diagrams attaching
to the Wilson lines.
\item For a non-abelian theory, if only two Wilson lines are present, webs are 
given by two-eikonal irreducible diagrams, {\it i.e.} diagrams that do not become 
disconnected upon cutting only the two Wilson lines. 
\item For general, multi-line correlators, webs are sets of diagrams that differ 
among themselves by the ordering of their gluon attachments to the Wilson lines. 
\end{itemize}
Clearly, by means of webs, one can directly compute the soft anomalous dimension 
matrix ${\bf \Gamma}_n$. It is important, however, to note that, for general multi-line 
correlators, multiplicative renormalisability and  exponentiation combine non-trivially, 
due to the non abelian nature of webs, and the calculation of renormalised correlators
includes commutators of counterterms and bare webs, as required~\cite{Gardi:2011yz}.
Not surprisingly, non-trivial simplifications of the final result for ${\bf \Gamma}_n$
appear only when the commutator terms are properly taken into account. We also
recall the special properties arising in the massless case, when the countours 
$\gamma_i$ are light-like, straight, semi-infinite Wilson lines: for such amplitudes,
scale invariance imposes strong constraints on the functional dependence of the 
soft anomalous dimension matrix. Up to two loops, ${\bf \Gamma}_n$ can only involve 
dipole colour correlations between Wilson lines~\cite{Aybat:2006mz,Aybat:2006wq,
Gardi:2009qi,Becher:2009cu,Becher:2009qa,Gardi:2009zv}; beyond two loops, scale-invariant conformal cross ratios of the form  $\rho_{ijkl} \equiv (\beta_i \cdot  \beta_j  \beta_k \cdot  
\beta_l)/(\beta_i \cdot  \beta_k  \beta_j \cdot  \beta_l)$ correlating four partons can also appear: the first such correlations 
arise at three loops, with at least four Wilson lines, and were computed in 
Ref.~\cite{Almelid:2015jia}. At four loops, one finds the first occurence of higher-order
Casimir operators of the gauge algebra, which have recently been computed in
the case of the cusp anomalous dimension~\cite{Henn:2019swt,vonManteuffel:2020vjv}:
the interplay of this class of contributions to the cusp with similar contributions
to multi-particle correlators is a very interesting open problem, with subtle connections
to the factorisation of collinear poles. We note that an interesting alternative approach
to IR exponentiation, focusing not on diagrammatics but on the symmetries 
and renormalisation properties of Wilson-line correlators, was developed in 
Refs.~\cite{Vladimirov:2014wga,Vladimirov:2015fea}.

The present paper focuses on the colour structure of the perturbative exponent
${\cal W}_n$ for the soft operator in \eq{softWLC}, mostly at the four-loop 
order. Specifically, we complete the task of computing the mixing matrices 
determining the exponentiated colour factors for all four-loop Cwebs, initiated 
in Ref.~\cite{Agarwal:2020nyc}, and we present some general properties of
mixing matrices, valid to all orders. We begin by reviewing existing  results on 
diagrammatic exponentiation in \secn{Webs}, where we discuss the concept of
Cweb and the main properties of webs and Cwebs, and we list the four-loop 
Cwebs connecting two and three Wilson lines. \secn{Repli} describes the replica 
method~\cite{Gardi:2010rn} for generating the Cweb mixing matrices, and its 
implementation in our code.  In \secn{Fourwe}, we give two detailed examples,
presenting the mixing matrices for one two-line Cweb and for one three-line Cweb. 
In \secn{direct-comp}, we present some observations on the structure of mixing 
matrices, including a derivation of two- and three-dimensional mixing matrices 
directly based on their combinatorial properties, bypassing the replica algorithm. 
Finally, we summarise our results and perspectives in \secn{Conclu}. In the 
Appendix, we list in detail the mixing matrices for all the Cwebs discussed in the 
main text.

%%%%%%%%%%%%%%%%%%%%%%%%%%%%%%%%%%%%%%%%%

\section{Webs and Cwebs for correlators of two and three Wilson lines}
\label{Webs}

In this section we introduce mixing matrices and their main properties, using the 
language of webs for multi-particle amplitudes, developed in ~\cite{Mitov:2010rp,
Gardi:2010rn,Gardi:2011wa,Gardi:2011yz,Dukes:2013wa, Gardi:2013ita,
Dukes:2013gea,Dukes:2016ger}. We then proceed to introduce correlator
webs, or Cwebs, following~\cite{Agarwal:2020nyc}, exploiting the fact that
the combinatorial problem associated with gluon attachments to Wilson lines
is structurally factored from the colour algebra associated with gluon subdiagrams.
Finally, as an example, we enumerate Cwebs with lowest-order contributions at 
three and four loops and connecting two or three Wilson lines.

A web is defined as a set of fixed-order Feynman diagrams contributing 
to a Wilson-line correlator, which differ only by the permutation of their gluon 
attachments to each Wilson line. If we write every Feynman diagram $D$ as 
the product of its kinematic factor  $K(D)$ and its colour 
factor $C(D)$, then the logarithm of the correlator, ${\cal W}_n$, can be written as 
\beq
  {\cal W}_n (\gamma_i) \, = \, \sum_D K(D) \, \widetilde{C}(D) \, ,
\label{ecfdefn} 
\eeq
where the {\it Exponentiated Colour Factor} (ECF) for diagram $D$, denoted
by $\widetilde{C}(D)$, is a linear combination of the colour factors of the diagrams 
which are present in the web. Thus we can write
\beq
  \widetilde{C}(D) \, = \, \sum_{D' \in w}  R_{w} (D,D') C(D') \, ,
\label{mixingmat}
\eeq
where the sum runs over all the diagrams present in web $w$ and $R_{w}$ is 
the {\it web mixing matrix}. Using \eq{ecfdefn} and \eq{mixingmat} we obtain
an expression for the correlator in \eq{diaxp},
\beq
  \mathcal{S}_n (\gamma_i) \, = \, \exp \bigg[ \sum_{D,D'} K(D) R(D,D') C(D) 
  \bigg] \, ,
\label{soft-defn}
\eeq
where the block-diagonal matrix $R$ is built using the web mixing matrices 
$R_w$ as blocks. In turn, each web $w$ can be written as
\beq
  w \, = \, \sum_{D \in  w} {\cal  K} (D) \, \widetilde{C} (D) \, = \,
  \sum_{D,D' \in  w} {\cal K} (D) \, R_w (D, D') \, C (D')  \, .
\label{eq:webredef}
\eeq
In this language, ${\cal W}_n = \sum w_n$, where the sum extends to all
webs arising in the presence of $n$ Wilson lines, order by order in perturbation
theory.
\begin{figure}[H]
	\centering
	\vspace{-3mm}
	\subfloat[$W_2^{(0,0,1)} (1,3)$]{\includegraphics[height=3.2cm,width=3.2cm]
	{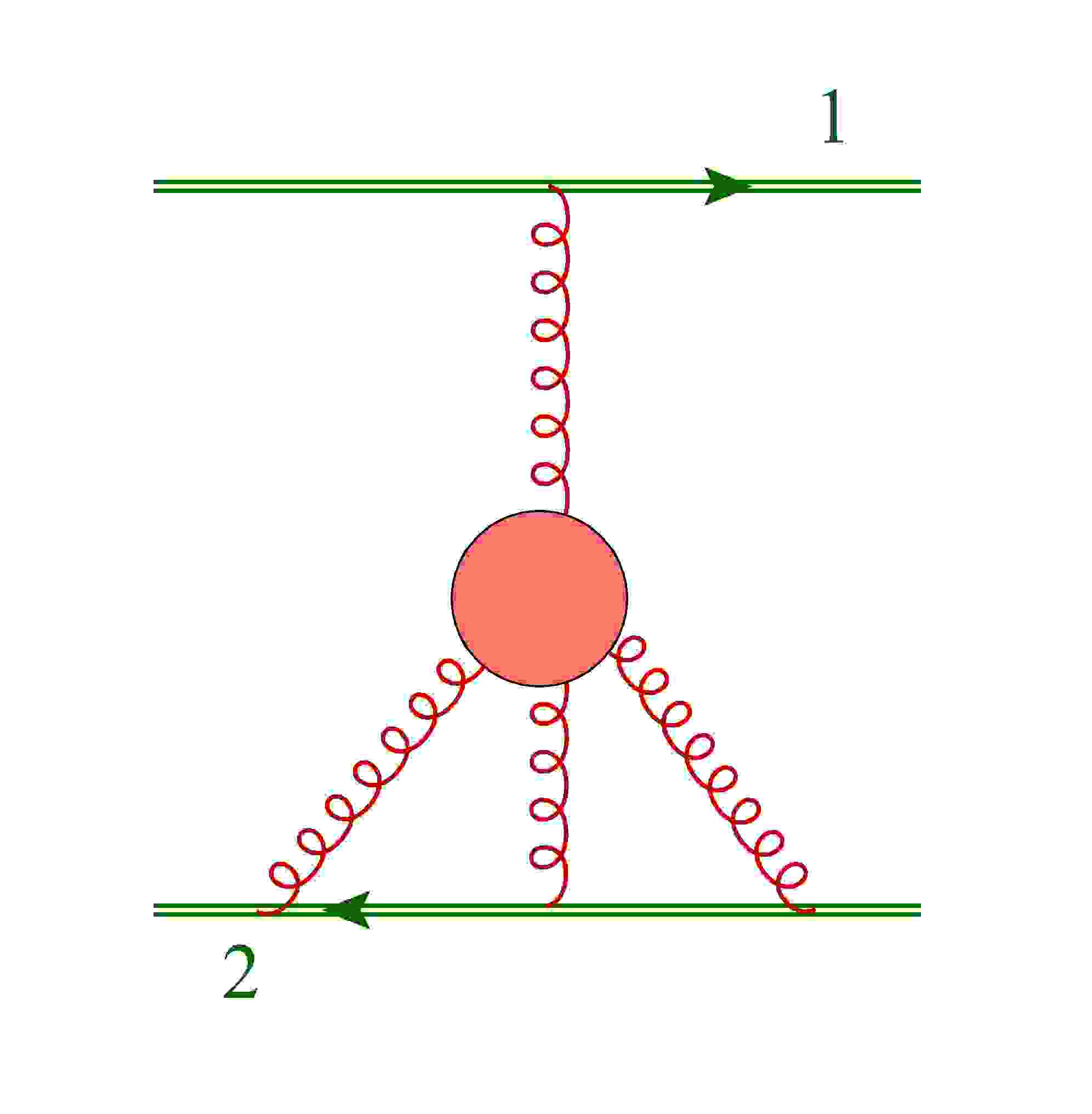} }
	\quad
	\subfloat[$W_2^{(0,0,1)} (2,2)$]{\includegraphics[height=3.2cm,width=3.2cm]
	{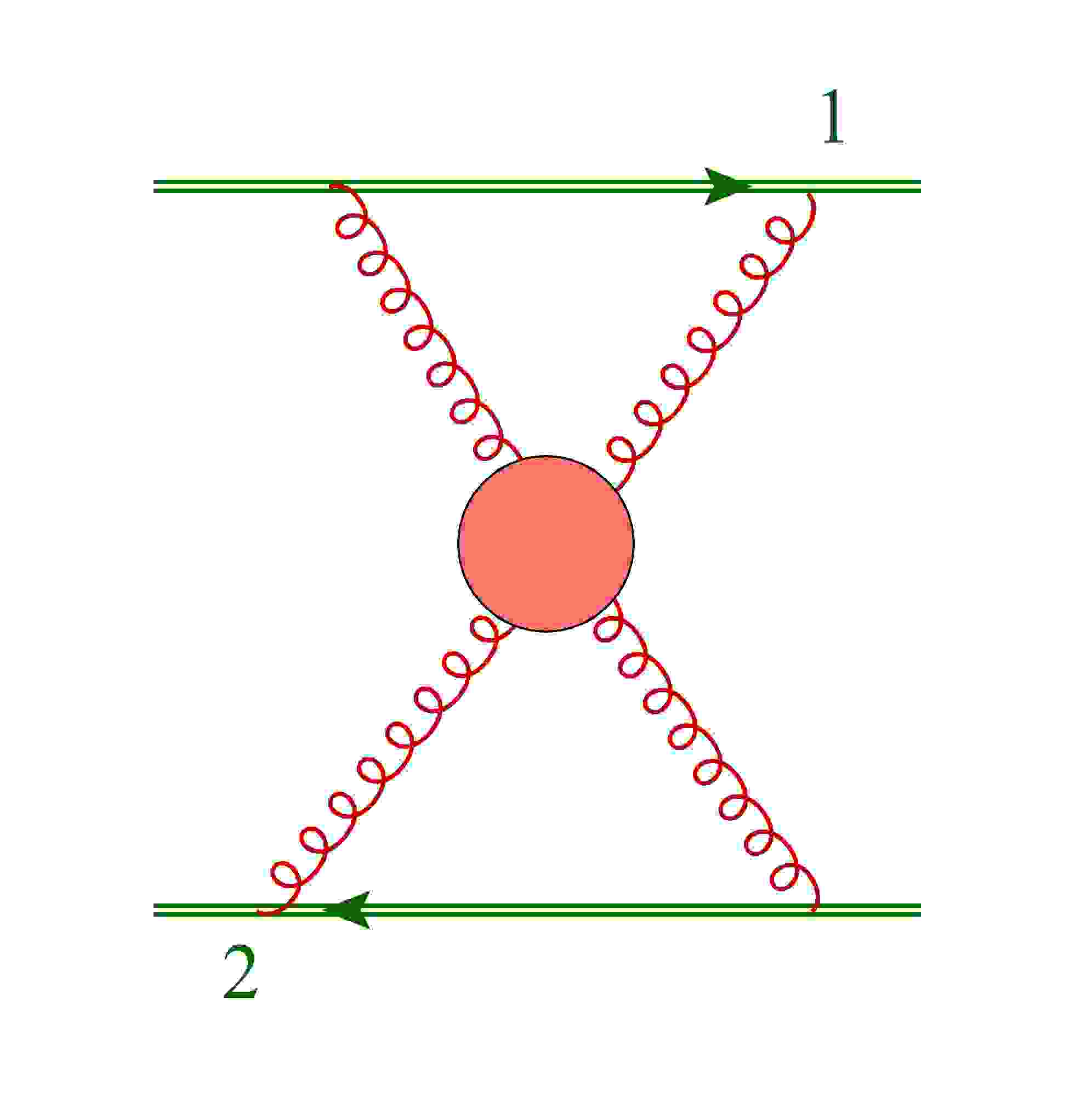} }
	\quad
	\subfloat[$W_2^{(1,1)} (2,3)$]{\includegraphics[height=3.2cm,width=3.2cm]
	{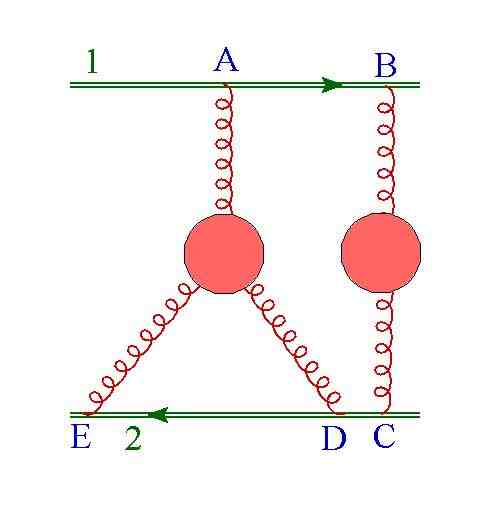} }
	\quad
	\subfloat[$W_2^{(3)} (3,3)$]{\includegraphics[height=3.2cm,width=3.2cm]
	{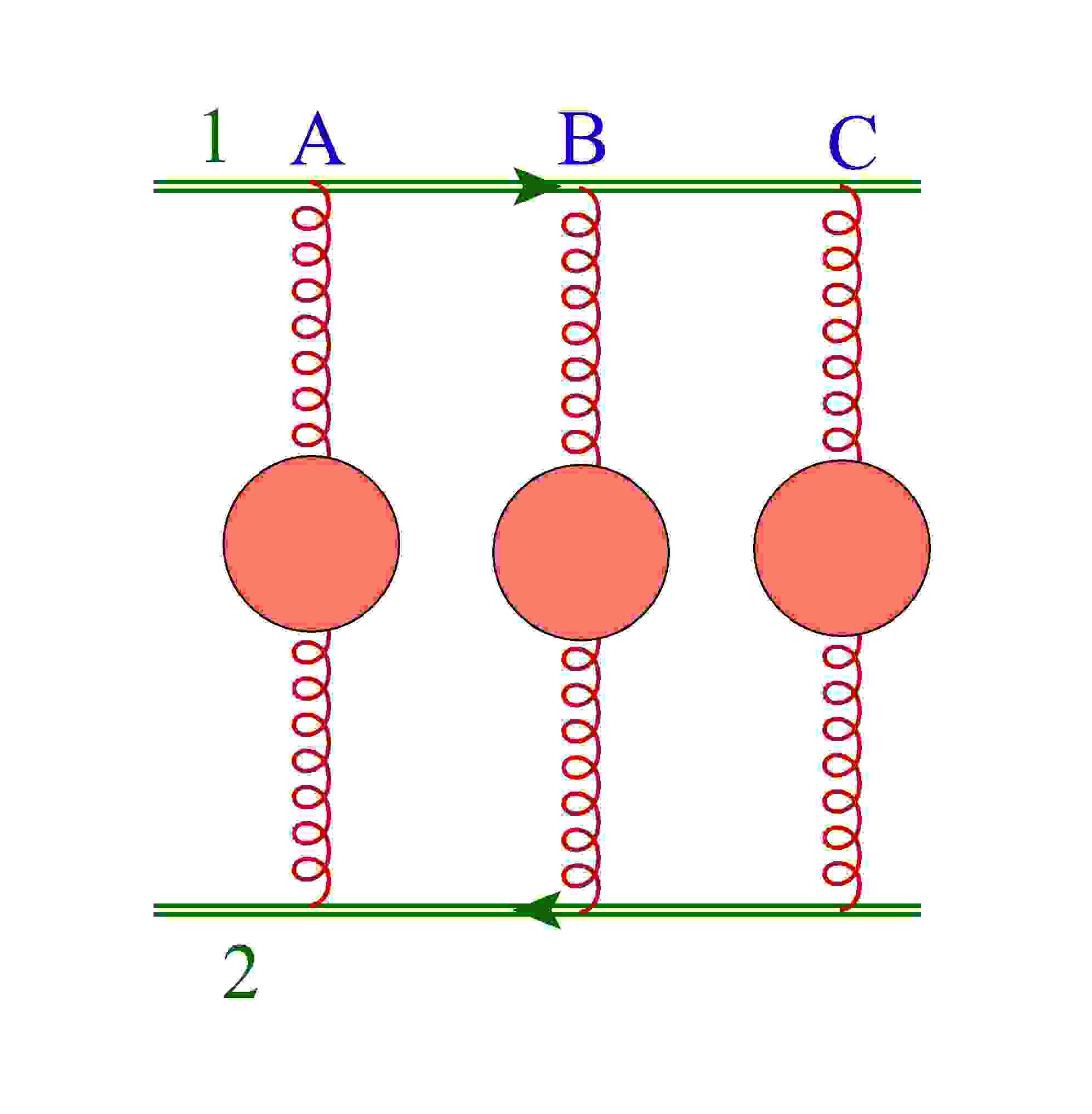} }
	\caption{Representative skeleton diagrams for the four three-loop Cwebs 
	connecting two Wilson lines in a massless theory. As clarified in the text, 
	the Cwebs in (a) and (b) have only one diagram, displayed here, while the
	Cwebs in (c) and (d) have six diagrams.}
	\label{Cwebg6_2}
\end{figure}
\begin{figure}[H]
	\centering
	\vspace{-3mm}
	\subfloat[$W_3^{(0,0,1)} (1,1,2)$]{\includegraphics[height=2.8cm,width=3.2cm]
	{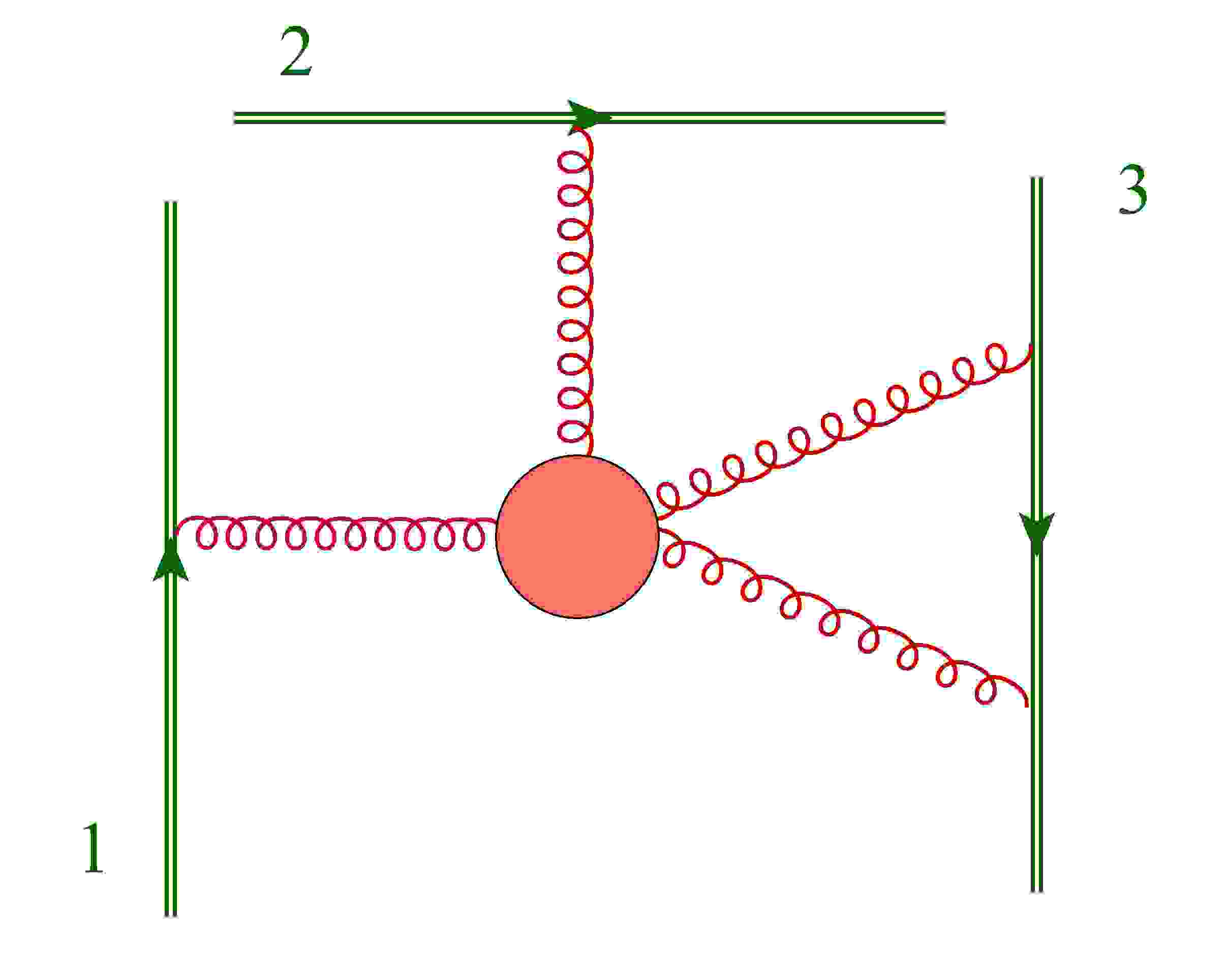} }
	\quad
	\subfloat[$W_3^{(1,1)} (1,1,3)$]{\includegraphics[height=2.8cm,width=3.2cm]
	{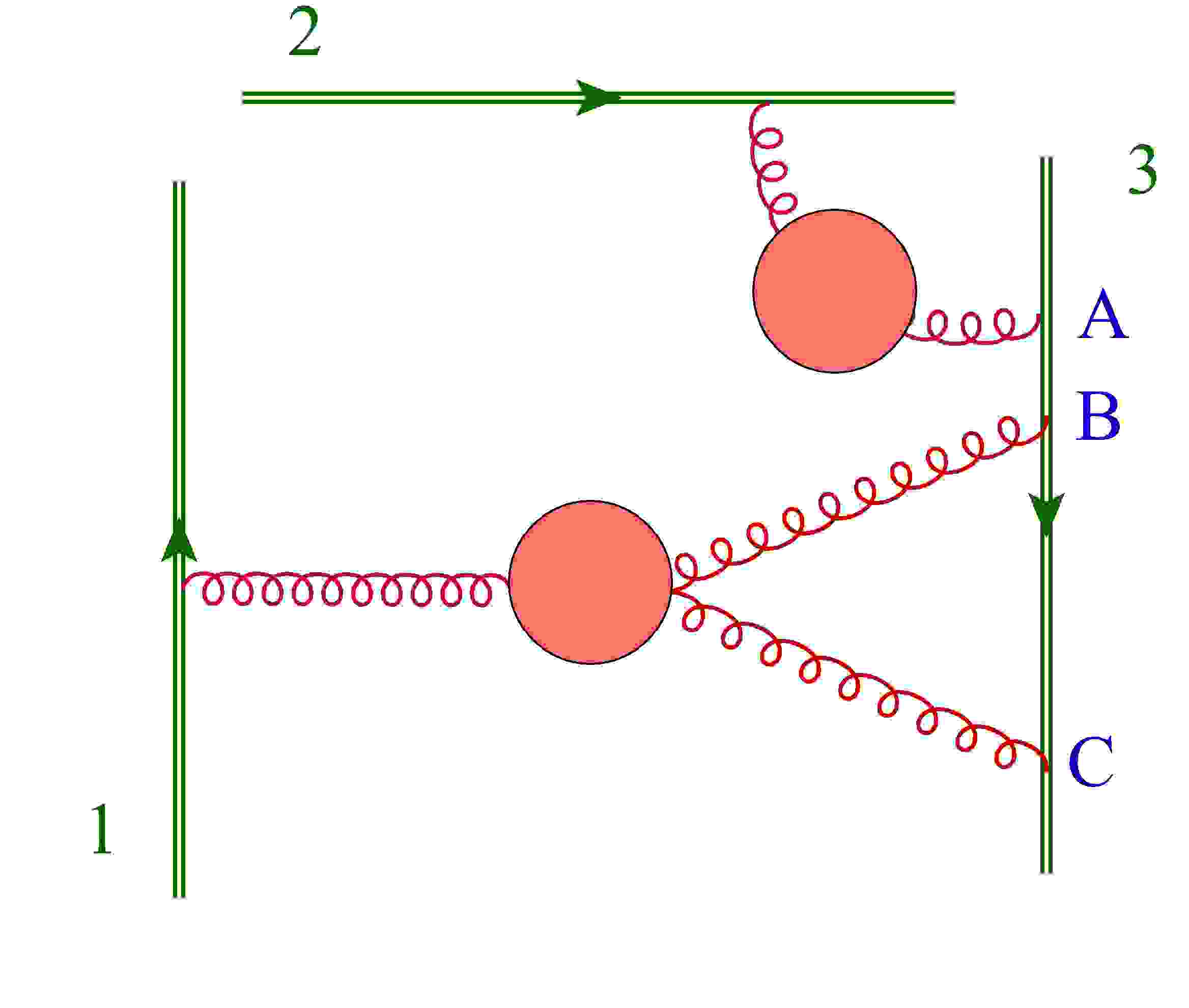} }
	\quad
	\subfloat[$W_{3, \, {\rm I}}^{(1,1)} (2,1,2)$]{\includegraphics[height=2.8cm,width=3.2cm]
	{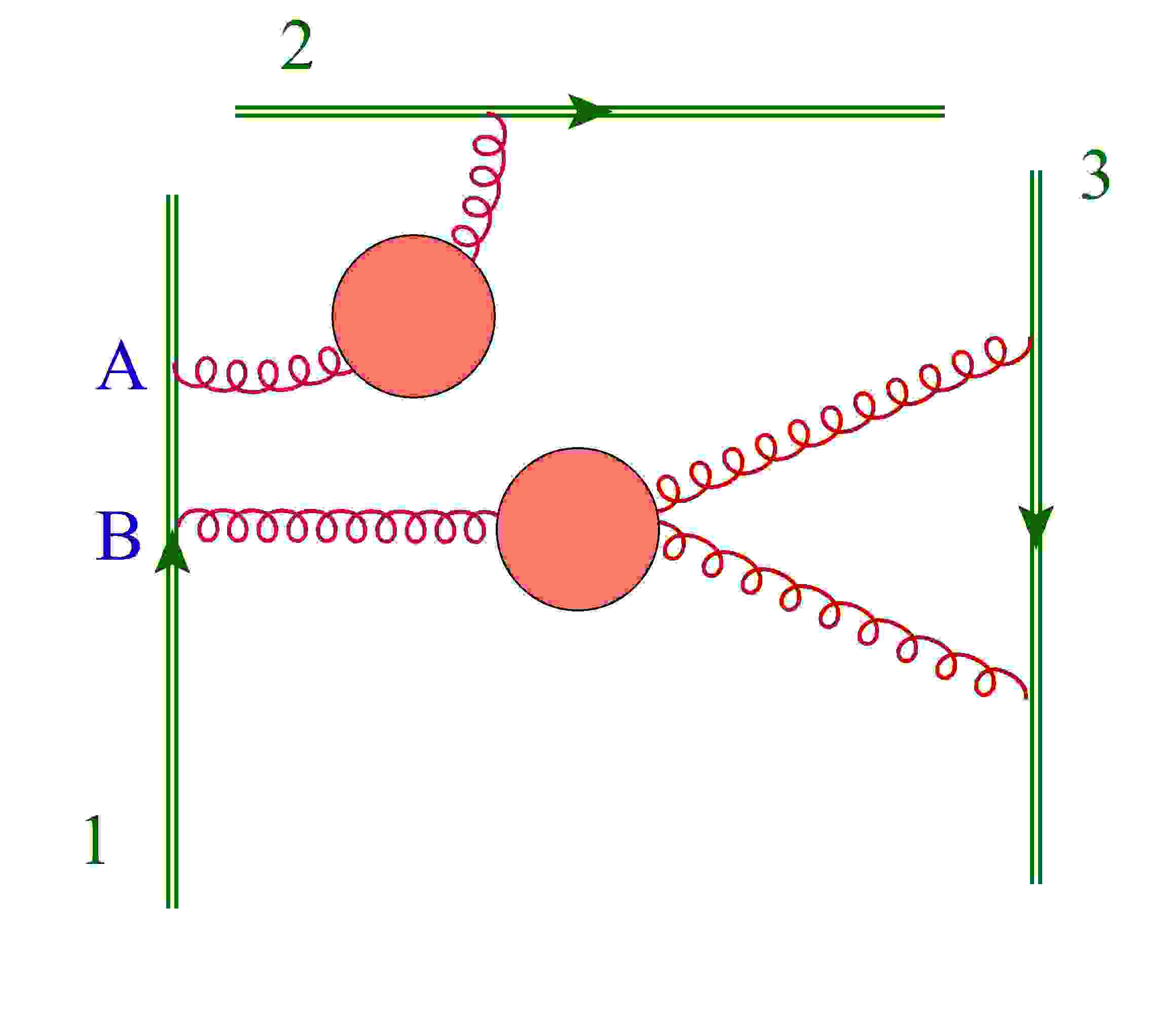} }
	\quad
	\subfloat[$W_{3, \, {\rm II}}^{(1,1)} (2,1,2)$]{\includegraphics[height=2.8cm,width=3.2cm]
	{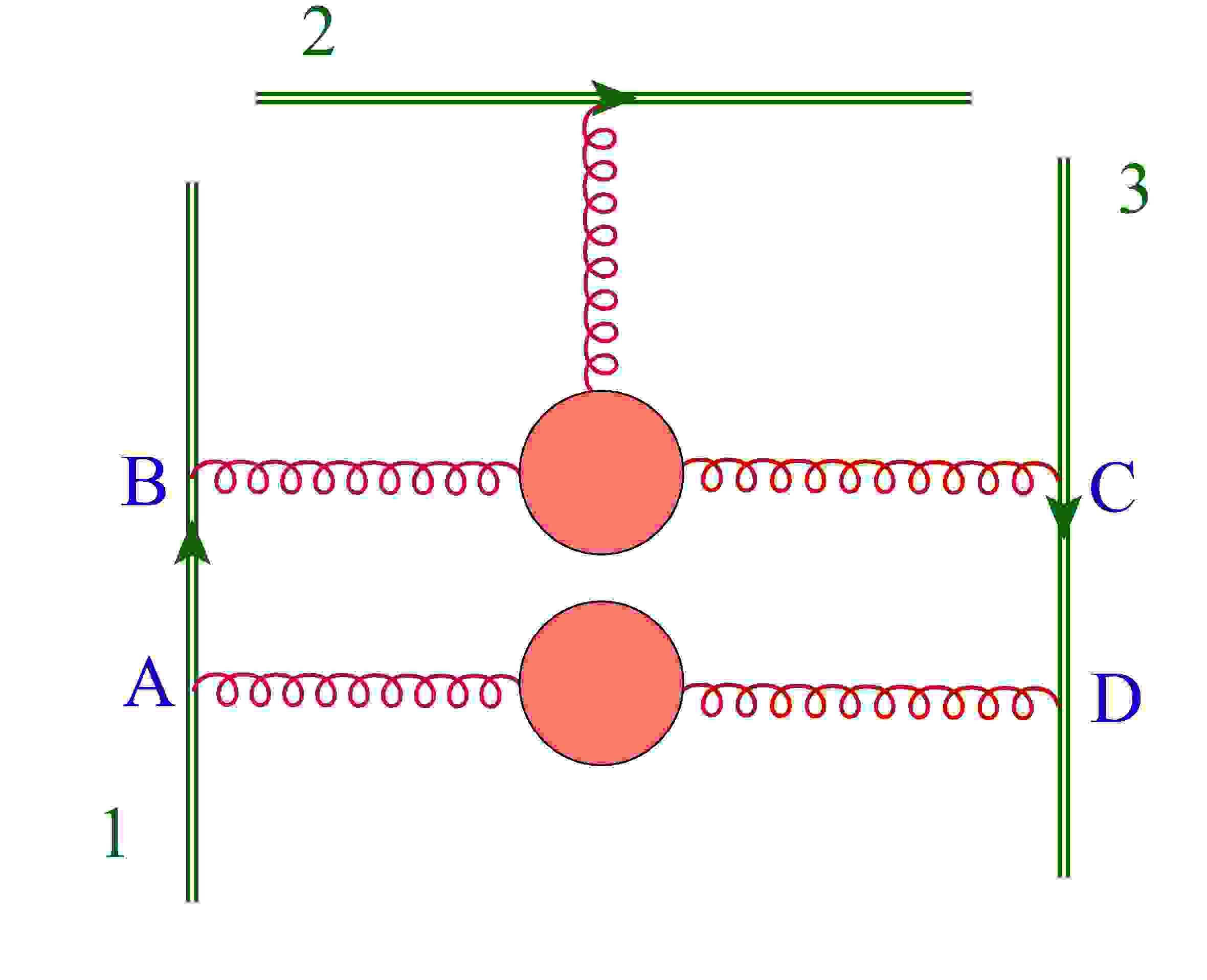} }
	\quad
	\subfloat[$W_3^{(3)} (3,1,2)$]{\includegraphics[height=2.8cm,width=3.2cm]
	{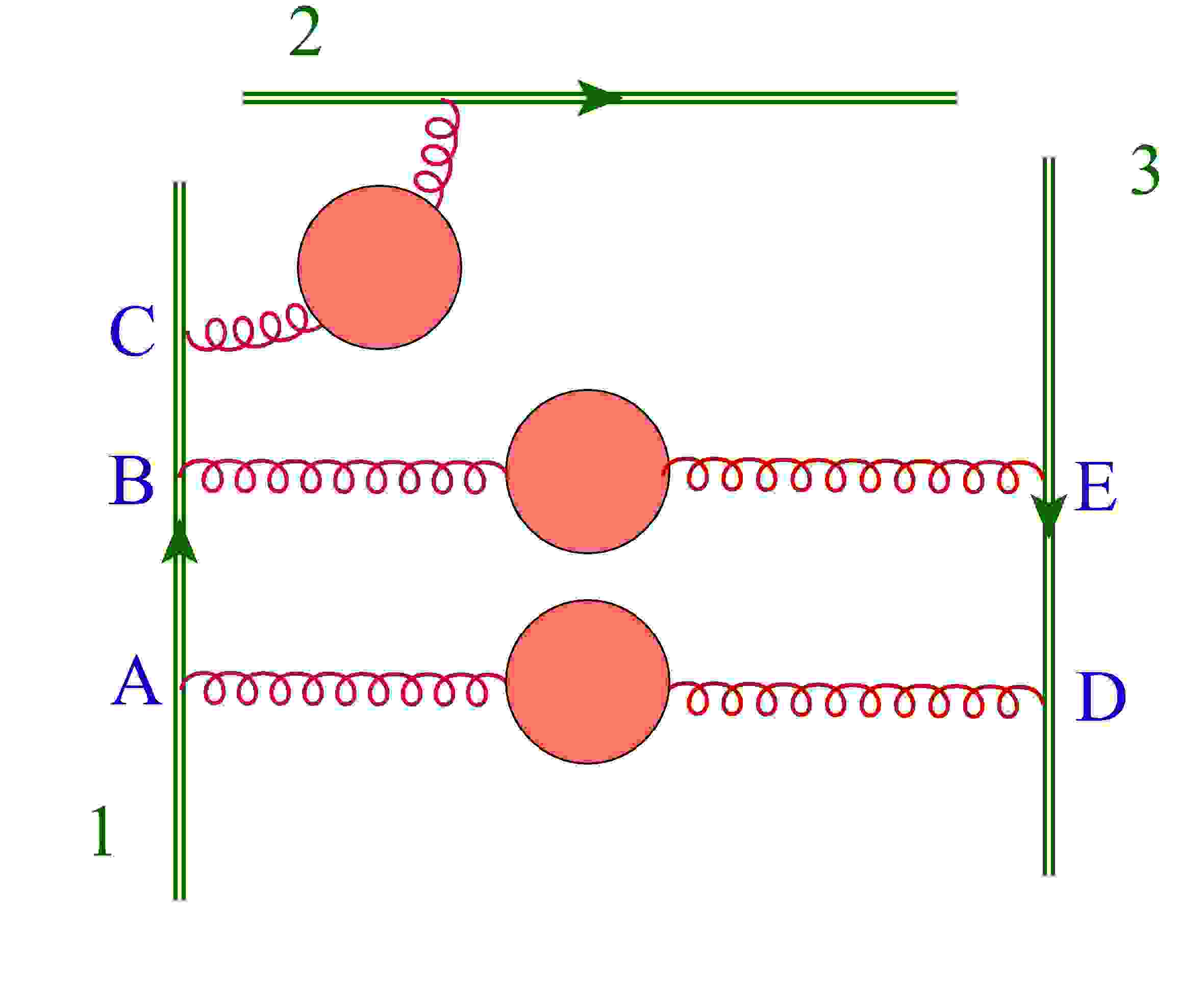} }
	\quad
	\subfloat[$W_3^{(3)} (2,2,2)$]{\includegraphics[height=2.8cm,width=3.2cm]
	{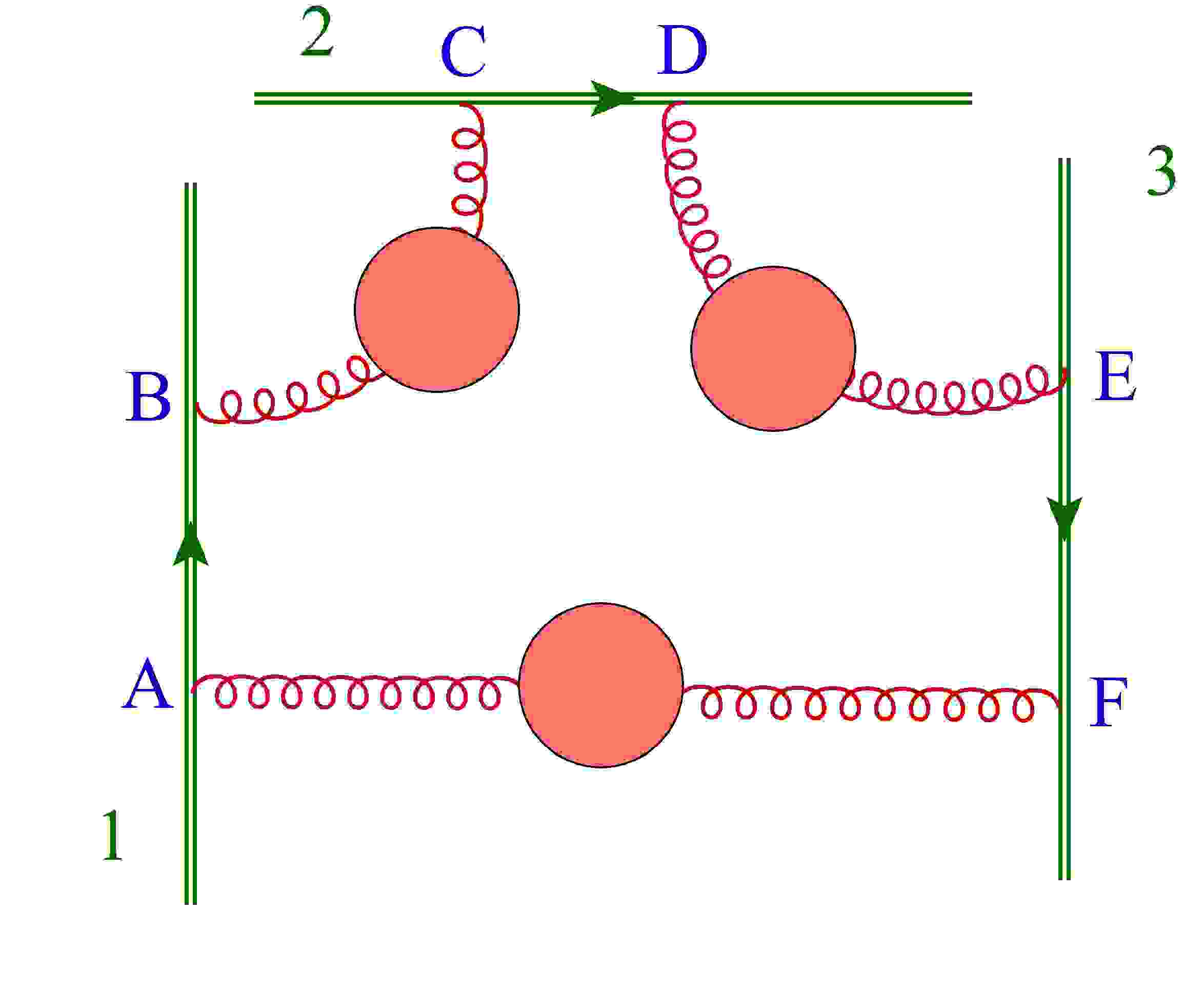} }
	\caption{Representative skeleton diagrams for the six three-loop Cwebs 
	connecting three Wilson lines in a massless theory. The dimensions of the 
	associated mixing matrices are $d_w = \{1,3,2,4,6,4\}$, in the order shown.}
\label{Cwebg6_3}
\end{figure} 
It is evident that a computation of the soft anomalous dimension matrix crucially 
involves these mixing matrices: they were studied systematically up to three loops, 
and several of their all-order properties were uncovered, in~\cite{Mitov:2010rp,
Gardi:2010rn,Gardi:2011wa,Gardi:2011yz,Dukes:2013wa,Gardi:2013ita,
Dukes:2013gea}. The first important point is that they are idempotent {\it i.e.} 
$R_w^2 = R_w$, and thus they are {\it projection operators}; in particular, their 
eigenvalues can either be 1 or 0. Acting on the right, according to \eq{eq:webredef},
they select a subset of the possible colour factors, while, acting on the left, 
they identify the set of linear combinations of kinematic factors that will contribute 
to the exponent. The rank $r_w$ of the mixing matrix $R_w$ equals the number 
of independent colour factors onto which it projects. Web mixing matrices, further,
enforce the {\it non-abelian exponentiation theorem}~\cite{Gardi:2013ita}, stating
that all exponentiated colour factors correspond to Feynman diagrams where the
gluon sub-diagram is completely connected. Two further combinatorial properties
have important physical consequences: first, web mixing matrices of dimension
$d_w > 1$ obey a {\it row sum rule}, stating that the sum of the elements of each 
row vanishes, $\sum_{D'} R_{D D'} \, = \, 0$; this ensures that diagrams contributing 
to the exponent are a proper subset of all diagrams contributing to the correlator. 
In addition, mixing matrices are conjectured~\cite{Gardi:2011yz} to obey a 
{\it column sum rule}, which can be written as $\sum_D s(D) R(D,D') = 0$, 
where $s(D)$ is called the column weight vector. For a diagram $D$, $s(D)$ 
denotes the number of different ways in which the connected gluon sub-diagrams 
can be sequentially shrunk to the common origin of the Wilson lines. This second
combinatorial property has important consequences for the renormalisation of
the correlator: it ensures that web mixing matrices project on kinematic factors 
which are free from ultraviolet sub-divergences; this, in turn, simplifies the 
evaluation of the exponent, implying that the UV counterterms are independent 
of the selected IR regulator.

A close examination of the properties of mixing matrices confirms the intuitive
notion that the internal structure of connected gluon sub-diagrams does not
affect the combinatorics of their attachments to the Wilson lines: a connected
multi-gluon correlator may generate a number of independent `internal' colour 
factors, but each one of them can be treated separately when considering the
connection to Wilson lines. This observation motivated the introduction, in
Ref.~\cite{Agarwal:2020nyc}, of the concept of Cwebs, building Wilson-line 
correlators in terms of connected gluon correlators instead of individual diagrams.

More precisely, a {\it Cweb} is a set of skeleton diagrams, built out of connected 
gluon correlators attached to Wilson lines, and closed under permutations of the 
gluon attachments to each Wilson line. Clearly, the main difference between 
webs and Cwebs is the fact that Cwebs are not fixed-order quantities, but admit 
their own perturbative expansion in powers of $g$. Cwebs strongly simplify the 
counting and organisation of contributions to the exponent, especially at high 
orders, where radiative corrections to gluon sub-diagrams become important 
and proliferate. They may also provide ingredients to generalise to multi-line 
correlators the fascinating arguments given in Ref.~\cite{Erdogan:2011yc}
for the two line case (see also Ref.~\cite{Falcioni:2019nxk}). Cweb mixing 
matrices are identical to those of their constituent webs, as the properties 
of mixing matrices are blind to gluon interactions away from the Wilson lines, 
and only depend on ordering of their attachments on the Wilson lines, which
is same in both the definitions.
\begin{figure}[H]
	\centering
	\subfloat[${W}_{2}^{(0,0,0,1)}(2,3)$]{\includegraphics[height=3.0cm,width=3.0cm]
	{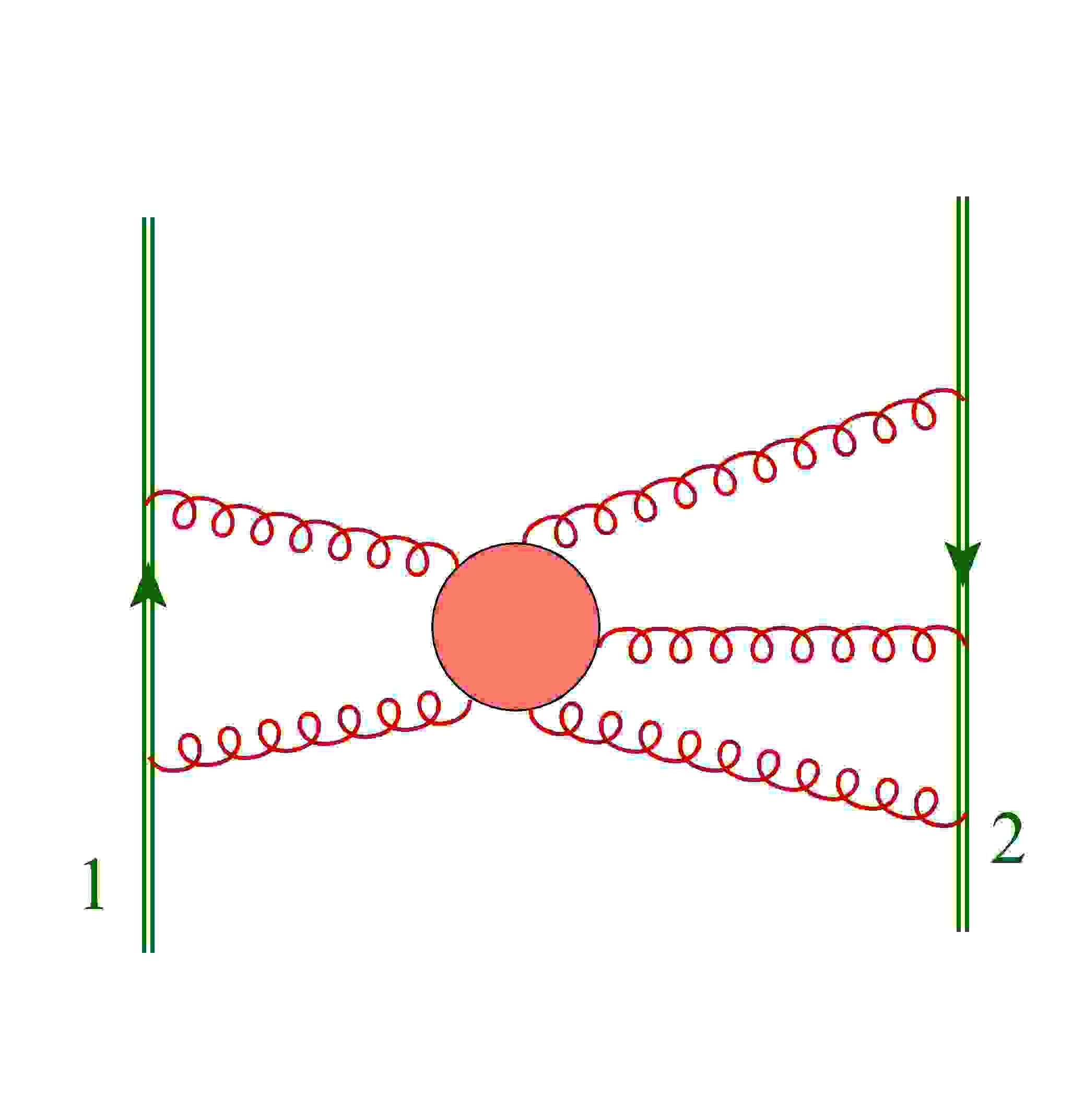} }
	\quad
	\subfloat[${W}_{2}^{(0,0,0,1)}(1,4)$]{\includegraphics[height=3.0cm,width=3.0cm]
	{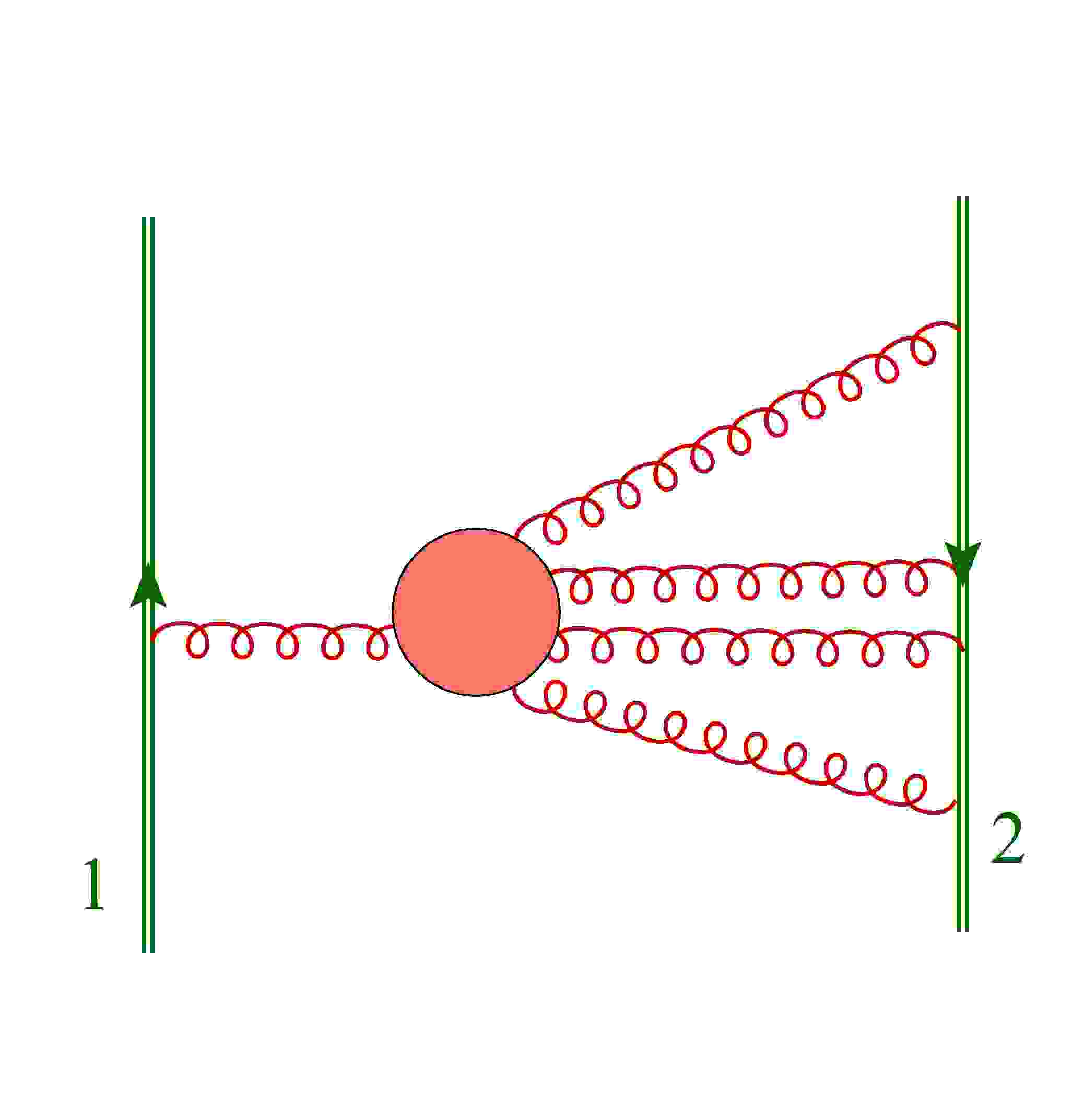} }
	\subfloat[${W}_{2}^{(1,0,1)}(2,4)$]{\includegraphics[height=3.0cm,width=3.0cm]
	{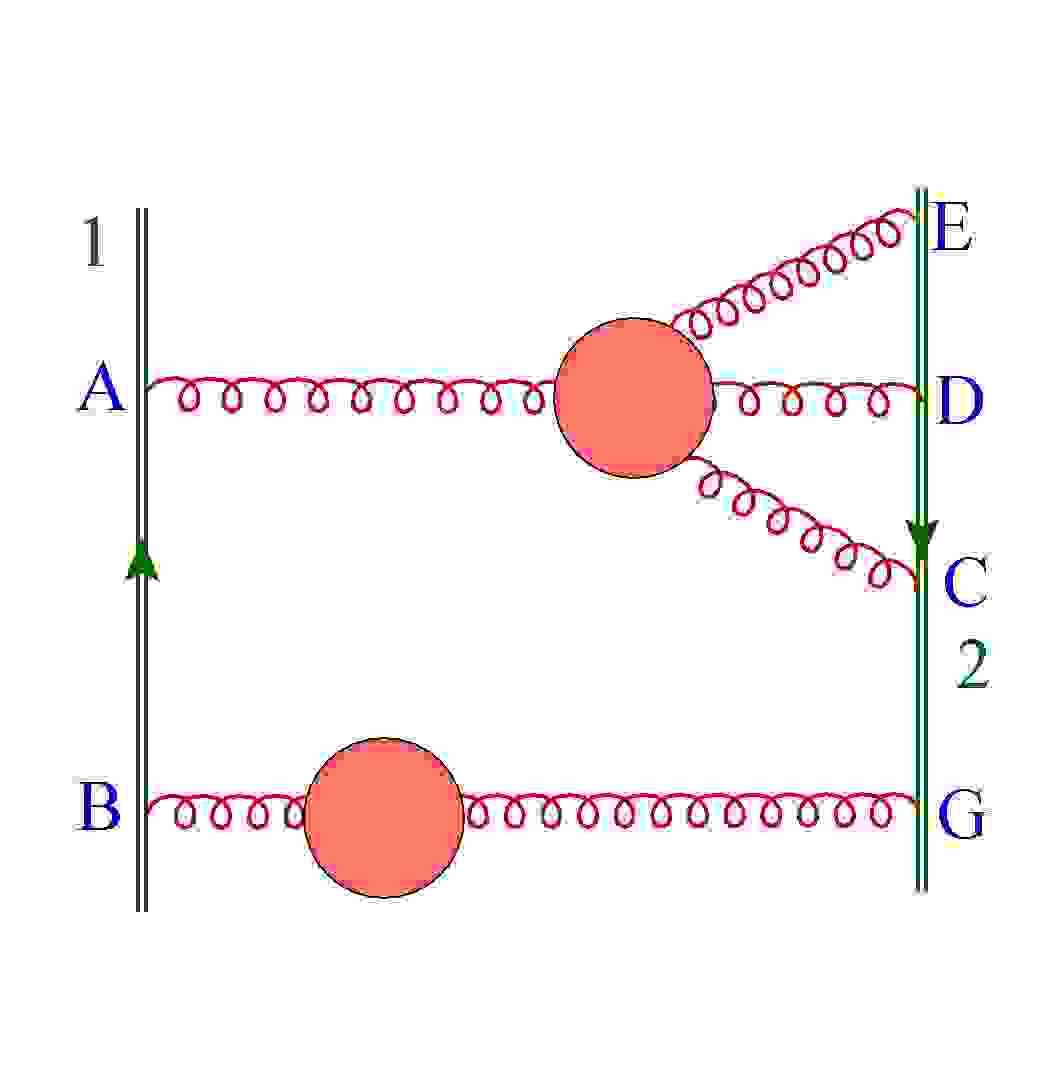} }
	\quad
	\subfloat[${W}_{2}^{(1,0,1)}(3,3)$]{\includegraphics[height=3.0cm,width=3.0cm]
	{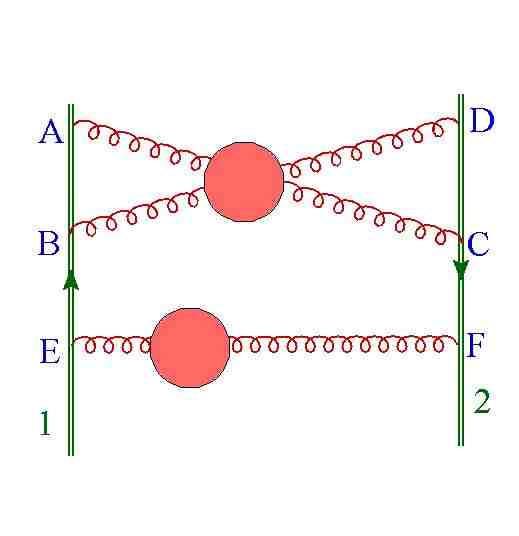} }
	\quad
	\subfloat[${W}_{2}^{(0,2)}(2,4)$]{\includegraphics[height=3.0cm,width=3.0cm]
	{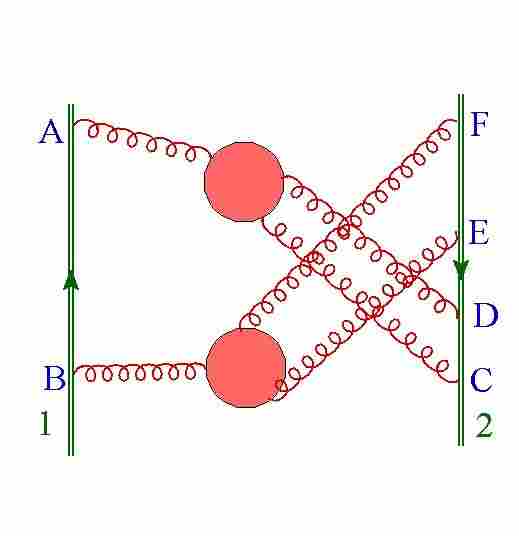} }
	\quad
	\subfloat[${W}_{2}^{(0,2)}(3,3)$]{\includegraphics[height=3.0cm,width=3.0cm]
	{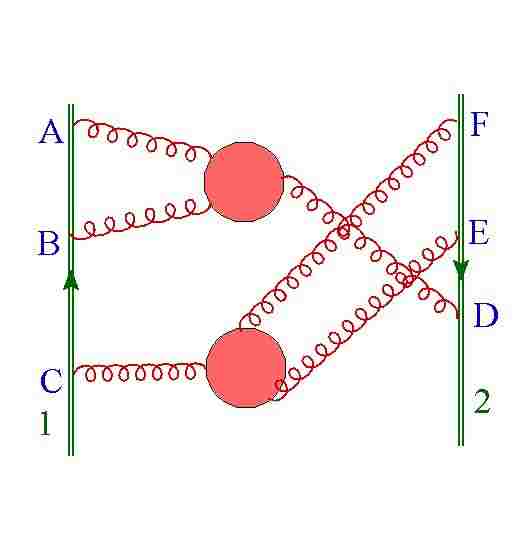} }
	\quad
	\subfloat[${W}_{2}^{(2,1)}(3,4)$]{\includegraphics[height=3.0cm,width=3.0cm]
	{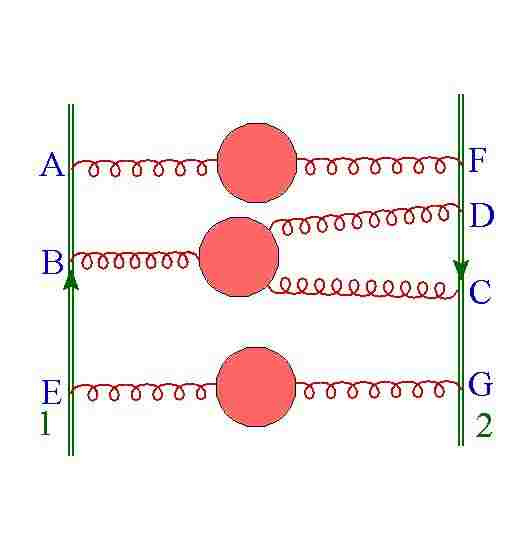} }
	\quad
	\subfloat[${W}_{2}^{(4)}(4,4)$]{\includegraphics[height=3.0cm,width=3.0cm]
	{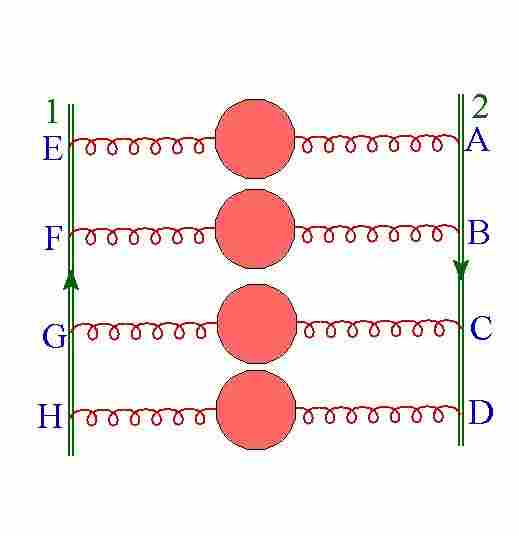} }
	\caption{Representative skeleton diagrams for the eight four-loop Cwebs 
	connecting two Wilson lines in a massless theory. The dimensions of the 
	associated mixing matrices are $d_w = \{1,1,8,9,6,9,36,24\}$, in the order 
	shown.}
\label{Cwebsg8_2}
\end{figure}
To identify Cwebs, we adopt the notation of Ref.~\cite{Agarwal:2020nyc},
writing $W_n^{(c_2, \ldots, c_p)} (k_1, \ldots, k_n)$, where $n$ denotes the 
number of Wilson lines, $c_m$ denotes the number of $m$-point correlators 
forming the Cweb, and $k_l$ denotes the number of attachments on each 
Wilson line, where, without loss of generality, we take $k_1 \leq k_2 \leq 
\ldots \leq k_n$. When a further degeneracy arises, we distinguish the 
Cwebs sharing the same notation by roman numerals.

A systematic way of generating all Cwebs recursively at any perturbative 
order was presented in~\cite{Agarwal:2020nyc}. Assuming one has already 
enumerated all Cwebs at $\mathcal{O}(g^{2 n - 2})$, the steps to generate 
all Cwebs at $\mathcal{O}(g^{2n})$ are listed below, keeping in mind the 
possibility of adding a Wilson line with no attachments at lower orders.
\begin{figure}[H]
	\vspace{-2cm}
	\centering
	\subfloat[${W}_{3}^{(0,0,0,1)}(1,1,3)$]{\includegraphics[height=3.0cm,width=3.0cm]
	{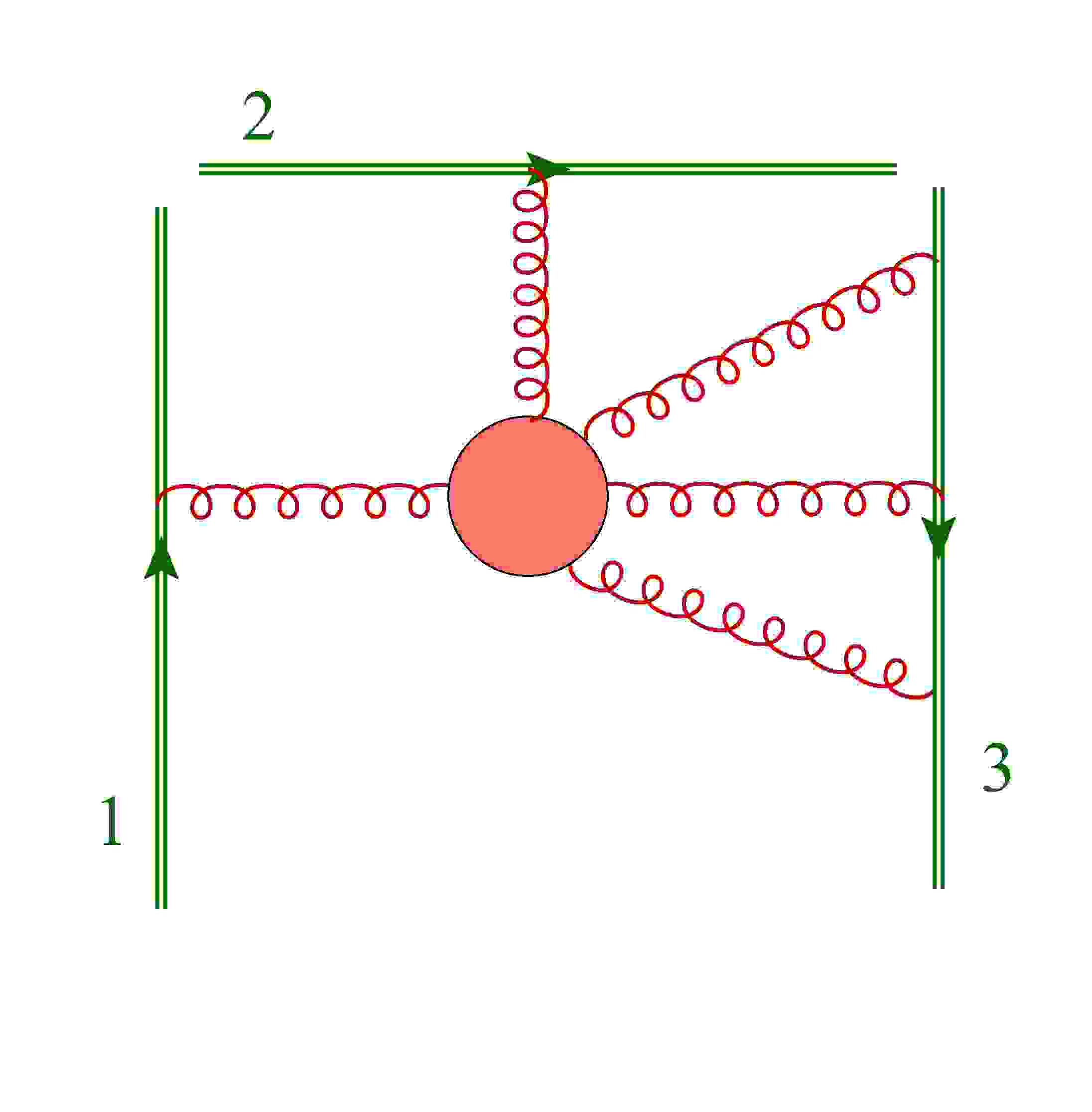} }
	\quad
	\subfloat[${W}_{3}^{(0,0,0,1)}(1,2,2)$]{\includegraphics[height=3.0cm,width=3.0cm]
	{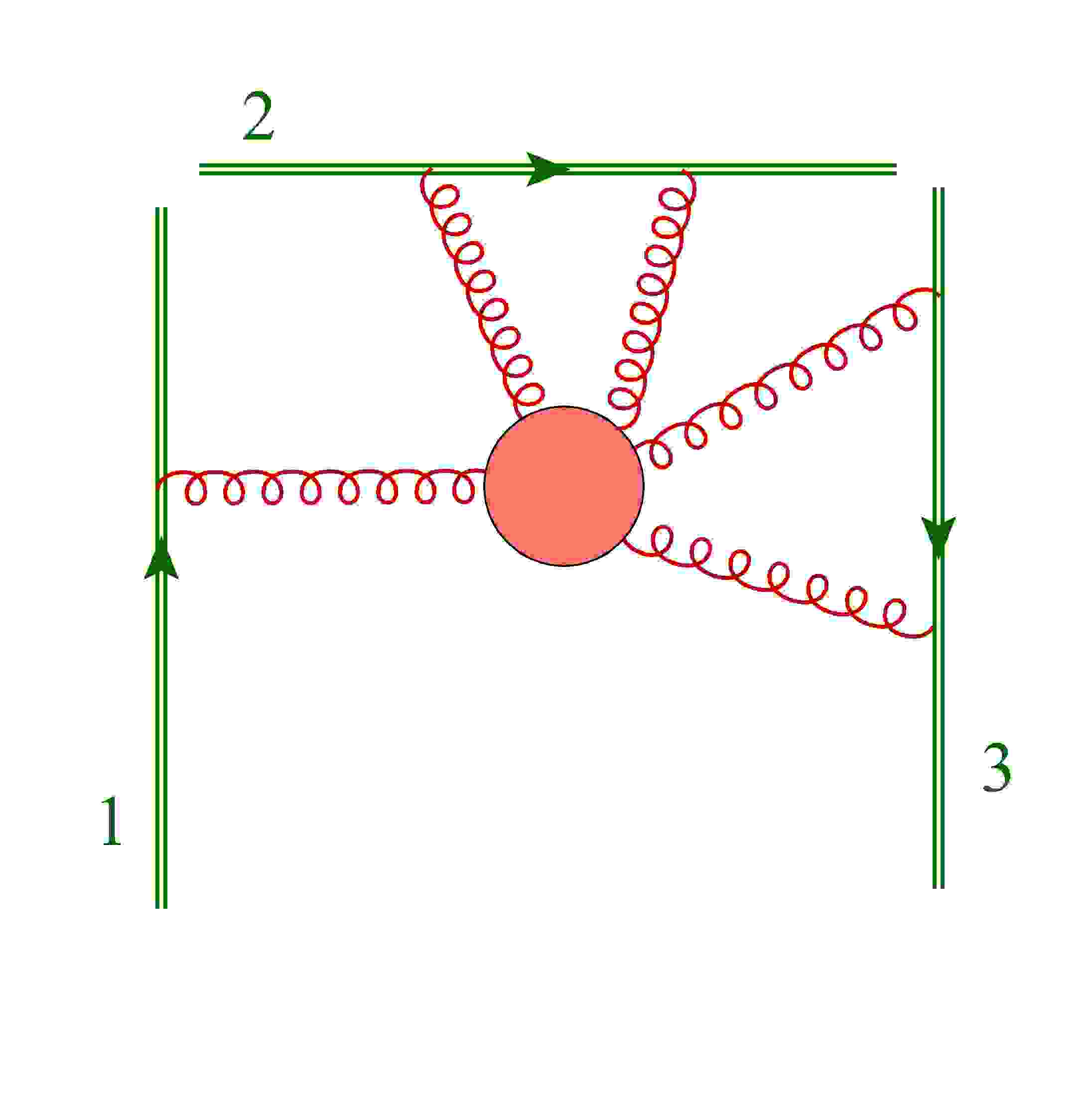} }
	\subfloat[${W}_{3, \rm{I}}^{(1,0,1)}(2,1,3)$]{\includegraphics[height=3.0cm,width=3.0cm]
	{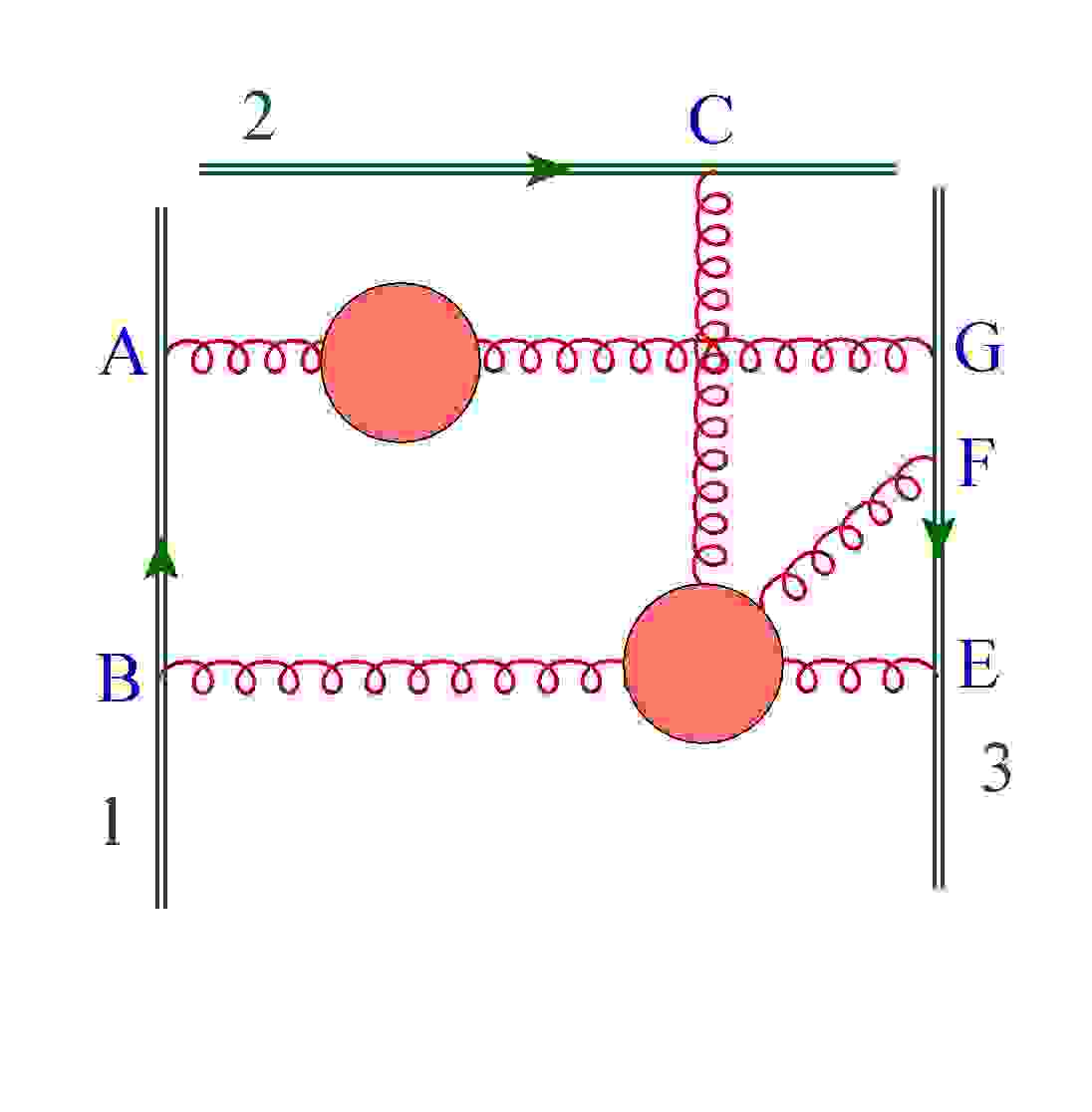} }
	\quad
	\subfloat[${W}_{3,\rm{II}}^{(1,0,1)}(2,1,3)$]{\includegraphics[height=3.0cm,width=3.0cm]
	{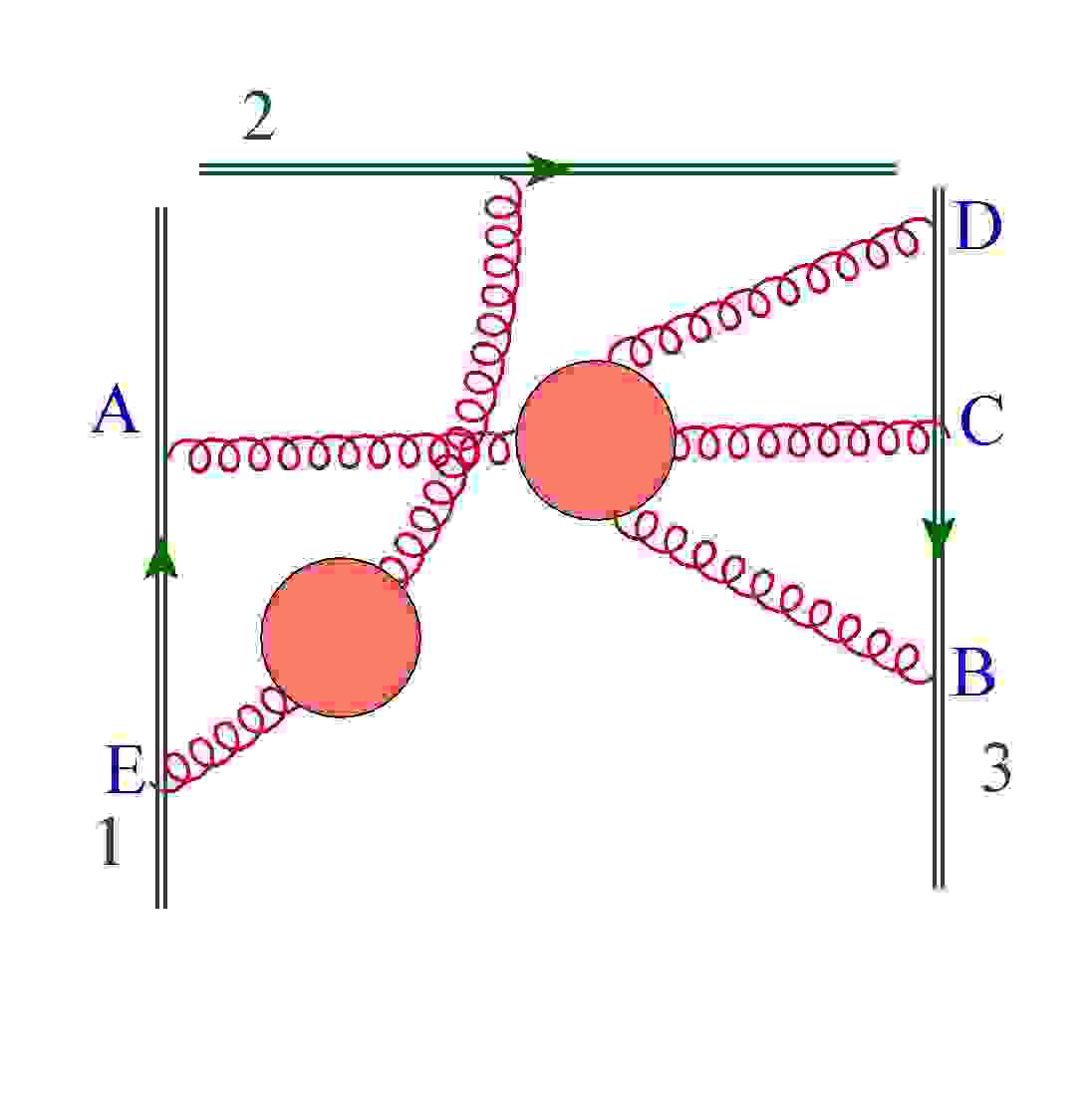} }
	\quad
	\subfloat[${W}_{3,\rm{I}}^{(0,2)}(3,1,2)$]{\includegraphics[height=3.0cm,width=3.0cm]
	{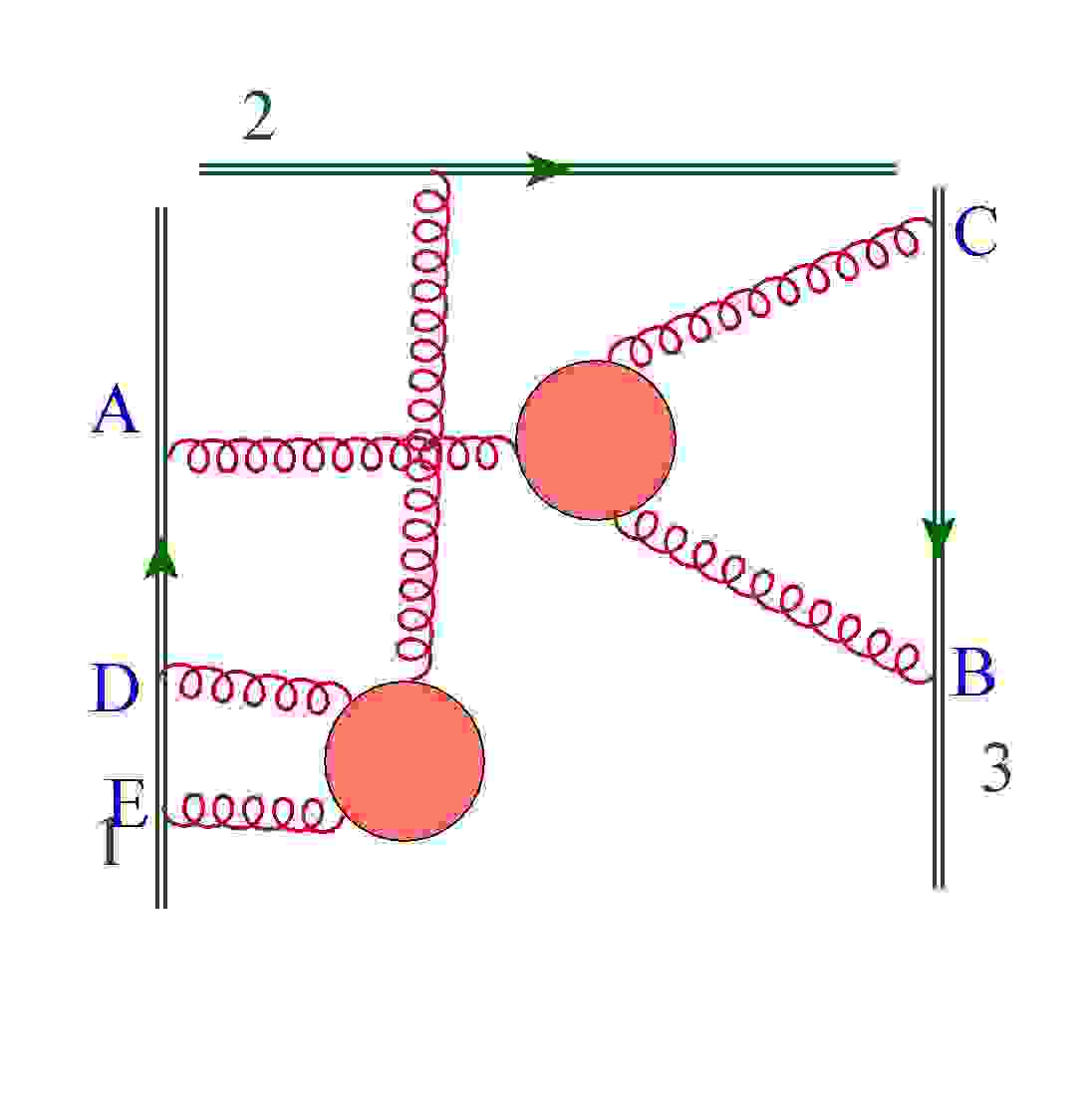} }
	\quad
	\subfloat[${W}_{3,\rm{II}}^{(0,2)}(2,1,3)$]{\includegraphics[height=3.0cm,width=3.0cm]
	{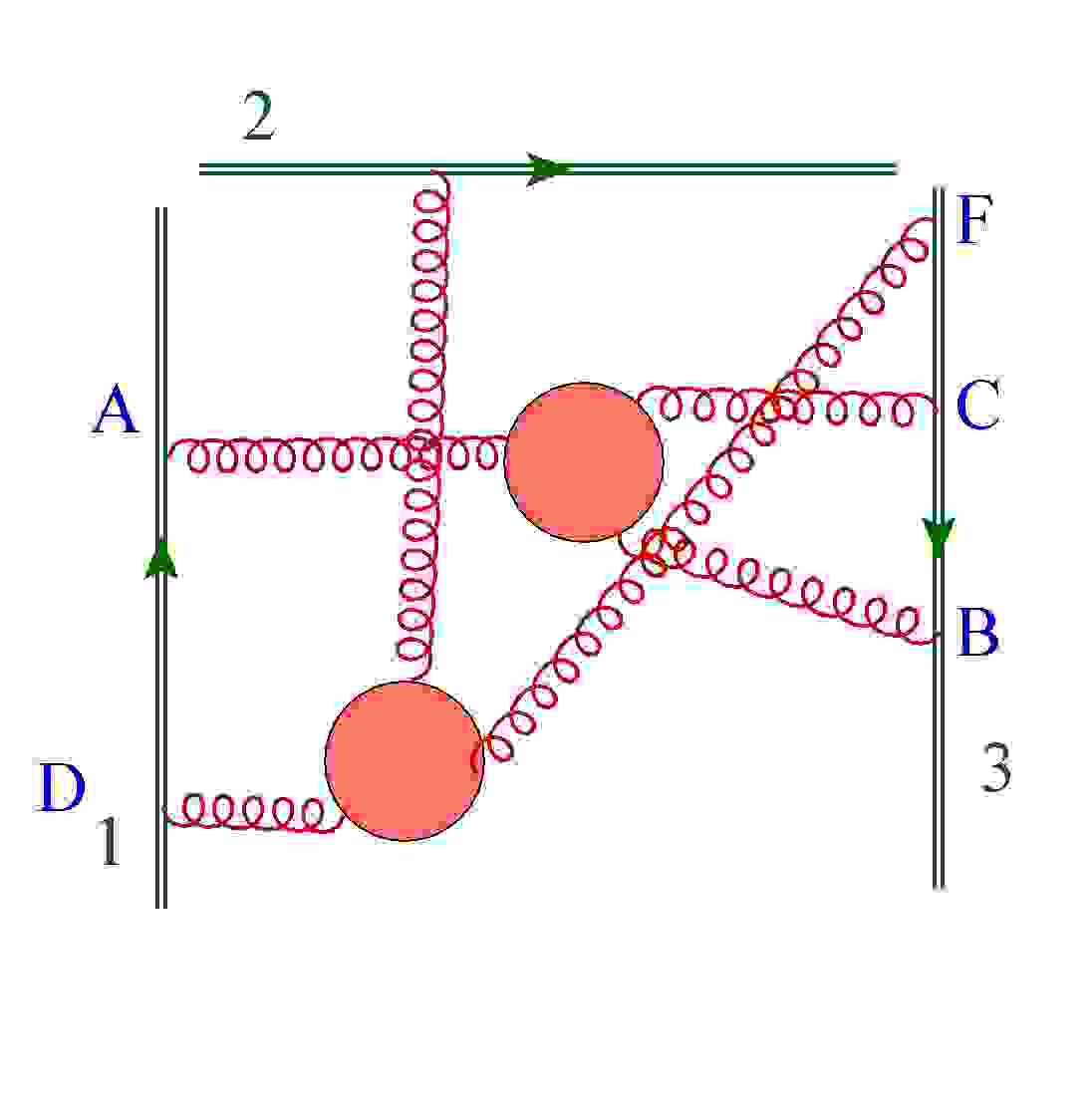} }
	\quad
	\subfloat[${W}_{3,\text{I}}^{(2,1)}(3,2,2)$]{\includegraphics[height=3.0cm,width=3.0cm]
	{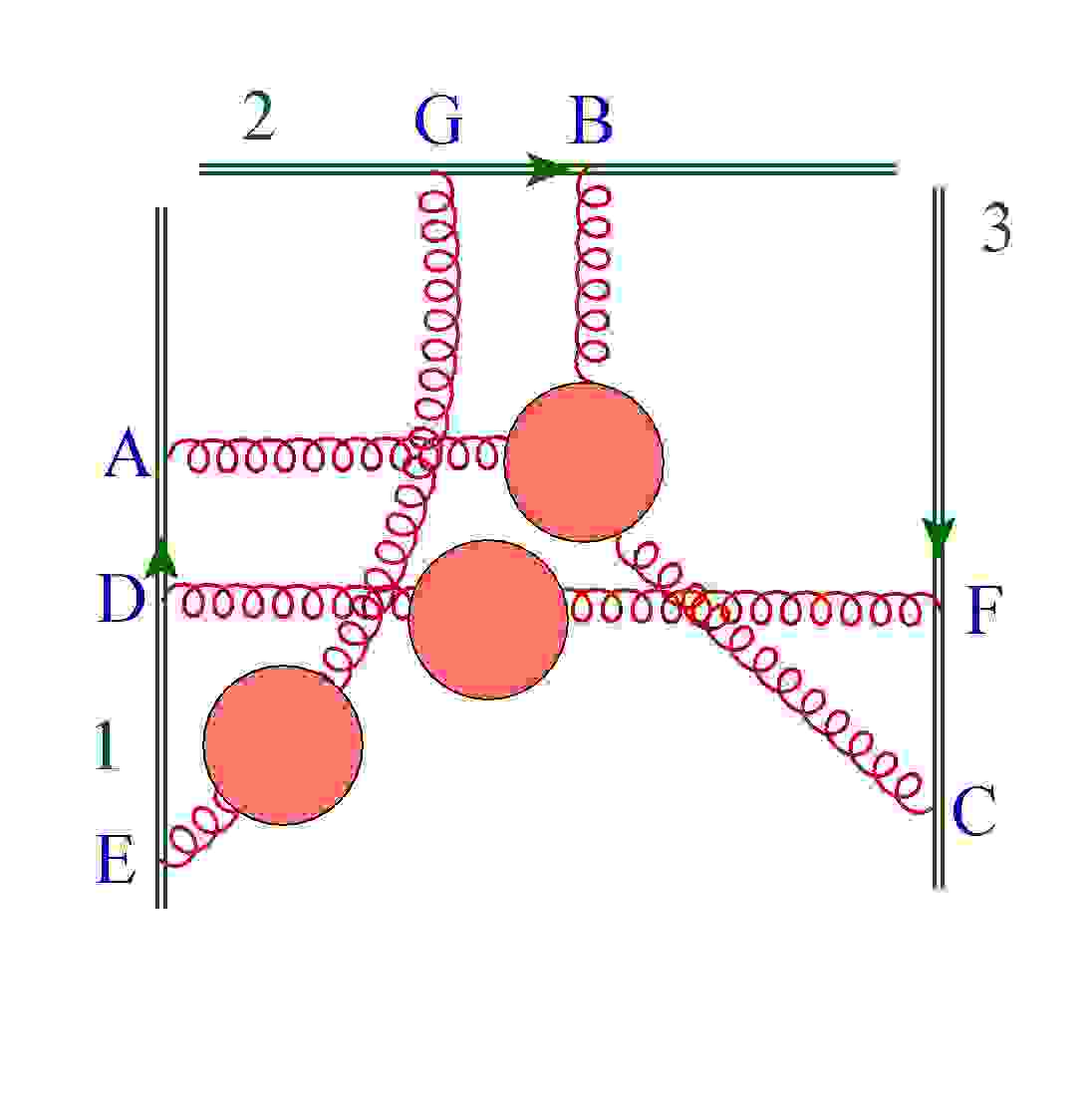} }
	\quad
	\subfloat[${W}_{3,\text{II}}^{(2,1)}(3,2,2)$]{\includegraphics[height=3.0cm,width=3.0cm]
	{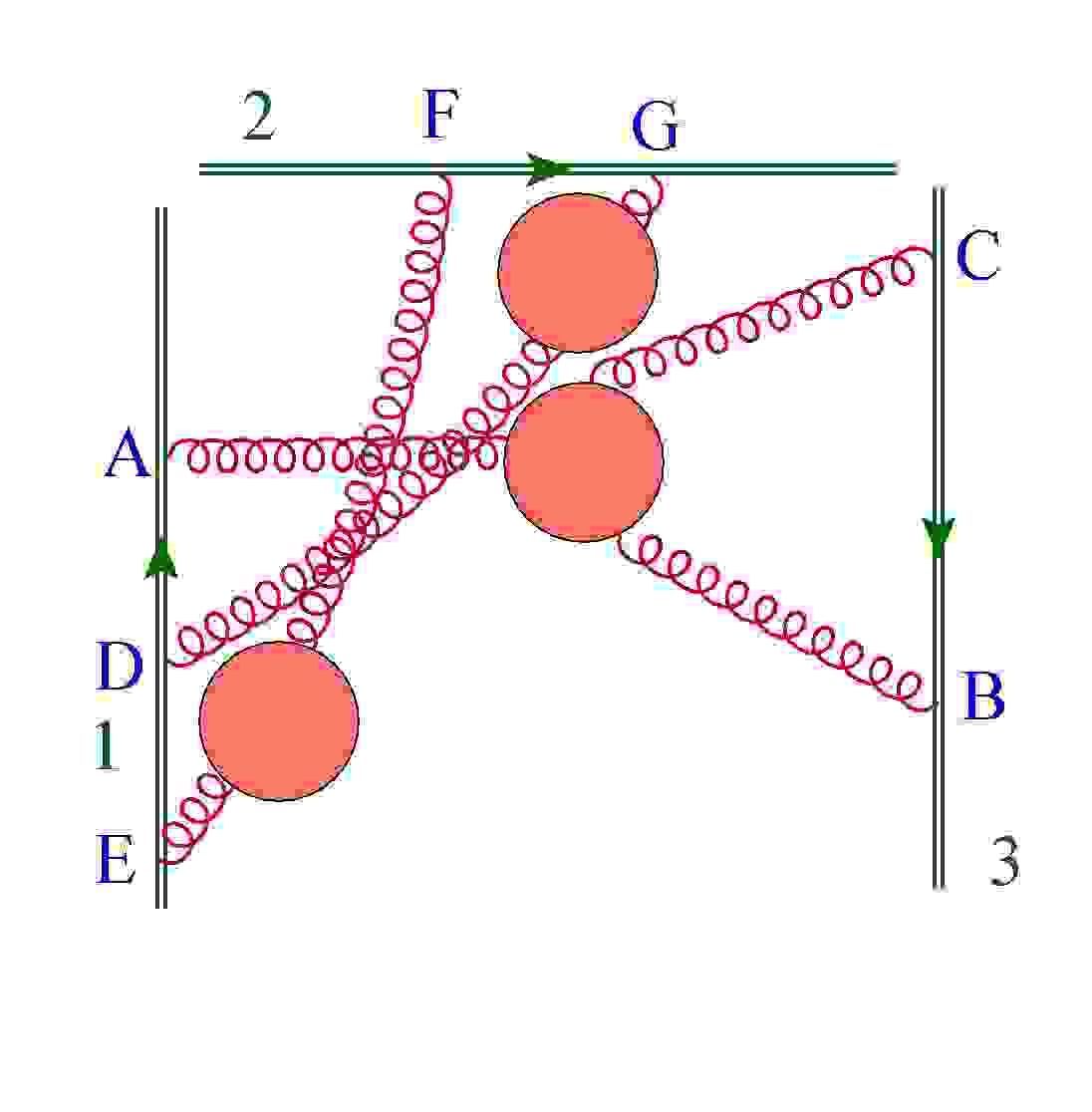} }
	\quad
	\subfloat[${W}_{3,\text{III}}^{(2,1)}(2,2,3)$]{\includegraphics[height=3.0cm,width=3.0cm]
	{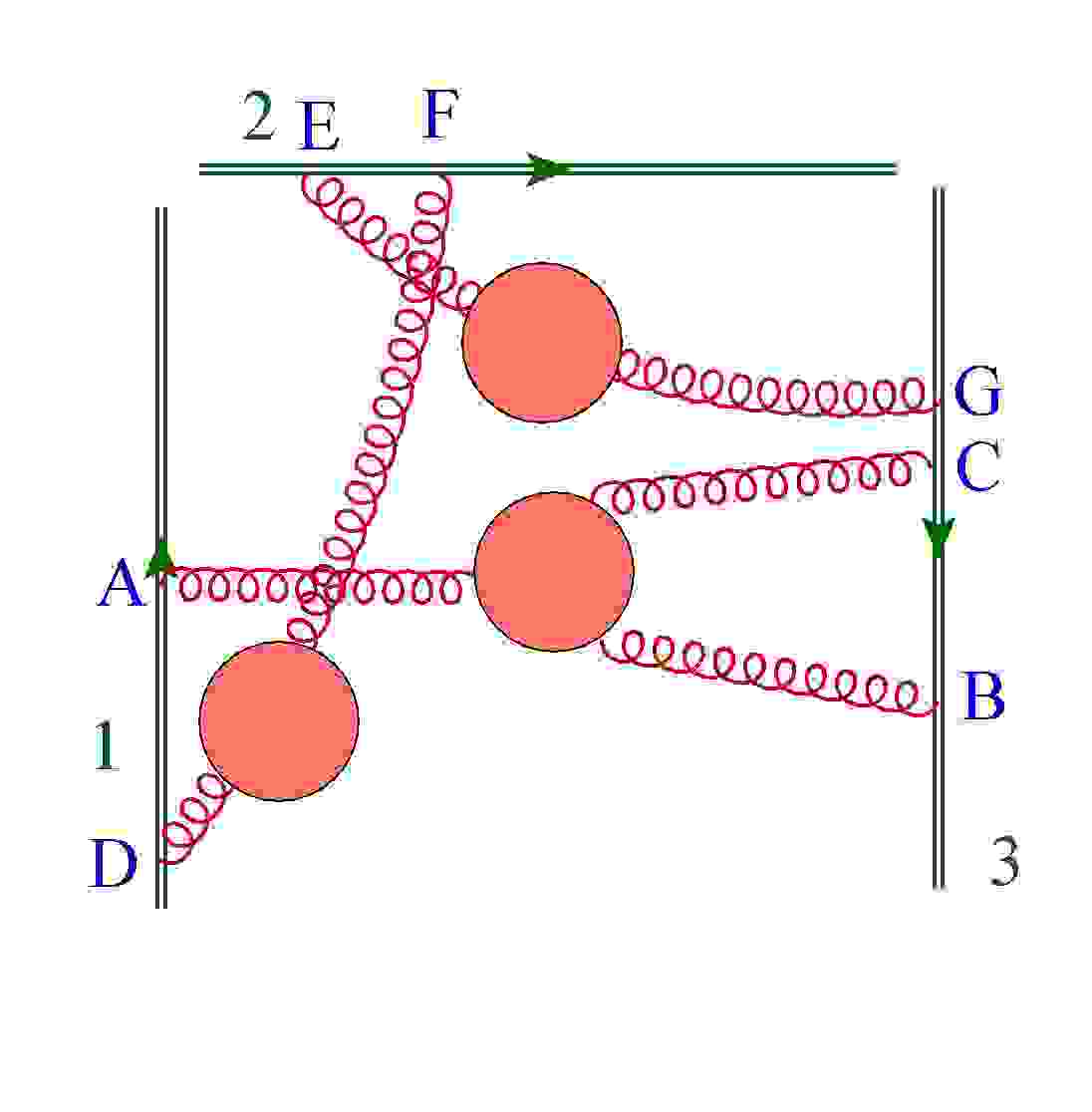} }
	\quad
	\subfloat[${W}_{3}^{(1,0,1)}(3,1,2)$]{\includegraphics[height=3.0cm,width=3.0cm]
	{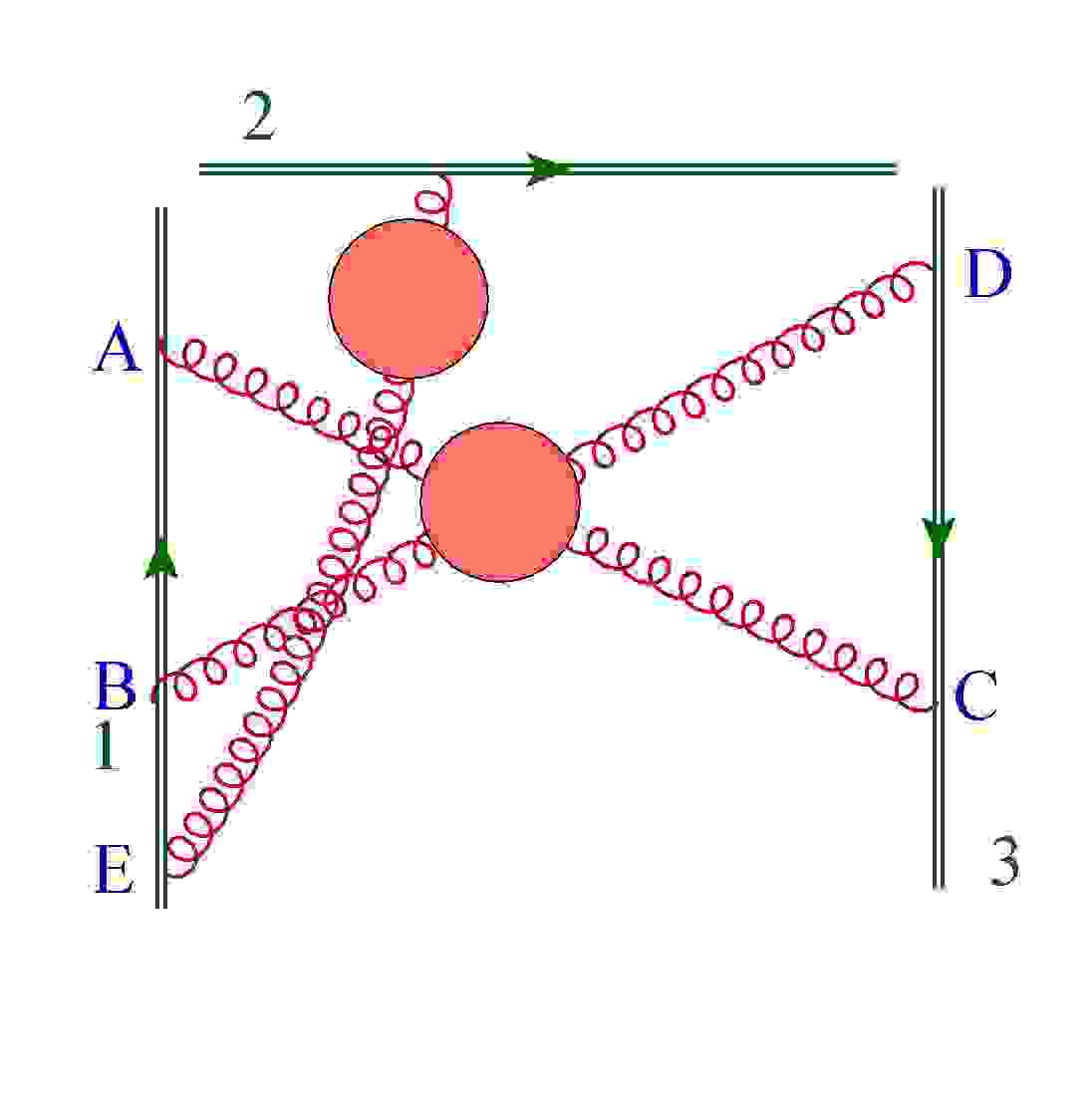} }
	\quad
	\subfloat[${W}_{3,\text{I}}^{(0,2)}(2,2,2)$]{\includegraphics[height=3.0cm,width=3.0cm]
	{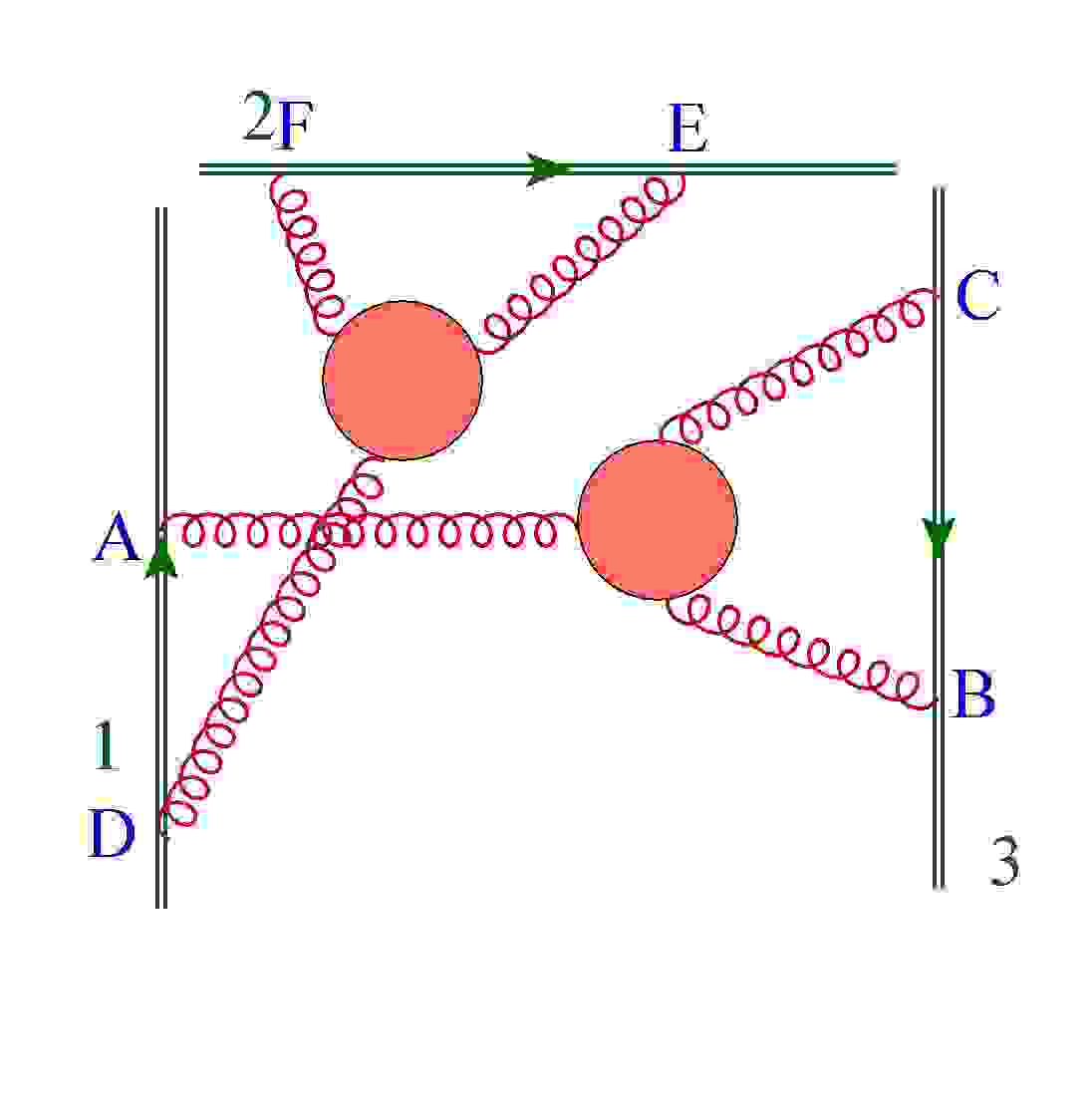} }
	\quad
	\subfloat[${W}_{3}^{(1,0,1)}(2,2,2)$]{\includegraphics[height=3.0cm,width=3.0cm]
	{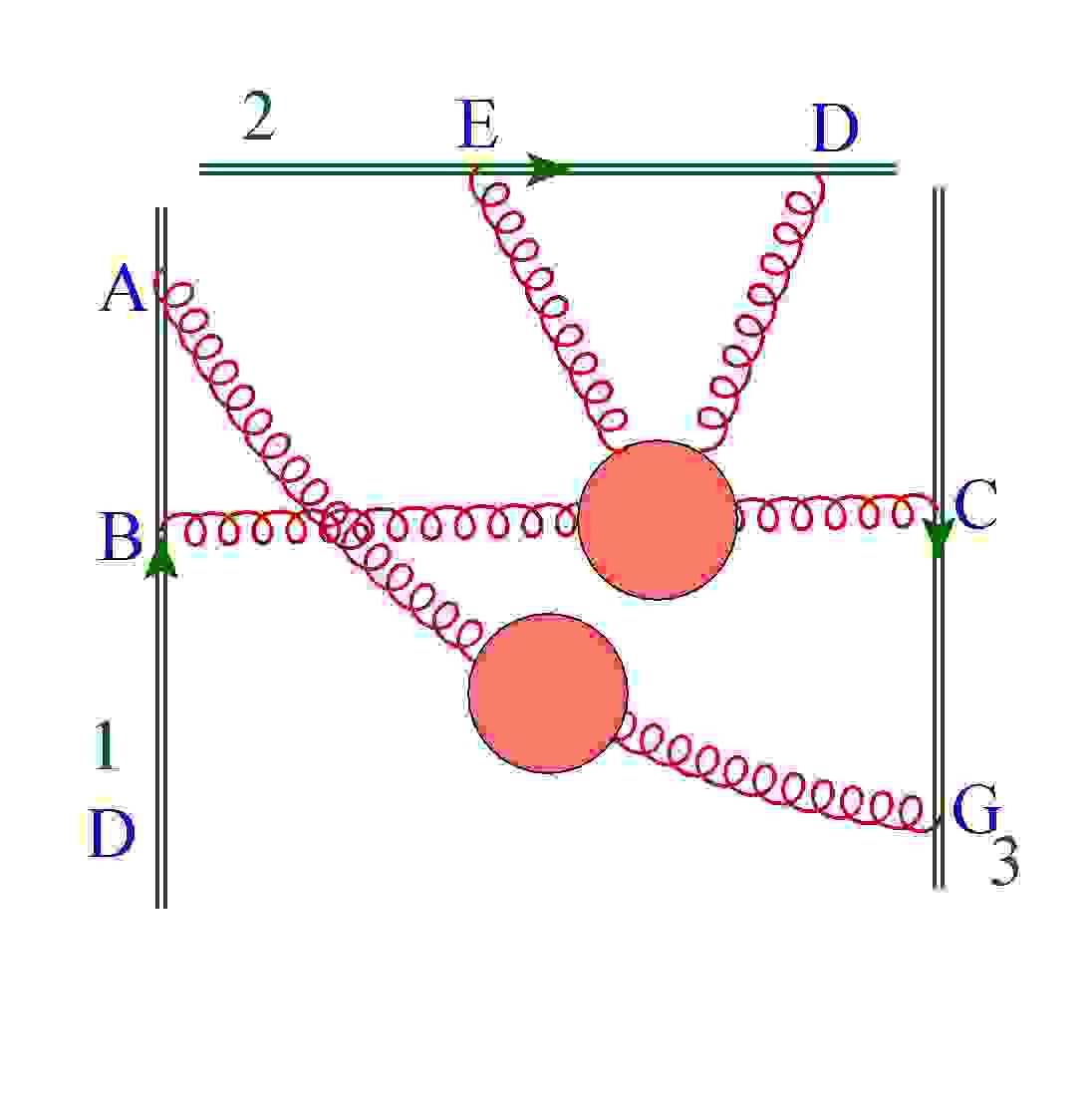} }
	\quad
	\subfloat[${W}_{3,\text{II}}^{(0,2)}(2,2,2)$]{\includegraphics[height=3.0cm,width=3.0cm]
	{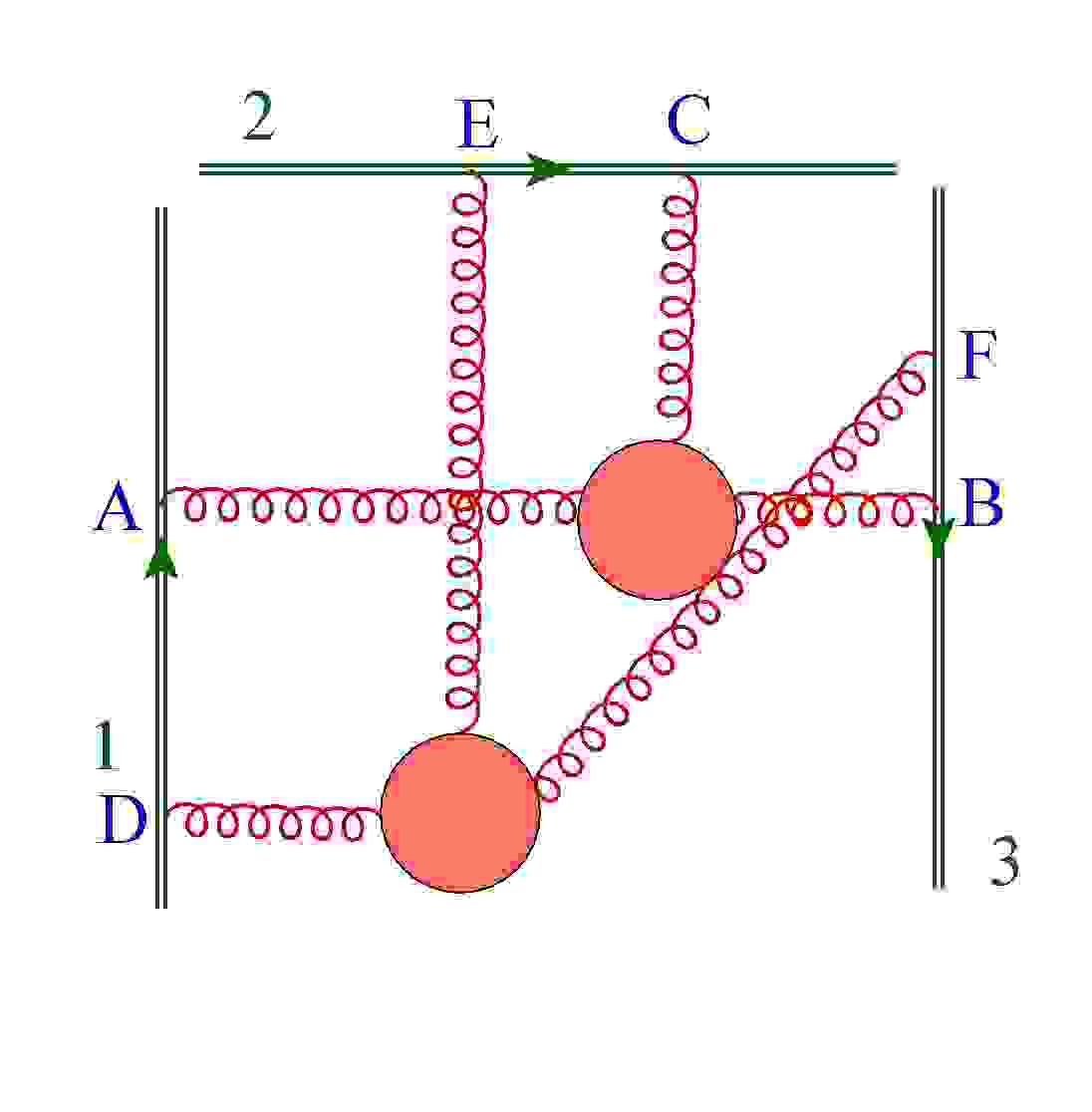} }
	\quad
	\subfloat[${W}_{3,\text{I}}^{(2,1)}(3,1,3)$]{\includegraphics[height=3.0cm,width=3.0cm]
	{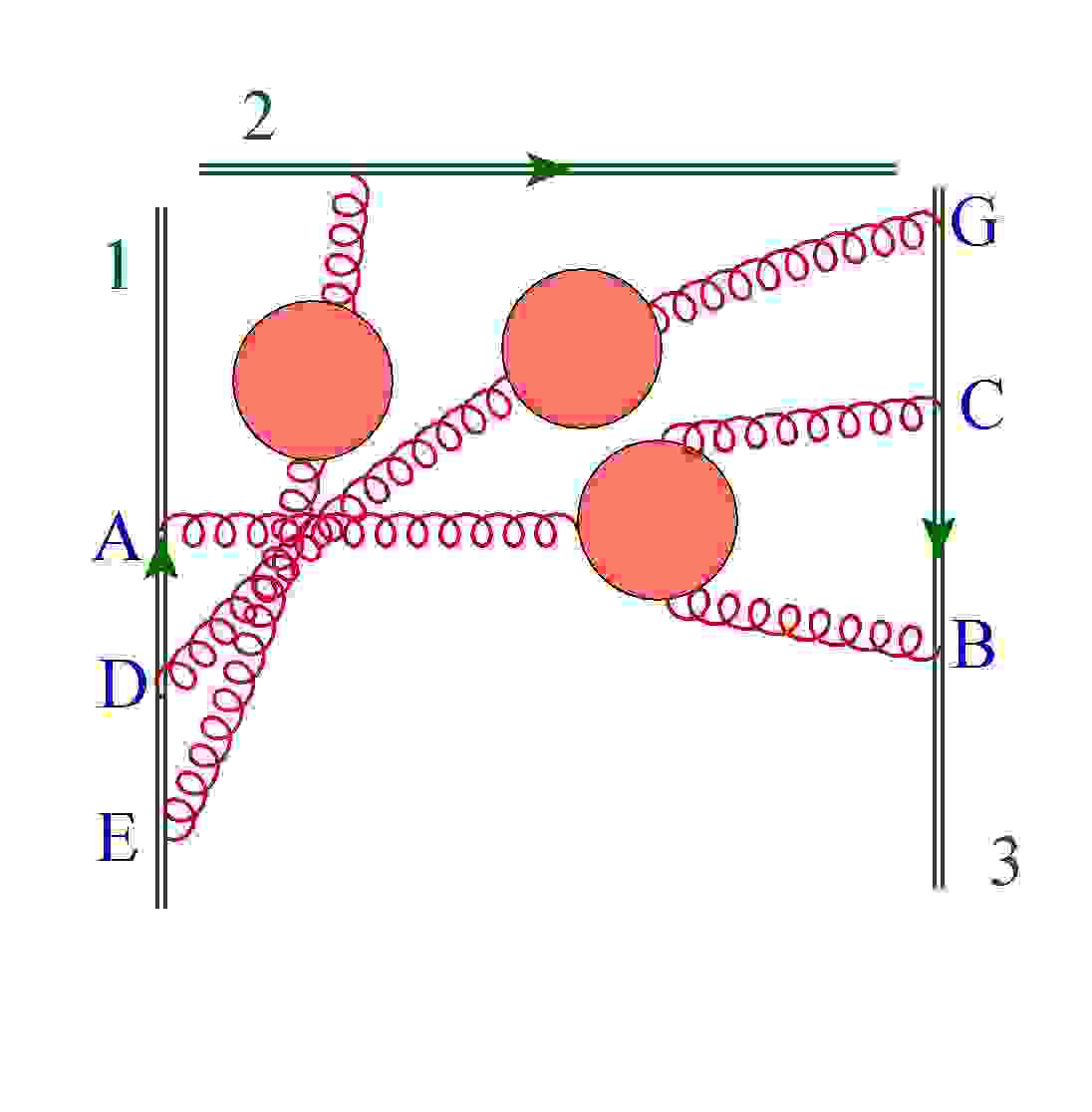} }
	\quad
	\subfloat[${W}_{3,\text{II}}^{(2,1)}(3,1,3)$]{\includegraphics[height=3.0cm,width=3.0cm]
	{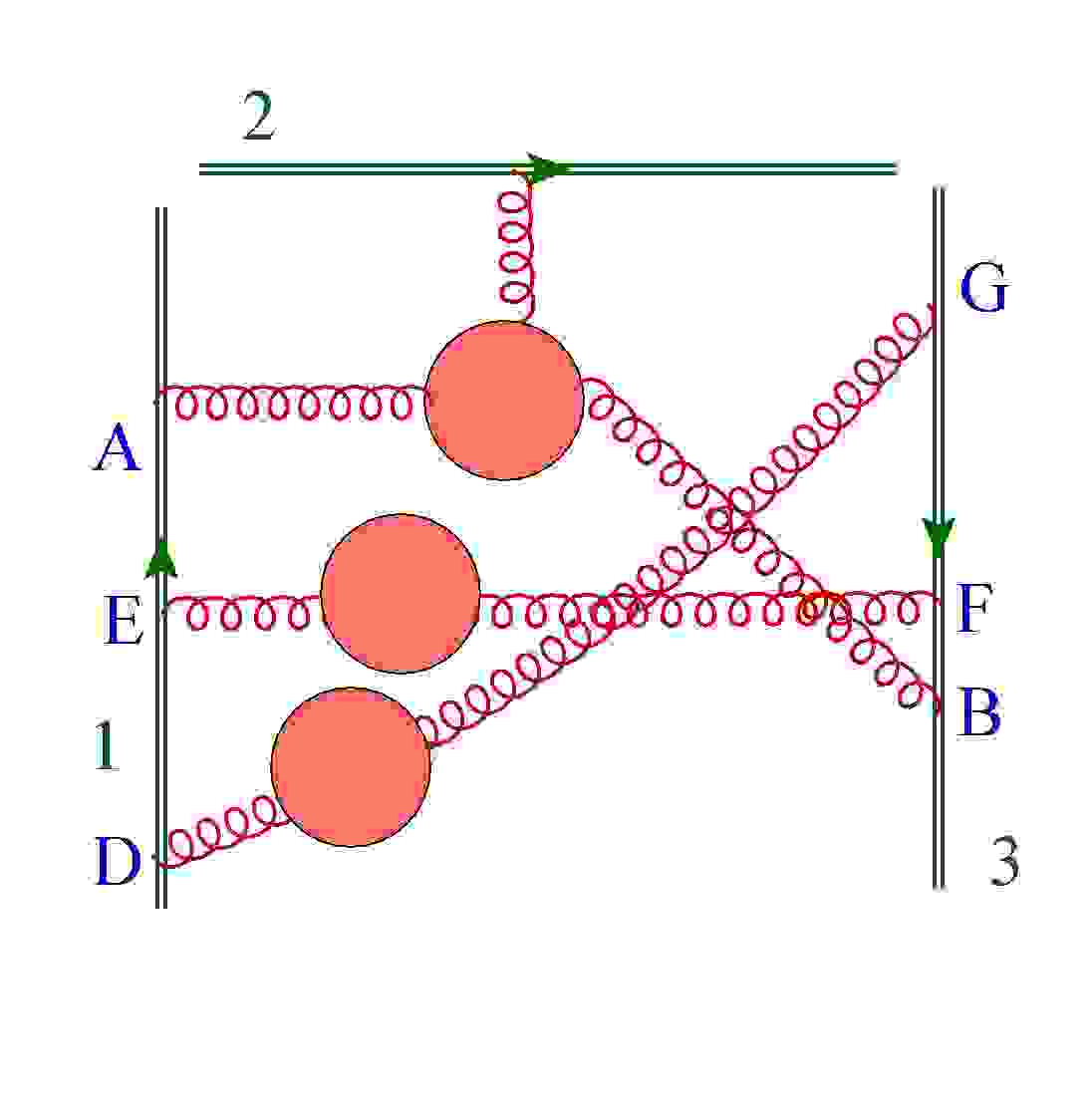} }
	\quad
	\subfloat[${W}_{3}^{(1,0,1)}(1,1,4)$]{\includegraphics[height=3.0cm,width=3.0cm]
	{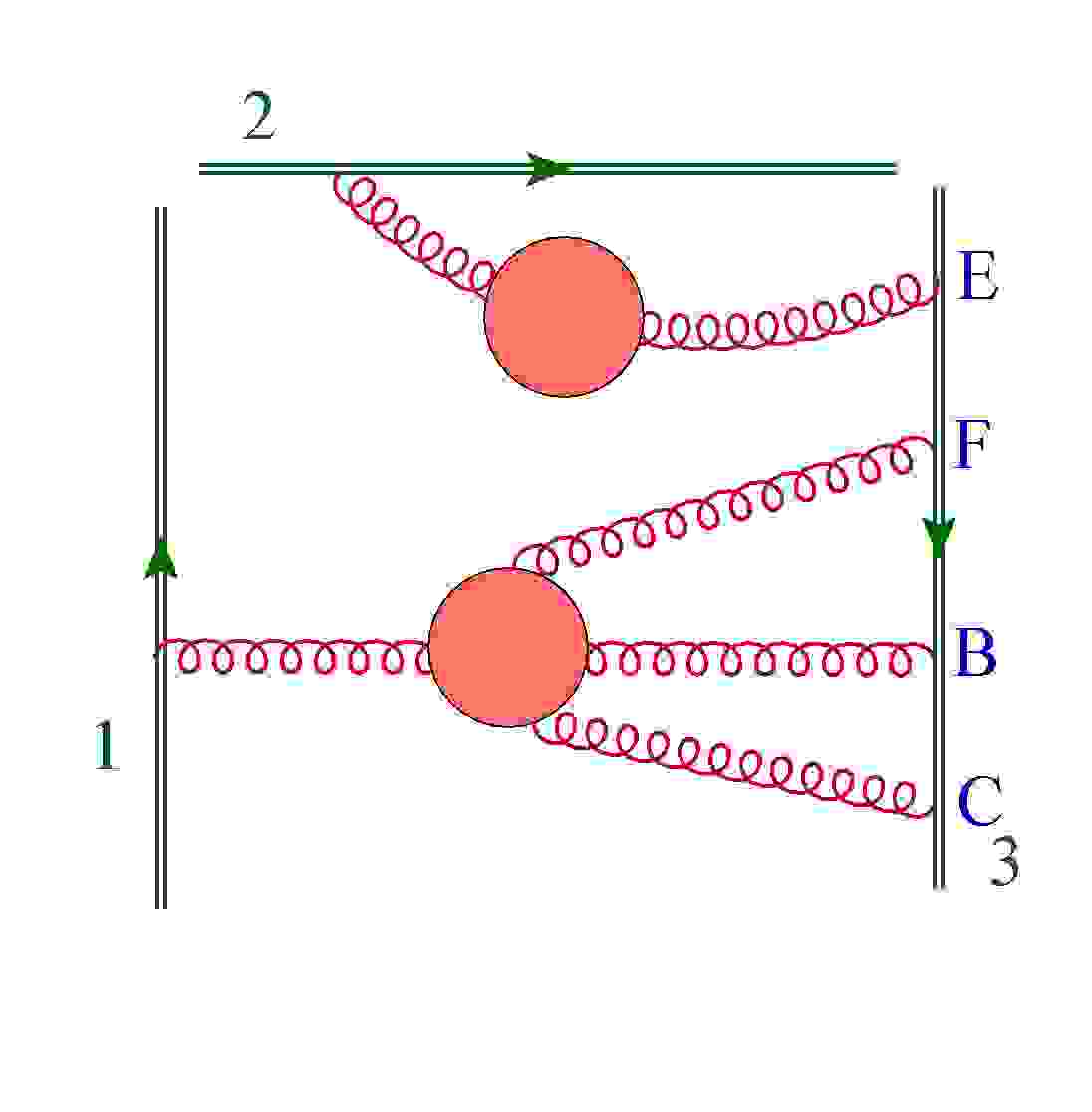} }
	\quad
	\subfloat[${W}_{3}^{(0,2)}(1,1,4)$]{\includegraphics[height=3.0cm,width=3.0cm]
	{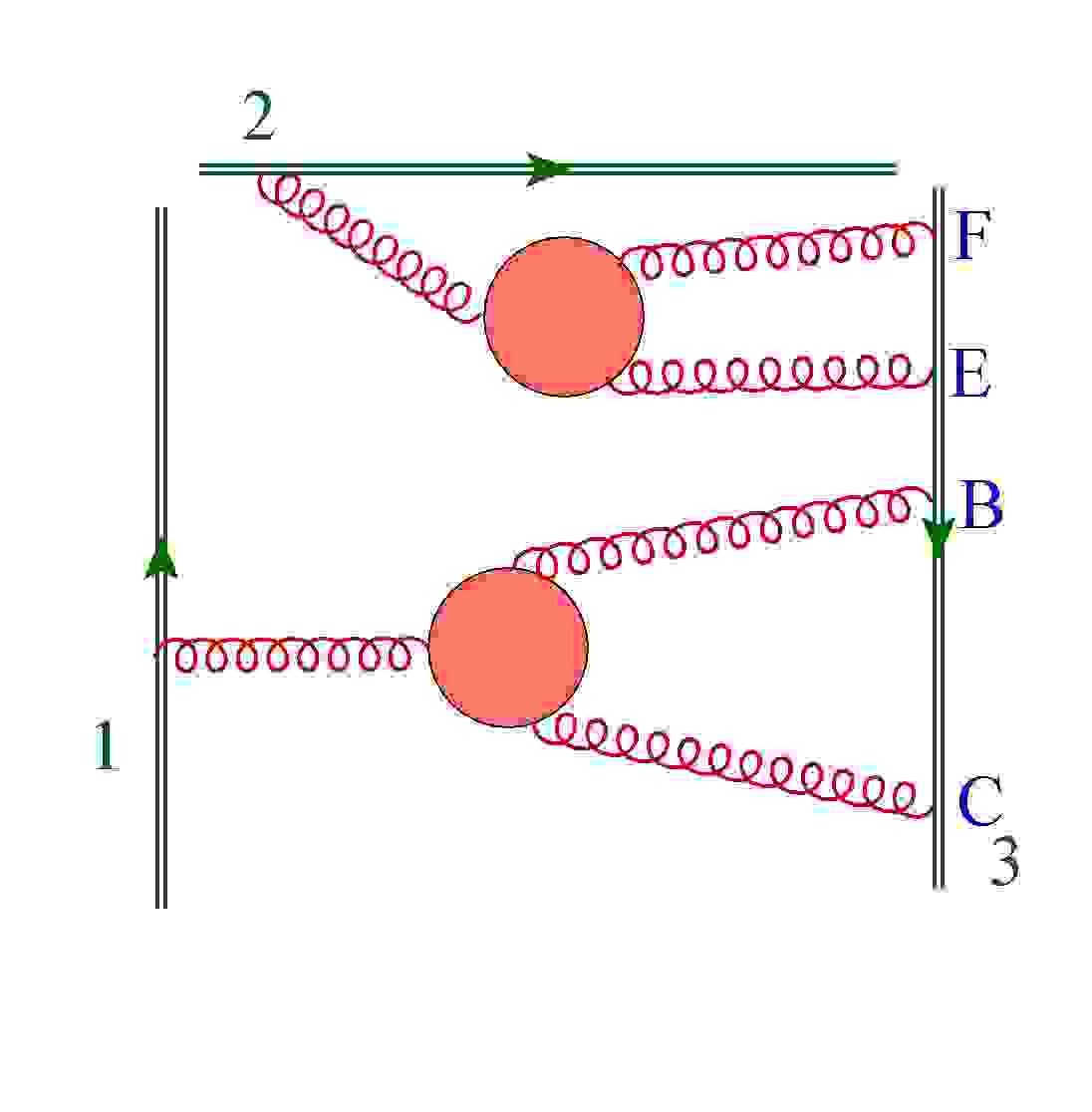} }
	\quad
	\subfloat[${W}_{3}^{(2,1)}(4,1,2)$]{\includegraphics[height=3.0cm,width=3.0cm]
	{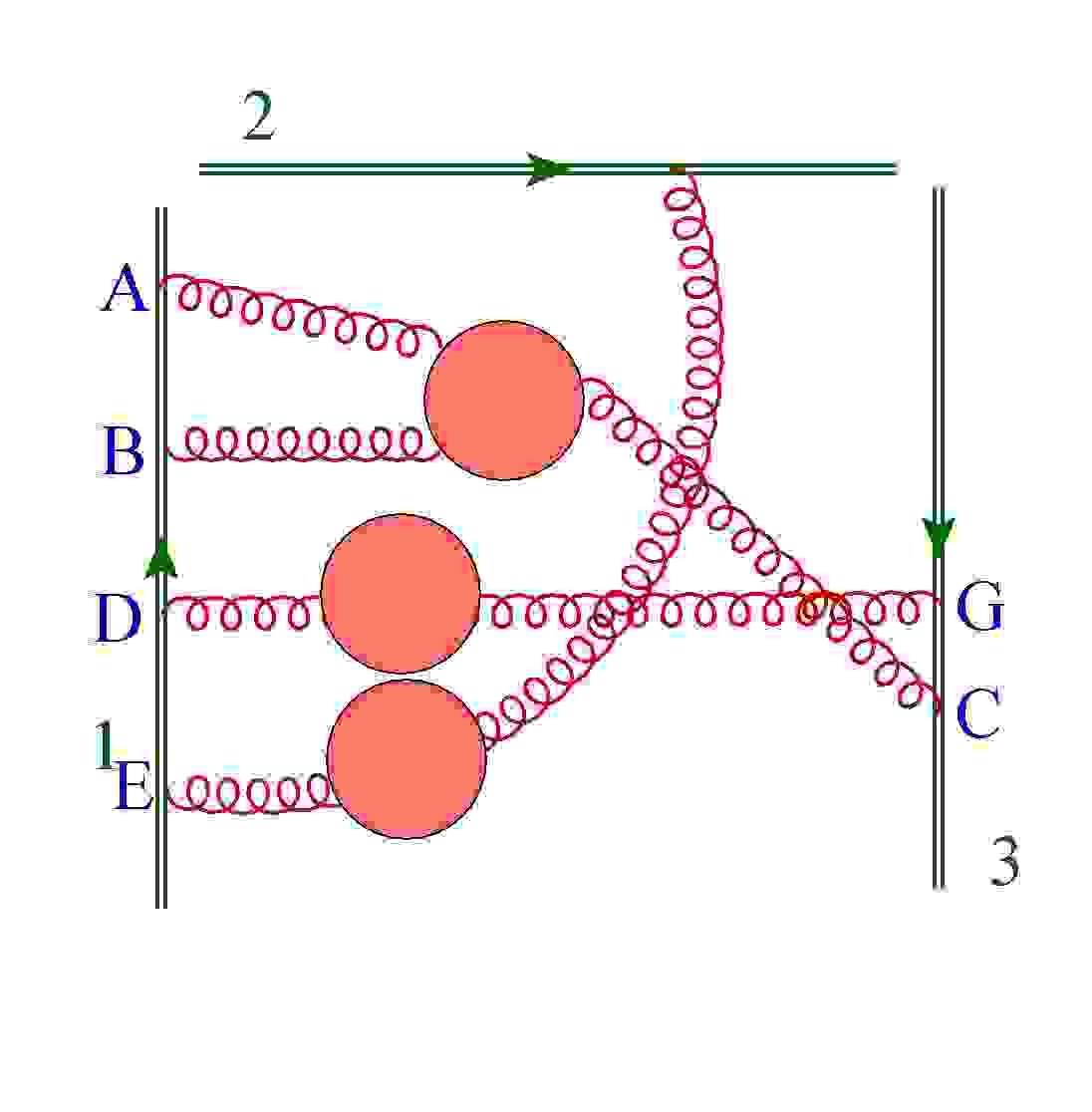} }
	\quad
	\subfloat[${W}_{3}^{(0,2)}(4,1,2)$]{\includegraphics[height=3.0cm,width=3.0cm]
	{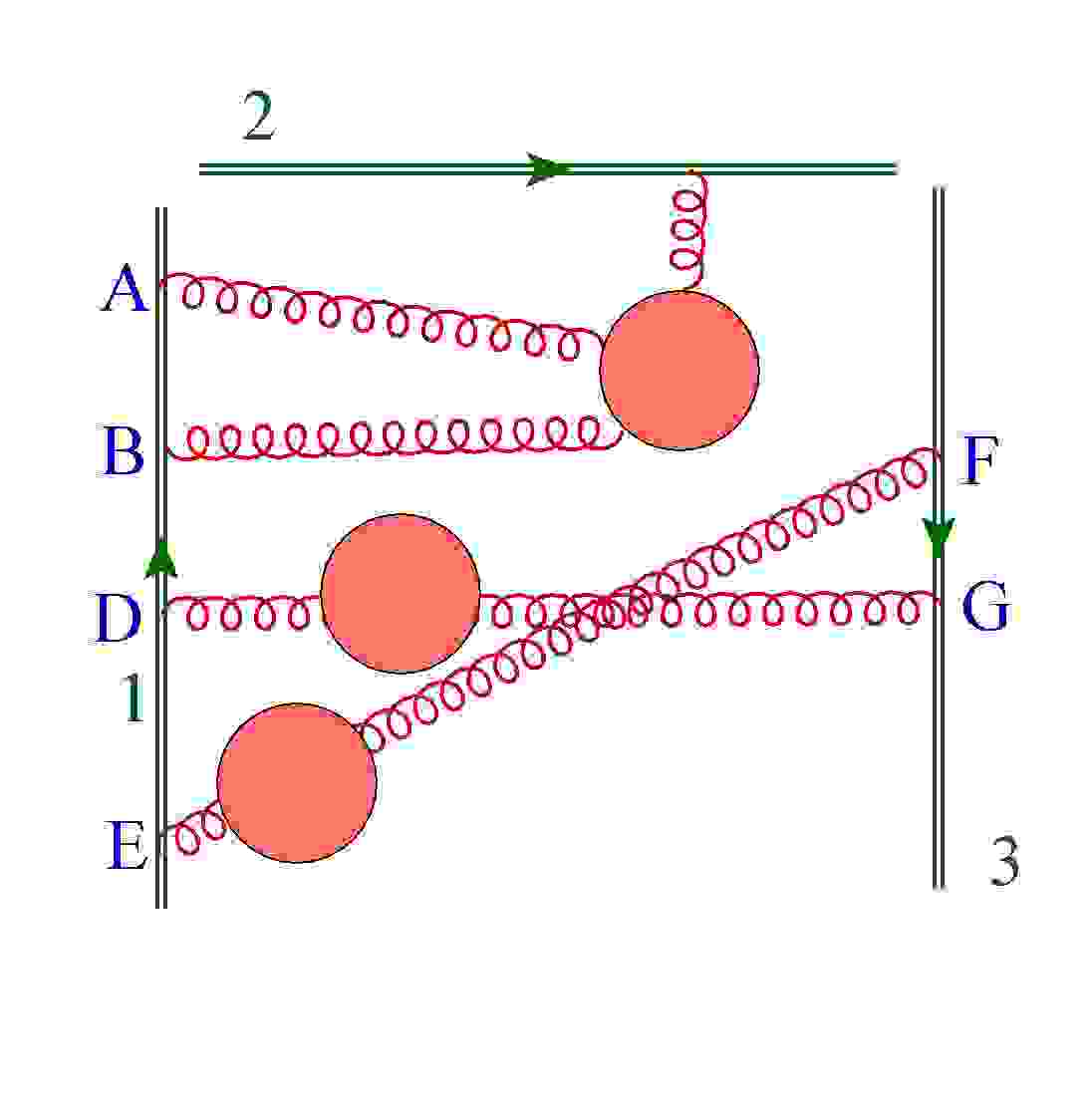} }
	\quad
	\subfloat[${W}_{3}^{(4)}(3,2,3)$]{\includegraphics[height=3.0cm,width=3.0cm]
	{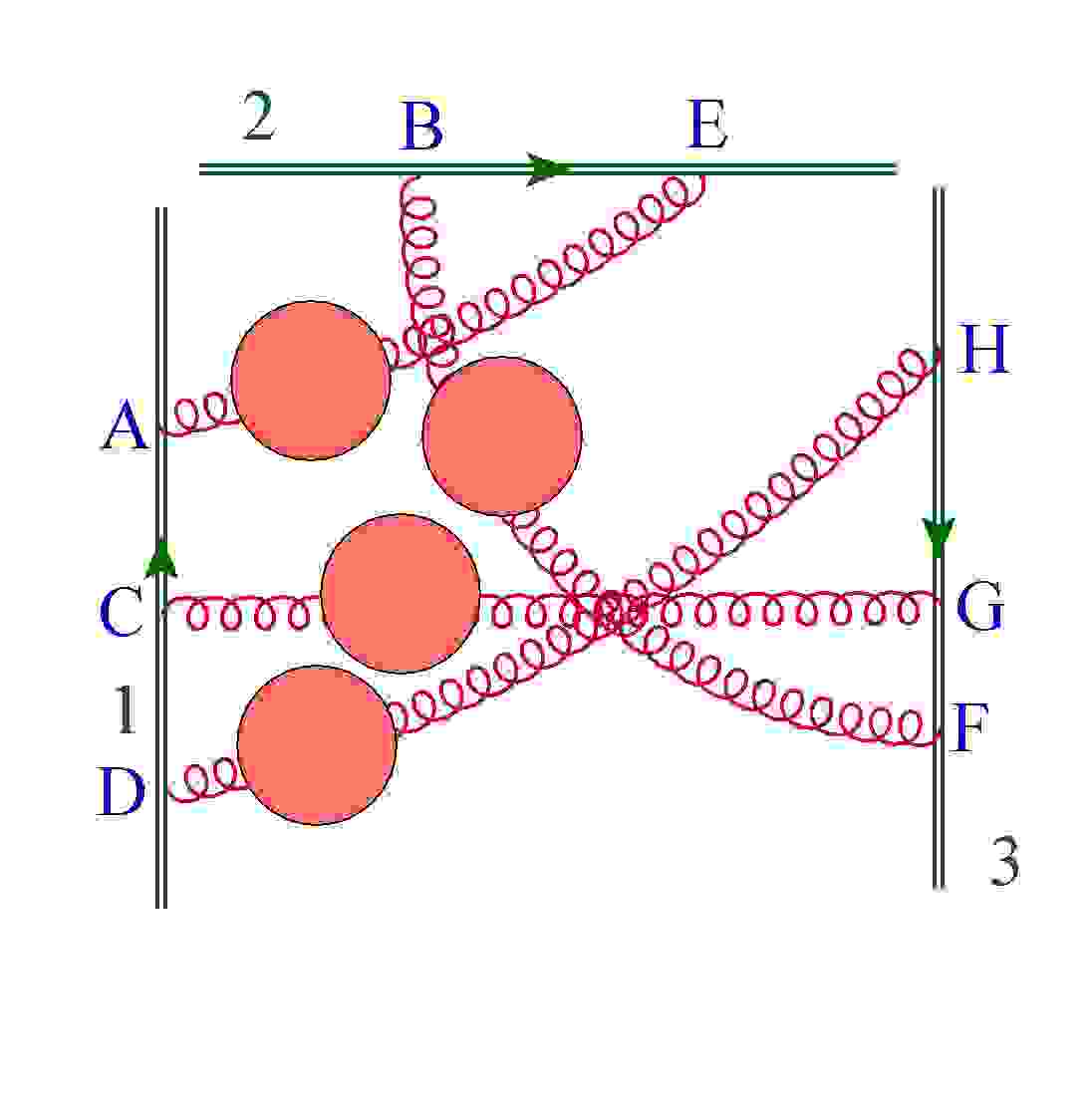} }
	\quad
	\subfloat[${W}_{3}^{(4)}(4,2,2)$]{\includegraphics[height=3.0cm,width=3.0cm]
	{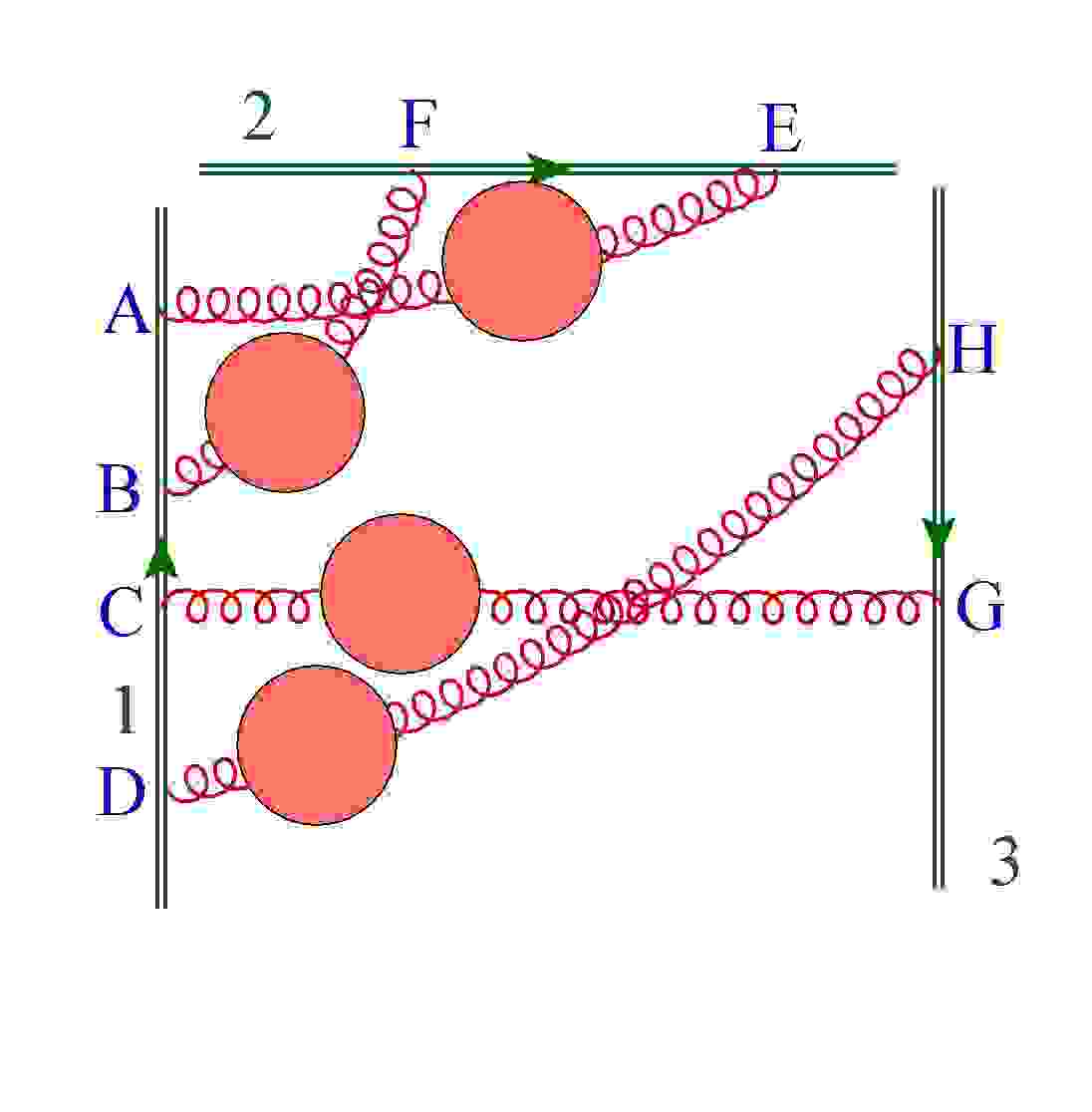} }
	\quad
	\subfloat[${W}_{3}^{(4)}(4,1,3)$]{\includegraphics[height=3.0cm,width=3.0cm]
	{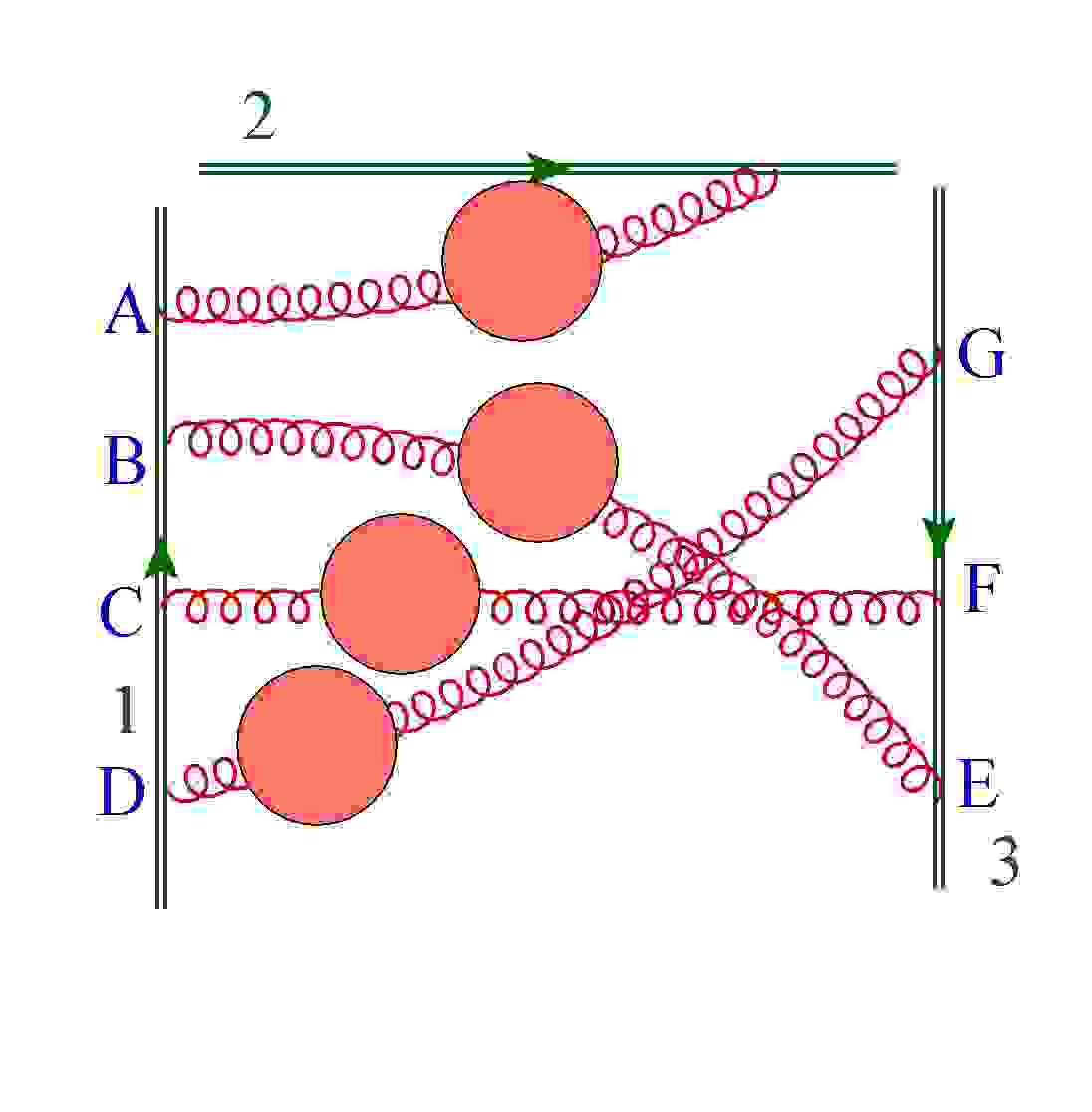} }
	\caption{Representative skeleton diagrams for the twenty-two four-loop Cwebs 
	connecting three Wilson lines in a massless theory. The dimensions of the 
	associated mixing matrices are $d_w = \{1,1,6,2,3,6,24,6,12,3,2,4,8,18,18,4,6,
	24,12,36,24,24\}$, in the order 
	shown.}
\label{Cwebsg8_3}
\end{figure}
\begin{enumerate}
\item Connect any two Wilson lines by introducing a two-gluon correlator.
\item Connect any existing $m$-point gluon correlator to a Wilson line, 
turning it into an  $(m+1)$-point gluon correlator.
\item Connect an existing $m$-point gluon correlator to an existing 
$n$-point gluon correlator by a single gluon, resulting in an $(n+m)$-point 
correlator.
\item Discard Cwebs that are given by the product of two or more
disconnected lower-order webs.
\item In a massless theory, discard all self-energy Cwebs, where all 
gluon lines attach to the same Wilson line, as they vanish as a 
consequence of the eikonal Feynman rules. 
\item Discard Cwebs that have been generated by the procedure more 
than once.
\end{enumerate}
Following the above steps, and focusing on the results for two and three
Wilson lines, we find four Cwebs connecting two Wilson lines at three 
loops, shown in Fig.~\ref{Cwebg6_2}, and six Cwebs connecting three 
Wilson lines at three loops, shown in Fig.~\ref{Cwebg6_3}. At four loops,
we find eight Cwebs connecting two lines, shown in Fig.~\ref{Cwebsg8_2},
and twenty-two Cwebs connecting threee lines, shown in Fig.~\ref{Cwebsg8_3}.
Cwebs  connecting four and five lines were studied in  ~\cite{Agarwal:2020nyc}.
Note that Cwebs that differ only by the permutation of the Wilson lines are 
identical in structure and we do not include them in our counting.

%%%%%%%%%%%%%%%%%%%%%%%%%%%%%%%%%%%%%%%%%
 
\section{Implementing a replica algorithm to generate Cweb mixing matrices}
\label{Repli} 

\begin{figure}[H]
	\centering
	\subfloat[]{\includegraphics[height=2.5cm,width=4.5cm]
	{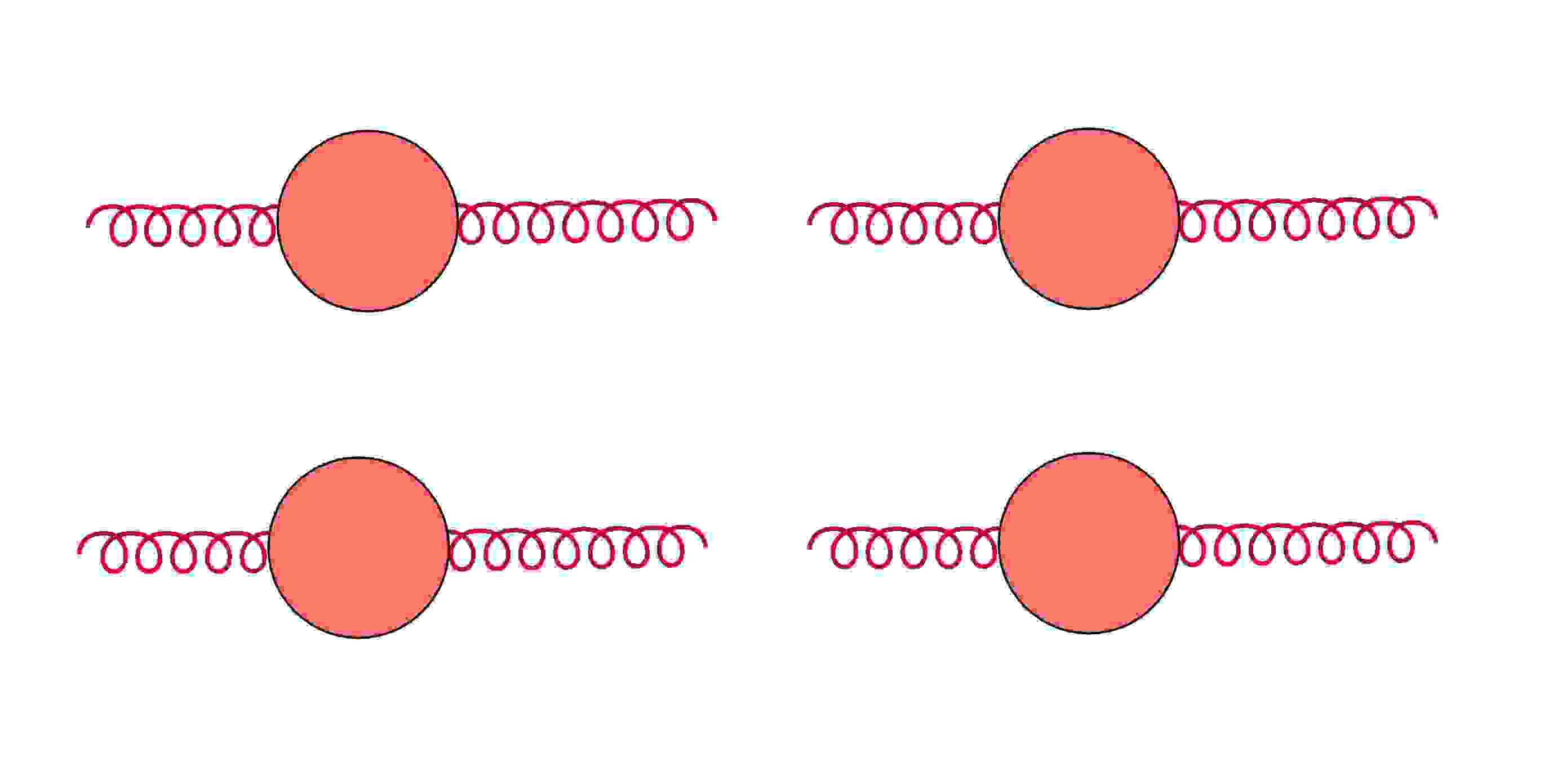} } 
	\qquad
	\qquad
	\qquad
	\subfloat[]{\includegraphics[height=2.5cm,width=4.5cm]
	{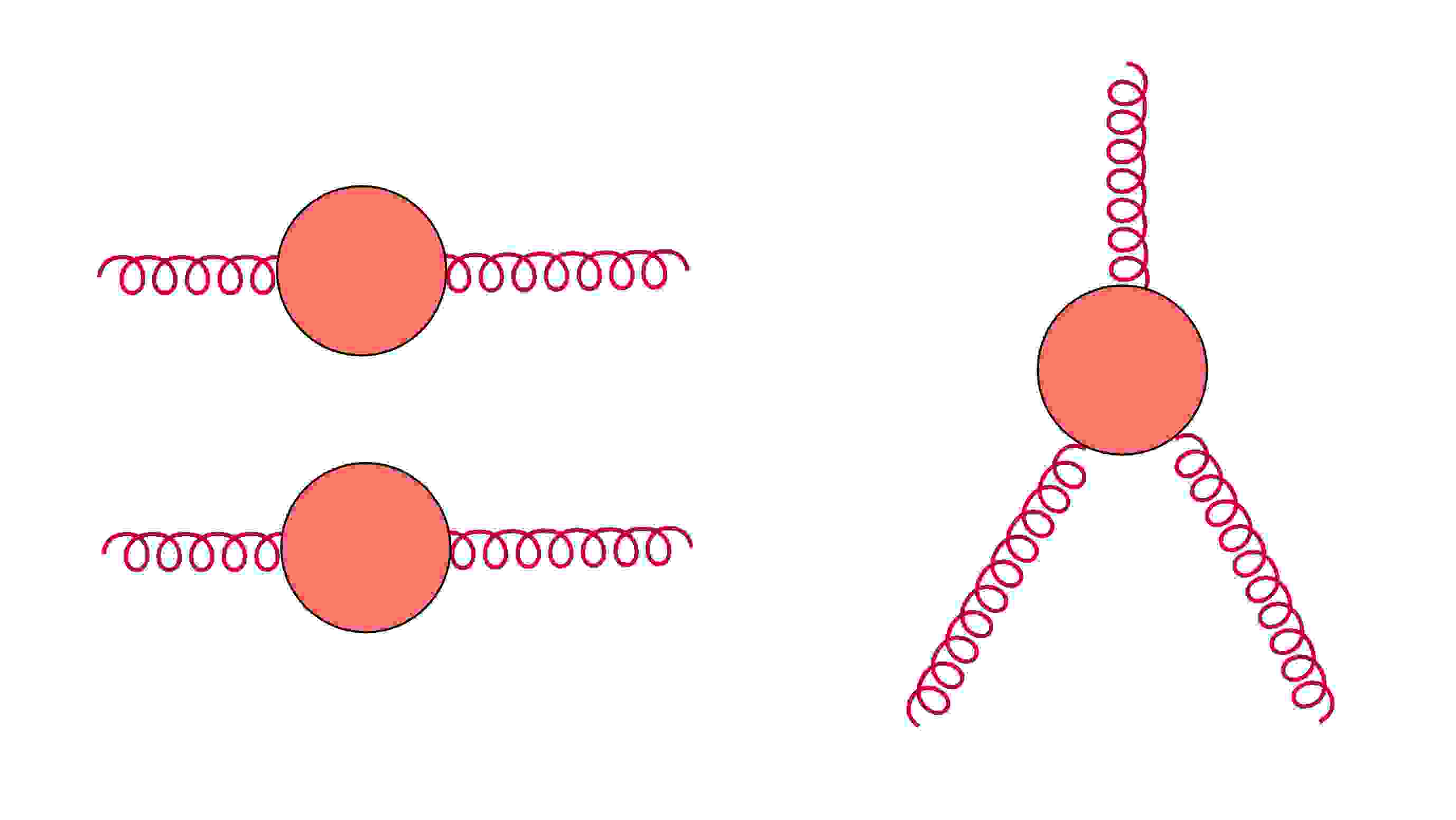} } \newline \\
	\subfloat[]{\includegraphics[height=2.5cm,width=4.5cm]
	{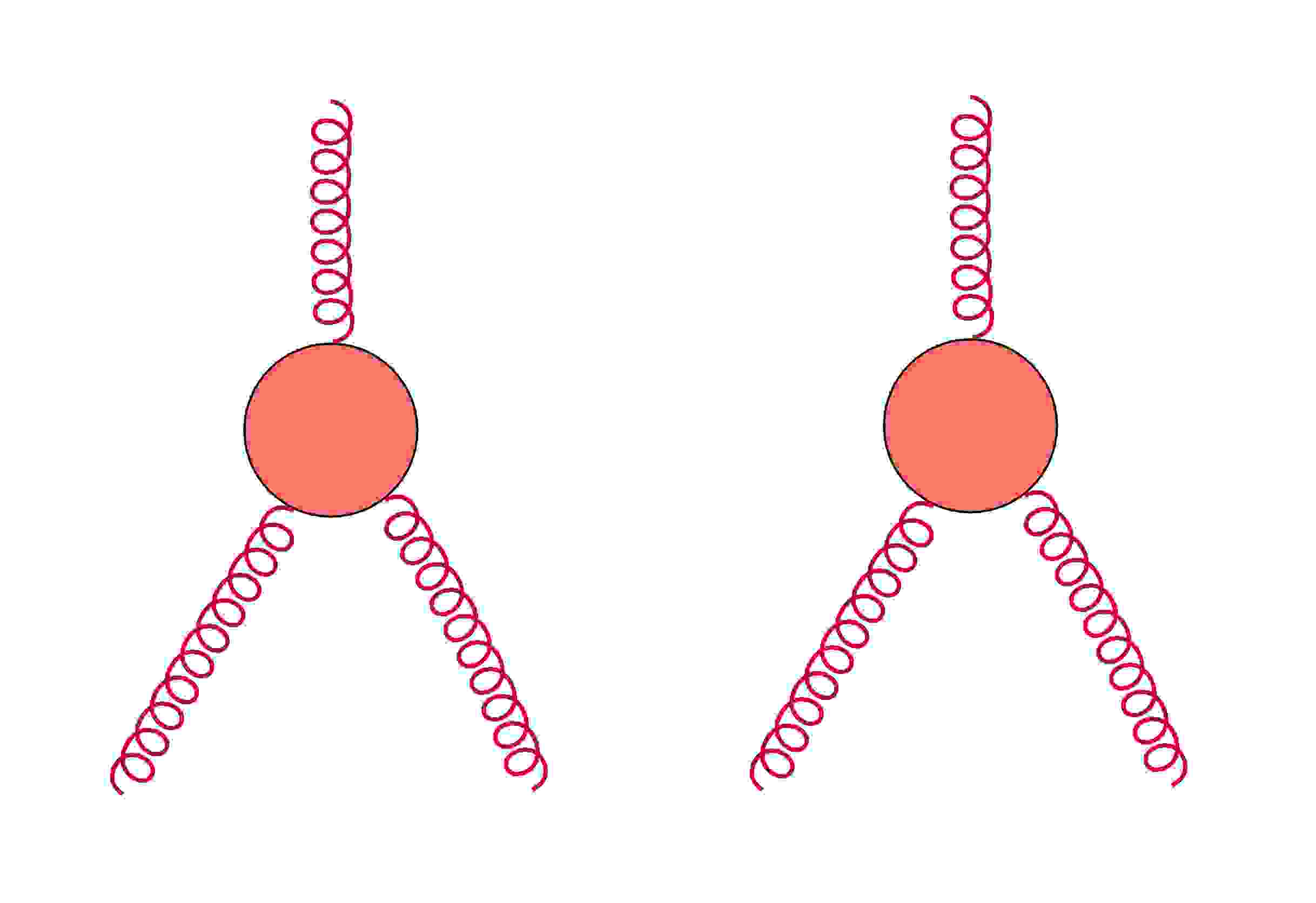} }
	\qquad
	\subfloat[]{\includegraphics[height=2.5cm,width=4.5cm]
	{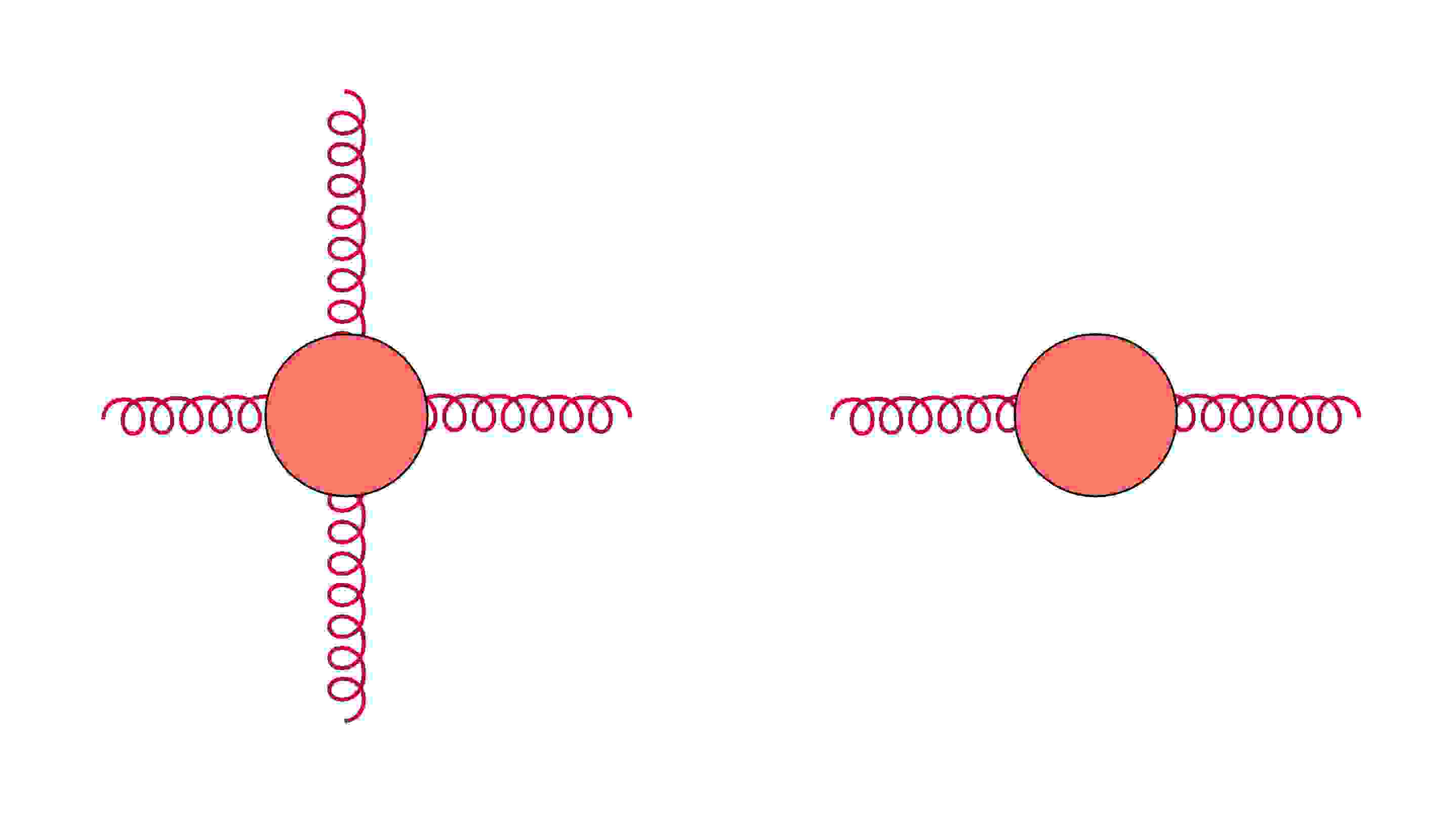} }
	\qquad
	\subfloat[]{\includegraphics[height=2.5cm,width=3.3cm]
	{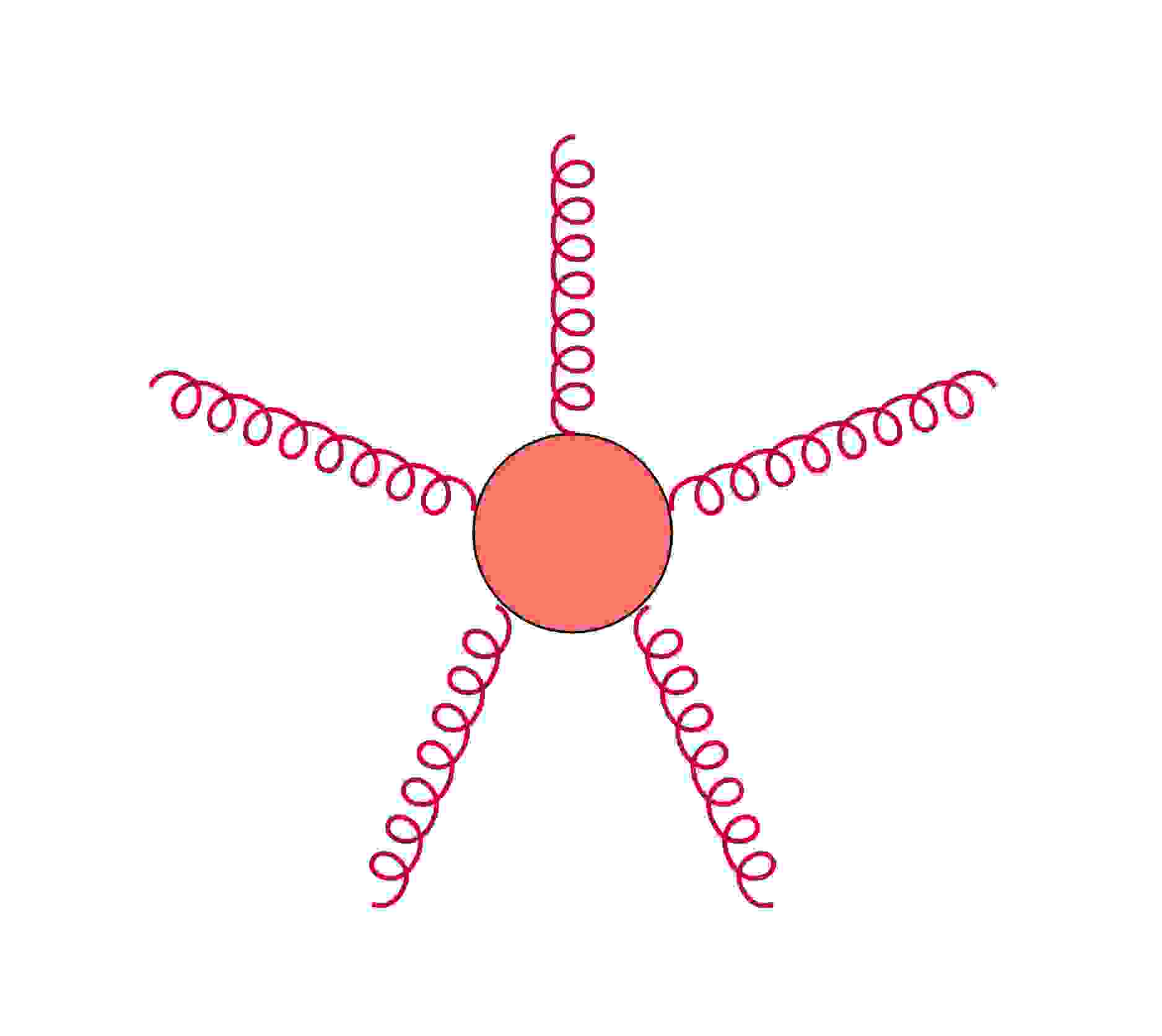} }\newline \\
	\caption{Combinations of connected correlators that can form Cwebs at 
	four loops.}
	\label{fig:connected}
\end{figure}

In this section, we will briefly describe our implementation of the replica method, 
introduced in~\cite{MezaPariVira}, and used in this context by~\cite{Laenen:2008gt,
Gardi:2010rn}, for the calculation of the web mixing matrices. The starting point 
is the path integral expression for the Wilson line correlator
\beq
  {\cal S}_n \left( \gamma_i \right) \, = \, \int {\cal D} A_\mu^a  \,\,
  {\rm e}^{{\rm i} S \left(  A_\mu^a \right)} \,  \prod_{k = 1}^n
  \Phi \left(  \gamma_k \right) \, = \, \exp \Big[ {\cal W}_n (\gamma_i) \Big] \, ,
\label{genWNCpath}
\eeq
where $S(A_\mu^a)$ is the classical action. The method is based on the 
construction of a {\it replicated theory}, where we replace the gluon field 
$A_\mu^a$ with $N_r$ identical copies, $A_\mu^{a,\,i}$ ($i = 1, \ldots, N_r$), 
which do not interact with each other. Further, we associate a copy of each 
Wilson line to each replica, effectively replacing each Wilson line in 
\eq{genWNCpath} with the product of $N_r$ Wilson lines. The correlator
in the replicated theory can then be written as
\beq
  {\cal S}_n^{\, {\rm repl.}} \left( \gamma_i \right) \, = \,   \Big[ 
  {\cal S}_n \left( \gamma_i \right) \Big]^{N_r} \, = \, \exp \Big[ N_r \,
  {\cal W}_n (\gamma_i) \Big] \, =  \, {\bf 1} + N_r \, {\cal W}_n (\gamma_i) 
  + {\cal O} (N_r^2) \, .
\label{exprepl}
\eeq
Using Eq. \ref{exprepl}, one can calculate $\mathcal{W}_n(\gamma_i)$ 
by computing the coefficient of the $\mathcal{O}(N_r)$ terms of the 
replicated Wilson line correlator. As there are no interaction vertices 
involving gluons belonging to different replicas, each connected gluon 
correlator in a Cweb is naturally assigned a unique replica number: this
means that the replica method designed for webs in~\cite{Gardi:2010rn,
Laenen:2008gt} immediately generalizes to Cwebs. The general structure
of the algorithm can be summarised as follows.
\begin{itemize}
\item Assign a replica number $i$, $1\leq i \leq N_r$, to each connected 
gluon correlator in a Cweb. 
\item Introduce a {\it replica ordering operator} $R$, which acts on the 
colour generators on each Wilson line ordering them according to their 
replica numbers. More precisely, if ${\bf T}_k^{(i)}$ denotes a generator 
associated with the emission of a gluon belonging to replica $i$ from 
Wilson line $k$, then R acts on a product of ${\bf T}_k^{(i)} {\bf T}_k^{(j)}$ 
by preserving the given order if $i \leq j$, and by reversing it if $i > j$.
The action of $R$ effectively replaces a skeleton diagram with another
one belonging to the same Cweb. 
\item In order to compute the colour factors in the replicated theory, one
then needs to determine the number of possible hierarchies of replica 
numbers occurring in a Cweb with $m$ connected pieces, which we 
call $h(m)$, and the number of occurrences of a particular hierarchy 
in the presence of $N_r$ replicas, which we call $M_{N_r}(h)$. $h(m)$ counts the  number of weak orderings on a set of $m$ elements, and is known in number theory and enumerative combinatorics as ordered Bell number
or Fubini number~\cite{IntSeq}. 
The first few ordered Bell numbers are given by $h(m) = \lbrace 1,1,
3,13,75,541 \rbrace$ for $m = {0,1,2,3,4,5}$, where we included
the case $m = 0$ to follow the mathematical convention. On the other 
hand, the multiplicity of a given hierarchy $h$, containing $n_r(h)$ distinct 
replicas, is easily determined, and it is given by 
\beq
  M_{N_r}(h) \, = \, \frac{N_r!}{\big( N_r - n_r(h) \big)! \,\, n_r(h)!}  \, .
\label{multhi}
\eeq
\item With these ingredients, one can compute the colour factor of a 
skeleton diagram $D$ in the replicated theory, which is given by
\beq
  C_{N_r}^{\, {\rm repl.}}  (D) \, = \, \sum_h M_{N_r} (h) \, R \big[ C(D) \big| h 
  \big]  \, ,
\label{expocolf}
\eeq
where $R \big[ C(D) \big|  h  \big]$ is the replica-ordered colour factor 
of diagram $D$ for a given hierarchy $h$. The exponentiated colour factor
for diagram $D$ is finally given by the $\mathcal{O}(N_r)$ terms in
\eq{expocolf}. 
\end{itemize} 
In order to compute the Cweb mixing matrices by applying the replica 
trick algorithm, we developed an in-house Mathematica code which 
was used to calculate the mixing matrices for Cwebs connecting four 
and five Wilson lines at four loops in Ref.~\cite{Agarwal:2020nyc}. 
Interestingly, lowering the number of Wilson lines increases the
combinatorial complexity of the problem, which may lead to a critical
slowing down of the algorithm for Cwebs involving many connected 
correlators attached to few Wilson lines. The basic reason is the fact
that, with few Wilson lines, at high orders the typical number of gluon
attachments to each Wilson line grows, and so does the size of a typical
mixing matrix. To keep the runtime under control, one has to take full 
advantage of the symmetries of the problem, including Bose symmetry
for each connected gluon correlator, and the symmetry under the exchange 
of identical gluon correlators connecting the same set of Wilson lines.
Below, we briefly present the main steps in our current implementation 
of the algorithm.
\begin{itemize}
\item The starting point is the building blocks for four-loop Cwebs, which
are the sets of connected gluon correlators shown in Fig.~\ref{fig:connected}. 
The code generates all four-loop Cwebs by combining the building blocks and 
attaching them to the Wilson lines in all possible ways. This is done starting
from the most intricate case, which is the attachment of a set of four two-point 
correlators in all possible ways to the Wilson lines. Proceeding with other
combinations of correlators, the combinatorial problem simplifies, and finally 
the five-point correlator shown in Fig.~\ref{fig:connected} only produces 
completely connected Cwebs with a mixing matrix $R=1$.
\item The above step generates Cwebs which are related to one another by 
the permutation of the Wilson lines. As far as the colour structure is concerned, 
they are all identical, so the duplicates are removed.   
\item The code then assigns a distinct replica number to each connected correlator. 
\item Starting from a single diagram, the code generates all diagrams for a 
Cweb by identifying the gluon attachments on each Wilson line, and placing 
the attachments originating from different gluon correlators in different packs. 
These packs are then shuffled.     
\item A subroutine generates all possible hierarchies $h$ for a Cweb. As an 
example, for a Cweb with two connected gluon correlators, the subroutine 
generates the three hierarchies $h=\lbrace i=j, i>j, i<j \rbrace$, where $i$ 
and $j$ denote the replica numbers. The code then determines the number 
of distinct replicas $n_r$ for each hierarchy. For example, for the hierarchy 
$i=j$ one has $n_r = 1$, whereas for hierarchies $i>j$ and $i<j$ one has 
$n_r = 2$. Using Eq. \ref{multhi}, one then computes the multiplicity 
$M_{N_r}(h)$.
\item At this stage, one can construct, for each Cweb, a table of the form of
Table 1 of Ref.~\cite{Gardi:2010rn}, and finally determine the mixing matrix 
$R_w$ for Cweb $w$. The code then diagonalizes $R_w$, constructing a 
matrix $Y_w$ such that $Y_w R_w Y_w^{-1}$ is of the form ${\bf 1}_{r_w}
\oplus {\bf 0}_{d_w - r_w}$, where $d_w$ is the dimension of the mixing 
matrix, and $r_w$ its rank. The matrix $Y_w$, acting from the left on the 
vector of colour factors for the diagrams in $w$, computes the exponentiated
colour factors.
\item In its present form, the code does not generate self-energy Cwebs, 
where a connected gluon correlator is attached only to a single Wilson 
line, since such Cwebs vanish in the massless theory. The code however
generates disconnected Cwebs (where the set of Wilson lines can be partitioned  
in subsets whose elements are not connected by any gluons): the vanishing 
of the mixing matrices for disconnected Cwebs works as a check on the code. 
Further checks are provided by verifying two known properties: the idempotence
of mixing matrices, and the row sum rule. The conjectured column sum rule, 
on the other hand, is verified a posteriori.
\end{itemize}
We note that, for a number of Cwebs with non-trivial symmetries, the procedure 
of generating all diagrams by shuffling the packs on each of the Wilson lines
will produce multiple duplicates of each diagram, which can be obtained from 
each other by simply relabelling some of the correlators. In the case at hand, 
this happens for the Cwebs:
${W}_{3,\text{II}}^{(2,1)}(2,2,3)$, 
${W}_{3,\text{II}}^{(2,1)}(1,3,3)$,  
${W}_{3}^{(0,2)}(1,2,4)$, 
${W}_{3}^{(4)}(2,3,3)$, 
${W}_{3}^{(4)}(2,2,4)$, 
${W}_{3}^{(4)}(1,3,4)$,
${W}_{2}^{(0,2)}(2,4)$,
${W}_{2}^{(2,1)}(3,4)$, 
${W}_{2}^{(4)}(4,4)$, shown, with all other Cwebs connecting three and four 
lines at four loops, in the Appendix. To contain the computation time, it is crucial 
to remove these duplicates before applying the replica algorithm. For example, 
the Cweb $W_2^{(4)}(4,4)$ has a total of 24 diagrams, but each diagram is 
generated in 24 duplicate copies: the present version of the code automatically
identifies and removes all duplicates.

\begin{figure}[H]
	\centering
	\subfloat[][$C_1$]{\includegraphics[height=3.0cm,width=3.0cm]{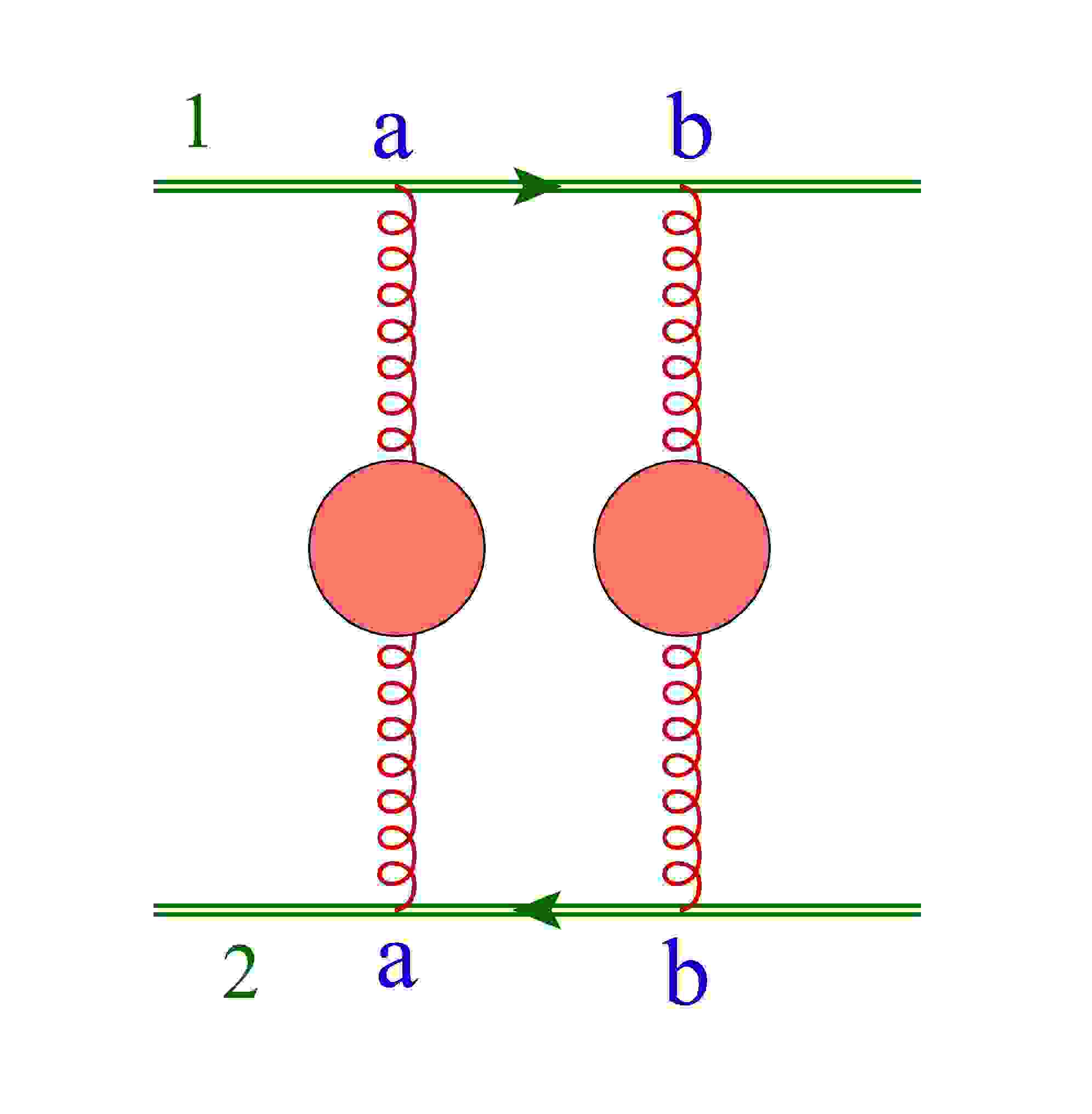} }
	\quad
	\subfloat[$C_2$]{\includegraphics[height=3.0cm,width=3.0cm]{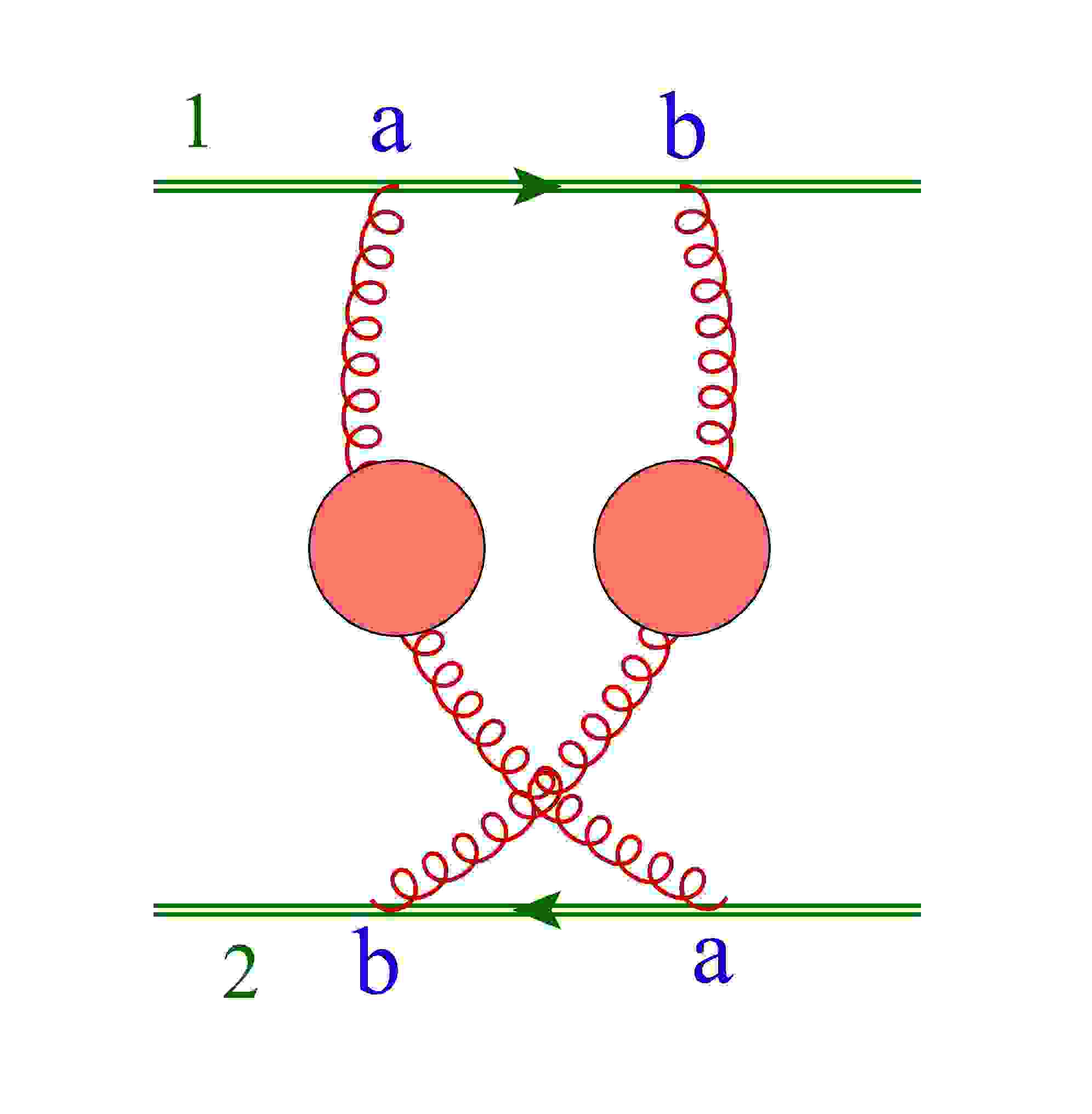} }
	\subfloat[$C_3$]{\includegraphics[height=3.0cm,width=3.0cm]{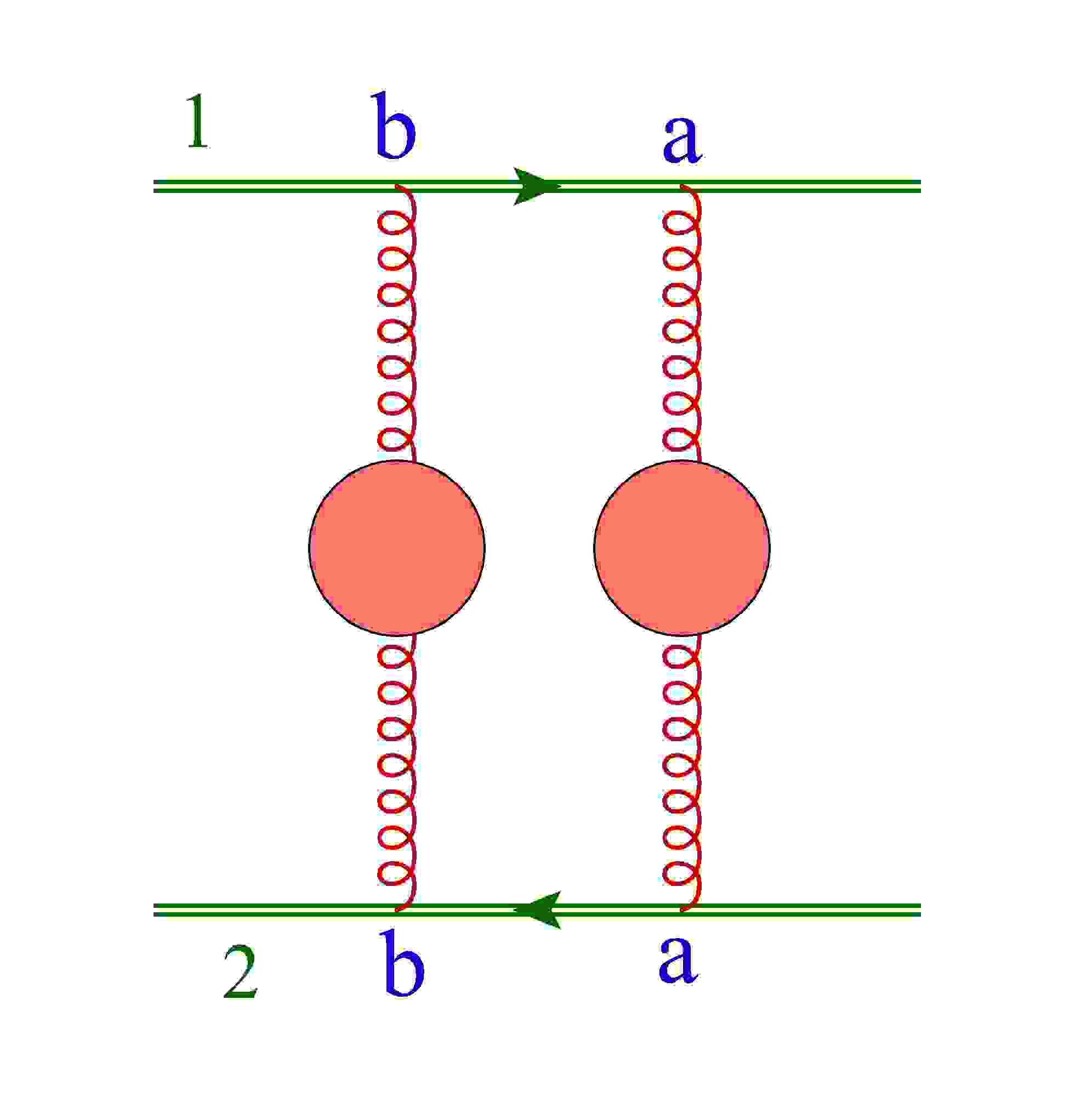} }
	\quad
	\subfloat[$C_4$]{\includegraphics[height=3.0cm,width=3.0cm]{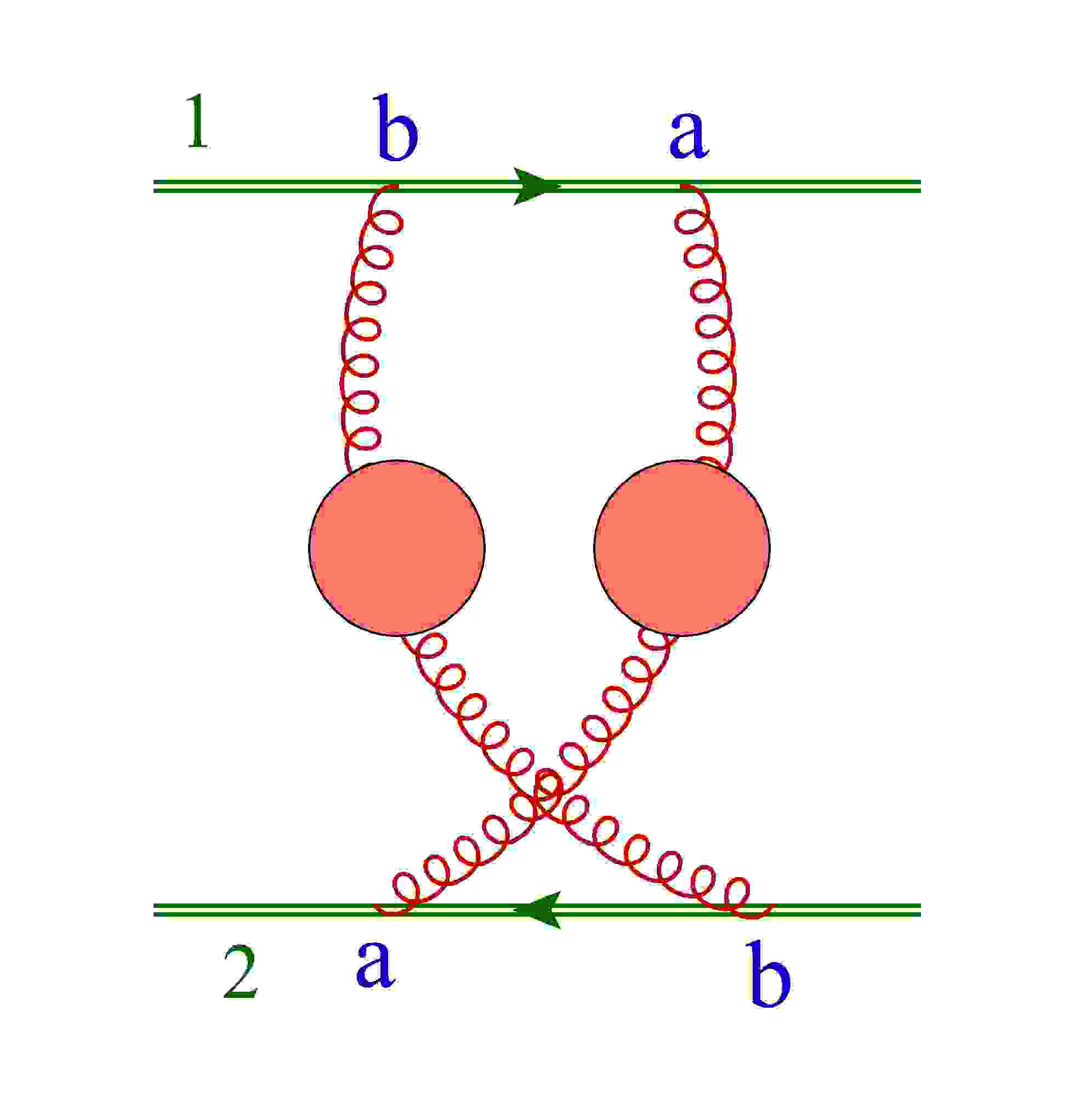} }
	\quad
	\caption{All diagrams generated by shuffling gluons on all Wilson lines, for 
	the case of $W_2^{(2)}(2,2)$.}
\label{double-replica}
\end{figure}

To further illustrate this issue consider the simple case of the two-loop Cweb  
$W_2^{(2)}(2,2)$, which has two attachments on both Wilson lines, and is shown 
in Fig.~\ref{double-replica}. 
The shuffle (which is just a single permutation in this case) of the correlators on 
each Wilson line generates the four diagrams shown in Fig.~\ref{double-replica}.   
As the colour indices $a$ and $b$ in Fig.~\ref{double-replica} are summed over, 
diagrams $C_1$ and $C_3$ are duplicates, and similarly $C_2$ and $C_4$. 
The code identifies the duplicates and deletes $C_3$ and $C_4$. 

The exponentiated colour factors generated by the code have open colour indices
for each gluon attachment to the Wilson lines: these open indices must then be
contracted with the colour structures arising from the connected gluon correlators.
These colour structures, in turn, are constrained by Bose symmetry and gauge 
invariance. For example, a two-gluon correlator joining any two lines $i$ and $j$ 
will generate only the dipole structure ${\bf T}_i \cdot {\bf T}_j$, while a three-point 
gluon correlator is conjectured to be proportional to the structure constants $\fabc$ 
to any perturbative order. For four-gluon correlators, three colour structures are 
available at tree level, while more complicated ones will arise when loop corrections 
are considered, including contributions proportional to quartic Casimir operators
of the gauge algebra~\cite{vanRitbergen:1998pn,Chetyrkin:2017bjc,Cvitanovic:2008zz}.

%%%%%%%%%%%%%%%%%%%%%%%%%%%%%%%%%%%%%%%%%

\section{Examples of four-loop Cwebs connecting two and three Wilson lines}
\label{Fourwe}

In this section, we present in some detail the calculation of two four-loop Cwebs 
connecting respectively two and three Wilson lines. We present their mixing matrices, 
the diagonalizing matrices Y, and we give explicit results for the exponentiated 
colour factors. The results for all other four-loop Cwebs connecting two and three 
Wilson lines are presented in the Appendix.

%%%%%%%%%%%%%%%%%%%%%%

\subsection{A two-line Cweb at four loops}
\label{four-two}

\begin{figure}[H]
  \begin{center}
  \includegraphics[height=4cm,width=4cm]{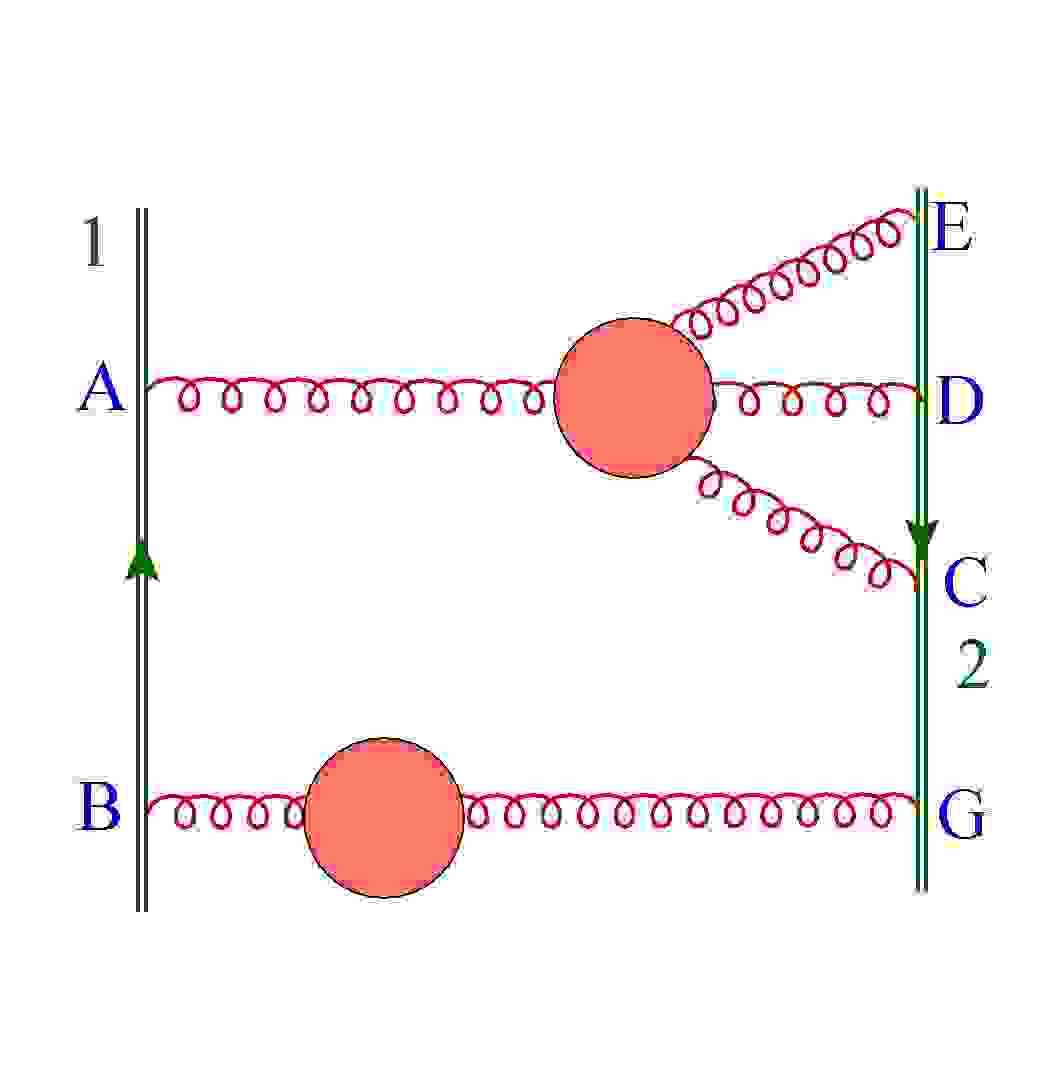}
  \caption{Skeleton diagram $C_1$ for the Cweb $W_{2}^{(1,0,1)}(2,4)$.} 
  \label{fig:twoweb}
  \end{center}
\end{figure}

As a first example, we consider the two-line Cweb $W_{2}^{(1,0,1)}(2,4)$, which 
contains one two-gluon correlator, no three-gluon correlators, and one four-gluon 
correlator; furthermore, there are two gluon attachments on line 1 and four gluon 
attachments on line 2. Clearly, the perturbative expansion for this Cweb starts 
at ${\cal O} (g^8)$, {\it  i.e.} at four loops. A representative skeleton diagram for 
the Cweb is shown in Fig.~\ref{fig:twoweb}. The available shuffles on line 1 and 
line 2 produce a total of eight skeleton diagrams for this Cweb which we denote
by the order of the gluon attachments on the two lines, and are given by
\beq
  \lbrace C_1,C_2,C_3,C_4,C_5,C_6,C_7,C_8 \rbrace & \equiv & 
  \Big\{ \lbrace \lbrace BA \rbrace, \lbrace EDCG \rbrace \rbrace, 
            \lbrace \lbrace BA \rbrace, \lbrace EDGC \rbrace \rbrace,
            \lbrace \lbrace BA \rbrace, \lbrace EGDC \rbrace \rbrace, \nn \\
  && \hspace{2mm}
            \lbrace \lbrace BA \rbrace, \lbrace GEDC \rbrace \rbrace, 
            \lbrace \lbrace AB \rbrace, \lbrace EDCG \rbrace \rbrace, 
            \lbrace \lbrace AB \rbrace, \lbrace EDGC \rbrace \rbrace, \nn \\
  && \hspace{2mm} 
            \lbrace \lbrace AB \rbrace, \lbrace EGDC \rbrace \rbrace, 
            \lbrace \lbrace AB \rbrace, \lbrace GEDC \rbrace \rbrace \Big\} \, .
\label{tabCw1}
\eeq
Keeping in mind the orientation of the Wilson lines, the diagram portrayed 
in Fig.~\ref{fig:twoweb} is diagram $C_1$. With this ordering, the column 
weights $s(C_i)$ for the diagrams in the Cweb are collected in the vector 
\beq
  \lbrace s(C_1), s(C_2), s(C_3), s(C_4), s(C_5), s(C_6), s(C_7), s(C_8) 
  \rbrace \, = \, \lbrace 0,0,0,1,1,0,0,0 \rbrace \, . \nn 
\label{svec1}
\eeq
\noindent Note that, only $C_4$ and $C_5$ have nonzero $s$ values.
We find that the mixing matrix $R$, the diagonalising matrix $Y$ and the 
diagonal matrix $D$ for this Cweb are given by
\begin{align}
\begin{split}
  &
  R \, = \, \left(
  \begin{array}{cccccccc}
  1 & 0 & 0 & -\frac{1}{2} & -\frac{1}{2} & 0 & 0 & 0 \\
  0 & 1 & 0 & -\frac{1}{2} & -\frac{1}{2} & 0 & 0 & 0 \\
  0 & 0 & 1 & -\frac{1}{2} & -\frac{1}{2} & 0 & 0 & 0 \\
  0 & 0 & 0 & \frac{1}{2} & -\frac{1}{2} & 0 & 0 & 0 \\
  0 & 0 & 0 & -\frac{1}{2} & \frac{1}{2} & 0 & 0 & 0 \\
  0 & 0 & 0 & -\frac{1}{2} & -\frac{1}{2} & 1 & 0 & 0 \\
  0 & 0 & 0 & -\frac{1}{2} & -\frac{1}{2} & 0 & 1 & 0 \\
  0 & 0 & 0 & -\frac{1}{2} & -\frac{1}{2} & 0 & 0 & 1 \\
  \end{array}
  \right) \, , \qquad \quad
  Y \, = \, \left(
  \begin{array}{cccccccc}
  -1 & 0 & 0 & 0 & 0 & 0 & 0 & 1 \\
  -1 & 0 & 0 & 0 & 0 & 0 & 1 & 0 \\
  -1 & 0 & 0 & 0 & 0 & 1 & 0 & 0 \\
  -1 & 0 & 0 & 0 & 1 & 0 & 0 & 0 \\
  -1 & 0 & 0 & 1 & 0 & 0 & 0 & 0 \\
  -1 & 0 & 1 & 0 & 0 & 0 & 0 & 0 \\
  -1 & 1 & 0 & 0 & 0 & 0 & 0 & 0 \\
  0 & 0 & 0 & 1 & 1 & 0 & 0 & 0 \\
  \end{array}
  \right)\, , \qquad \quad
  D \,= \, \D{7} \, .
\end{split}
\label{eq:web1app}
\end{align}
One may easily verify the properties of the mixing matrix: $R$ is idempotent, 
the matrix elements in each row sums to zero, and the column sum rule is 
obeyed. After diagonalising the mixing matrix, we find that the rank of the mixing 
matrix is $r_w = 7$, which means that there will be 7 independent exponentiated 
colour factors for this Cweb, which are given by
\beq
  (YC)_1 & = &
  i f^{acg} f^{deg} f^{ebh} \tb 1 \ta 1 \tc 2 \td 2 \thh 2  
  + i f^{acg} f^{bdj} f^{deg} \tb 1 \ta 1 \tc 2 \tj 2 \te 2   \nnn \\&&
  + i f^{acg} f^{cbm} f^{deg} \tb 1 \ta 1 \tm 2 \td 2 \te 2  
  - i f^{abu} f^{acg} f^{deg} \tu 1 \tb 2 \tc 2 \td 2 \te 2 \, ,
  \nnn \\ \nnn \\
  (YC)_2 & = &
  i f^{acg} f^{bdj} f^{deg} \tb 1 \ta 1 \tc 2 \tj 2 \te 2  
  + i f^{acg} f^{cbm} f^{deg} \tb 1 \ta 1 \tm 2 \td 2 \te 2   \nnn \\&&
  - i f^{abu} f^{acg} f^{deg} \tu 1 \tb 2 \tc 2 \td 2 \te 2 \, ,
  \nnn \\ \nnn \\
  (YC)_3 & = &
  i f^{acg} f^{cbm} f^{deg} \tb 1 \ta 1 \tm 2 \td 2 \te 2  
  - i f^{abu} f^{acg} f^{deg} \tu 1 \tb 2 \tc 2 \td 2 \te 2 \, ,
  \nnn \\ \nnn \\
  (YC)_4 & = &
  - i f^{abu} f^{acg} f^{deg} \tu 1 \tb 2 \tc 2 \td 2 \te 2 \, ,
  \nnn \\ \nnn \\
  (YC)_5 &=&
  i f^{acg} f^{deg} f^{ebh} \ta 1 \tb 1 \tc 2 \td 2 \thh 2  
  + i f^{acg} f^{bdj} f^{deg} \ta 1 \tb 1 \tc 2 \tj 2 \te 2   \nnn \\&&
  + i f^{acg} f^{cbm} f^{deg} \ta 1 \tb 1 \tm 2 \td 2 \te 2 \, ,
\nnn \\ \nnn \\
\label{ecf11}
\eeq
\beq
  (YC)_6 &=&
  i f^{acg} f^{bdj} f^{deg} \ta 1 \tb 1 \tc 2 \tj 2 \te 2  
  + i f^{acg} f^{cbm} f^{deg} \ta 1 \tb 1 \tm 2 \td 2 \te 2 \, ,
  \nnn \\ \nnn \\
  (YC)_7 &=&
  i f^{acg} f^{cbm} f^{deg} \ta 1 \tb 1 \tm 2 \td 2 \te 2 \, .  
\label{ecf12}
\eeq
We observe that all the exponentiated colour factors correspond to completely 
connected Feynman diagrams, which verifies, as expected, the non-abelian 
exponentiation theorem~\cite{Gardi:2013ita}. 

%%%%%%%%%%%%%%%%%%%%%%

\subsection{A three-line Cweb at four loops}
\label{four-three}

\begin{figure}[H] 
  \begin{center}
  \includegraphics[height=4cm,width=4cm]{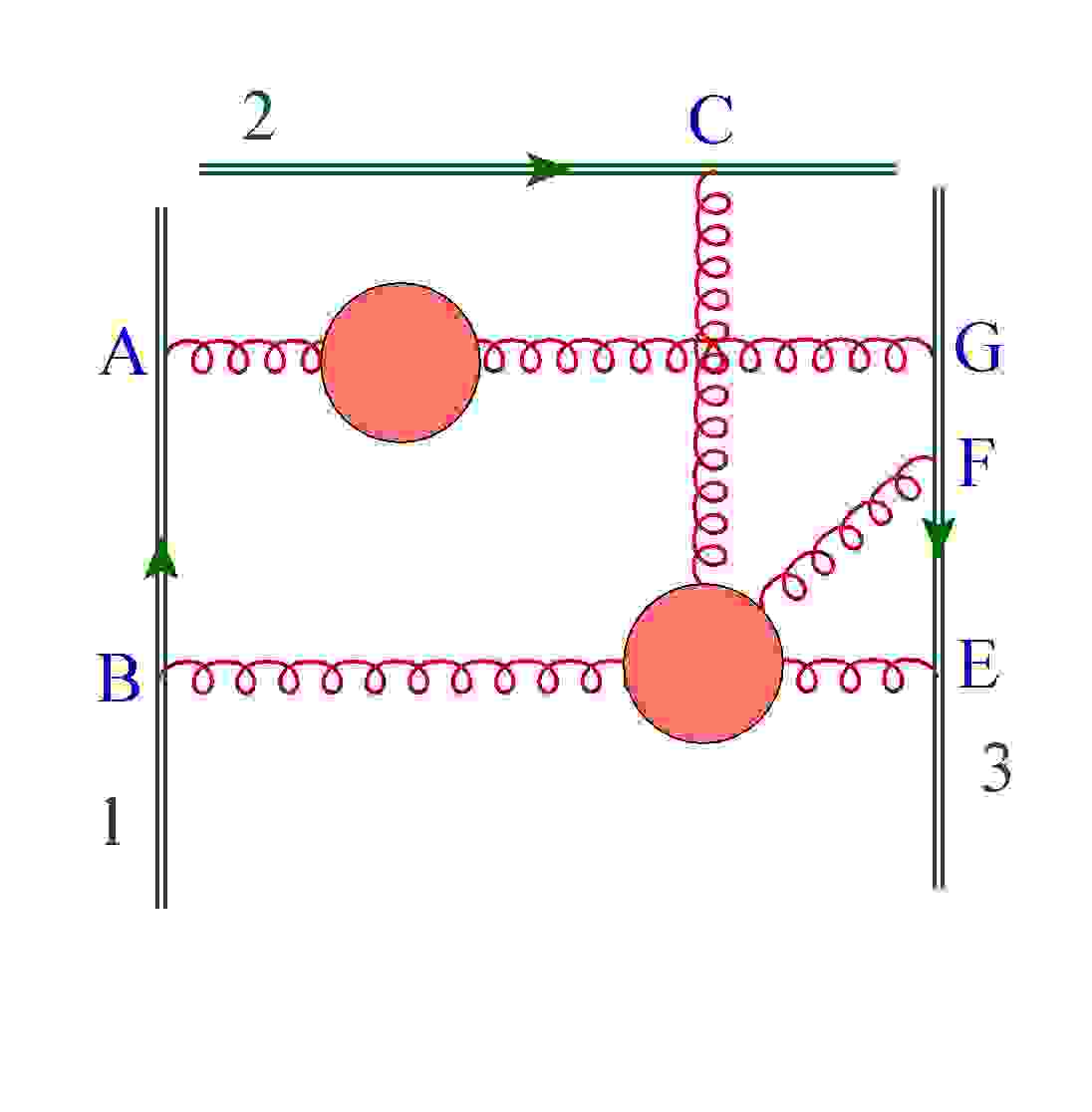}
  \caption{Skeleton diagram $C_1$ for the Cweb $W_{3,\text{I}}^{(1,0,1)}(1,2,3)$.}
  \label{fig:threeweb}
  \end{center}
\end{figure}

As a second example, we select a three-line Cweb at four loops, labelled as 
$W_{3, \rm{I}}^{(1,0,1)}(2,1,3)$, with a two-gluon correlator and a four-gluon 
correlator. Here the roman numeral I appears to distinguish this Cweb from a 
second one, shown in the Appendix, and denoted by $W_{3, \rm{II}}^{(1,0,1)}$, 
which shares the same number of attachments to the Wilson lines and the
same number of two-, three- and four-gluon correlators. A representative 
skeleton diagram is displayed in Fig.~\ref{fig:threeweb}.
\noindent This Cweb has 6 diagrams which are denoted by \\ 
\beq
  \lbrace C_1, C_2, C_3, C_4, C_5, C_6 \rbrace \, = \, 
  \Big\{ \lbrace\lbrace BA \rbrace, \lbrace GFE \rbrace \rbrace,
           \lbrace\lbrace BA \rbrace, \lbrace FGE \rbrace \rbrace,
           \lbrace\lbrace BA \rbrace, \lbrace FEG \rbrace \rbrace \, , \nn \\ 
           \lbrace \lbrace AB \rbrace, \lbrace GFE \rbrace \rbrace,
           \lbrace \lbrace AB \rbrace, \lbrace FGE \rbrace \rbrace,
           \lbrace \lbrace AB \rbrace, \lbrace FEG \rbrace \rbrace \Big\} \, . \nn
\label{tabCw2} 
\eeq
With this ordering, the skeleton diagram depicted in Fig.~\ref{fig:threeweb} is
diagram $C_1$. The weight factors $s(C_i)$ for the diagrams in this Cwebs 
are given by 
\beq
  \lbrace s(C_1), s(C_2), s(C_3), s(C_4), s(C_5), s(C_6) \rbrace \, = \,
  \lbrace 0,0,1,1,0,0,0,0 \rbrace \, . \nn
\label{svec2}
\eeq
We find the mixing matrix $R$, the diagonalising matrix $Y$, and the diagonal 
matrix $D$, as
\begin{align}
\begin{split}
  &
  R \, = \, \left(
  \begin{array}{cccccc}
  1 & 0 & -\frac{1}{2} & -\frac{1}{2} & 0 & 0 \\
  0 & 1 & -\frac{1}{2} & -\frac{1}{2} & 0 & 0 \\
  0 & 0 & \frac{1}{2} & -\frac{1}{2} & 0 & 0 \\
  0 & 0 & -\frac{1}{2} & \frac{1}{2} & 0 & 0 \\
  0 & 0 & -\frac{1}{2} & -\frac{1}{2} & 1 & 0 \\
  0 & 0 & -\frac{1}{2} & -\frac{1}{2} & 0 & 1 \\
  \end{array} 
  \right) \, ,  \qquad \quad
  Y \, = \, \left( 
  \begin{array}{cccccc}
  -1 & 0 & 0 & 0 & 0 & 1 \\
  -1 & 0 & 0 & 0 & 1 & 0 \\
  -1 & 0 & 0 & 1 & 0 & 0 \\
  -1 & 0 & 1 & 0 & 0 & 0 \\
  -1 & 1 & 0 & 0 & 0 & 0 \\
  0 & 0 & 1 & 1 & 0 & 0 \\
  \end{array}
  \right) \, , \qquad \quad
  D \, = \, \D{5} \, ,
\end{split}
\label{eq:web1app}
\end{align}
so that the rank in this case is $r_w = 5$. Once again, this mixing matrix satisfies 
all the desired properties: idempotence, row sum rule, and column sum rule. 
The exponentiated colour factors are
\beq
  (YC)_1 & = &  i \fafk \fbcg \fefg \tb 1 \ta 1 \tc 2 \te 3 \tkk 3  
  + i \faeh \fbcg \fefg \tb 1 \ta 1 \tc 2 \thh 3   \tf 3 \nonumber \\ \nonumber
  & & - i \fabm \fbcg \fefg \tm 1 \tc 2 \te 3 \tf 3 \ta 3 \, , \\ \nonumber \\
  (YC)_2 & = &i \fafk \fbcg \fefg \tb 1 \ta 1 \tc 2 \te 3 \tkk 3 
  - i \fabm \fbcg \fefg \tm 1 \tc 2 \te 3 \tf 3 \ta 3 \, , \nonumber \\ \nonumber \\
  (YC)_3 & = & -i \fabm \fbcg \fefg \tm 1 \tc 2 \te 3 \tf 3 \ta 3 \, , \\ \nonumber \\ 
  (YC)_4 & = & i \fafk \fbcg \fefg \ta 1 \tb 1 \tc 2 \te 3 \tkk 3  
  + i \faeh \fbcg \fefg \ta 1 \tb 1 \tc 2 \thh 3 \tf 3 \, , \nonumber \\ \nonumber \\
  (YC)_5 & = &  i \fafk \fbcg \fefg \ta 1 \tb 1 \tc 2 \te 3 \tkk 3 \, . \nonumber
\label{three-ecf}
\eeq
As expected, they all correspond to connected Feynman diagrams.

%%%%%%%%%%%%%%%%%%%%%%%%%%%%%%%%%%%%%%%%%

\section{Observations on mixing matrices and their direct construction}
\label{direct-comp}

The present work, together with the results presented in \cite{Agarwal:2020nyc},
as well as earlier work in Refs.~\cite{Gardi:2010rn,Gardi:2011wa,Gardi:2011yz,
Gardi:2013ita}, provides a considerable amount of empirical data about mixing 
matrices, partly summarised in Table~\ref{statistics} and in Table~\ref{ranks}.
In a massless theory, the total number of non-vanishing Cwebs having 
lowest-order contributions at ${\cal O} (g^8)$ or below is 79. There is just 
one Cweb at ${\cal O} (g^2)$, four at ${\cal O} (g^4)$, fourteen at ${\cal O} (g^6)$
and sixty at ${\cal O} (g^8)$. Of these Cwebs, thirteen are fully connected, so 
they consist of only one skeleton diagram and have a trivial mixing matrix, 
$R = 1$. Setting those aside, the dimensions of the remaining sixty-six
mixing matrices range between $d_w = 2$ and $d_w = 36$, while their ranks 
range between $r_w = 1$ and $r_w = 29$. %
\begin{table}[h]
	\begin{tabular}
	{|>{\centering}p{2cm}|>{\centering}p{2.5cm}
	|>{\centering}p{2.5cm}|>{\centering}p{2.5cm}|>{\centering}p{2.5cm}|c|}
        \hline 
	Dimension of the Mixing matrix & No. of Cwebs connecting 5 lines & 
	No. of Cwebs connecting 4 lines & No. of Cwebs connecting 3 lines & 
	No. of Cwebs connecting 2 lines & Total
	\tabularnewline
		\hline 
		$1$ & 1 & 1 & 2 & 2 & 6\tabularnewline
		\hline 
		$2$ & 2 & 2 & 2 & 0 & 6\tabularnewline
		\hline 
		$3$ & 0 & 2 & 2 & 0 & 4\tabularnewline
		\hline 
		$4$ & 2 & 3 & 2 & 0 & 7\tabularnewline
		\hline 
		$6$ & 1 & 4 & 4 & 1 & 10\tabularnewline
		\hline 
		$8$ & 1 & 2 & 1 & 1 & 5\tabularnewline
		\hline 
		$9$ & 0 & 0 & 0 & 2 & 2\tabularnewline
		\hline 
		$12$ & 1 & 3 & 2 & 0 & 6\tabularnewline
		\hline 
		$16$ & 0 & 1 & 0 & 0 & 1\tabularnewline
		\hline 
		$18$ & 0 & 1 & 2 & 0 & 3\tabularnewline
		\hline 
		$24$ & 1 & 2 & 4 & 1 & 8\tabularnewline
		\hline 
		$36$ & 0 & 0 & 1 & 1 & 2\tabularnewline
		\hline 
		Total & 9 & 21 & 22 &8 & 60 \\
		\hline
	\end{tabular}
	\caption{Distribution of four-loop Cwebs according to the dimension of their
	mixing matrices and the number of their Wilson lines. \label{statistics}}
\end{table}
To give a flavour of the distribution of mixing matrices, in Table~\ref{statistics}
we present the dimensions of mixing matrices appearing at four loops,
distributed according to the number of Wilson lines occurring in the corresponding 
Cwebs. In Table~\ref{ranks} we present the ranks of all mixing matrices appearing
up to four loops, for different matrix sizes. It is clear that the combinatorial problems 
associated with mixing matrices are non-trivial, and indeed interesting connections 
to abstract combinatorics have already been uncovered and exploited in 
Refs.~\cite{Gardi:2011wa,Dukes:2013gea,Dukes:2013wa,Dukes:2016ger}. 
In particular, while the dimensions of mixing matrices are easily computable 
from the diagrammatic structure of their Cwebs, their ranks are not in general
predictable with current knowledge, and Table~\ref{ranks} does not display a 
discernible pattern.

In what follows, we provide some simple results emerging from the empirical 
data up to four loops, which in some cases allow for all-order generalisations. 
In particular, low-dimensional mixing matrices are highly constrained and can 
be uniquely determined from their general properties up to dimension $d_w = 3$, 
and partially at $d_w = 4$.
\begin{table}[h]
	\begin{center}
		\begin{tabular}{|c|c|c|c|}
			\hline 
			Dimension of mixing matrix & Ranks at 2 loops & Ranks at 3 loops & Ranks at 4 loops\tabularnewline
			\hline 
			2 & 1 & 1 & 1\tabularnewline
			\hline 
			3 & - & 2 & 2\tabularnewline
			\hline 
			4 & - & 1,3 & 1,3\tabularnewline
			\hline 
			6 & - & 2,3,4 & 2,3,5\tabularnewline
			\hline 
			8 & - & 4 & 1,3,4,7\tabularnewline
			\hline 
			9 & - & - & 8\tabularnewline
			\hline 
			12 & - & - & 2,3,6,7,8\tabularnewline
			\hline 
			16 & - & - & 5\tabularnewline
			\hline 
			18 & - & - & 6,10,13\tabularnewline
			\hline 
			24 & - & - & 6,8,9,12,13,15,16,17\tabularnewline
			\hline 
			36 & - & - & 19,29\tabularnewline
			\hline 
		\end{tabular}\caption{Dimension and the corresponding ranks of mixing matrices at different
			perturbative orders\label{ranks}}
	\end{center}
\end{table}

A first natural question to ask is how many different mixing matrices can be 
generated with a given dimensionality $d_w$. This requires establishing when
two mixing matrices should be considered different: given their definition, 
we take the viewpoint that two matrices are in the same equivalence class
if they are connected by a permutation of their rows or of their columns: this
corresponds to permuting the diagrams in the web and the components of the 
vector of colour factors; row and column sum rules are preserved by these
permutations, provided one permutes the components of the $s$ vectors
appropriately. On the other hand, taking general linear combinations of rows 
or columns of mixing matrices does not have a diagrammatic interpretation
and we do not consider it. With this definition of equivalence, we note that
matrices of dimension $d_w = 2$ occur ten times up to four loops, but there
are only two different matrices, one of which occurs only once. This is
easily understood and is discussed in \secn{twodimma}. Matrices of dimension
$d_w = 3$ occur six times up to four loops, but actually these are six occurrences
of the same matrix: this is discussed in \secn{threedimma}. At $d_w = 4$,
for the first time we have matrices of different rank ($r_w = 1,3$), which 
occur both at three and at four loops: it turns out that the rank-three matrix
is uniquely determined, while in principle different $d_w = 4$ matrices of lower 
rank could appear at higher orders. For $d_w = \{6,8,9,12,16,18,24,36\}$,
the numbers of inequivalent mixing matrices arising up to four loops is 
$n_i = \{6,4,1,6,1,3,8,2\}$, and in several cases inequivalent matrices of 
the same rank are found. Given these data, we now examine in more detail
the low-dimensional cases.

%%%%%%%%%%%%%%%%%%%%%%

\subsection{On two-dimensional mixing matrices}
\label{twodimma}

\begin{figure}[H]
	\centering
		\subfloat[][$C_1$]{\includegraphics[height=3cm,width=3cm]{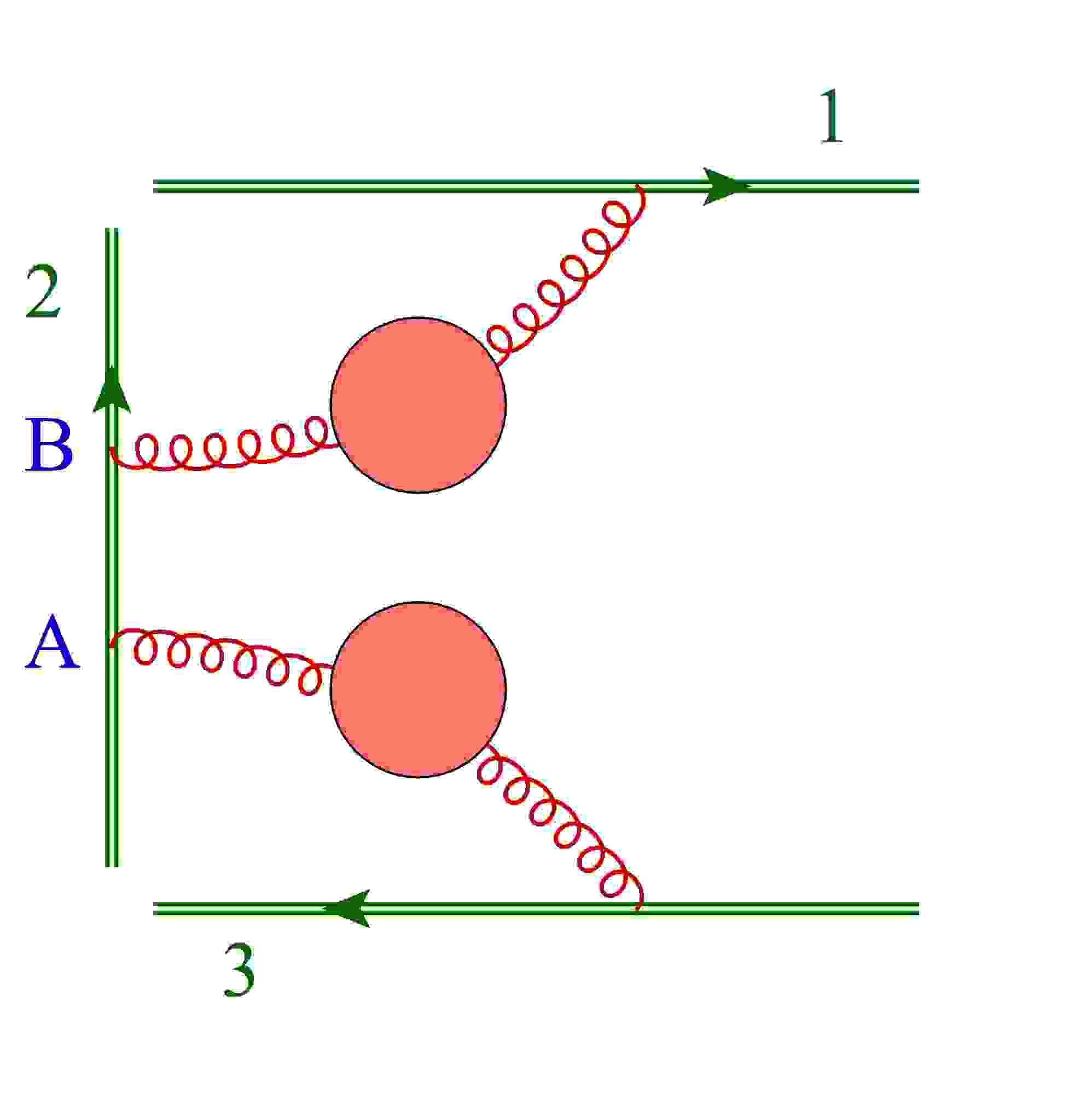} }
		\qquad
		\subfloat[][$C_2$]{\includegraphics[height=3cm,width=3cm]{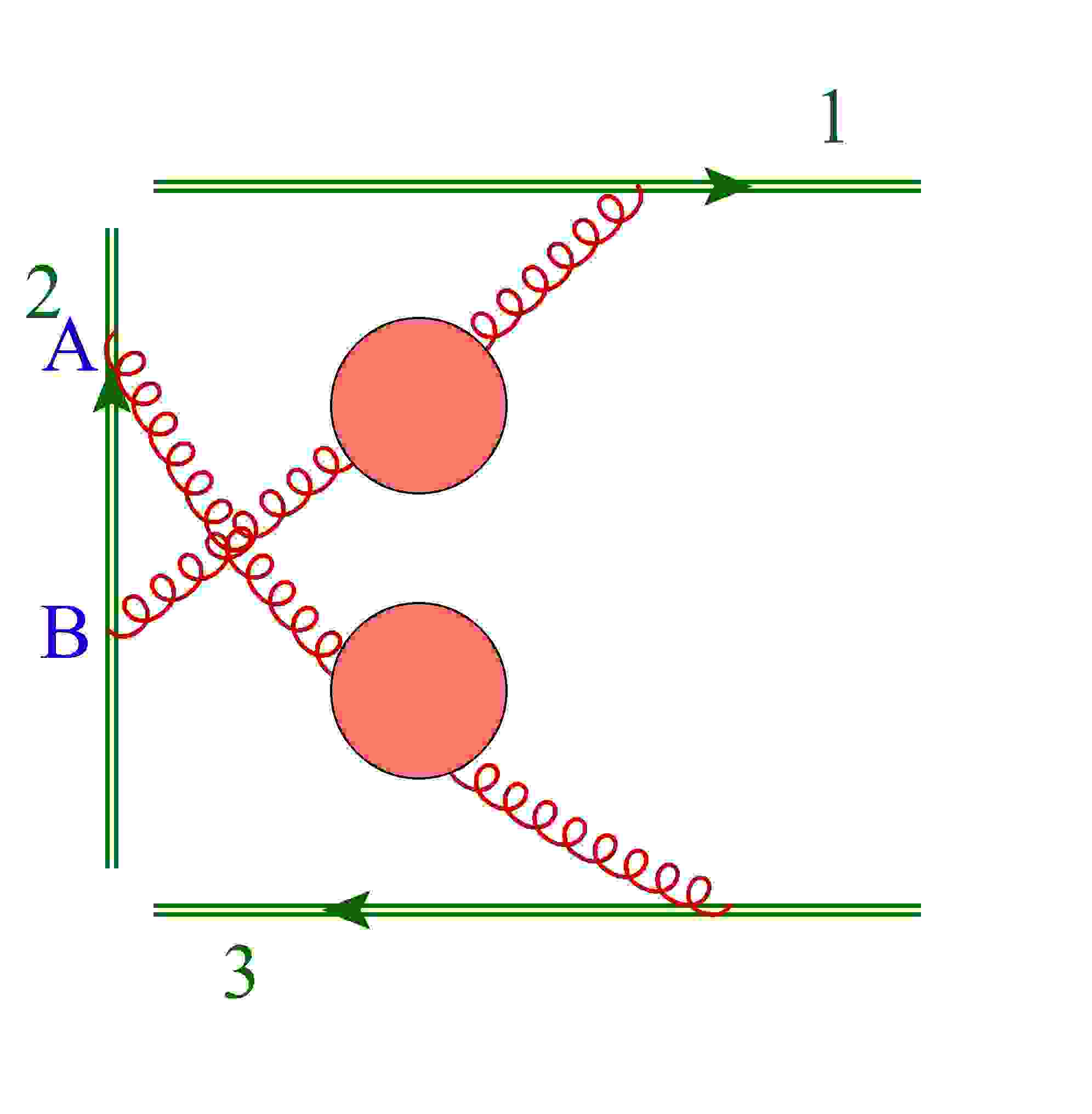} }
		\caption{The two skeleton diagrams of Cweb $W_3^{(2)}(1,2,1)$.}
		\label{2by2type1}
\end{figure}

All matrices of dimension two appearing at three and at four loops are the same 
and have the form 
\begin{align}
        R \, = \, \left(
        \begin{array}{cc}
        \frac{1}{2} & -\frac{1}{2} \\
        -\frac{1}{2} & \frac{1}{2} \\
        \end{array}
        \right) \, .
\label{2by2}
\end{align}
At two loops, however, the matrix in \eq{2by2} appears once, but, in addition, 
one more two dimensional matrix appears, in the case of Cweb $W_2^{(2)}(2,2)$.
It is
\begin{align}
  R \, = \, \left(
  \begin{array}{cc}
       1 & -1 \\
       0 &  0  \\
   \end{array}
              \right) \, .
\label{2by2type2}
\end{align}
It is easy to show that indeed these are the only two possible two-dimensional
mixing matrices at any perturbative order, and furthermore \eq{2by2type2} occurs 
only once, at two loops. With $n \geq 3$ Wilson lines, a two-dimensional mixing 
matrix requires a Cweb with precisely two gluon attachments from distinct 
correlators to a specific Wilson line, while all other Wilson lines must allow 
no shuffles, {\it i.e.} they must each be attached to gluons from a single 
correlator. The $s$ vector for such a configuration is always $s = \{1, 1\}$,
as seen from the example in Fig.~\ref{2by2type1}, since in each contributing 
diagram one of the two correlators involved can always be shrunk to the 
origin without affecting the other.
\noindent Armed with the knowledge of $s$, we can impose the row and column sum 
rules on a generic $2 \times 2$ matrix, obtaining
\begin{align}
  \begin{split}
  R \, = \, \displaystyle{\left(
	\begin{array}{cc}
	a & -a \\
	-a & a \\
	\end{array}
	\right) } \, .
  \label{2by2ex}
  \end{split}
\end{align}
The normalisation can be fixed by observing that the matrix has rank $r = 1$, 
and, since it is a projection operator, the trace of the matrix must equal the rank.
This leads uniquely to \eq{2by2}.

\begin{figure}[H]
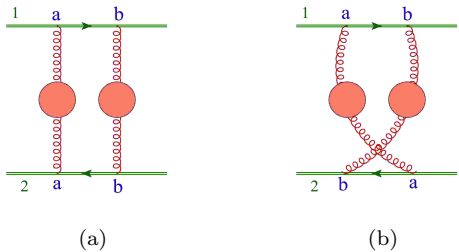

	\centering
	\subfloat[]{\includegraphics[height=3cm,width=3cm]{Cwebg4-1.jpg} }
	\qquad
	\subfloat[]{\includegraphics[height=3cm,width=3cm]{Cwebg4-2.jpg} }
	\caption{The two inequivalent skeleton diagrams of Cweb 
	$W_2^{(2)}(2,2)$.}
	\label{2by2type4}
\end{figure}

The only exception to the above reasoning occurs at two loops, when there
are only two Wilson lines, and two dressed gluon propagators attach to
both lines, as depicted in Fig.~\ref{2by2type4}. The dimension of the mixing 
matrix is reduced from 4 to 2 by the symmetry of the correlators, but in this 
case the $s$ vector is given by $s = \{0, 1\}$, since for diagram (a) there
are no possibilities to shrink sequentially the two correlators.
\noindent Once again, imposing the row and column sum rules, with the new $s$ vector,
and normalising the trace of the matrix to its rank, we find uniquely \eq{2by2type2}.
We conclude that, at any perturbative order, all two-dimensional mixing matrices 
will have the form of \eq{2by2}, with the sole exception \eq{2by2type2} in the
two-loop, two-line case we just examined.    

%%%%%%%%%%%%%%%%%%%%%%

\subsection{The three-dimensional mixing matrix}
\label{threedimma}

\begin{figure}[H]
	\centering
	\subfloat[]{\includegraphics[height=3cm,width=3cm]{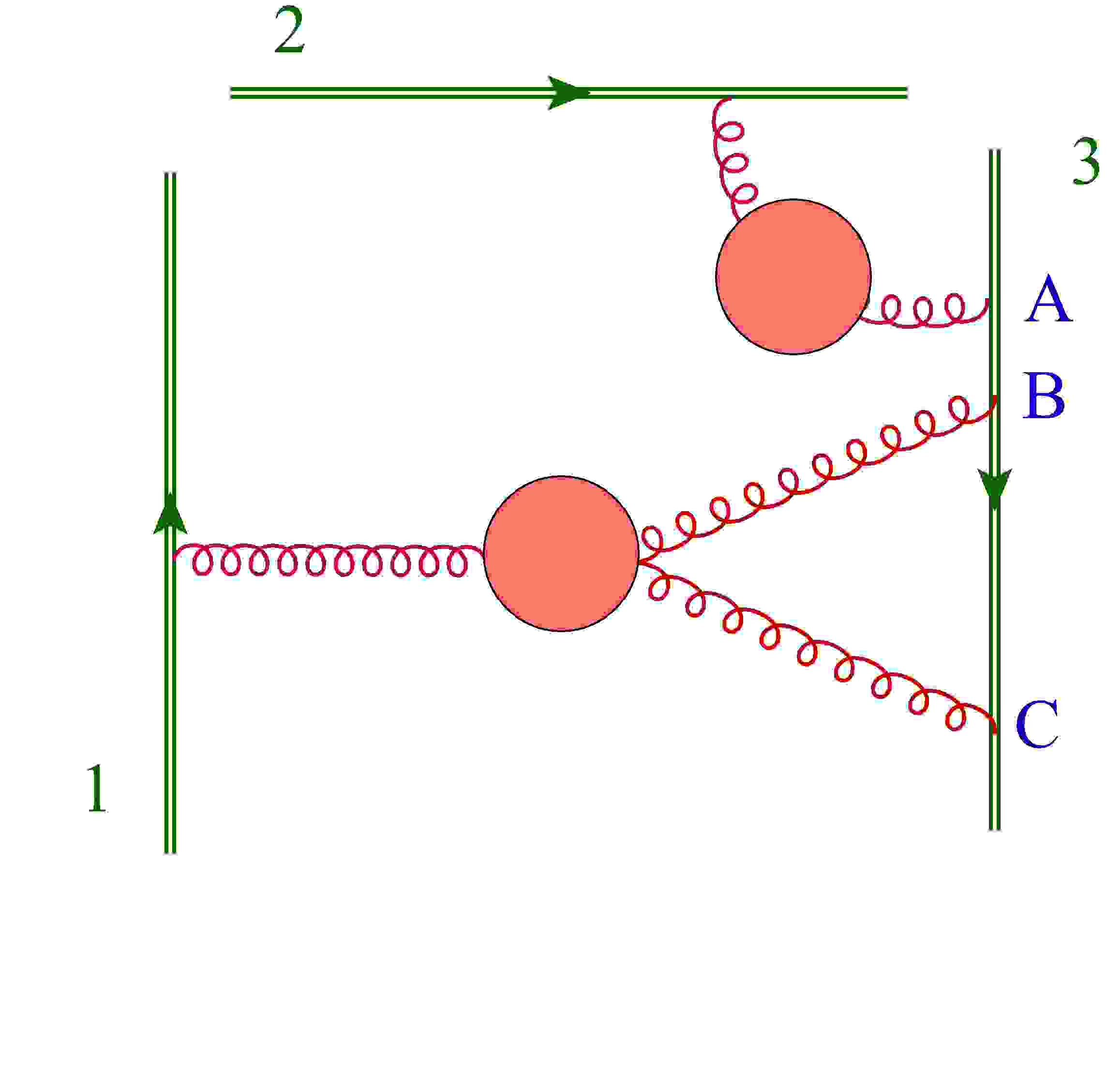} }
	\qquad
	\subfloat[]{\includegraphics[height=3cm,width=3cm]{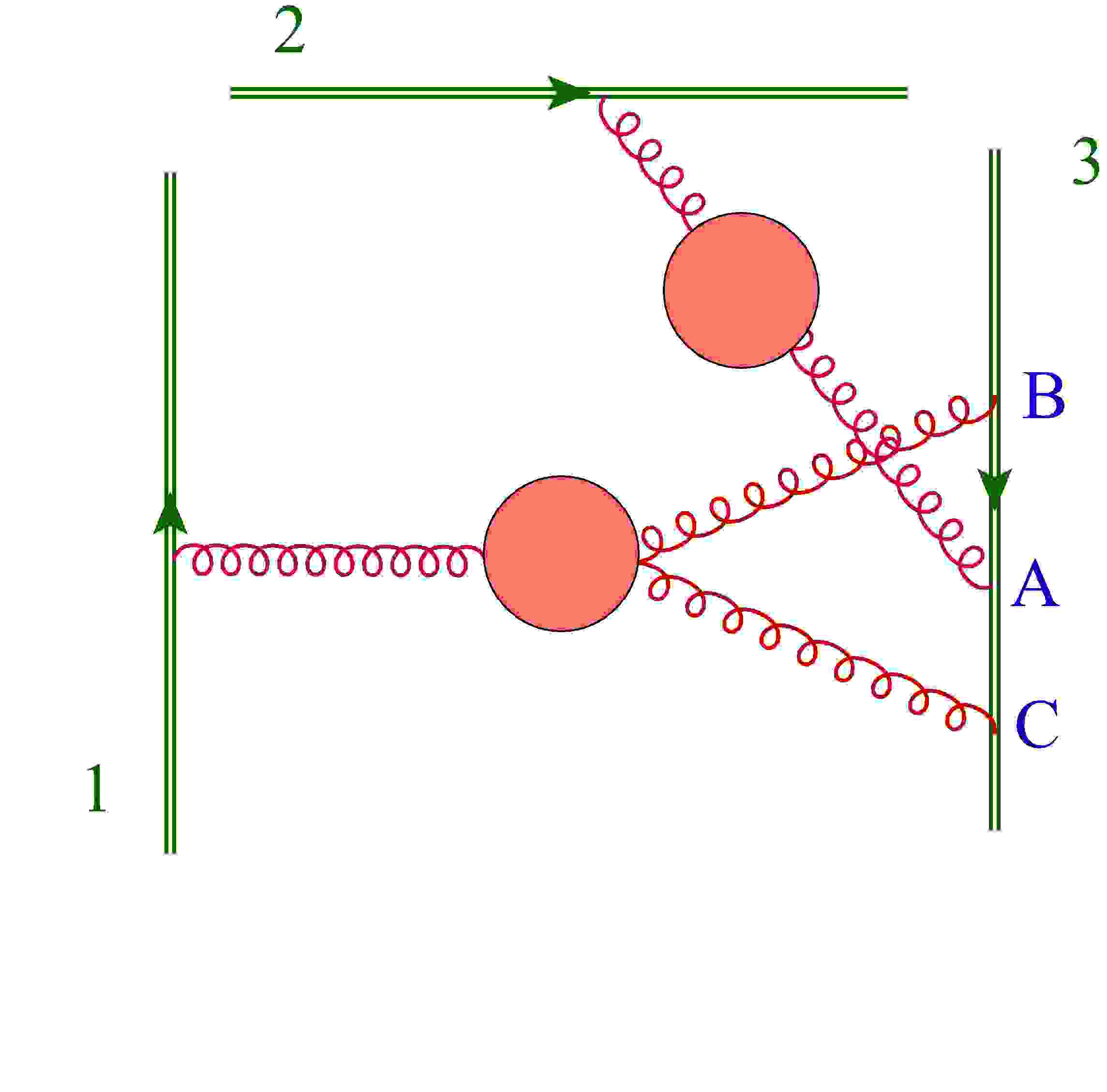} }
	\qquad
	\subfloat[]{\includegraphics[height=3cm,width=3cm]{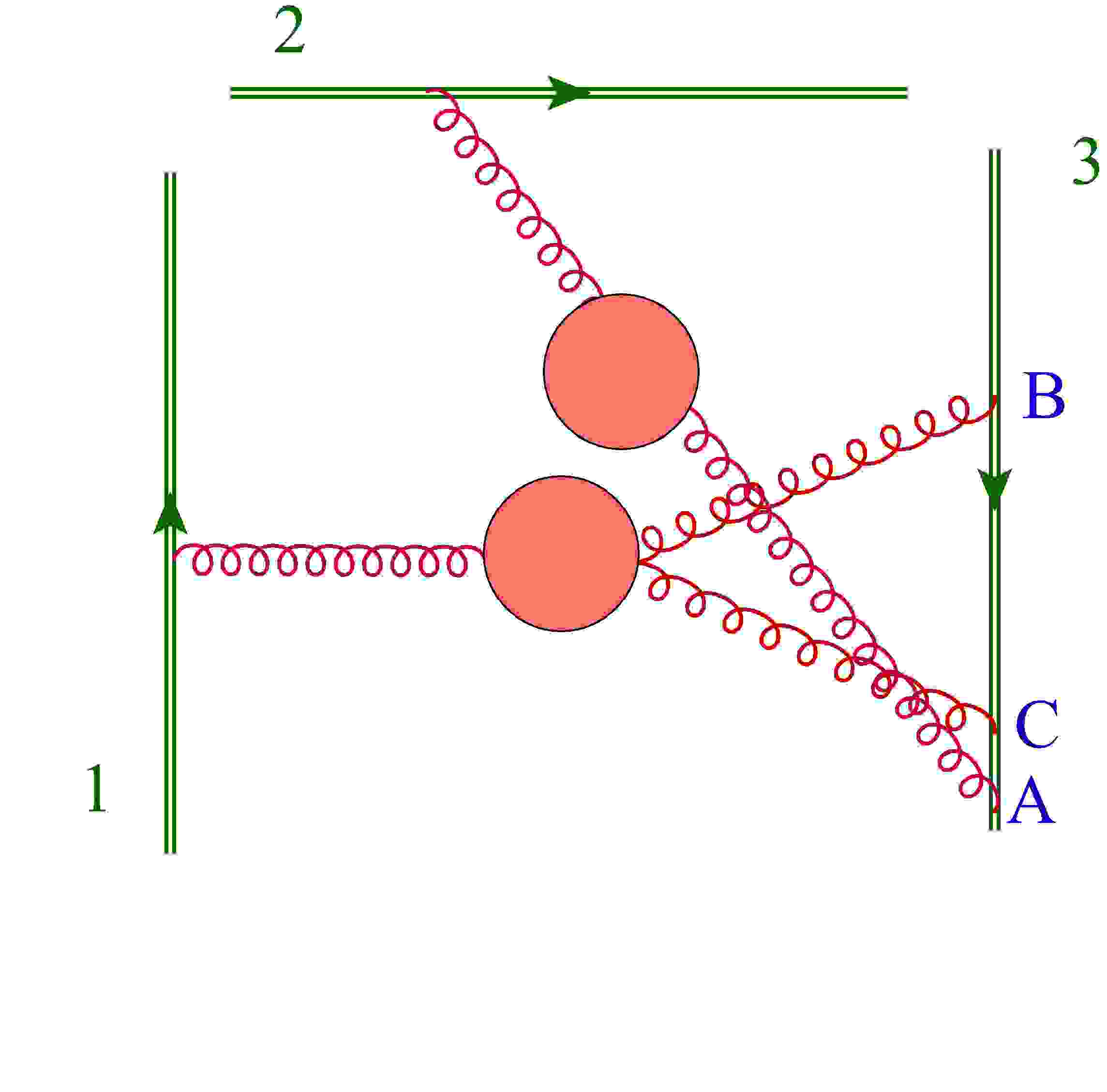} }
	\caption{The three skeleton diagrams contributing to $W_3^{(1,1)}(1,3,1)$.}
\label{3by3fig}
\end{figure}

There are a total of six three-dimensional mixing matrices up to ${\cal O}(g^8)$:
two arising at three loops and four at four loops. It turns out that these are six
occurrences of the same matrix, which is
\begin{align}
\begin{split}
  &
  R \, = \, \left(
  \begin{array}{ccc}
  \frac{1}{2} & 0 & -\frac{1}{2} \\
  - \frac{1}{2} & 1 & -\frac{1}{2} \\
  - \frac{1}{2} & 0 & \frac{1}{2} \\
  \end{array}
  \right) \, .
\label{3by3}
\end{split}
\end{align}
It is not difficult to prove that this is a general result, and \eq{3by3} is the only
possible three-dimensional mixing matrix at any order. To see it, note that, 
since $d_w = 3$ is a prime number, it can only arise for Cwebs where three 
shuffles are possible on a single Wilson line, and no shuffles are available 
on any other Wilson line. The only configuration of gluon attachments on a 
Wilson line which leads to three shuffles involves two gluon correlators, one 
attaching to the Wilson line via a single gluon, while the second one attaches 
with two gluons. An example is the Cweb $W_3^{(1,1)}(1,1,3)$ at three loops,
shown in Fig.~\ref{3by3fig}. 
\noindent The column weight vector for this Cweb is $s = \{1,0,1\}$. Importantly, 
we notice that diagram $(b)$, which has $s=0$, cannot be generated from diagrams 
$(a)$ and $(c)$, which have $s=1$, by the action of the replica ordering operator:  
under that action, all attachments belonging to a single gluon correlator (and 
thus to a single replica) move together. In other words, diagram $(b)$ is a singlet
under the mixing operation, whereas diagrams $(a)$ and $(c)$ form a doublet. 
Taking this into account, and applying the row and column sum rules, one sees 
that the mixing matrix must be of the form
\begin{align}
\begin{split}
  &
  R \, = \, \displaystyle{\left(
  \begin{array}{ccc}
  a & 0 & -a \\
  b & 1  & - 1 - b \\
  - a & 0 & a  \\
  \end{array}
  \right) \, . }\
\end{split}
\end{align}
Finally, using the idempotence property $R^2=R$ and imposing that the trace 
of the matrix must equal its rank (which is $r_w = 2$ in this case) one can fix the
two remaining parameters and obtain precisely \eq{3by3}. We conclude that
all three-dimensional mixing matrices at any perturbative order will be given by
\eq{3by3}.  

%%%%%%%%%%%%%%%%%%%%%%

\subsection{Results for higher-dimensional mixing matrices}
\label{highdimma}

The reasoning of the previous two sections generalises in two different directions.
On the one hand, one may consider mixing matrices whose dimension is a 
prime number, $d_w = p$. As in the case $p=3$, such matrices can only arise
from Cwebs where one Wilson line has a single attachment from a gluon correlator
$G_1$, together with $p-1$ attachments from a second gluon correlator $G_2$, 
with no other Wilson lines providing further shuffles. The column weight vector 
for such a web is again of the form $s = \{1, 0, \ldots, 0, 1\}$, with $p-2$ zeroes. 
Furthermore, once again, the $p-2$ diagrams where the $G_1$ gluon is inserted
between attachments of $G_2$ gluons cannot be reached by the action of the
replica ordering operator, and are singlets under the mixing operation, while
the two diagrams where the $G_1$ gluon is on one or the other side of the 
$G_2$ cluster form a doublet. This information, together with the row sum rule, 
is sufficient to conclude that the rank of the mixing matrix is $p-1$. Imposing
idempotency, and normalising by setting the trace of the matrix equal to the rank,
one finds that mixing matrices of prime dimension $p$ are unique, and they can 
be written as
\begin{align}
\begin{split}
  &
  R \, = \, \displaystyle{\left(
  \begin{array}{ccc}
  \frac{1}{2} & 0 \,\, 0 \, \ldots \, 0 & -\frac{1}{2} \\
  - \frac{1}{2} & \!1 \,\, 0  \ldots 0 & -\frac{1}{2} \\
  &\,\,\, \ldots & \\
  - \frac{1}{2} & 0 \,\, 0 \, \ldots \, 1 & - \frac{1}{2} \\
  - \frac{1}{2} & 0 \,\, 0 \, \ldots \, 0 & \frac{1}{2}
  \end{array}
  \right) \, . }\
\label{dimprime}  
\end{split}
\end{align}
This reasoning explains the fact that the largest prime number appearing as 
the dimension of a mixing matrix at four loops is $p=3$. The first mixing matrix
of dimension $d_w = p = 5$ will appear at ${\cal O}(g^{10})$, arising when four 
gluons from a five-point gluon correlator are shuffled with a single gluon from
a two-point correlator attaching on the same Wilson line.
 
Conversely, we can consider any Cweb where a single gluon from a correlator
$G_1$ is shuffled on a Wilson line with a set of $n-1$ gluons from a second 
correlator $G_2$, and no other shuffles are available on other lines, even 
when $n$ is not a prime number. The reasoning above still holds, so that the 
mixing matrix will have dimension $n$ and rank $n-1$, and it will be of the
form of \eq{dimprime}. In this case, however, we cannot argue that the matrix
is unique, since other sets of correlators and attachments will likely lead to
other matrices with the same dimension. A case in point is $d_w = 4$, where
we find two distinct mixing matrices, occurring in two three-loop Cwebs
and in seven four-loop Cwebs. One of these two matrices, as expected, 
has rank $r = 3$ and is of the form of \eq{dimprime}: it arises, among others,
from the Cweb $W_3^{1,0,1}(1,1,4)$, depicted in Fig.~\ref{Cwebsg8_3}(p), 
which is of the form just described. The second type of $d_w = 4$ mixing 
matrix emerging from the algorithm has rank $r_w = 1$. It arises for 
example in the case of Cweb $W_{4, \, {\rm I}} ^{(2,1)} (1,2,2,2)$, 
discussed in Ref.~\cite{Agarwal:2020nyc}, and it has a column weight 
vector given by $s = \{1,2,2,1\}$. In this case, there are no `singlet' 
skeleton diagrams in the set, so the available constraints are not 
sufficient to determine this second matrix uniquely. Clearly, we are 
barely scratching the surface of this combinatorial problem, and further
dedicated studies are likely to bring about a much deeper understanding.

%%%%%%%%%%%%%%%%%%%%%%%%%%%%%%%%%%%%%%%%%

\section{Summary and Outlook}
\label{Conclu}

The study of diagrammatic exponentiation of Wilson-line correlators has a 
long history, and has provided many important insights concerning the 
infrared structure of perturbative gauge amplitudes. In this paper, we developed
the idea of correlator webs, or Cwebs, introduced in Ref.~\cite{Agarwal:2020nyc},
and we completed the study of all Cweb mixing matrices appearing up to 
four loops. Specifically, we have listed all four-loop Cwebs connecting 
two and three Wilson lines, following the recursive algorithm developed 
in~\cite{Agarwal:2020nyc}. Using an improved version of our in-house 
Mathematica code, we have computed the mixing matrices for these Cwebs 
using the replica method~\cite{Gardi:2010rn,Laenen:2008gt}. In all cases,
the mixing matrices we have found verified both the proven and the
conjectured properties listed in the literature~\cite{Gardi:2010rn,Gardi:2011wa,
Gardi:2011yz,Dukes:2013wa,Gardi:2013ita,Dukes:2013gea}, in
particular the general form of the non-abelian exponentiation theorem:
all the exponentiated colour factors correspond to fully connected gluon
sub-diagrams~\cite{Gardi:2013ita}. 

As shown in the examples reported in \secn{four-two} and in \secn{four-three}, 
once the mixing matrix is determined, exponentiated colour factors are readily 
computed by multiplying the diagonalising matrix $Y_w$ for the given Cweb 
times the column vector formed out of the colour factors of individual skeleton 
diagrams. One must then use the commutation relations of the gauge algebra 
to simplify products of colour generators attached to each Wilson line. All mixing
matrices for four-loop Cwebs connecting two and three Wilson lines are presented
in the Appendix. When the number of Wilson lines is small, as in this case, at high 
orders many attachments are present on each Wilson line, and the commutation
relations must be applied repeatedly, generating a large number of ECFs, 
many of which have lengthy expressions. As a consequence, we refrain from 
presenting explictly the ECFs in the Appendix: given the mixing matrices, the 
steps required to derive the ECFs are straightforward. In any case, ECFs can 
be obtained from the authors upon request in the form of Mathematica code. 

Given the considerable accumulated data about mixing matrices, from the
knowledge of the 66 non-trivial Cwebs arising up to four loops, we have taken
the opportunity to examine their emerging properties. We noted that, while
the dimension $d_w$ of mixing matrices is directly computable, given
the set of gluon correlators and attachments in the selected Cweb, the 
rank $r_w$ of the resulting matrices is not easy to predict in general.
On the other hand, we have observed that low-dimensional mixing 
matrices are highly constrained, and we have derived some general 
results. We have shown that two- and three-dimensional mixing matrices
are uniquely determined to all orders in perturbation theory by their
general properties, given the limited possibilities available for their
column weight vectors. We have also shown that mixing matrices whose
dimension is a prime number are unique, and we have given their form; 
furthermore, we have uniquely determined the mixing matrices for an 
infinite series of simple Cwebs, consistently with the low-order examples
that we have explicitly computed.

Together with the results presented in \cite{Agarwal:2020nyc}, the results 
for three- and two-line Cwebs at four loops presented here complete the 
computation of the mixing matrices for all four-loop Cwebs. A direct calculation 
of mixing matrices using posets (partial ordered sets) was also explored 
in~\cite{Dukes:2013gea,Dukes:2013wa,Dukes:2016ger} for a class of 
webs: our results provide additional useful data for the direct construction 
of the mixing matrices, and for the study of their interesting algebraic and 
combinatorial properties.

We emphasize that exponentiated colour factors are not all independent, 
and it is necessary to reduce them to a basis, implementing the constraint
of colour conservation, which implies that, when acting on a physical amplitude,
the colour-insertion operators must satisfy $\sum_i \textbf{T}_i=0$, where the
sum runs over all hard particles, represented here by the Wilson lines.
The consequences of colour conservation were implemented at three
loops in~\cite{Almelid:2015jia}, and studied at four loops in~\cite{Ahrens:2012qz,
Becher:2019avh}. Together with Bose symmetry, they allow to reduce 
the form of the soft anomalous dimension matrix to a relatively simple
parametrisation in terms of a few scalar functions. Of particular interest is the 
appearance, at four loops, of contributions proportional to quartic Casimir
operators of the gauge algebra, which are fully known for the simple case
of two Wilson lines~\cite{Moch:2017uml,Moch:2018wjh,Henn:2019swt,
vonManteuffel:2020vjv}, but, as yet, undetermined in the general case, 
although partial results have begun to emerge~\cite{Catani:2019nqv,
Falcioni:2020lvv}. Our results provide all the necessary colour ingredients 
for the complete four-loop calculation, and other tools are available for
the study of colour structures at high orders, such as the effective vertex 
analysis of Ref.~\cite{Gardi:2013ita}, and the generating functional approach 
of Ref.~\cite{Vladimirov:2017ksc}. That being said, of course by far the most 
difficult challenge remains the calculation of the kinematic contributions, in 
particular for four-loop connected diagrams. The interplay of colour and 
kinematics which emerges with striking power and simplicity in the soft 
anomalous dimension matrix remains one of the most challenging and 
interesting topics in the study of perturbative non-abelian gauge theories.

%%%%%%%%%%%%%%%%%%%%%%%%%%%%%%%%%%%%%%%%%

\section*{Acknowledgments}

\noindent AT and LM and SP would like to thank MHRD Govt. of India for 
the GIAN grant (171008M01), ``The Infrared Structure of Perturbative Gauge 
Theories'' and for the SPARC grant (P578) ``Perturbative QCD for Precision 
Physics at the LHC'', which were crucial to the completion of the present 
research. SP would also like to thank MHRD Govt. of India for an SRF 
fellowship, and the University of Turin and INFN Turin for warm hospitality 
during the course of this work. 

%%%%%%%%%%%%%%%%%%%%%%%%%%%%%%%%%%%%%%%%%
\pagebreak
\appendix

%%%%%%%%%%%%%%%%%%%%%%%%%%%%%%%%%%%%%%%%%

\section{All four-loop mixing matrices connecting two and three Wilson lines}
\label{App}

In this appendix we give results for all the Cwebs that appear at 4 loops in the 
scattering amplitude, that can connect two or three Wilson lines. Throughout 
the list,  $R$ and $D$ denote the mixing matrix and the diagonalized matrix 
respectively. Clearly, $D$ is just a representation of the rank of $R$, and 
we write it as $D = \D{r}$, where $r$ is the rank. We display only one skeleton 
diagram per web, and we explicitly give the order of the shuffles that generate 
the other diagrams, which is tied to the order the columns of the mixing matrix 
in the chosen basis. The tables of shuffles also give the components of 
the column weight vector for each Cweb.
\begin{table}[h]
	\begin{center}
		\begin{tabular}{|c|l|c|c|c|}
			\hline 
			Sl. No. & Name & No. of diagrams & No. of hierarchies & Rank of R\tabularnewline
			\hline 
			1 & $\ensuremath{{W}_{3,{\rm {I}}}^{(1,0,1)}(1,2,3)}$ & 6 & 3 & 5\tabularnewline
			\hline 
			2 & $\ensuremath{{W}_{3,{\rm {II}}}^{(1,0,1)}(1,2,3)}$ & 2 & 3 & 1\tabularnewline
			\hline 
			3 & ${W}_{3,{\rm {I}}}^{(0,2)}(1,2,3)$ & 3 & 3 & 2\tabularnewline
			\hline 
			4 & ${W}_{3,{\rm {II}}}^{(0,2)}(1,2,3)$ & 6 & 3 & 5\tabularnewline
			\hline 
			5 & ${W}_{3,\text{I}}^{(2,1)}(2,2,3)$ & 24 & 13 & 16\tabularnewline
			\hline 
			6 & ${W}_{3,\text{II}}^{(2,1)}(2,2,3)$ & 6 & 13 & 3\tabularnewline
			\hline 
			7 & ${W}_{3,\text{III}}^{(2,1)}(2,2,3)$ & 12 & 13 & 7\tabularnewline
			\hline 
			8 & ${W}_{3}^{(1,0,1)}(1,2,3)$ & 3 & 3 & 2\tabularnewline
			\hline 
			9 & ${W}_{3,\text{I}}^{(0,2)}(2,2,2)$ & 2 & 3 & 1\tabularnewline
			\hline 
			10 & ${W}_{3}^{(1,0,1)}(2,2,2)$ & 4 & 3 & 3\tabularnewline
			\hline 
			11 & ${W}_{3,\text{II}}^{(0,2)}(2,2,2)$ & 8 & 3 & 7\tabularnewline
			\hline 
			12 & ${W}_{3,\text{I}}^{(2,1)}(1,3,3)$ & 18 & 13 & 10\tabularnewline
			\hline 
			13 & ${W}_{3,\text{II}}^{(2,1)}(1,3,3)$ & 18 & 13 & 13\tabularnewline
			\hline 
			14 & ${W}_{3}^{(1,0,1)}(1,1,4)$ & 4 & 3 & 3\tabularnewline
			\hline 
			15 & ${W}_{3}^{(0,2)}(1,1,4)$ & 6 & 3 & 5\tabularnewline
			\hline 
			16 & ${W}_{3}^{(2,1)}(1,2,4)$ & 24 & 13 & 15\tabularnewline
			\hline 
			17 & ${W}_{3}^{(0,2)}(1,2,4)$ & 12 & 3 & 8\tabularnewline
			\hline 
			18 & ${W}_{3}^{(4)}(2,3,3)$ & 36 & 75 & 19\tabularnewline
			\hline 
			19 & ${W}_{3}^{(4)}(2,2,4)$ & 24 & 75 & 12\tabularnewline
			\hline 
			20 & ${W}_{3}^{(4)}(1,3,4)$ & 24 & 75 & 13\tabularnewline
			\hline 
		\end{tabular}
		\caption{Summary of the results for four-loop Cwebs connecting three Wilson
			lines, \label{table-threelines}}
	\end{center}
\end{table}
\begin{table}[h]
	\begin{center}
		\begin{tabular}{|c|c|c|c|c|}
			\hline 
			Sl. No. & Name & No. of diagrams & No. of hierarchies & Rank of R\tabularnewline
			\hline 
			1 & ${W}_{2}^{(1,0,1)}(2,4)$ & 8 & 3 & 7\tabularnewline
			\hline 
			2 & ${W}_{2}^{(1,0,1)}(3,3)$ & 9 & 3 & 8\tabularnewline
			\hline 
			3 & ${W}_{2}^{(0,2)}(2,4)$ & 6 & 3 & 5\tabularnewline
			\hline 
			4 & ${W}_{2}^{(0,2)}(3,3)$ & 9 & 3 & 8\tabularnewline
			\hline 
			5 & ${W}_{2}^{(2,1)}(3,4)$ & 36 & 13 & 29\tabularnewline
			\hline 
			6 & ${W}_{2}^{(4)}(4,4)$ & 24 & 75 & 17\tabularnewline
			\hline 
		\end{tabular}
		\caption{Summary of the results for four-loop Cwebs connecting two Wilson
			lines, \label{table-twolegs}}
		
	\end{center}
\end{table}
In Tables~\ref{table-threelines} and~\ref{table-twolegs}, we present the list
of Cwebs, the total number of skeleton diagrams for each Cweb, the number 
of replica hierarchies generated in the application of the replica method, and 
the rank of the resulting mixing matrices.  We omit from the list the Cwebs 
that are composed of a single skeleton diagram, whose mixing matrix is just 
a number, $R = 1$.

\vspace{2mm}

%%%%%%%%%%%%%%%%%%

\subsection{Cwebs connecting three Wilson lines}

\vspace{2mm}

%%%%%%%%%%%%%%%%%%

\begin{itemize}
	
	%%%%%%%%%%%%%%%%%%
	
	%%%%%%%%%%%%%%%%%%%%%%%%%%%%%%%%%%%
	%end-of-one-web%
	%%%%%%%%%%%%%%%%%%%%%%%%%%%%%%%%%%%

	\item[{\bf 1}.]  $\textbf{W}_{3, \rm{I}}^{(1,0,1)}(1,2,3)$	\\
	This Cweb has six diagrams, one of which is displayed below. The table gives
	the chosen order of the six shuffles of the gluon attachments, and the corresponding
	$s$ factors.
	\begin{minipage}{0.5\textwidth}
		\begin{figure}[H]
			\vspace{-2mm}
			\includegraphics[height=4cm,width=4cm]{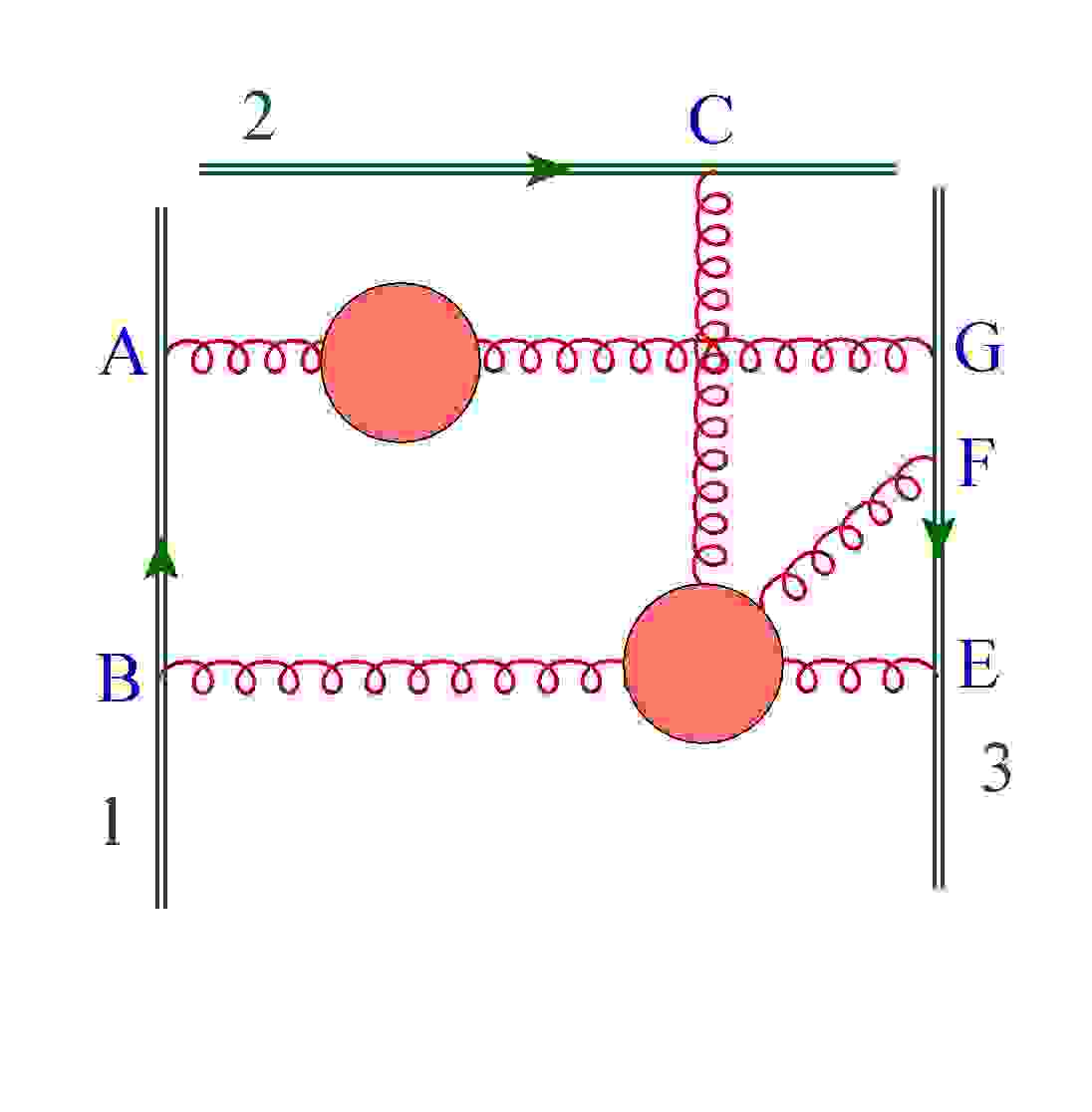}
		\end{figure}
	\end{minipage}
	\hspace{-2cm}
	\begin{minipage}{0.46\textwidth}
		\vspace{2cm}
		\begin{tabular}{ | c | c | c |}
			\hline
			\textbf{Diagrams} & \textbf{Sequences} & \textbf{s-factors} \\ \hline
			$C_1$ & $\lbrace\lbrace BA\rbrace,\lbrace GFE \rbrace \rbrace$ & 0 \\ \hline
			$C_2$ & $\lbrace\lbrace BA\rbrace,\lbrace FGE\rbrace \rbrace$ & 0 \\ \hline
			$C_3$ & $\lbrace\lbrace BA\rbrace,\lbrace FEG \rbrace \rbrace$ & 1 \\ \hline
			$C_4$ & $\lbrace\lbrace AB\rbrace,\lbrace GFE \rbrace \rbrace$ & 1 \\ \hline
			$C_5$ & $\lbrace\lbrace AB\rbrace,\lbrace FGE \rbrace \rbrace$ & 0 \\ \hline
			$C_6$ & $\lbrace\lbrace AB\rbrace,\lbrace FEG \rbrace \rbrace$ & 0 \\ \hline
		\end{tabular}
		\label{tab:abcd1}	
	\end{minipage}
	
	\noindent The $R$ matrix is given by
	\begin{align}
	\begin{split}
	&
	R=\left(
	\begin{array}{cccccc}
	1 & 0 & -\frac{1}{2} & -\frac{1}{2} & 0 & 0 \\
	0 & 1 & -\frac{1}{2} & -\frac{1}{2} & 0 & 0 \\
	0 & 0 & \frac{1}{2} & -\frac{1}{2} & 0 & 0 \\
	0 & 0 & -\frac{1}{2} & \frac{1}{2} & 0 & 0 \\
	0 & 0 & -\frac{1}{2} & -\frac{1}{2} & 1 & 0 \\
	0 & 0 & -\frac{1}{2} & -\frac{1}{2} & 0 & 1 \\
	\end{array}
	\right)\,, \qquad \qquad
	D = \D{5} \, .
	\end{split}
	\label{eq:web1app}
	\end{align}
	\item[{\bf 2}.]  $\textbf{W}_{3,\rm{II}}^{(1,0,1)}(1,2,3)$
	\\
	This Cweb is a second kind of Cweb with notation $\textbf{W}_{3}^{(1,0,1)}(1,2,3)$ and has two diagrams, one of which is displayed below. The table gives
	the chosen order of the two shuffles of the gluon attachments, and the corresponding
	$s$ factors. \\
	\begin{minipage}{0.5\textwidth}
		\begin{figure}[H]
			\vspace{-2mm}
			\includegraphics[height=4cm,width=4cm]{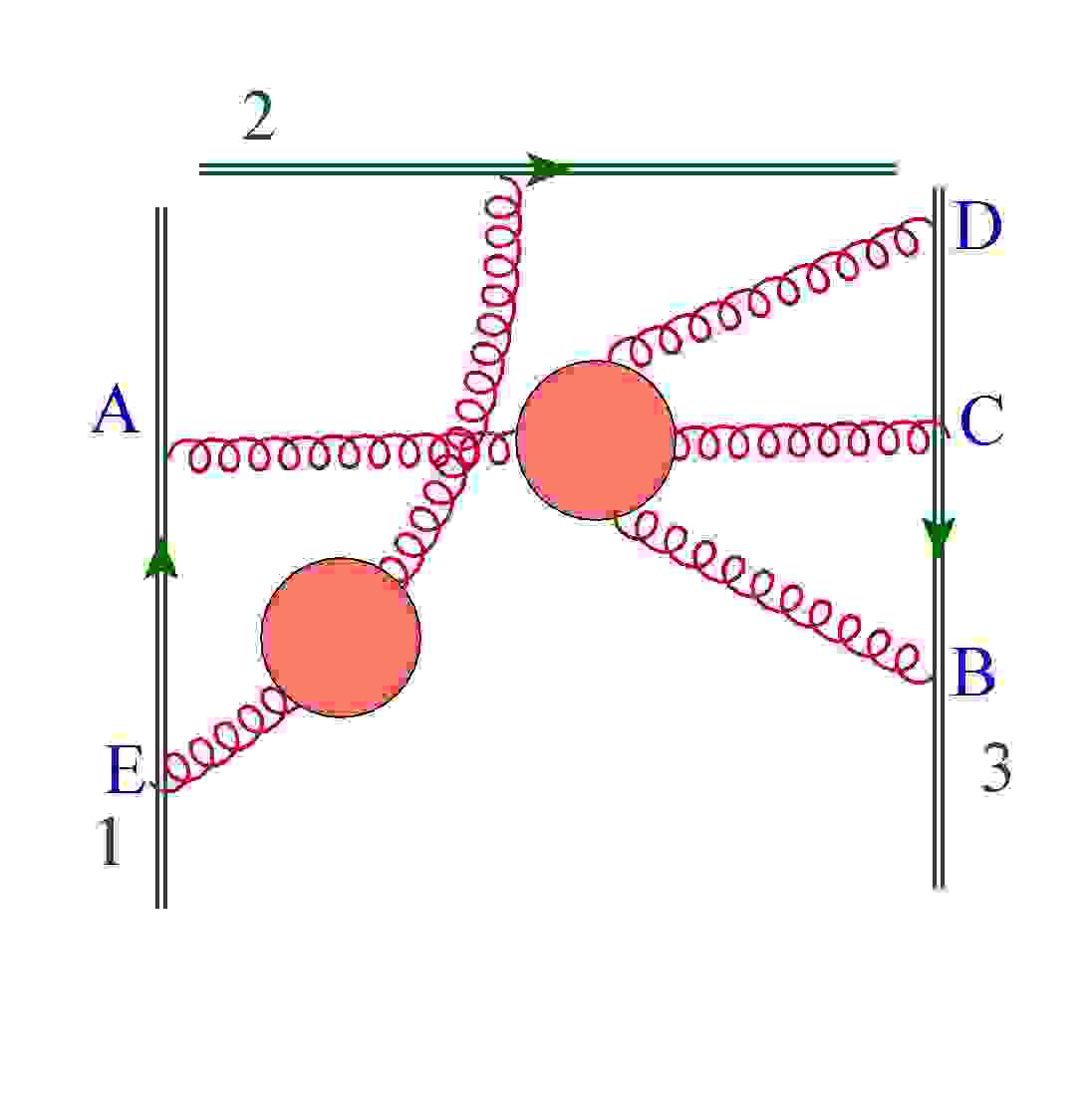}
		\end{figure}
	\end{minipage} 
	\hspace{-2cm}
	\begin{minipage}{0.46\textwidth}
		\vspace{2cm}
\begin{tabular}{ | c | c | c |}
			\hline
			\textbf{Diagrams} & \textbf{Sequences} & \textbf{s-factors} \\ \hline
			$C_1$ & $\lbrace\lbrace EA\rbrace,\lbrace DCB \rbrace \rbrace$ & 1 \\ \hline
			$C_2$ & $\lbrace\lbrace AE\rbrace,\lbrace DCB\rbrace \rbrace$ & 1 \\ \hline
		\end{tabular}
		\label{tab:abcd1}	
	\end{minipage}
	
	\noindent The $R$ matrix is given by
	\begin{align}
	\begin{split}
	&
	R=\left(
	\begin{array}{cc}
	\frac{1}{2} & -\frac{1}{2} \\
	-\frac{1}{2} & \frac{1}{2} \\
	\end{array}
	\right)\,, \qquad \qquad
	D = \D{1} \, .
	\end{split}
	\label{eq:web1app}
	\end{align}
	\item[{\bf 3}.]  $\textbf{W}_{3,\rm{I}}^{(0,2)}(1,2,3)$
	\\
	This Cweb has three diagrams, one of which is displayed below. The table gives
	the chosen order of the three shuffles of the gluon attachments, and the corresponding
	$s$ factors.
	\begin{minipage}{0.5\textwidth}
		\begin{figure}[H]
			\vspace{-2mm}
			\includegraphics[height=4cm,width=4cm]{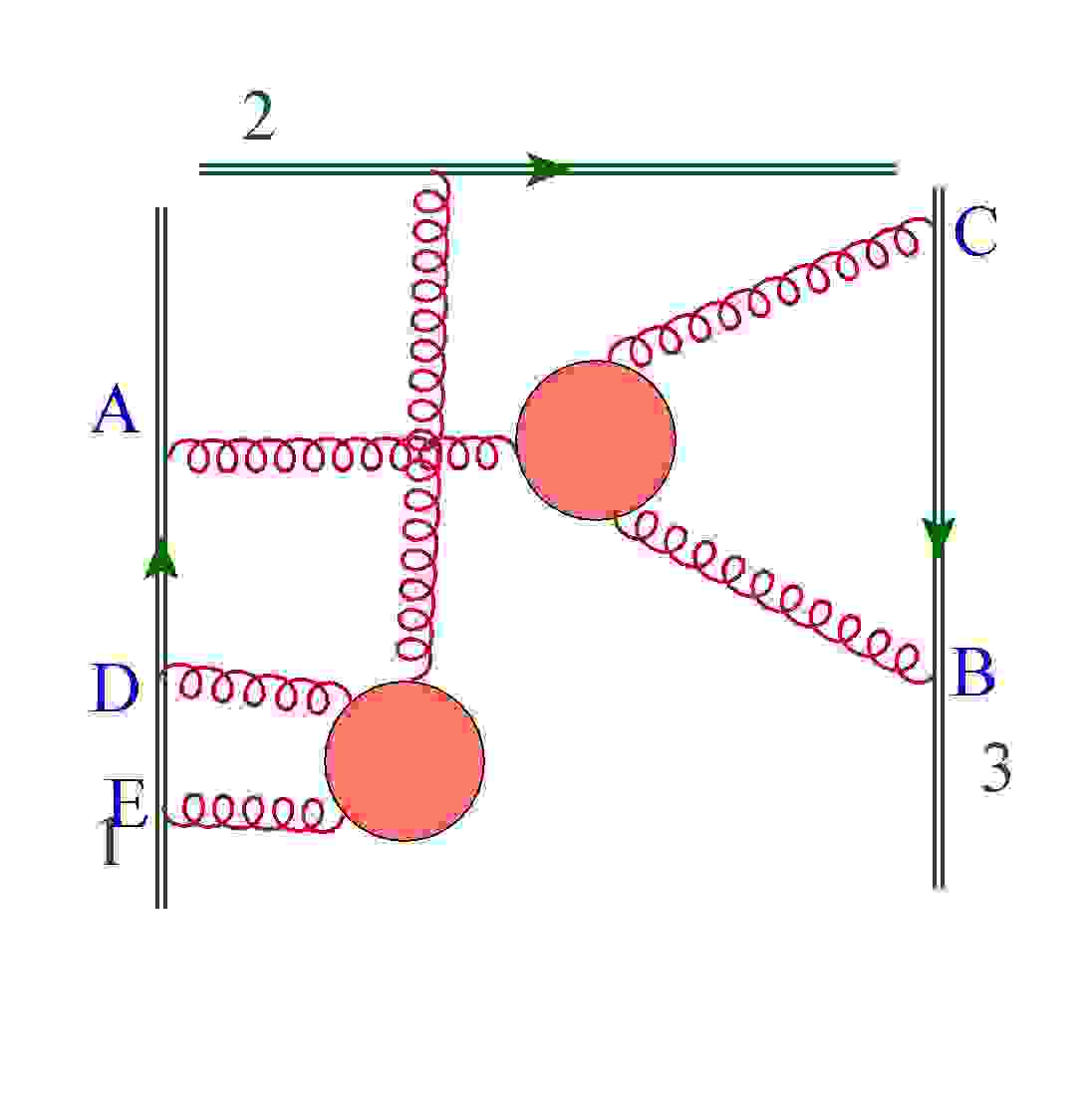}
		\end{figure}
	\end{minipage} 
	\hspace{-2cm}
	\begin{minipage}{0.46\textwidth}
		\vspace{2cm}
		\begin{tabular}{ | c | c | c |}
			\hline
			\textbf{Diagrams} & \textbf{Sequences} & \textbf{s-factors} \\ \hline
			$C_1$ & $\lbrace\lbrace EDA\rbrace,\lbrace CB \rbrace \rbrace$ & 1 \\ \hline
			$C_2$ & $\lbrace\lbrace EAD\rbrace,\lbrace CB\rbrace \rbrace$ & 0 \\ \hline
			$C_3$ & $\lbrace\lbrace EAD\rbrace,\lbrace CB\rbrace \rbrace$ & 1 \\ \hline
		\end{tabular}
		\label{tab:abcd1}	
	\end{minipage}
	
	\noindent The $R$ matrix is given by
	\begin{align}
	\begin{split}
	&
	R=\left(
	\begin{array}{ccc}
	\frac{1}{2} & 0 & -\frac{1}{2} \\
	-\frac{1}{2} & 1 & -\frac{1}{2} \\
	-\frac{1}{2} & 0 & \frac{1}{2} \\
	\end{array}
	\right)\,, \qquad \qquad
	D = \D{2} \, .
	\end{split}
	\label{eq:web1app}
	\end{align}
	\item[{\bf 4}.]  $\textbf{W}_{3,\rm{II}}^{(0,2)}(1,2,3)$
	\\
	This Cweb has six diagrams, one of which is displayed below. The table gives
	the chosen order of the six shuffles of the gluon attachments, and the corresponding
	$s$ factors.
	\begin{minipage}{0.5\textwidth}
		\begin{figure}[H]
			\vspace{-2mm}
			\includegraphics[height=4cm,width=4cm]{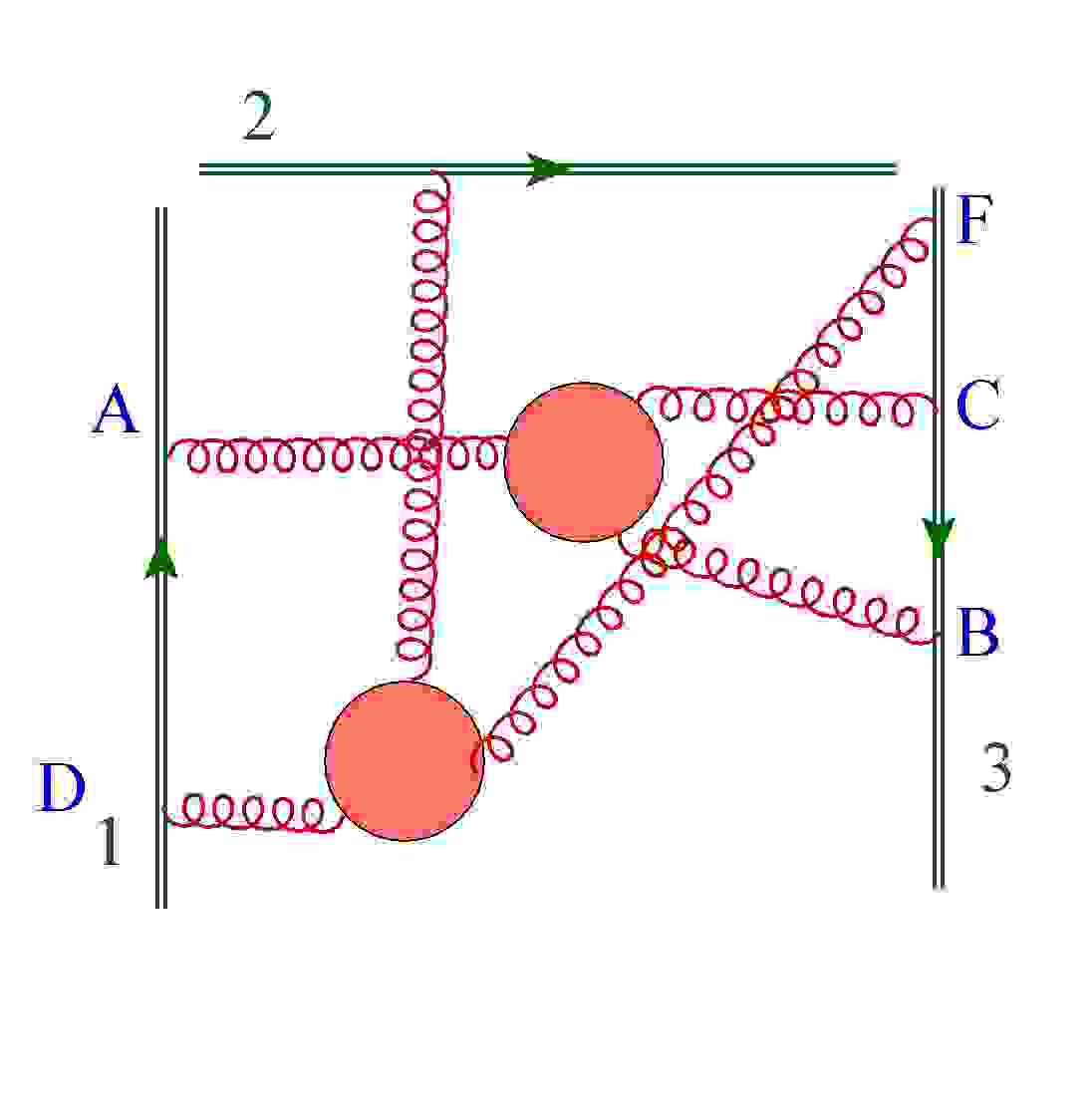}
		\end{figure}
	\end{minipage} 
	\hspace{-2cm}
	\begin{minipage}{0.46\textwidth}
		\vspace{2cm}
		\begin{tabular}{ | c | c | c |}
			\hline
			\textbf{Diagrams} & \textbf{Sequences} & \textbf{s-factors} \\ \hline
			$C_1$ & $\lbrace\lbrace DA\rbrace,\lbrace FCB \rbrace \rbrace$ & 1 \\ \hline
			$C_2$ & $\lbrace\lbrace DA\rbrace,\lbrace CFB\rbrace \rbrace$ & 0 \\ \hline
			$C_3$ & $\lbrace\lbrace DA\rbrace,\lbrace CBF\rbrace \rbrace$ & 0 \\ \hline
			$C_4$ & $\lbrace\lbrace AD\rbrace,\lbrace FCB \rbrace \rbrace$ & 0 \\ \hline
			$C_5$ & $\lbrace\lbrace AD\rbrace,\lbrace CFB\rbrace \rbrace$ & 0 \\ \hline
			$C_6$ & $\lbrace\lbrace AD\rbrace,\lbrace CBF\rbrace \rbrace$ & 1\\ \hline
		\end{tabular}
		\label{tab:abcd1}	
	\end{minipage}
	
	\noindent The $R$ matrix is given by
	\begin{align}
	\begin{split}
	&
	R=\left(
	\begin{array}{cccccc}
	\frac{1}{2} & 0 & 0 & 0 & 0 & -\frac{1}{2} \\
	-\frac{1}{2} & 1 & 0 & 0 & 0 & -\frac{1}{2} \\
	-\frac{1}{2} & 0 & 1 & 0 & 0 & -\frac{1}{2} \\
	-\frac{1}{2} & 0 & 0 & 1 & 0 & -\frac{1}{2} \\
	-\frac{1}{2} & 0 & 0 & 0 & 1 & -\frac{1}{2} \\
	-\frac{1}{2} & 0 & 0 & 0 & 0 & \frac{1}{2} \\
	\end{array}
	\right)\,, \qquad \qquad
	D = \D{5} \, .
	\end{split}
	\label{eq:web1app}
	\end{align}
	\item[{\bf 5}.]  $\textbf{W}_{3,\text{I}}^{(2,1)}(2,2,3)$
	\\
	This Cweb has 24 diagrams, one of which is displayed below. The table gives
	the chosen order of the 24 shuffles of the gluon attachments, and the corresponding
	$s$ factors.
	\begin{minipage}{0.5\textwidth}
		\begin{figure}[H]
			\vspace{-2mm}
			\includegraphics[height=4cm,width=4cm]{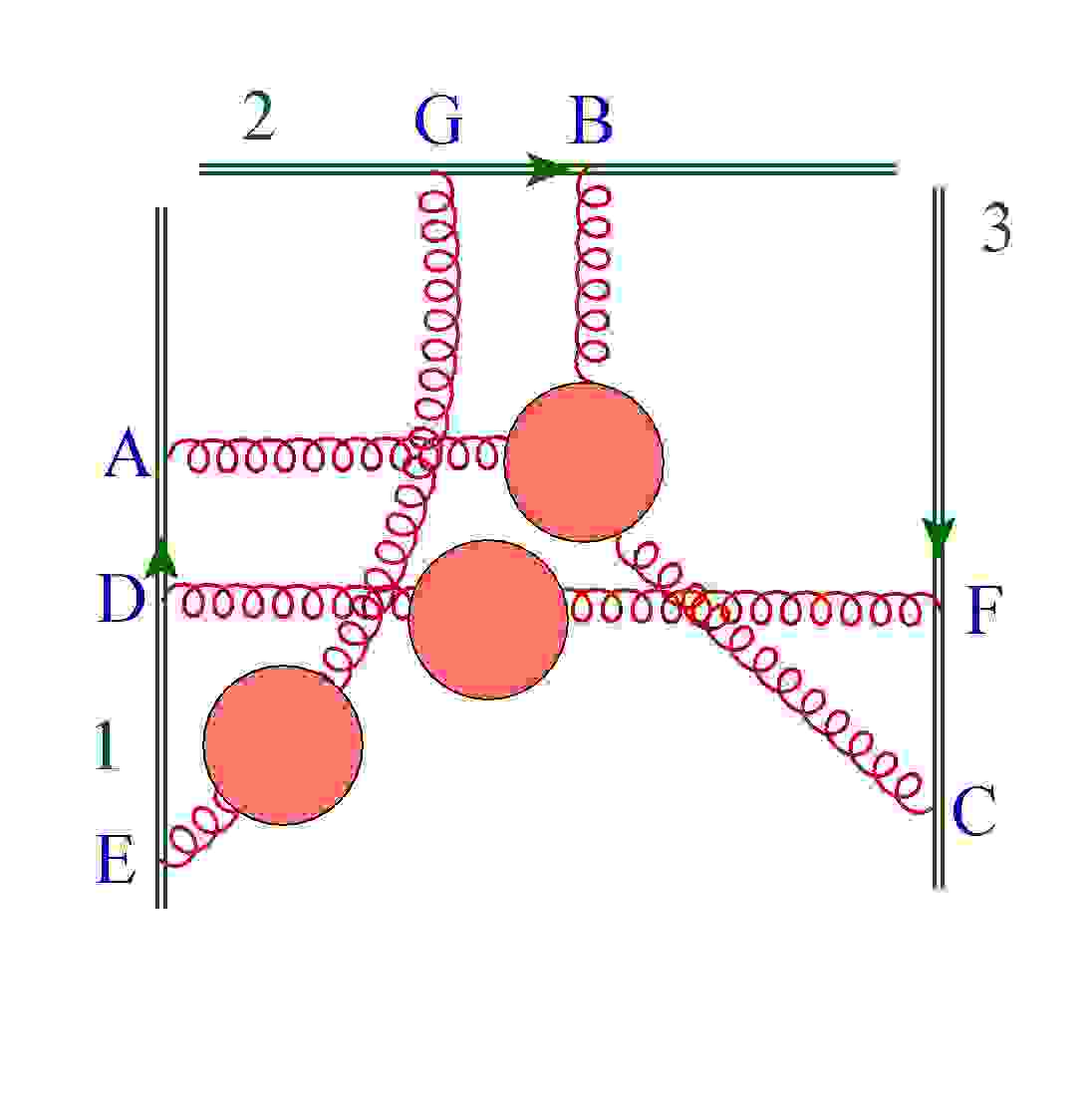}
		\end{figure}
	\end{minipage} 
	\hspace{-2cm}
	\begin{minipage}{0.46\textwidth}
		\vspace{2cm}
		\begin{tabular}{ | c | c | c |}
			\hline
			\textbf{Diagrams} & \textbf{Sequences} & \textbf{s-factors} \\ \hline
			$C_1$ & $\lbrace\lbrace EDA\rbrace,\lbrace GB \rbrace, \lbrace FC \rbrace\rbrace$ & 1 \\ \hline
			$C_2$ & $\lbrace\lbrace EDA\rbrace,\lbrace GB \rbrace, \lbrace CF \rbrace\rbrace$ & 0 \\ \hline
			$C_3$ & $\lbrace\lbrace EDA\rbrace,\lbrace BG \rbrace, \lbrace FC \rbrace\rbrace$ & 0 \\ \hline
			$C_4$ & $\lbrace\lbrace EDA\rbrace,\lbrace BG \rbrace, \lbrace CF \rbrace\rbrace$ & 0 \\ \hline
			$C_5$ & $\lbrace\lbrace DEA\rbrace,\lbrace GB \rbrace, \lbrace FC \rbrace\rbrace$ & 1 \\ \hline
			$C_6$ & $\lbrace\lbrace DEA\rbrace,\lbrace GB \rbrace, \lbrace CF \rbrace\rbrace$ & 0 \\ \hline
			$C_7$ & $\lbrace\lbrace DEA\rbrace,\lbrace BG \rbrace, \lbrace FC \rbrace\rbrace$ & 0 \\
\hline
			
		\end{tabular}
		\label{tab:abcd1}	
	\end{minipage}
	
	\noindent The $R$ matrix is given by
	\begin{align}
	\begin{split}
	&
	R=\left(
	\begin{array}{cccccccccccccccccccccccc}
	\frac{1}{3} & 0 & 0 & 0 & -\frac{1}{6} & 0 & 0 & 0 & 0 & -\frac{1}{6} & 0 & 0 & 0 & 0 & 0 & -\frac{1}{6} & 0 & 0 & -\frac{1}{6} & 0 & 0 & 0 & 0 & \frac{1}{3} \\
	-\frac{1}{6} & \frac{1}{2} & 0 & 0 & -\frac{1}{6} & 0 & 0 & 0 & 0 & -\frac{1}{6} & 0 & 0 & 0 & 0 & 0 & -\frac{1}{6} & 0 & 0 & \frac{1}{3} & -\frac{1}{2} & 0 & 0 & 0 & \frac{1}{3} \\
	-\frac{2}{3} & 0 & 1 & 0 & \frac{1}{3} & 0 & -\frac{1}{2} & 0 & 0 & \frac{1}{3} & 0 & -\frac{1}{2} & 0 & 0 & 0 & -\frac{1}{6} & 0 & 0 & -\frac{1}{6} & 0 & 0 & 0 & 0 & \frac{1}{3} \\
	-\frac{1}{6} & -\frac{1}{2} & 0 & 1 & \frac{1}{3} & 0 & -\frac{1}{2} & 0 & 0 & \frac{1}{3} & 0 & -\frac{1}{2} & 0 & 0 & 0 & -\frac{1}{6} & 0 & 0 & \frac{1}{3} & -\frac{1}{2} & 0 & 0 & 0 & \frac{1}{3} \\
	-\frac{1}{6} & 0 & 0 & 0 & \frac{1}{3} & 0 & 0 & 0 & 0 & -\frac{1}{6} & 0 & 0 & 0 & 0 & 0 & \frac{1}{3} & 0 & 0 & -\frac{1}{6} & 0 & 0 & 0 & 0 & -\frac{1}{6} \\
	\frac{1}{3} & -\frac{1}{2} & 0 & 0 & -\frac{2}{3} & 1 & 0 & 0 & 0 & -\frac{1}{6} & 0 & 0 & 0 & 0 & 0 & \frac{1}{3} & 0 & 0 & \frac{1}{3} & -\frac{1}{2} & 0 & 0 & 0 & -\frac{1}{6} \\
	-\frac{1}{6} & 0 & 0 & 0 & -\frac{1}{6} & 0 & \frac{1}{2} & 0 & 0 & \frac{1}{3} & 0 & -\frac{1}{2} & 0 & 0 & 0 & \frac{1}{3} & 0 & 0 & -\frac{1}{6} & 0 & 0 & 0 & 0 & -\frac{1}{6} \\
	\frac{1}{3} & -\frac{1}{2} & 0 & 0 & -\frac{1}{6} & 0 & -\frac{1}{2} & 1 & 0 & \frac{1}{3} & 0 & -\frac{1}{2} & 0 & 0 & 0 & \frac{1}{3} & 0 & 0 & \frac{1}{3} & -\frac{1}{2} & 0 & 0 & 0 & -\frac{1}{6} \\
\end{array}
	\right)\,, \qquad \qquad \\ &
	D = \D{16} \, .
	\end{split}
	\label{eq:web1app}
	\end{align}
	
	\item[{\bf 6}.]  $\textbf{W}_{3,\text{II}}^{(2,1)}(2,2,3)$
	\\
	This Cweb has six diagrams, one of which is displayed below. The table gives
	the chosen order of the six shuffles of the gluon attachments, and the corresponding
	$s$ factors.
	\begin{minipage}{0.5\textwidth}
		\begin{figure}[H]
			\vspace{-2mm}
			\includegraphics[height=4cm,width=4cm]{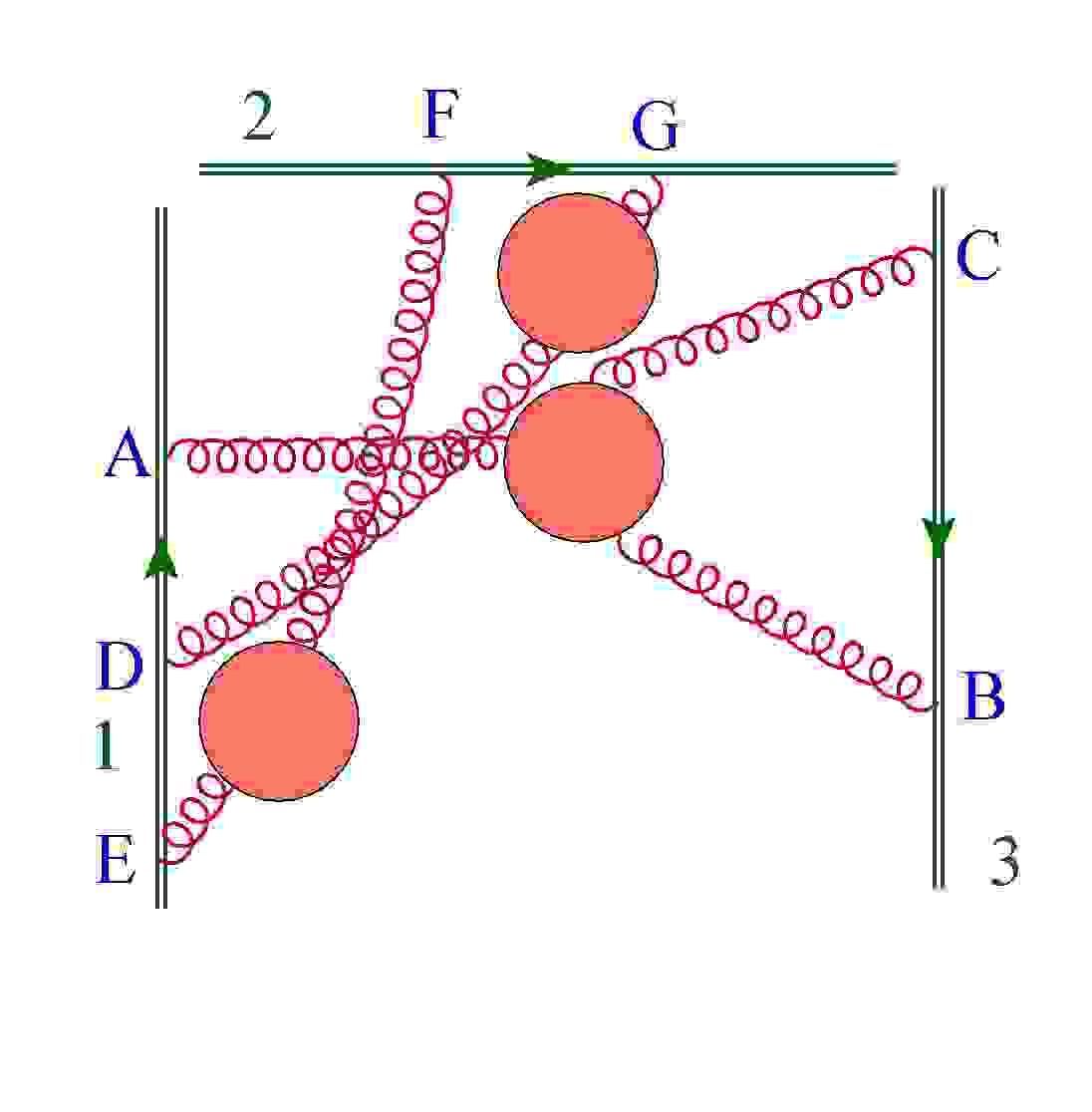}
		\end{figure}
	\end{minipage} 
	\hspace{-2cm}
	\begin{minipage}{0.46\textwidth}
		\vspace{2cm}
		\begin{tabular}{ | c | c | c |}
			\hline
			\textbf{Diagrams} & \textbf{Sequences} & \textbf{s-factors} \\ \hline
			$C_1$ & $\lbrace\lbrace EDA\rbrace,\lbrace FG \rbrace,\lbrace CB \rbrace \rbrace$ & 1 \\ \hline
			$C_2$ & $\lbrace\lbrace EDA\rbrace,\lbrace GF \rbrace,\lbrace CB \rbrace \rbrace$ & 0\\ \hline
			$C_3$ & $\lbrace\lbrace EAD\rbrace,\lbrace FG \rbrace,\lbrace CB \rbrace \rbrace$ & 1 \\ \hline
			$C_4$ & $\lbrace\lbrace EAD\rbrace,\lbrace GF \rbrace,\lbrace CB \rbrace \rbrace $ & 0 \\ \hline
			$C_5$ & $\lbrace\lbrace AED\rbrace,\lbrace FG \rbrace,\lbrace CB \rbrace \rbrace$ & 1 \\ \hline
			$C_6$ & $\lbrace\lbrace AED\rbrace,\lbrace GF \rbrace,\lbrace CB \rbrace \rbrace$ & 0 \\ \hline
			
		\end{tabular}
		\label{tab:abcd1}	
	\end{minipage}
	
	\noindent The $R$ matrix is given by
	\begin{align}
	\begin{split}
	&
	R=\left(
	\begin{array}{cccccc}
	\frac{1}{6} & 0 & -\frac{1}{3} & 0 & \frac{1}{6} & 0 \\
	-\frac{1}{3} & \frac{1}{2} & -\frac{1}{3} & 0 & \frac{2}{3} & -\frac{1}{2} \\
	-\frac{1}{3} & 0 & \frac{2}{3} & 0 & -\frac{1}{3} & 0 \\
	\frac{1}{6} & -\frac{1}{2} & -\frac{1}{3} & 1 & \frac{1}{6} & -\frac{1}{2} \\
	\frac{1}{6} & 0 & -\frac{1}{3} & 0 & \frac{1}{6} & 0 \\
	\frac{2}{3} & -\frac{1}{2} & -\frac{1}{3} & 0 & -\frac{1}{3} & \frac{1}{2} \\
	\end{array}
	\right)\,, \qquad \qquad
	D = \D{3} \, .
	\end{split}
	\label{eq:web1app}
	\end{align}
	\item[{\bf 7}.]  $\textbf{W}_{3,\text{III}}^{(2,1)}(2,2,3)$
	\\
	This Cweb has twelve diagrams, one of which is displayed below. The table gives
	the chosen order of the twelve shuffles of the gluon attachments, and the corresponding
	$s$ factors.
	\begin{minipage}{0.5\textwidth}
		\begin{figure}[H]
			\vspace{-2mm}
			\includegraphics[height=4cm,width=4cm]{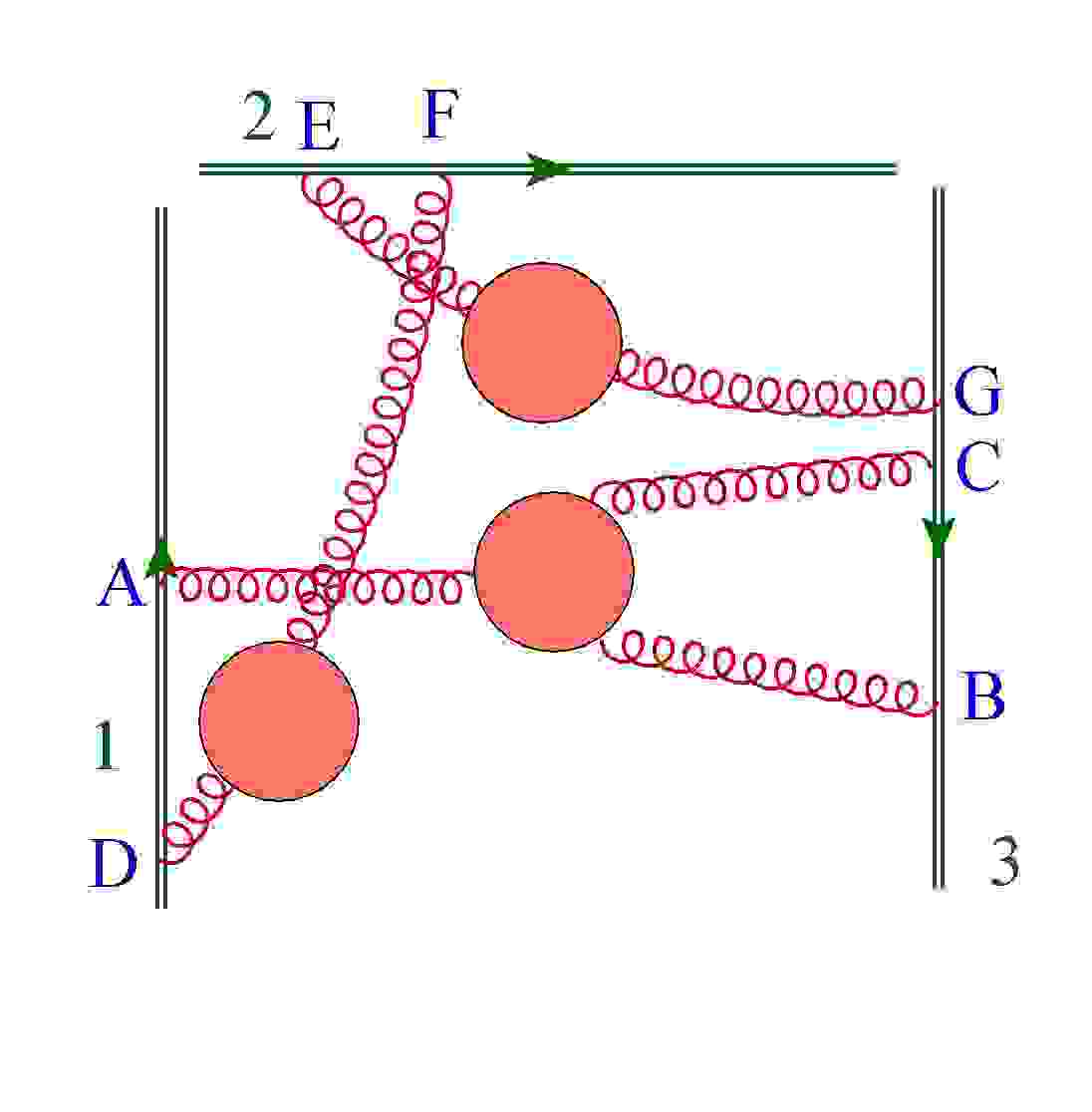}
		\end{figure}
	\end{minipage} 
	\hspace{-2cm}
	\begin{minipage}{0.46\textwidth}
		\vspace{2cm}
		\begin{tabular}{ | c | c | c |}
			\hline
			\textbf{Diagrams} & \textbf{Sequences} & \textbf{s-factors} \\ \hline
			$C_1$ & $\lbrace\lbrace DA\rbrace,\lbrace EF \rbrace,\lbrace GCB \rbrace \rbrace$ & 1 \\ \hline
			$C_2$ & $\lbrace\lbrace DA\rbrace,\lbrace EF \rbrace,\lbrace CGB \rbrace \rbrace$ & 0 \\ \hline
			$C_3$ & $\lbrace\lbrace DA\rbrace,\lbrace EF \rbrace,\lbrace CBG \rbrace \rbrace$ & 0 \\ \hline
			$C_4$ & $\lbrace\lbrace DA\rbrace,\lbrace FE \rbrace,\lbrace GCB \rbrace \rbrace$ & 1 \\ \hline
			$C_5$ & $\lbrace\lbrace DA\rbrace,\lbrace FE \rbrace,\lbrace CGB \rbrace \rbrace$ & 0 \\ \hline
			$C_6$ & $\lbrace\lbrace DA\rbrace,\lbrace FE \rbrace,\lbrace CBG \rbrace \rbrace$ & 1 \\ \hline
			$C_7$ & $\lbrace\lbrace AD\rbrace,\lbrace EF \rbrace,\lbrace GCB \rbrace \rbrace$ & 1 \\ \hline
			$C_8$ & $\lbrace\lbrace AD\rbrace,\lbrace EF \rbrace,\lbrace CGB \rbrace \rbrace$ & 0 \\ \hline
			$C_9$ & $\lbrace\lbrace AD\rbrace,\lbrace EF \rbrace,\lbrace CBG \rbrace \rbrace$ & 1 \\ \hline
			$C_{10}$ & $\lbrace\lbrace AD\rbrace,\lbrace FE \rbrace,\lbrace GCB \rbrace \rbrace$ & 0 \\ \hline
			$C_{11}$ & $\lbrace\lbrace AD\rbrace,\lbrace FE \rbrace,\lbrace CGB \rbrace \rbrace$ & 0 \\ \hline
			$C_{12}$ & $\lbrace\lbrace AD\rbrace,\lbrace FE \rbrace,\lbrace CBG \rbrace \rbrace$ & 1 \\ \hline
		\end{tabular}
		\label{tab:abcd1}	
	\end{minipage}
	
	\noindent The $R$ matrix is given by
	\begin{align}
	\begin{split}
	&
	R=\left(
	\begin{array}{cccccccccccc}
	\frac{1}{3} & 0 & 0 & -\frac{1}{6} & 0 & -\frac{1}{6} & -\frac{1}{6} & 0 & -\frac{1}{6} & 0 & 0 & \frac{1}{3} \\
	-\frac{2}{3} & 1 & 0 & \frac{1}{3} & -\frac{1}{2} & -\frac{1}{6} & \frac{1}{3} & -\frac{1}{2} & -\frac{1}{6} & 0 & 0 & \frac{1}{3} \\
	-\frac{2}{3} & 0 & 1 & \frac{1}{3} & 0 & -\frac{2}{3} & \frac{1}{3} & 0 & -\frac{2}{3} & 0 & 0 & \frac{1}{3} \\
	-\frac{1}{6} & 0 & 0 & \frac{1}{3} & 0 & -\frac{1}{6} & -\frac{1}{6} & 0 & \frac{1}{3} & 0 & 0 & -\frac{1}{6} \\
	-\frac{1}{6} & 0 & 0 & -\frac{1}{6} & \frac{1}{2} & -\frac{1}{6} & \frac{1}{3} & -\frac{1}{2} & \frac{1}{3} & 0 & 0 & -\frac{1}{6} \\
	-\frac{1}{6} & 0 & 0 & -\frac{1}{6} & 0 & \frac{1}{3} & \frac{1}{3} & 0 & -\frac{1}{6} & 0 & 0 & -\frac{1}{6} \\
	-\frac{1}{6} & 0 & 0 & -\frac{1}{6} & 0 & \frac{1}{3} & \frac{1}{3} & 0 & -\frac{1}{6} & 0 & 0 & -\frac{1}{6} \\
	-\frac{1}{6} & 0 & 0 & \frac{1}{3} & -\frac{1}{2} & \frac{1}{3} & -\frac{1}{6} & \frac{1}{2} & -\frac{1}{6} & 0 & 0 & -\frac{1}{6} \\
	-\frac{1}{6} & 0 & 0 & \frac{1}{3} & 0 & -\frac{1}{6} & -\frac{1}{6} & 0 & \frac{1}{3} & 0 & 0 & -\frac{1}{6} \\
	\frac{1}{3} & 0 & 0 & -\frac{2}{3} & 0 & \frac{1}{3} & -\frac{2}{3} & 0 & \frac{1}{3} & 1 & 0 & -\frac{2}{3} \\
	\frac{1}{3} & 0 & 0 & -\frac{1}{6} & -\frac{1}{2} & \frac{1}{3} & -\frac{1}{6} & -\frac{1}{2} & \frac{1}{3} & 0 & 1 & -\frac{2}{3} \\
	\frac{1}{3} & 0 & 0 & -\frac{1}{6} & 0 & -\frac{1}{6} & -\frac{1}{6} & 0 & -\frac{1}{6} & 0 & 0 & \frac{1}{3} \\
	\end{array}
	\right)\,, \qquad \qquad \\ &
	D = \D{7} \, 
	\end{split}
	\label{eq:web1app}
	\end{align}
	
	\item[{\bf 8}.]  $\textbf{W}_{3}^{(1,0,1)}(1,2,3)$
		\\
	This Cweb has three diagrams, one of which is displayed below. The table gives
	the chosen order of the three shuffles of the gluon attachments, and the corresponding
	$s$ factors.
	\begin{minipage}{0.5\textwidth}
		\begin{figure}[H]
			\vspace{-2mm}
			\includegraphics[height=4cm,width=4cm]{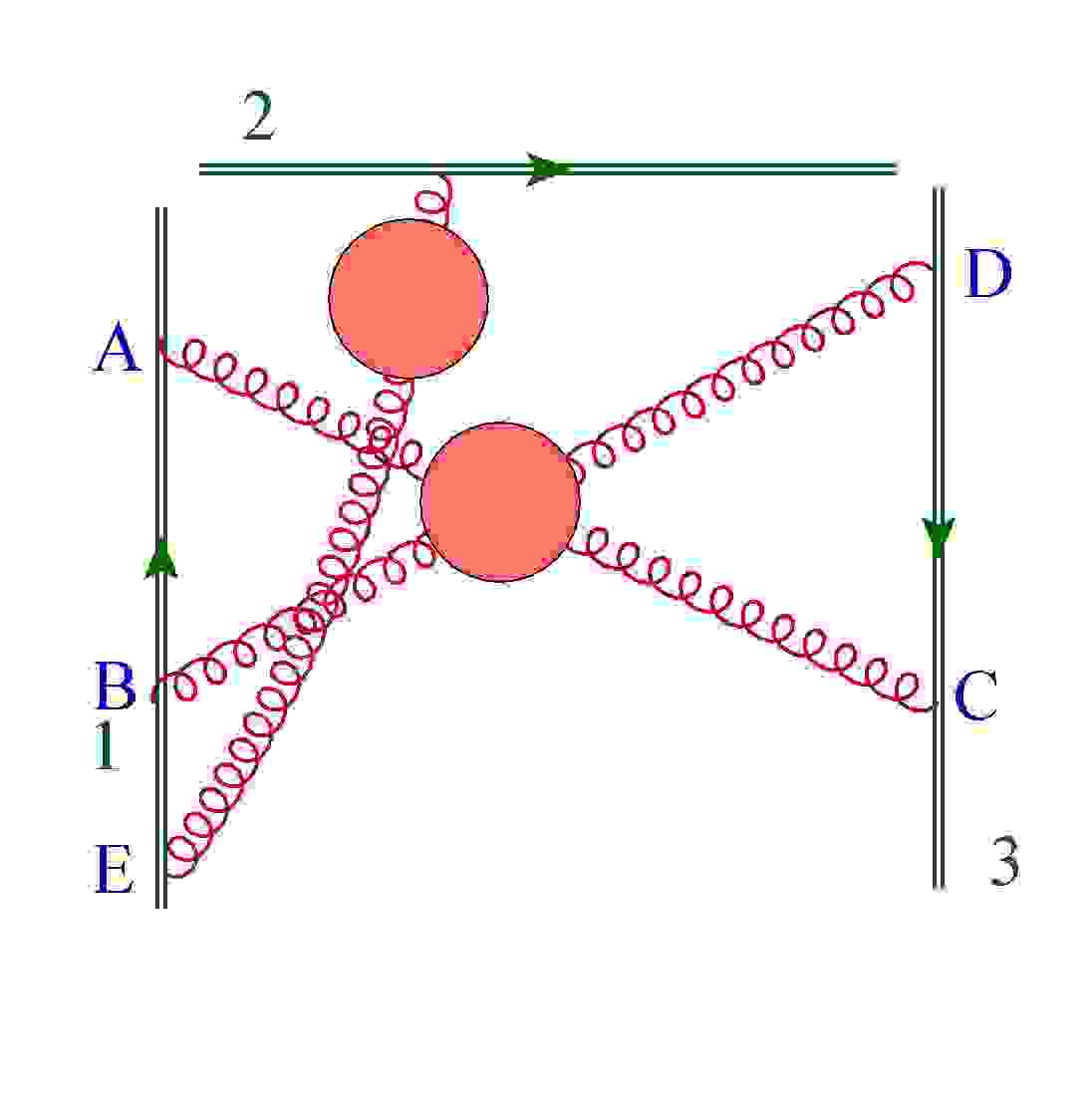}
		\end{figure}
	\end{minipage} 
	\hspace{-2cm}
	\begin{minipage}{0.46\textwidth}
		\vspace{2cm}
		\begin{tabular}{ | c | c | c |}
			\hline
			\textbf{Diagrams} & \textbf{Sequences} & \textbf{s-factors} \\ \hline
			$C_1$ & $\lbrace\lbrace EBA\rbrace,\lbrace DC \rbrace \rbrace$ & 1 \\ \hline
			$C_2$ & $\lbrace\lbrace BEA\rbrace,\lbrace DC \rbrace \rbrace$ & 0 \\ \hline
			$C_3$ & $\lbrace\lbrace AEB\rbrace,\lbrace DC \rbrace \rbrace$ & 1 \\ \hline
		\end{tabular}
		\label{tab:abcd1}	
	\end{minipage}
	
	\noindent The $R$ matrix is given by
	\begin{align}
	\begin{split}
	&
	R=\left(
	\begin{array}{ccc}
	\frac{1}{2} & 0 & -\frac{1}{2} \\
	-\frac{1}{2} & 1 & -\frac{1}{2} \\
	-\frac{1}{2} & 0 & \frac{1}{2} \\
	\end{array}
	\right)\,, \qquad \qquad 
	D = \D{2} \, 
	\end{split}
	\label{eq:web1app}
	\end{align}
	\item[{\bf 9}.]  $\textbf{W}_{3, \rm{I}}^{(0,2)}(2,2,2)$
	\\
	This Cweb has two diagrams, one of which is displayed below. The table gives
	the chosen order of the two shuffles of the gluon attachments, and the corresponding
	$s$ factors.
	\begin{minipage}{0.5\textwidth}
		\begin{figure}[H]
			\vspace{-2mm}
			\includegraphics[height=4cm,width=4cm]{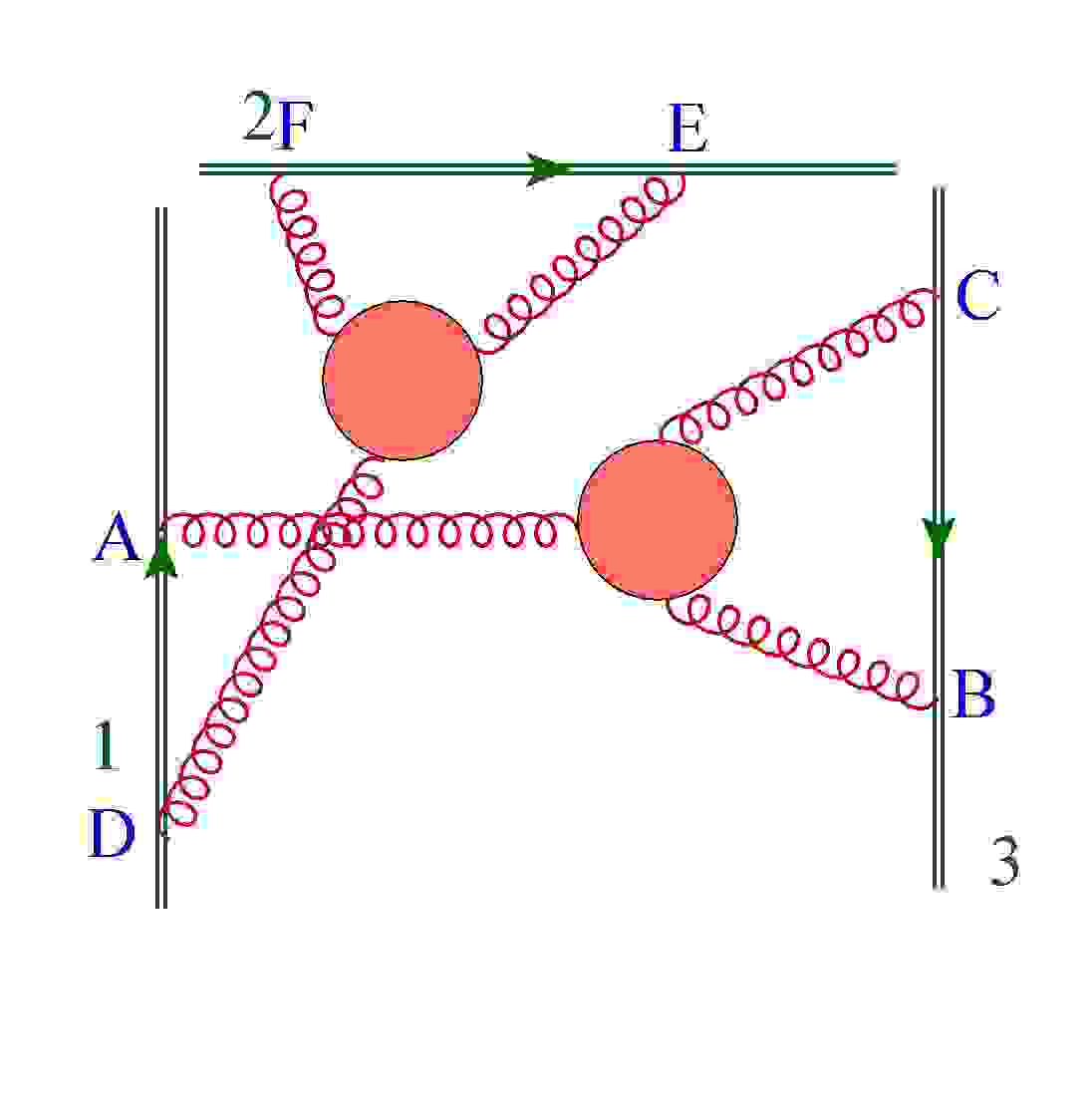}
		\end{figure}
	\end{minipage} 
	\hspace{-2cm}
	\begin{minipage}{0.46\textwidth}
		\vspace{2cm}
		\begin{tabular}{ | c | c | c |}
			\hline
			\textbf{Diagrams} & \textbf{Sequences} & \textbf{s-factors} \\ \hline
			$C_1$ & $\lbrace\lbrace DA\rbrace,\lbrace FE \rbrace, \lbrace CB \rbrace \rbrace$ & 1 \\ \hline
			$C_2$ & $\lbrace\lbrace AD\rbrace,\lbrace FE \rbrace, \lbrace  CB \rbrace \rbrace$ & 1 \\ \hline
		\end{tabular}
		\label{tab:abcd1}	
	\end{minipage}
	
	\noindent The $R$ matrix is given by
	\begin{align}
	\begin{split}
	&
	R=\left(
	\begin{array}{cc}
	\frac{1}{2} & -\frac{1}{2} \\
	-\frac{1}{2} & \frac{1}{2} \\
	\end{array}
	\right)\,, \qquad \qquad 
%	Y=\left(
%	\begin{array}{cc}
%	-1 & 1 \\
%	1 & 1 \\
%	\end{array}
%	\right),\qquad \qquad 
	D = \D{1} \, 
	\end{split}
	\label{eq:web1app}
	\end{align}
	\item[{\bf 10}.]  $\textbf{W}_{3}^{(1,0,1)}(2,2,2)$
	\\
	This Cweb has four diagrams, one of which is displayed below. The table gives
	the chosen order of the four shuffles of the gluon attachments, and the corresponding
	$s$ factors.
	\begin{minipage}{0.5\textwidth}
		\begin{figure}[H]
			\vspace{-2mm}
			\includegraphics[height=4cm,width=4cm]{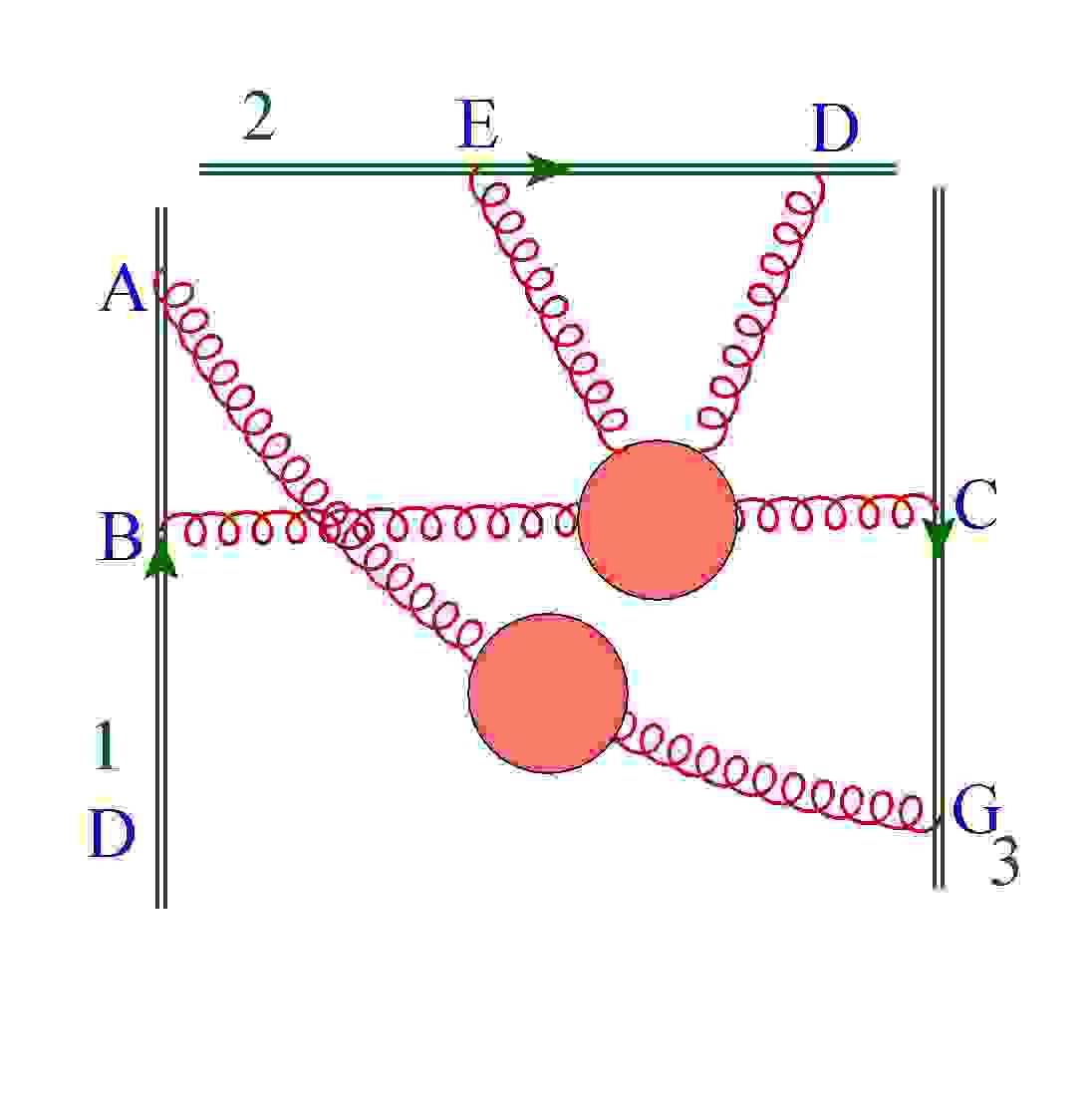}
		\end{figure}
	\end{minipage} 
	\hspace{-2cm}
	\begin{minipage}{0.46\textwidth}
		\vspace{2cm}
		\begin{tabular}{ | c | c | c |}
			\hline
			\textbf{Diagrams} & \textbf{Sequences} & \textbf{s-factors} \\ \hline
			$C_1$ & $\lbrace\lbrace BA\rbrace,\lbrace ED \rbrace, \lbrace CG \rbrace \rbrace$ & 1 \\ \hline
			$C_2$ & $\lbrace\lbrace BA\rbrace,\lbrace ED \rbrace, \lbrace GC \rbrace \rbrace$ & 0 \\ \hline
			$C_3$ & $\lbrace\lbrace AB\rbrace,\lbrace ED \rbrace, \lbrace CG \rbrace \rbrace$ & 0 \\ \hline
			$C_4$ & $\lbrace\lbrace AB\rbrace,\lbrace ED \rbrace, \lbrace GC \rbrace \rbrace$ & 1 \\ \hline
		\end{tabular}
		\label{tab:abcd1}	
	\end{minipage}
	
	\noindent The $R$ matrix is given by
	\begin{align}
	\begin{split}
	&
	R=\left(
	\begin{array}{cccc}
	\frac{1}{2} & 0 & 0 & -\frac{1}{2} \\
	-\frac{1}{2} & 1 & 0 & -\frac{1}{2} \\
	-\frac{1}{2} & 0 & 1 & -\frac{1}{2} \\
	-\frac{1}{2} & 0 & 0 & \frac{1}{2} \\
	\end{array}
	\right)\,, \qquad \qquad 
	D = \D{3} \, 
	\end{split}
	\label{eq:web1app}
	\end{align}
	\item[{\bf 11}.]  $\textbf{W}_{3, \rm{II}}^{(0,2)}(2,2,2)$
	\\
	This Cweb has eight diagrams, one of which is displayed below. The table gives
	the chosen order of the eight shuffles of the gluon attachments, and the corresponding
	$s$ factors.
	\begin{minipage}{0.5\textwidth}
		\begin{figure}[H]
			\vspace{-2mm}
			\includegraphics[height=4cm,width=4cm]{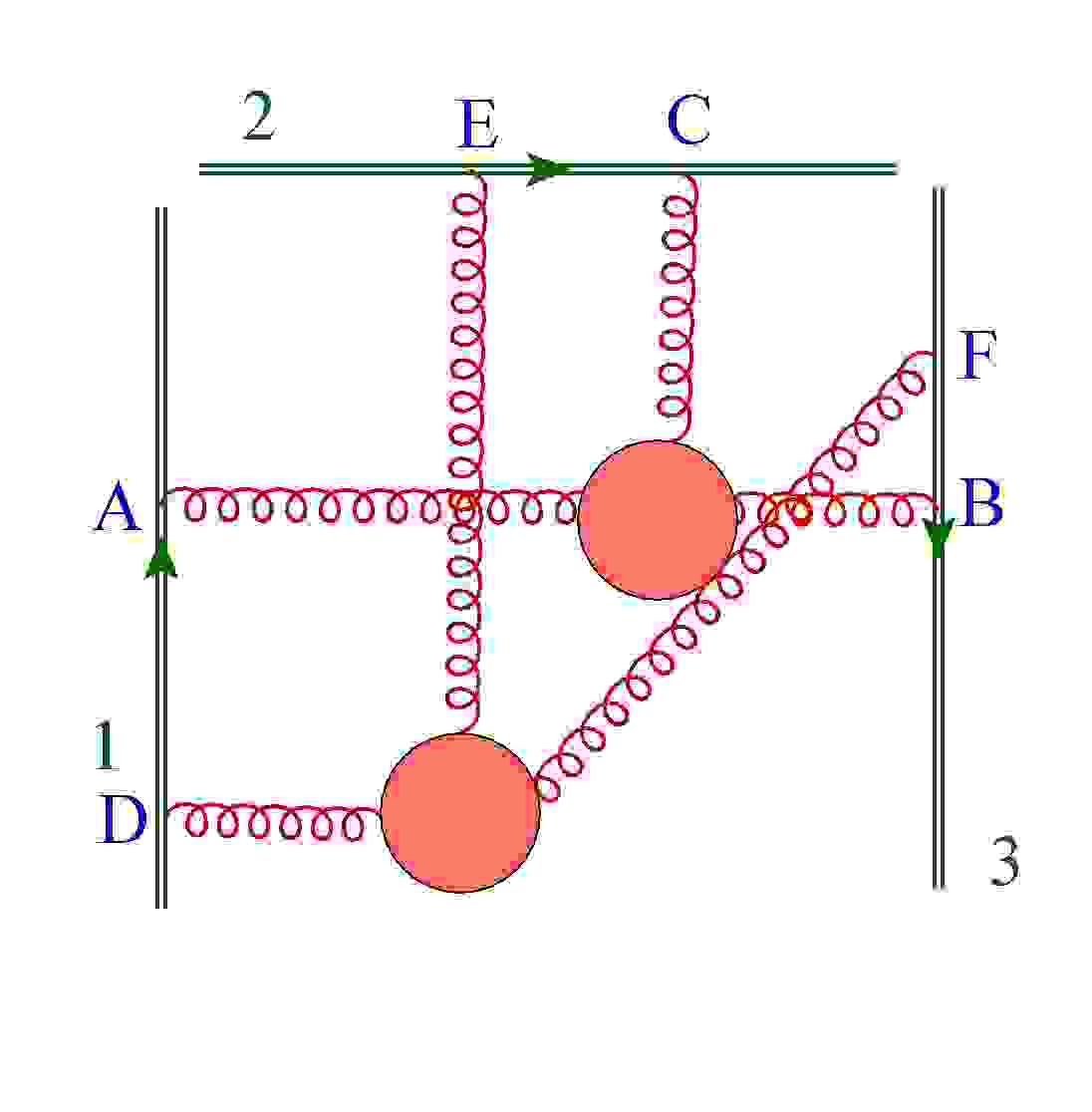}
		\end{figure}
	\end{minipage} 
	\hspace{-2cm}
	\begin{minipage}{0.46\textwidth}
		\vspace{2cm}
		\begin{tabular}{ | c | c | c |}
			\hline
			\textbf{Diagrams} & \textbf{Sequences} & \textbf{s-factors} \\ \hline
			$C_1$ & $\lbrace\lbrace DA\rbrace,\lbrace EC \rbrace, \lbrace FB \rbrace \rbrace$ & 1 \\ \hline
			$C_2$ & $\lbrace\lbrace DA\rbrace,\lbrace EC \rbrace, \lbrace BF \rbrace \rbrace$ & 0 \\ \hline
			$C_3$ & $\lbrace\lbrace DA\rbrace,\lbrace CE \rbrace, \lbrace FB \rbrace \rbrace$ & 0 \\ \hline
			$C_4$ & $\lbrace\lbrace DA\rbrace,\lbrace CE \rbrace, \lbrace BF \rbrace \rbrace$ & 0 \\ \hline
			$C_5$ & $\lbrace\lbrace AD\rbrace,\lbrace EC \rbrace, \lbrace FB \rbrace \rbrace$ & 0 \\ \hline
			$C_6$ & $\lbrace\lbrace AD\rbrace,\lbrace EC \rbrace, \lbrace BF \rbrace \rbrace$ & 0 \\ \hline
			$C_7$ & $\lbrace\lbrace AD\rbrace,\lbrace CE \rbrace, \lbrace FB \rbrace \rbrace$ & 0 \\ \hline
			$C_8$ & $\lbrace\lbrace AD\rbrace,\lbrace CE \rbrace, \lbrace BF \rbrace \rbrace$ & 1 \\ \hline
		\end{tabular}
		\label{tab:abcd1}	
	\end{minipage}
	
	\noindent The $R$ matrix is given by
	\begin{align}
	\begin{split}
	&
	R=\left(
	\begin{array}{cccccccc}
	\frac{1}{2} & 0 & 0 & 0 & 0 & 0 & 0 & -\frac{1}{2} \\
	-\frac{1}{2} & 1 & 0 & 0 & 0 & 0 & 0 & -\frac{1}{2} \\
	-\frac{1}{2} & 0 & 1 & 0 & 0 & 0 & 0 & -\frac{1}{2} \\
	-\frac{1}{2} & 0 & 0 & 1 & 0 & 0 & 0 & -\frac{1}{2} \\
	-\frac{1}{2} & 0 & 0 & 0 & 1 & 0 & 0 & -\frac{1}{2} \\
	-\frac{1}{2} & 0 & 0 & 0 & 0 & 1 & 0 & -\frac{1}{2} \\
	-\frac{1}{2} & 0 & 0 & 0 & 0 & 0 & 1 & -\frac{1}{2} \\
	-\frac{1}{2} & 0 & 0 & 0 & 0 & 0 & 0 & \frac{1}{2} \\
	\end{array}
	\right)\,, \qquad \qquad 
	D = \D{7} \, 
	\end{split}
	\label{eq:web1app}
	\end{align}
	\item[{\bf 12}.]  $\textbf{W}_{3,\text{I}}^{(2,1)}(1,3,3)$
	%\begin{figure}[H]
	%	\centering
	%	\subfloat[]{\includegraphics[scale=0.5]{Diagrams-mathematica/3-legs/Web14.jpg} }
	%	\label{fig:Web-14}
	%\end{figure} 
	\\
	This Cweb has eighteen diagrams, one of which is displayed below. The table gives
	the chosen order of the eighteen shuffles of the gluon attachments, and the corresponding
	$s$ factors.
	\begin{minipage}{0.5\textwidth}
		\begin{figure}[H]
			\vspace{-2mm}
			\includegraphics[height=4cm,width=4cm]{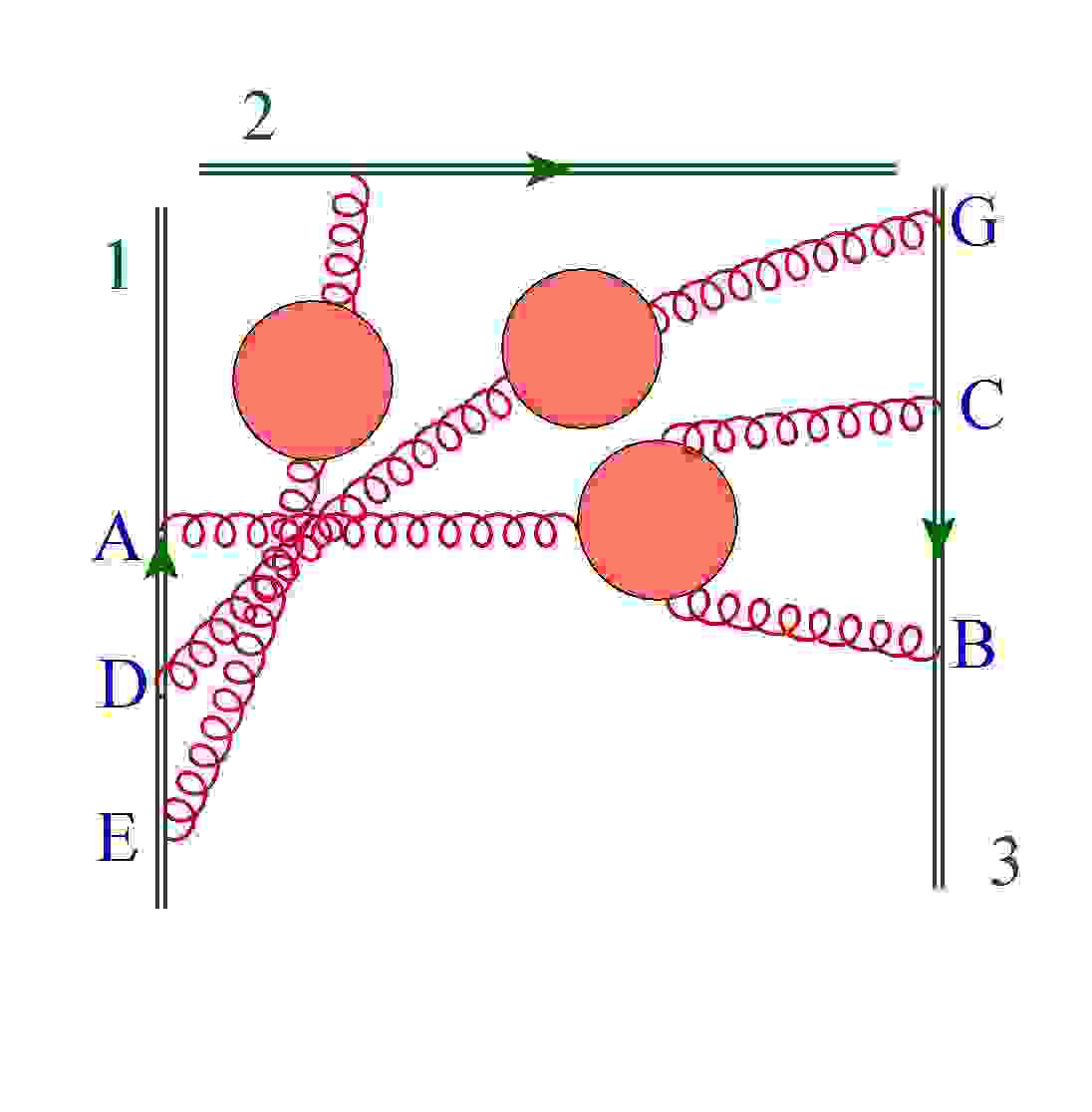}
		\end{figure}
	\end{minipage} 
	\hspace{-2cm}
	\begin{minipage}{0.46\textwidth}
		\vspace{2cm}
\begin{tabular}{ | c | c | c |}
			\hline
			\textbf{Diagrams} & \textbf{Sequences} & \textbf{s-factors} \\ \hline
			$C_1$ & $\lbrace\lbrace EDA\rbrace,\lbrace GCB \rbrace \rbrace$ & 1 \\ \hline
			$C_2$ & $\lbrace\lbrace EDA\rbrace,\lbrace CGB \rbrace \rbrace$ & 0 \\ \hline
			$C_3$ & $\lbrace\lbrace EDA\rbrace,\lbrace CBG \rbrace \rbrace$ & 0 \\ \hline
			$C_4$ & $\lbrace\lbrace DEA\rbrace,\lbrace GCB \rbrace \rbrace$ & 1 \\ \hline
			$C_5$ & $\lbrace\lbrace DEA\rbrace,\lbrace CGB \rbrace \rbrace$ & 0 \\ \hline
			$C_6$ & $\lbrace\lbrace DEA\rbrace,\lbrace CBG \rbrace \rbrace$ & 0 \\ \hline
			$C_7$ & $\lbrace\lbrace EAD\rbrace,\lbrace GCB \rbrace \rbrace$ & 0 \\ \hline
			$C_8$ & $\lbrace\lbrace EAD\rbrace,\lbrace CGB \rbrace \rbrace$ & 0 \\ \hline
			$C_9$ & $\lbrace\lbrace EAD\rbrace,\lbrace CBG \rbrace \rbrace$ & 1 \\ \hline
			$C_{10}$ & $\lbrace\lbrace AED\rbrace,\lbrace GCB \rbrace \rbrace$ & 0 \\ \hline
			$C_{11}$ & $\lbrace\lbrace AED\rbrace,\lbrace CGB \rbrace \rbrace$ & 0 \\ \hline
			$C_{12}$ & $\lbrace\lbrace AED\rbrace,\lbrace CBG \rbrace \rbrace$ & 1 \\ \hline
			$C_{13}$ & $\lbrace\lbrace DAE\rbrace,\lbrace GCB \rbrace \rbrace$ & 1 \\ \hline
			$C_{14}$ & $\lbrace\lbrace DAE\rbrace,\lbrace CGB \rbrace \rbrace$ & 0 \\ \hline
			$C_{15}$ & $\lbrace\lbrace DAE\rbrace,\lbrace CBG \rbrace \rbrace$ & 0 \\ \hline
			$C_{16}$ & $\lbrace\lbrace ADE\rbrace,\lbrace GCB \rbrace \rbrace$ & 0 \\ \hline
			$C_{17}$ & $\lbrace\lbrace ADE\rbrace,\lbrace CGB \rbrace \rbrace$ & 0 \\ \hline
			$C_{18}$ & $\lbrace\lbrace ADE\rbrace,\lbrace CBG \rbrace \rbrace$ & 1 \\ \hline
		\end{tabular}
		\label{tab:abcd1}	
	\end{minipage}
	
	\noindent The $R$ matrix is given by
	\begin{align}
	\begin{split}
	&
	R=\left(
	\begin{array}{cccccccccccccccccc}
	\frac{1}{3} & 0 & 0 & -\frac{1}{6} & 0 & 0 & 0 & 0 & -\frac{1}{6} & 0 & 0 & -\frac{1}{6} & -\frac{1}{6} & 0 & 0 & 0 & 0 & \frac{1}{3} \\
	-\frac{1}{6} & \frac{1}{2} & 0 & -\frac{1}{6} & 0 & 0 & 0 & 0 & -\frac{1}{6} & 0 & 0 & -\frac{1}{6} & \frac{1}{3} & -\frac{1}{2} & 0 & 0 & 0 & \frac{1}{3} \\
	-\frac{1}{6} & 0 & \frac{1}{2} & -\frac{1}{6} & 0 & 0 & 0 & 0 & -\frac{1}{6} & 0 & 0 & -\frac{1}{6} & \frac{1}{3} & 0 & -\frac{1}{2} & 0 & 0 & \frac{1}{3} \\
	-\frac{1}{6} & 0 & 0 & \frac{1}{3} & 0 & 0 & 0 & 0 & -\frac{1}{6} & 0 & 0 & \frac{1}{3} & -\frac{1}{6} & 0 & 0 & 0 & 0 & -\frac{1}{6} \\
	\frac{1}{3} & -\frac{1}{2} & 0 & -\frac{2}{3} & 1 & 0 & 0 & 0 & -\frac{1}{6} & 0 & 0 & \frac{1}{3} & \frac{1}{3} & -\frac{1}{2} & 0 & 0 & 0 & -\frac{1}{6} \\
	\frac{1}{3} & 0 & -\frac{1}{2} & -\frac{2}{3} & 0 & 1 & 0 & 0 & -\frac{1}{6} & 0 & 0 & \frac{1}{3} & \frac{1}{3} & 0 & -\frac{1}{2} & 0 & 0 & -\frac{1}{6} \\
	-\frac{1}{6} & 0 & 0 & -\frac{1}{6} & 0 & 0 & \frac{1}{2} & 0 & -\frac{1}{6} & 0 & 0 & -\frac{1}{6} & \frac{1}{3} & 0 & 0 & -\frac{1}{2} & 0 & \frac{1}{3} \\
	-\frac{1}{6} & 0 & 0 & -\frac{1}{6} & 0 & 0 & 0 & \frac{1}{2} & -\frac{1}{6} & 0 & 0 & -\frac{1}{6} & \frac{1}{3} & 0 & 0 & 0 & -\frac{1}{2} & \frac{1}{3} \\
\end{array}
	\right)\,, \qquad \qquad \\ &
	D = \D{10} \, 
	\end{split}
	\label{eq:web1app}
	\end{align}
	\item[{\bf 13}.]  $\textbf{W}_{3,\text{II}}^{(2,1)}(1,3,3)$
	\\
	This Cweb has 18 diagrams, one of which is displayed below. The table gives
	the chosen order of the 18 shuffles of the gluon attachments, and the corresponding
	$s$ factors.
	\begin{minipage}{0.5\textwidth}
		\begin{figure}[H]
			\vspace{-2mm}
			\includegraphics[height=4cm,width=4cm]{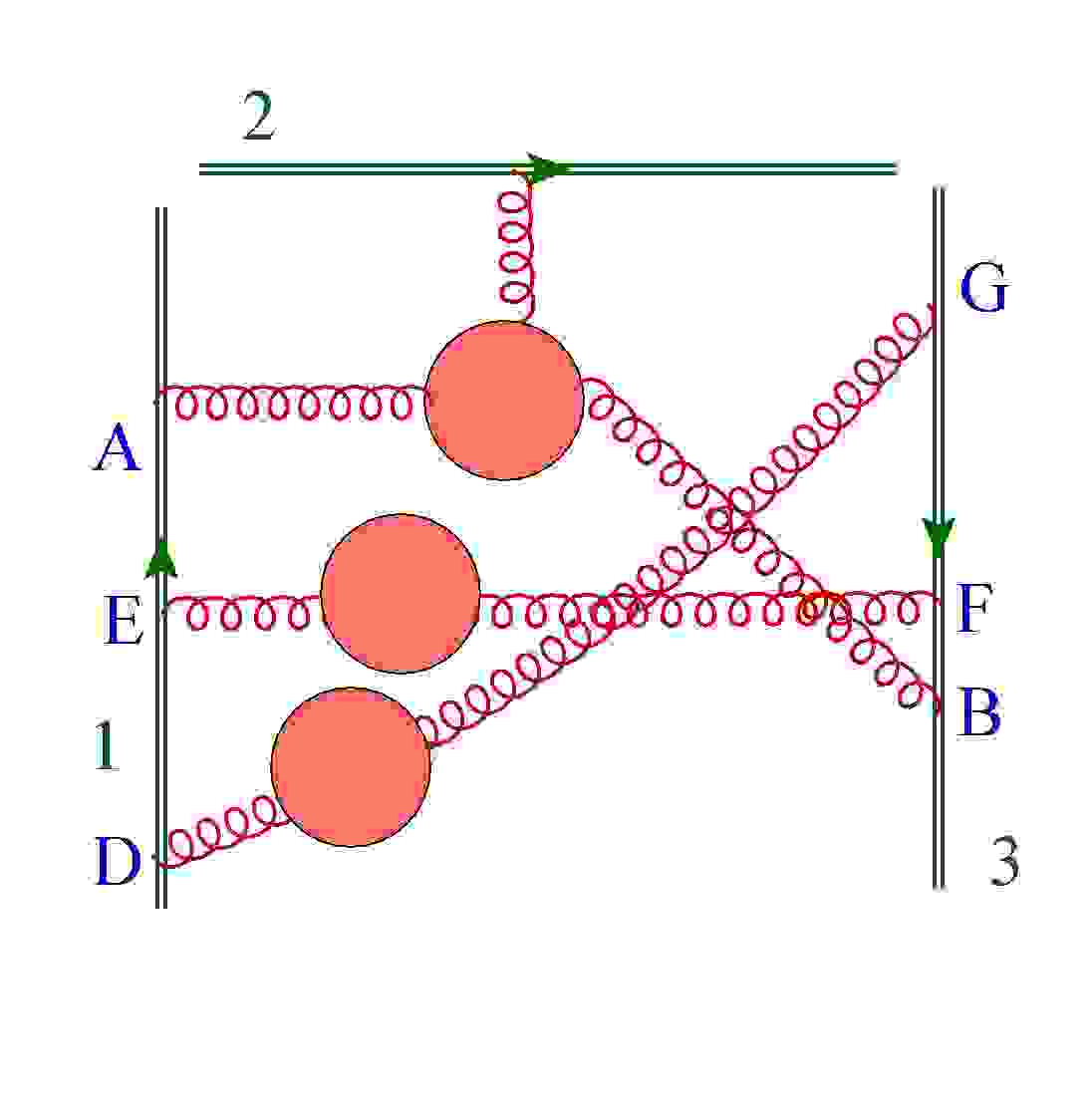}
		\end{figure}
	\end{minipage} 
	\hspace{-2cm}
	\begin{minipage}{0.46\textwidth}
		\vspace{2cm}
		\begin{tabular}{ | c | c | c |}
			\hline
			\textbf{Diagrams} & \textbf{Sequences} & \textbf{s-factors} \\ \hline
			$C_1$ & $\lbrace\lbrace DEA\rbrace,\lbrace GFB \rbrace \rbrace$ & 1 \\ \hline
			$C_2$ & $\lbrace\lbrace DEA\rbrace,\lbrace FGB \rbrace \rbrace$ & 0 \\ \hline
			$C_3$ & $\lbrace\lbrace DEA\rbrace,\lbrace GBF \rbrace \rbrace$ & 0 \\ \hline
			$C_4$ & $\lbrace\lbrace DEA\rbrace,\lbrace BGF \rbrace \rbrace$ & 0 \\ \hline
			$C_5$ & $\lbrace\lbrace DEA\rbrace,\lbrace FBG \rbrace \rbrace$ & 0 \\ \hline
			$C_6$ & $\lbrace\lbrace DEA\rbrace,\lbrace BFG \rbrace \rbrace$ & 0 \\ \hline
			$C_7$ & $\lbrace\lbrace DAE\rbrace,\lbrace GFB \rbrace \rbrace$ & 0 \\ \hline
			$C_8$ & $\lbrace\lbrace DAE\rbrace,\lbrace FGB \rbrace \rbrace$ & 0 \\ \hline
			$C_9$ & $\lbrace\lbrace DAE\rbrace,\lbrace GBF \rbrace \rbrace$ & 1 \\ \hline
			$C_{10}$ & $\lbrace\lbrace DAE\rbrace,\lbrace BGF \rbrace \rbrace$ & 0 \\ \hline
			$C_{11}$ & $\lbrace\lbrace DAE\rbrace,\lbrace FBG \rbrace \rbrace$ & 0 \\ \hline
			$C_{12}$ & $\lbrace\lbrace DAE\rbrace,\lbrace BFG \rbrace \rbrace$ & 0 \\ \hline
			$C_{13}$ & $\lbrace\lbrace ADE\rbrace,\lbrace GFB \rbrace \rbrace$ & 0\\ \hline
			$C_{14}$ & $\lbrace\lbrace ADE\rbrace,\lbrace FGB \rbrace \rbrace$ & 0 \\ \hline
			$C_{15}$ & $\lbrace\lbrace ADE\rbrace,\lbrace GBF \rbrace \rbrace$ & 0 \\ \hline
			$C_{16}$ & $\lbrace\lbrace ADE\rbrace,\lbrace BGF \rbrace \rbrace$ & 1 \\ \hline
			$C_{17}$ & $\lbrace\lbrace ADE\rbrace,\lbrace FBG \rbrace \rbrace$ & 0 \\ \hline
			$C_{18}$ & $\lbrace\lbrace ADE\rbrace,\lbrace BFG \rbrace \rbrace$ & 0 \\ \hline
		\end{tabular}
		\label{tab:abcd1}	
	\end{minipage}
	
	\noindent The $R$ matrix is given by
	\begin{align}
	\begin{split}
	&
	R=\left(
	\begin{array}{cccccccccccccccccc}
	\frac{1}{6} & 0 & 0 & 0 & 0 & 0 & 0 & 0 & -\frac{1}{3} & 0 & 0 & 0 & 0 & 0 & 0 & \frac{1}{6} & 0 & 0 \\
	-\frac{1}{3} & \frac{1}{2} & 0 & 0 & 0 & 0 & 0 & 0 & -\frac{1}{3} & 0 & 0 & 0 & 0 & 0 & 0 & \frac{2}{3} & 0 & -\frac{1}{2} \\
	-\frac{1}{3} & 0 & \frac{1}{2} & 0 & 0 & 0 & 0 & 0 & \frac{1}{6} & -\frac{1}{2} & 0 & 0 & 0 & 0 & 0 & \frac{1}{6} & 0 & 0 \\
	\frac{1}{6} & 0 & -1 & 1 & 0 & 0 & 0 & 0 & \frac{2}{3} & -1 & 0 & 0 & 0 & 0 & 0 & \frac{1}{6} & 0 & 0 \\
	\frac{1}{6} & -\frac{1}{2} & -\frac{1}{2} & 0 & 1 & 0 & 0 & 0 & \frac{1}{6} & -\frac{1}{2} & 0 & 0 & 0 & 0 & 0 & \frac{2}{3} & 0 & -\frac{1}{2} \\
	\frac{2}{3} & -\frac{1}{2} & -1 & 0 & 0 & 1 & 0 & 0 & \frac{2}{3} & -1 & 0 & 0 & 0 & 0 & 0 & \frac{2}{3} & 0 & -\frac{1}{2} \\
	-\frac{1}{3} & 0 & 0 & 0 & 0 & 0 & \frac{1}{2} & 0 & \frac{1}{6} & 0 & 0 & 0 & 0 & 0 & -\frac{1}{2} & \frac{1}{6} & 0 & 0 \\
	\frac{1}{6} & -\frac{1}{2} & 0 & 0 & 0 & 0 & -\frac{1}{2} & 1 & \frac{1}{6} & 0 & 0 & 0 & 0 & 0 & -\frac{1}{2} & \frac{2}{3} & 0 & -\frac{1}{2} \\
\end{array}
	\right)\,, \qquad \qquad  \\&
	D = \D{13} \, 
	\end{split}
	\label{eq:web1app}
	\end{align}
	\item[{\bf 14}.]  $\textbf{W}_{3}^{(1,0,1)}(1,1,4)$
	\\
	This Cweb has four diagrams, one of which is displayed below. The table gives
	the chosen order of the four shuffles of the gluon attachments, and the corresponding
	$s$ factors.
	\begin{minipage}{0.5\textwidth}
		\begin{figure}[H]
			\vspace{-2mm}
			\includegraphics[height=4cm,width=4cm]{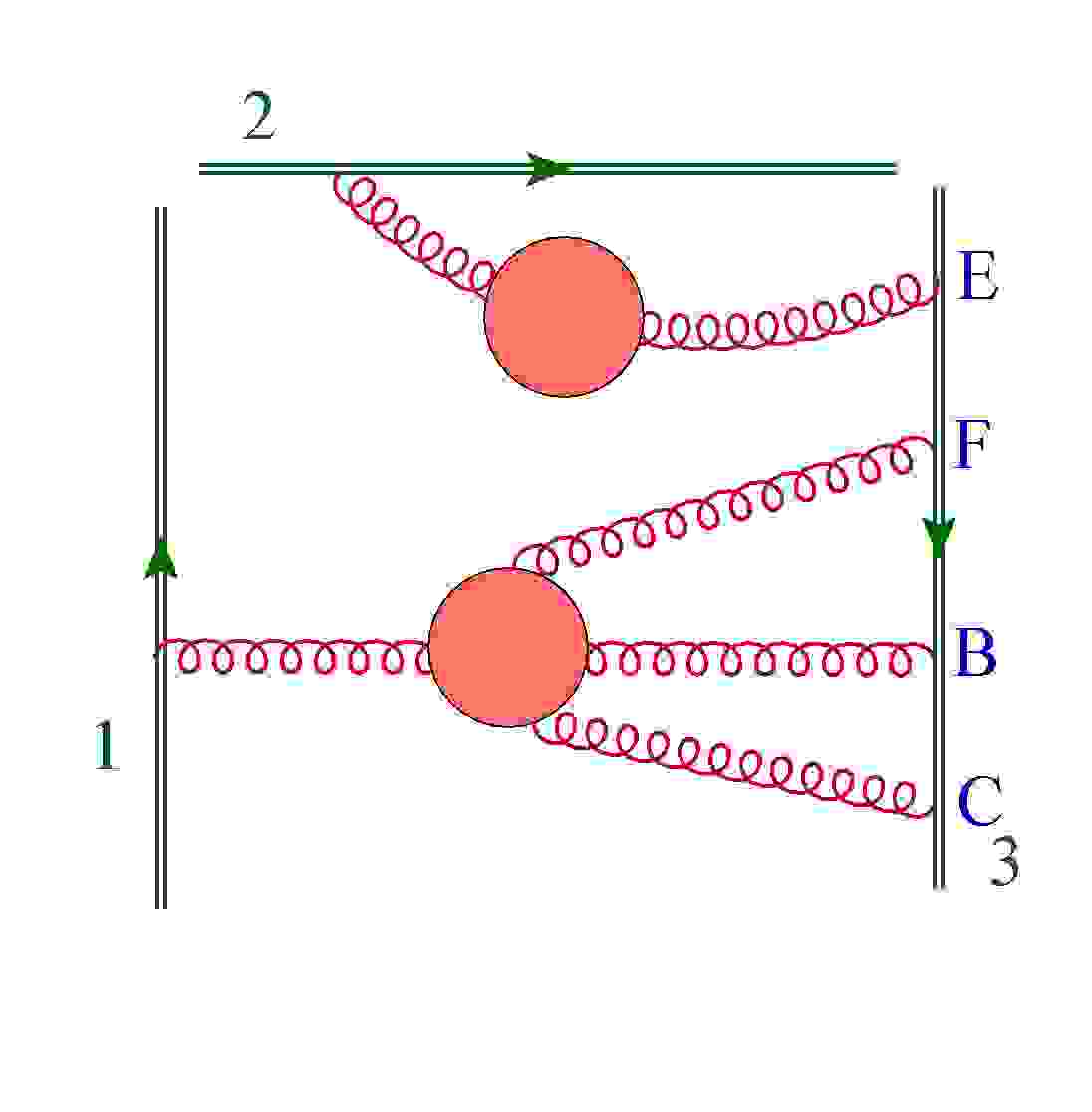}
		\end{figure}
	\end{minipage} 
	\hspace{-2cm}
	\begin{minipage}{0.46\textwidth}
		\vspace{2cm}
		\begin{tabular}{ | c | c | c |}
			\hline
			\textbf{Diagrams} & \textbf{Sequences} & \textbf{s-factors} \\ \hline
			$C_1$ & $\lbrace\lbrace EFBC \rbrace \rbrace$ & 1 \\ \hline
			$C_2$ & $\lbrace\lbrace FEBC \rbrace \rbrace$ & 0 \\ \hline
			$C_3$ & $\lbrace\lbrace FBEC \rbrace \rbrace$ & 0 \\ \hline
			$C_4$ & $\lbrace\lbrace FBCE \rbrace \rbrace$ & 1 \\ \hline
		\end{tabular}
		\label{tab:abcd1}	
	\end{minipage}
	
	\noindent The $R$ matrix is given by
	\begin{align}
	\begin{split}
	&
	R=\left(
	\begin{array}{cccc}
	\frac{1}{2} & 0 & 0 & -\frac{1}{2} \\
	-\frac{1}{2} & 1 & 0 & -\frac{1}{2} \\
	-\frac{1}{2} & 0 & 1 & -\frac{1}{2} \\
	-\frac{1}{2} & 0 & 0 & \frac{1}{2} \\
	\end{array}
	\right)\,, \qquad \qquad 
	D = \D{3} \, 
	\end{split}
	\label{eq:web1app}
	\end{align}
	\item[{\bf 15}.]  $\textbf{W}_{3}^{(0,2)}(1,1,4)$
	\\
	This Cweb has six diagrams, one of which is displayed below. The table gives
	the chosen order of the six shuffles of the gluon attachments, and the corresponding
	$s$ factors.
	\begin{minipage}{0.5\textwidth}
		\begin{figure}[H]
			\vspace{-2mm}
			\includegraphics[height=4cm,width=4cm]{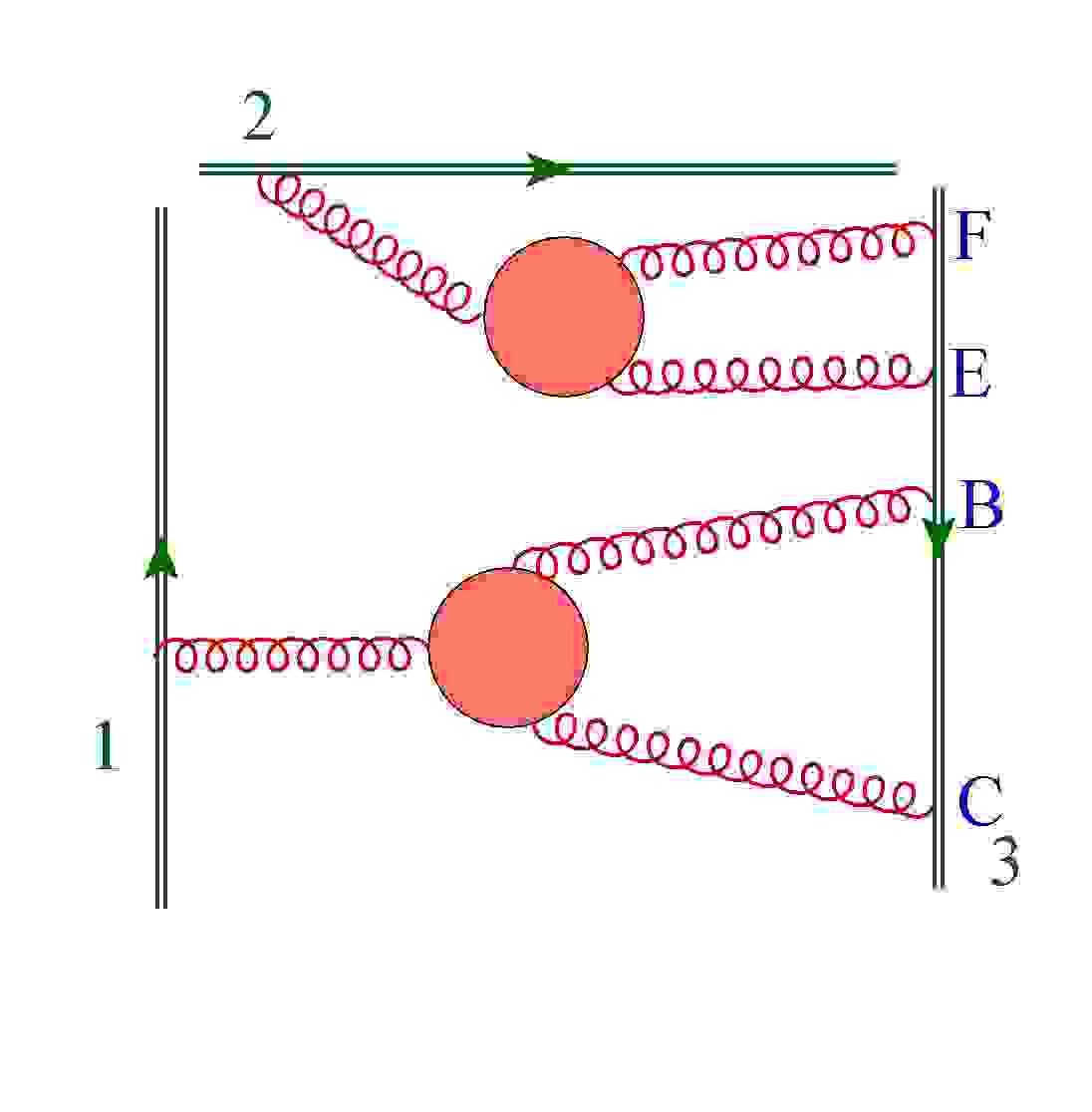}
		\end{figure}
	\end{minipage} 
	\hspace{-2cm}
	\begin{minipage}{0.46\textwidth}
		\vspace{2cm}
		\begin{tabular}{ | c | c | c |}
			\hline
			\textbf{Diagrams} & \textbf{Sequences} & \textbf{s-factors} \\ \hline
			$C_1$ & $\lbrace\lbrace FEBC \rbrace \rbrace$ & 1 \\ \hline
			$C_2$ & $\lbrace\lbrace FBEC \rbrace \rbrace$ & 0 \\ \hline
			$C_3$ & $\lbrace\lbrace BFEC \rbrace \rbrace$ & 0 \\ \hline
			$C_4$ & $\lbrace\lbrace FBCE \rbrace \rbrace$ & 0 \\ \hline
			$C_5$ & $\lbrace\lbrace BFCE \rbrace \rbrace$ & 0 \\ \hline
			$C_6$ & $\lbrace\lbrace BCFE \rbrace \rbrace$ & 1 \\ \hline
		\end{tabular}
		\label{tab:abcd1}	
	\end{minipage}
	
	\noindent The $R$ matrix is given by
	\begin{align}
	\begin{split}
	&
	R=\left(
	\begin{array}{cccccc}
	\frac{1}{2} & 0 & 0 & 0 & 0 & -\frac{1}{2} \\
	-\frac{1}{2} & 1 & 0 & 0 & 0 & -\frac{1}{2} \\
	-\frac{1}{2} & 0 & 1 & 0 & 0 & -\frac{1}{2} \\
	-\frac{1}{2} & 0 & 0 & 1 & 0 & -\frac{1}{2} \\
	-\frac{1}{2} & 0 & 0 & 0 & 1 & -\frac{1}{2} \\
	-\frac{1}{2} & 0 & 0 & 0 & 0 & \frac{1}{2} \\
	\end{array}
	\right)\,, \qquad \qquad 
	D = \D{5} \, 
	\end{split}
	\label{eq:web1app}
	\end{align}
	\item[{\bf 16}.]  $\textbf{W}_{3}^{(2,1)}(1,2,4)$
	\\
	This Cweb has 24 diagrams, one of which is displayed below. The table gives
	the chosen order of the 24 shuffles of the gluon attachments, and the corresponding
	$s$ factors.
	\begin{minipage}{0.5\textwidth}
		\begin{figure}[H]
			\vspace{-2mm}
			\includegraphics[height=4cm,width=4cm]{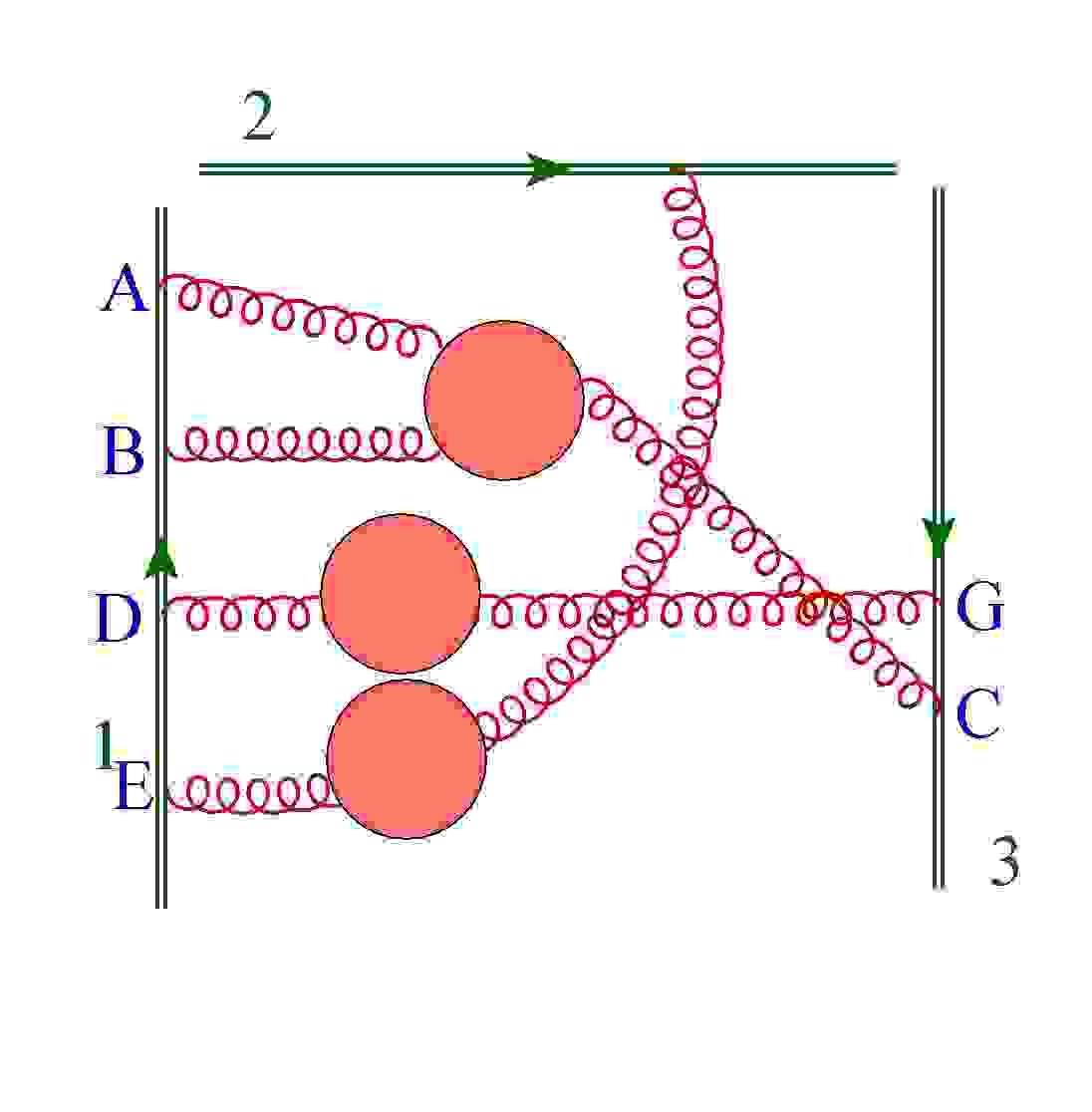}
		\end{figure}
	\end{minipage} 
	\hspace{-2cm}
	\begin{minipage}{0.46\textwidth}
		\vspace{2cm}
		\begin{tabular}{ | c | c | c |}
			\hline
			\textbf{Diagrams} & \textbf{Sequences} & \textbf{s-factors} \\ \hline
			$C_1$ & $\lbrace\lbrace EDBA \rbrace, \lbrace GC\rbrace \rbrace$ & 1 \\ \hline
			$C_2$ & $\lbrace\lbrace EDBA \rbrace, \lbrace CG\rbrace \rbrace$ & 0 \\ \hline
			$C_3$ & $\lbrace\lbrace DEBA\rbrace, \lbrace GC\rbrace \rbrace$ & 1 \\ \hline
			$C_4$ & $\lbrace\lbrace DEBA \rbrace, \lbrace CG\rbrace \rbrace$ & 0 \\ \hline
			$C_5$ & $\lbrace\lbrace EBDA \rbrace, \lbrace GC\rbrace \rbrace$ & 0 \\ \hline
			$C_6$ & $\lbrace\lbrace EBDA \rbrace, \lbrace CG\rbrace \rbrace$ & 0 \\ \hline
			$C_7$ & $\lbrace\lbrace BEDA \rbrace, \lbrace GC\rbrace \rbrace$ & 0 \\
\hline
		\end{tabular}
		\label{tab:abcd1}	
	\end{minipage}
	
	\noindent The $R$ matrix is given by
	\begin{align}
	\begin{split}
	&
	R=\left(
	\begin{array}{cccccccccccccccccccccccc}
	\frac{1}{3} & 0 & -\frac{1}{6} & 0 & 0 & 0 & 0 & 0 & 0 & 0 & 0 & 0 & 0 & -\frac{1}{6} & 0 & 0 & 0 & -\frac{1}{6} & -\frac{1}{6} & 0 & 0 & 0 & 0 & \frac{1}{3} \\
	-\frac{1}{6} & \frac{1}{2} & -\frac{1}{6} & 0 & 0 & 0 & 0 & 0 & 0 & 0 & 0 & 0 & 0 & -\frac{1}{6} & 0 & 0 & 0 & -\frac{1}{6} & \frac{1}{3} & -\frac{1}{2} & 0 & 0 & 0 & \frac{1}{3} \\
	-\frac{1}{6} & 0 & \frac{1}{3} & 0 & 0 & 0 & 0 & 0 & 0 & 0 & 0 & 0 & 0 & -\frac{1}{6} & 0 & 0 & 0 & \frac{1}{3} & -\frac{1}{6} & 0 & 0 & 0 & 0 & -\frac{1}{6} \\
	\frac{1}{3} & -\frac{1}{2} & -\frac{2}{3} & 1 & 0 & 0 & 0 & 0 & 0 & 0 & 0 & 0 & 0 & -\frac{1}{6} & 0 & 0 & 0 & \frac{1}{3} & \frac{1}{3} & -\frac{1}{2} & 0 & 0 & 0 & -\frac{1}{6} \\
	-\frac{1}{6} & 0 & -\frac{1}{6} & 0 & \frac{1}{2} & 0 & 0 & 0 & 0 & 0 & 0 & 0 & 0 & -\frac{1}{6} & 0 & 0 & 0 & -\frac{1}{6} & \frac{1}{3} & 0 & -\frac{1}{2} & 0 & 0 & \frac{1}{3} \\
	-\frac{1}{6} & 0 & -\frac{1}{6} & 0 & 0 & \frac{1}{2} & 0 & 0 & 0 & 0 & 0 & 0 & 0 & -\frac{1}{6} & 0 & 0 & 0 & -\frac{1}{6} & \frac{1}{3} & 0 & 0 & -\frac{1}{2} & 0 & \frac{1}{3} \\
	-\frac{1}{6} & 0 & \frac{1}{3} & 0 & -\frac{1}{2} & 0 & 1 & 0 & -\frac{1}{2} & 0 & 0 & 0 & 0 & \frac{1}{3} & 0 & -\frac{1}{2} & 0 & -\frac{1}{6} & \frac{1}{3} & 0 & -\frac{1}{2} & 0 & 0 & \frac{1}{3} \\
	-\frac{1}{6} & 0 & \frac{1}{3} & 0 & 0 & -\frac{1}{2} & 0 & 1 & -\frac{1}{2} & 0 & 0 & 0 & 0 & \frac{1}{3} & 0 & -\frac{1}{2} & 0 & -\frac{1}{6} & \frac{1}{3} & 0 & 0 & -\frac{1}{2} & 0 & \frac{1}{3} \\
\end{array}
	\right)\,, \qquad \qquad \\&
	D = \D{15} \, 
	\end{split}
	\label{eq:web1app}
	\end{align}
	\item[{\bf 17}.]  $\textbf{W}_{3}^{(0,2)}(1,2,4)$
	\\
	This Cweb has 12 diagrams, one of which is displayed below. The table gives
	the chosen order of the 12 shuffles of the gluon attachments, and the corresponding
	$s$ factors.
	\begin{minipage}{0.5\textwidth}
		\begin{figure}[H]
			\vspace{-2mm}
			\includegraphics[height=4cm,width=4cm]{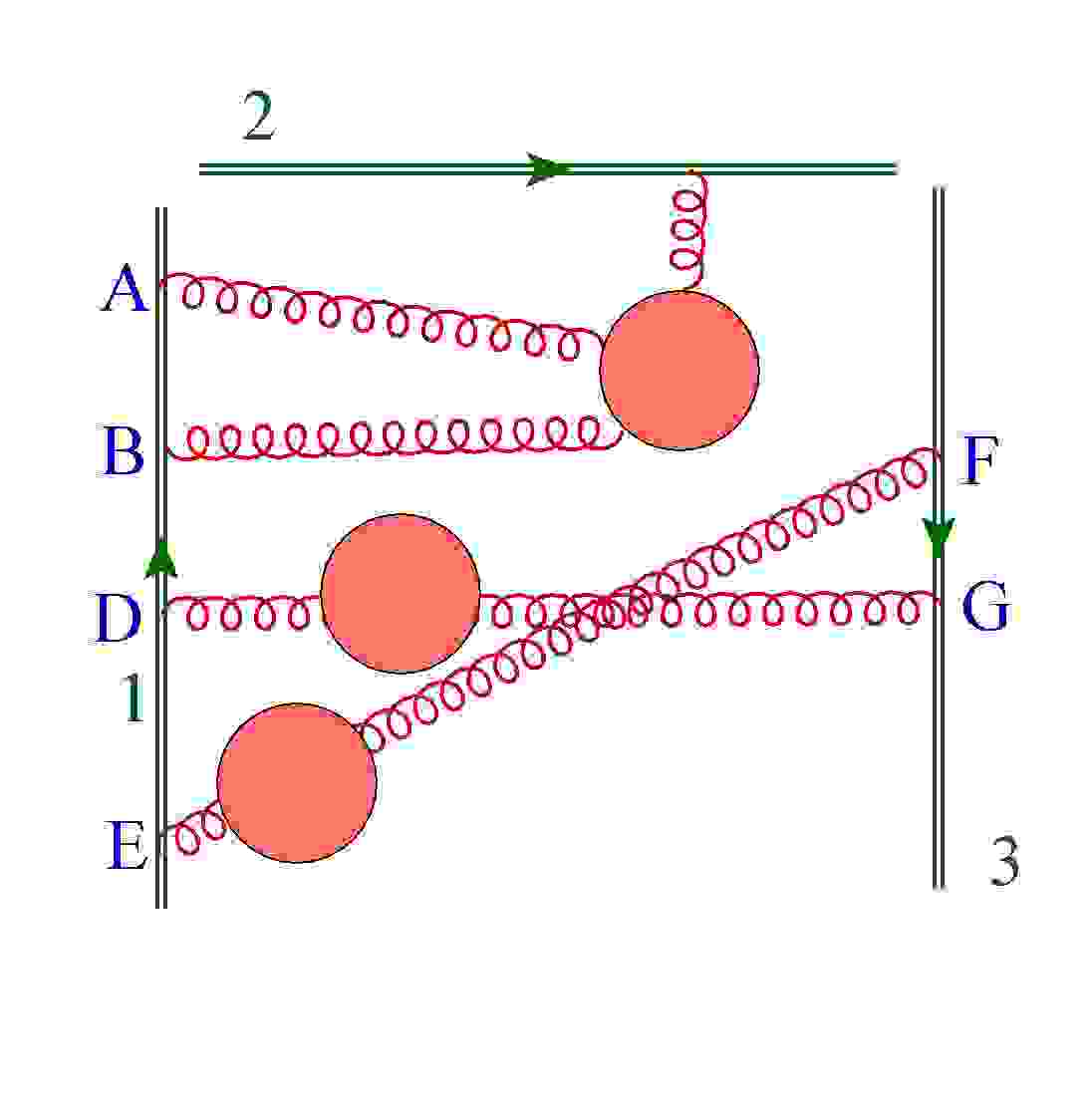}
		\end{figure}
	\end{minipage} 
	\hspace{-2cm}
	\begin{minipage}{0.46\textwidth}
		\vspace{2cm}
		\begin{tabular}{ | c | c | c |}
			\hline
			\textbf{Diagrams} & \textbf{Sequences} & \textbf{s-factors} \\ \hline
			$C_1$ & $\lbrace\lbrace EDBA\rbrace,\lbrace FG \rbrace \rbrace$ & 1 \\ \hline
			$C_2$ & $\lbrace\lbrace EDBA\rbrace,\lbrace GF \rbrace \rbrace$ & 0 \\ \hline
			$C_3$ & $\lbrace\lbrace EBDA\rbrace,\lbrace FG \rbrace \rbrace$ & 0 \\ \hline
			$C_4$ & $\lbrace\lbrace EBDA\rbrace,\lbrace GF \rbrace \rbrace$ & 0 \\ \hline
			$C_5$ & $\lbrace\lbrace BEDA\rbrace,\lbrace FG \rbrace \rbrace$ & 0 \\ \hline
			$C_6$ & $\lbrace\lbrace BEDA\rbrace,\lbrace GF \rbrace \rbrace$ & 0 \\ \hline
			$C_7$ & $\lbrace\lbrace EBAD\rbrace,\lbrace FG \rbrace \rbrace$ & 1 \\ \hline
			$C_8$ & $\lbrace\lbrace EBAD\rbrace,\lbrace GF \rbrace \rbrace$ & 0 \\ \hline
			$C_9$ & $\lbrace\lbrace BEAD\rbrace,\lbrace FG \rbrace \rbrace$ & 0 \\ \hline
			$C_{10}$ & $\lbrace\lbrace BEAD\rbrace,\lbrace GF \rbrace \rbrace$ & 0 \\ \hline
			$C_{11}$ & $\lbrace\lbrace BEAD\rbrace,\lbrace FG \rbrace \rbrace$ & 1 \\ \hline
			$C_{12}$ & $\lbrace\lbrace BEAD\rbrace,\lbrace GF \rbrace \rbrace$ & 0 \\ \hline
		\end{tabular}
		\label{tab:abcd1}	
	\end{minipage}
	
	\noindent The $R$ matrix is given by
	\begin{align}
	\begin{split}
	&
	R=\left(
	\begin{array}{cccccccccccc}
	\frac{1}{6} & 0 & 0 & 0 & 0 & 0 & -\frac{1}{3} & 0 & 0 & 0 & \frac{1}{6} & 0 \\
	-\frac{1}{3} & \frac{1}{2} & 0 & 0 & 0 & 0 & -\frac{1}{3} & 0 & 0 & 0 & \frac{2}{3} & -\frac{1}{2} \\
	-\frac{1}{3} & 0 & \frac{1}{2} & 0 & 0 & 0 & \frac{1}{6} & 0 & -\frac{1}{2} & 0 & \frac{1}{6} & 0 \\
	\frac{1}{6} & -\frac{1}{2} & -\frac{1}{2} & 1 & 0 & 0 & \frac{1}{6} & 0 & -\frac{1}{2} & 0 & \frac{2}{3} & -\frac{1}{2} \\
	\frac{1}{6} & 0 & -1 & 0 & 1 & 0 & \frac{2}{3} & 0 & -1 & 0 & \frac{1}{6} & 0 \\
	\frac{2}{3} & -\frac{1}{2} & -1 & 0 & 0 & 1 & \frac{2}{3} & 0 & -1 & 0 & \frac{2}{3} & -\frac{1}{2} \\
	-\frac{1}{3} & 0 & 0 & 0 & 0 & 0 & \frac{2}{3} & 0 & 0 & 0 & -\frac{1}{3} & 0 \\
	\frac{1}{6} & -\frac{1}{2} & 0 & 0 & 0 & 0 & -\frac{1}{3} & 1 & 0 & 0 & \frac{1}{6} & -\frac{1}{2} \\
	\frac{1}{6} & 0 & -\frac{1}{2} & 0 & 0 & 0 & \frac{1}{6} & 0 & \frac{1}{2} & 0 & -\frac{1}{3} & 0 \\
	\frac{2}{3} & -\frac{1}{2} & -\frac{1}{2} & 0 & 0 & 0 & \frac{1}{6} & 0 & -\frac{1}{2} & 1 & \frac{1}{6} & -\frac{1}{2} \\
	\frac{1}{6} & 0 & 0 & 0 & 0 & 0 & -\frac{1}{3} & 0 & 0 & 0 & \frac{1}{6} & 0 \\
	\frac{2}{3} & -\frac{1}{2} & 0 & 0 & 0 & 0 & -\frac{1}{3} & 0 & 0 & 0 & -\frac{1}{3} & \frac{1}{2} \\
	\end{array}
	\right), \qquad \qquad  \\&
	D = \D{8} \, 
	\end{split}
	\label{eq:web1app}
	\end{align}
	\item[{\bf 18}.]  $\textbf{W}_{3}^{(4)}(2,3,3)$
	\\
	This Cweb has 36 diagrams, one of which is displayed below. This is the biggest web mixing matrix, the table gives
	the chosen order of the 36 shuffles of the gluon attachments, and the corresponding
	$s$ factors. \\
	\begin{minipage}{0.5\textwidth}
		\begin{figure}[H]
			\vspace{-2mm}
			\includegraphics[height=4cm,width=4cm]{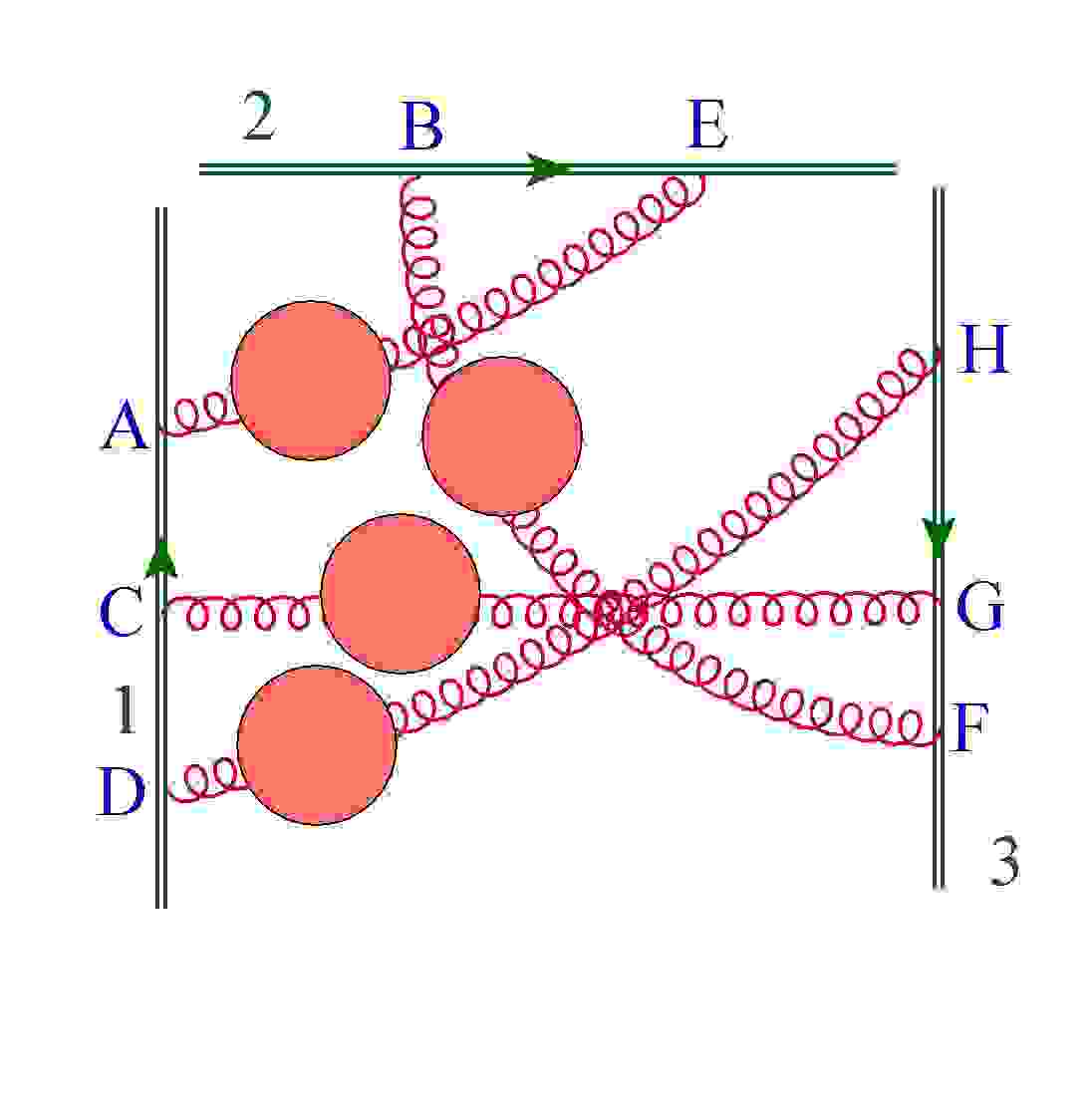}
		\end{figure}
	\end{minipage} 
	\hspace{-2cm}
	\begin{minipage}{0.46\textwidth}
		\vspace{2cm}
		\begin{tabular}{ | c | c | c |}
			\hline
			\textbf{Diagrams} & \textbf{Sequences} & \textbf{s-factors} \\ \hline
			$C_1$ & $\lbrace\lbrace DCA\rbrace,\lbrace BE\rbrace,\lbrace HGF \rbrace \rbrace$ & 1 \\ \hline
			$C_2$ & $\lbrace\lbrace DCA\rbrace,\lbrace BE\rbrace,\lbrace GHF \rbrace \rbrace$ & 0 \\ \hline
			$C_3$ & $\lbrace\lbrace DCA\rbrace,\lbrace BE\rbrace,\lbrace HFG \rbrace \rbrace$ & 1 \\ \hline
			$C_4$ & $\lbrace\lbrace DCA\rbrace,\lbrace BE\rbrace,\lbrace FGH \rbrace \rbrace$ & 1 \\ \hline
			$C_5$ & $\lbrace\lbrace DCA\rbrace,\lbrace BE\rbrace,\lbrace GFH \rbrace \rbrace$ & 0 \\ \hline
			$C_6$ & $\lbrace\lbrace DCA\rbrace,\lbrace BE\rbrace,\lbrace FGH \rbrace \rbrace$ & 0 \\ \hline	
			$C_7$ & $\lbrace\lbrace DCA\rbrace,\lbrace EB\rbrace,\lbrace HGF \rbrace \rbrace$ & 1 \\
\hline
		\end{tabular}
		\label{tab:abcd1}	
	\end{minipage}
	
	\noindent The $R$ matrix is given by
	
	\resizebox{.8\linewidth}{!}{
		\begin{minipage}{\linewidth}
			\begin{align}
			\begin{split}
			&
			R=\left(
			\begin{array}{cccccccccccccccccccccccccccccccccccc}
			\frac{1}{6} & 0 & -\frac{1}{6} & 0 & 0 & 0 & 0 & 0 & 0 & 0 & 0 & 0 & 0 & 0 & -\frac{1}{6} & \frac{1}{6} & 0 & 0 & -\frac{1}{6} & 0 & \frac{1}{6} & 0 & 0 & 0 & 0 & 0 & 0 & 0 & 0 & 0 & 0 & 0 & \frac{1}{6} & -\frac{1}{6} & 0 & 0 \\
			-\frac{1}{6} & \frac{1}{3} & -\frac{1}{6} & \frac{1}{6} & 0 & -\frac{1}{6} & \frac{1}{6} & -\frac{1}{6} & 0 & 0 & 0 & 0 & 0 & 0 & -\frac{1}{6} & \frac{1}{6} & 0 & 0 & -\frac{1}{6} & 0 & \frac{1}{6} & 0 & 0 & 0 & 0 & 0 & 0 & \frac{1}{6} & 0 & -\frac{1}{6} & \frac{1}{6} & -\frac{1}{6} & \frac{1}{6} & -\frac{1}{2} & 0 & \frac{1}{3} \\
			-\frac{1}{6} & 0 & \frac{1}{3} & -\frac{1}{6} & 0 & 0 & 0 & 0 & 0 & 0 & 0 & 0 & 0 & 0 & 0 & 0 & 0 & 0 & 0 & 0 & 0 & 0 & 0 & 0 & 0 & 0 & 0 & 0 & 0 & 0 & \frac{1}{6} & 0 & -\frac{1}{3} & \frac{1}{6} & 0 & 0 \\
			0 & 0 & -\frac{1}{6} & \frac{1}{6} & 0 & 0 & 0 & 0 & 0 & 0 & 0 & 0 & 0 & 0 & \frac{1}{6} & -\frac{1}{6} & 0 & 0 & \frac{1}{6} & 0 & -\frac{1}{6} & 0 & 0 & 0 & 0 & 0 & 0 & 0 & 0 & 0 & -\frac{1}{6} & 0 & \frac{1}{6} & 0 & 0 & 0 \\
			0 & -\frac{1}{6} & -\frac{1}{6} & 0 & \frac{1}{2} & -\frac{1}{6} & \frac{1}{6} & -\frac{1}{6} & 0 & 0 & 0 & 0 & 0 & 0 & 0 & 0 & 0 & 0 & 0 & 0 & 0 & 0 & 0 & 0 & 0 & 0 & 0 & \frac{1}{6} & 0 & -\frac{1}{6} & -\frac{1}{6} & \frac{1}{3} & \frac{1}{6} & -\frac{1}{6} & -\frac{1}{2} & \frac{1}{3} \\
			\frac{1}{6} & -\frac{1}{6} & -\frac{1}{6} & -\frac{1}{6} & 0 & \frac{1}{3} & \frac{1}{6} & -\frac{1}{6} & 0 & 0 & 0 & 0 & 0 & 0 & \frac{1}{6} & -\frac{1}{6} & 0 & 0 & \frac{1}{6} & 0 & -\frac{1}{6} & 0 & 0 & 0 & 0 & 0 & 0 & \frac{1}{6} & 0 & -\frac{1}{6} & -\frac{1}{2} & \frac{1}{3} & \frac{1}{6} & \frac{1}{6} & 0 & -\frac{1}{6} \\
			0 & 0 & -\frac{1}{6} & 0 & 0 & 0 & \frac{1}{6} & 0 & 0 & 0 & 0 & 0 & 0 & 0 & \frac{1}{6} & \frac{1}{6} & 0 & 0 & -\frac{1}{6} & 0 & -\frac{1}{6} & 0 & 0 & 0 & 0 & 0 & 0 & -\frac{1}{6} & 0 & 0 & 0 & 0 & \frac{1}{6} & 0 & 0 & 0 \\
\end{array}
			\right)\,,
			\end{split}
			\label{eq:web1app}
			\end{align}
		\end{minipage}
	}
	\\
	$D = \D{19}$ 
	\item[{\bf 19}.]  $\textbf{W}_{3}^{(4)}(2,2,4)$
	\\
	This Cweb has 24 diagrams, one of which is displayed below. The table gives
	the chosen order of the 24 shuffles of the gluon attachments, and the corresponding
	$s$ factors.
	\begin{minipage}{0.5\textwidth}
		\begin{figure}[H]
			\vspace{-2mm}
			\includegraphics[height=4cm,width=4cm]{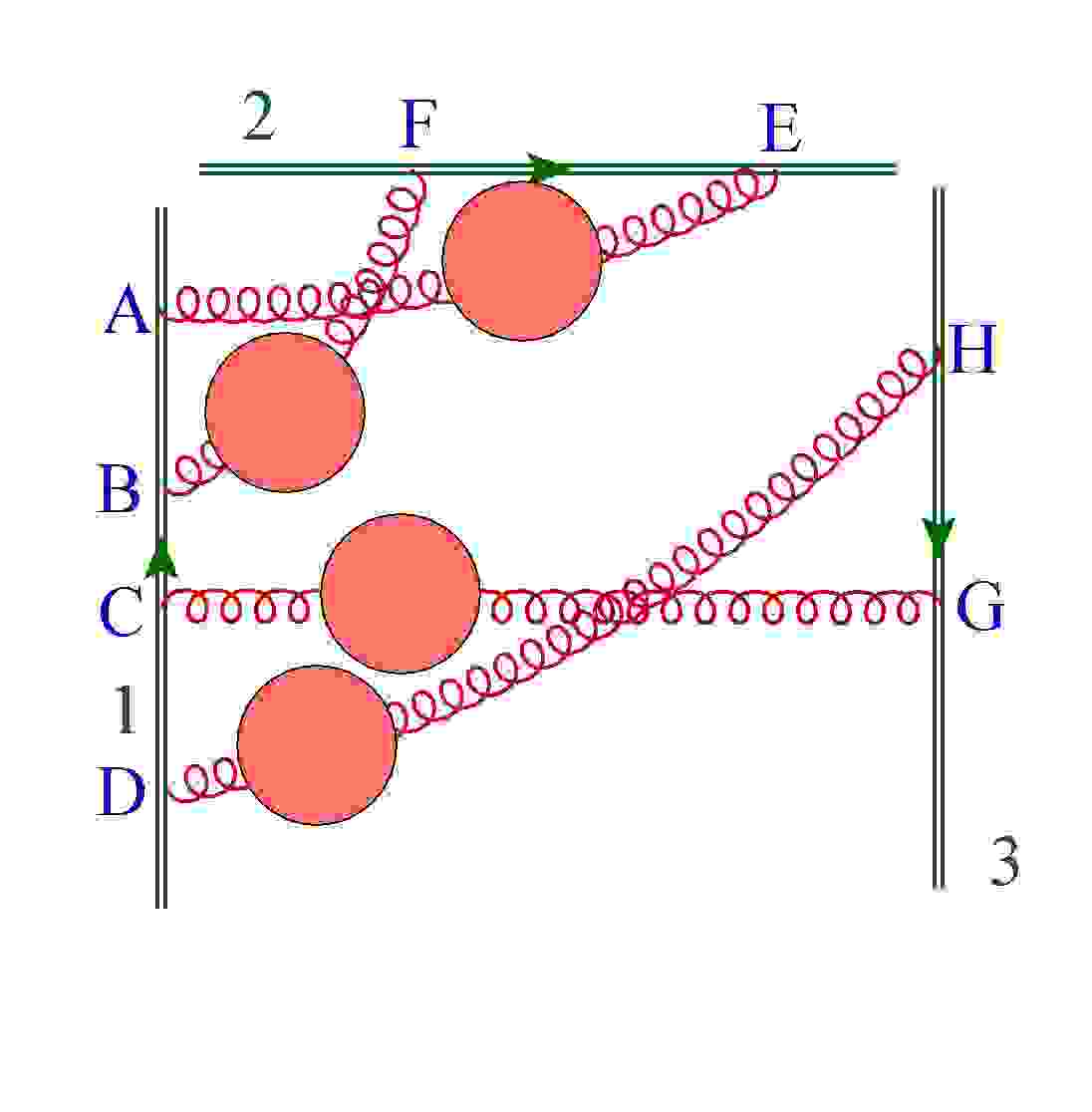}
		\end{figure}
	\end{minipage} 
	\hspace{-2cm}
	\begin{minipage}{0.46\textwidth}
		\vspace{2cm}
		\begin{tabular}{ | c | c | c |}
			\hline
			\textbf{Diagrams} & \textbf{Sequences} & \textbf{s-factors} \\ \hline
			$C_1$ & $\lbrace\lbrace DCBA\rbrace, \lbrace EF\rbrace , \lbrace HG\rbrace \rbrace$ & 1 \\ \hline
			$C_2$ & $\lbrace\lbrace DCBA\rbrace, \lbrace EF\rbrace , \lbrace GH\rbrace \rbrace$ & 0 \\ \hline
			$C_3$ & $\lbrace\lbrace DCBA\rbrace, \lbrace FE\rbrace , \lbrace HG\rbrace \rbrace$ & 0 \\ \hline
			$C_4$ & $\lbrace\lbrace DCBA\rbrace, \lbrace FE\rbrace , \lbrace GH\rbrace \rbrace$ & 0 \\ \hline
			$C_5$ & $\lbrace\lbrace DBCA\rbrace, \lbrace EF\rbrace , \lbrace HG\rbrace \rbrace$ & 1 \\ \hline
			$C_6$ & $\lbrace\lbrace DBCA\rbrace, \lbrace EF\rbrace , \lbrace GH\rbrace \rbrace$ & 0 \\ \hline
			$C_7$ & $\lbrace\lbrace DBCA\rbrace, \lbrace FE\rbrace , \lbrace HG\rbrace \rbrace$ & 0 \\
\hline
		\end{tabular}
		\label{tab:abcd1}	
	\end{minipage}
	
	\noindent The $R$ matrix is given by
	\begin{align}
	\begin{split}
	&
	R=\left(
	\begin{array}{cccccccccccccccccccccccc}
	\frac{1}{6} & 0 & 0 & 0 & -\frac{1}{3} & 0 & 0 & 0 & 0 & 0 & 0 & 0 & 0 & 0 & 0 & 0 & \frac{1}{3} & 0 & 0 & 0 & -\frac{1}{6} & 0 & 0 & 0 \\
	0 & \frac{1}{6} & 0 & 0 & -\frac{1}{3} & 0 & 0 & 0 & \frac{1}{3} & -\frac{1}{3} & 0 & 0 & 0 & 0 & 0 & 0 & \frac{1}{3} & 0 & 0 & 0 & -\frac{1}{3} & \frac{1}{6} & 0 & 0 \\
	0 & 0 & \frac{1}{6} & 0 & -\frac{1}{3} & 0 & 0 & 0 & 0 & 0 & 0 & 0 & \frac{1}{3} & 0 & -\frac{1}{3} & 0 & \frac{1}{3} & 0 & 0 & 0 & -\frac{1}{3} & 0 & \frac{1}{6} & 0 \\
	\frac{1}{3} & -\frac{1}{3} & -\frac{1}{3} & \frac{1}{2} & -\frac{1}{3} & 0 & 0 & 0 & \frac{1}{3} & -\frac{1}{3} & 0 & 0 & \frac{1}{3} & 0 & -\frac{1}{3} & 0 & \frac{1}{3} & 0 & 0 & 0 & -1 & \frac{2}{3} & \frac{2}{3} & -\frac{1}{2} \\
	-\frac{1}{6} & 0 & 0 & 0 & \frac{1}{3} & 0 & 0 & 0 & 0 & 0 & 0 & 0 & 0 & 0 & 0 & 0 & -\frac{1}{3} & 0 & 0 & 0 & \frac{1}{6} & 0 & 0 & 0 \\
	\frac{1}{6} & -\frac{1}{3} & 0 & 0 & -\frac{1}{6} & \frac{1}{2} & 0 & 0 & -\frac{1}{6} & \frac{1}{6} & 0 & 0 & 0 & 0 & 0 & 0 & \frac{1}{6} & -\frac{1}{2} & 0 & 0 & 0 & \frac{1}{6} & 0 & 0 \\
	\frac{1}{6} & 0 & -\frac{1}{3} & 0 & -\frac{1}{6} & 0 & \frac{1}{2} & 0 & 0 & 0 & 0 & 0 & -\frac{1}{6} & 0 & \frac{1}{6} & 0 & \frac{1}{6} & 0 & -\frac{1}{2} & 0 & 0 & 0 & \frac{1}{6} & 0 \\
	0 & \frac{1}{6} & \frac{1}{6} & -\frac{1}{2} & \frac{1}{3} & -\frac{1}{2} & -\frac{1}{2} & 1 & -\frac{1}{6} & \frac{1}{6} & 0 & 0 & -\frac{1}{6} & 0 & \frac{1}{6} & 0 & \frac{2}{3} & -\frac{1}{2} & -\frac{1}{2} & 0 & -\frac{2}{3} & \frac{2}{3} & \frac{2}{3} & -\frac{1}{2} \\
\end{array}
	\right)\,, \qquad \qquad \\&
	D = \D{12} \, 
	\end{split}
	\label{eq:web1app}
	\end{align}
	\item[{\bf 20}.]  $\textbf{W}_{3}^{(4)}(1,3,4)$
	\\
	This Cweb has 24 diagrams, one of which is displayed below. The table gives
	the chosen order of the 24 shuffles of the gluon attachments, and the corresponding
	$s$ factors.
	\begin{minipage}{0.5\textwidth}
		\begin{figure}[H]
			\vspace{-2mm}
			\includegraphics[height=4cm,width=4cm]{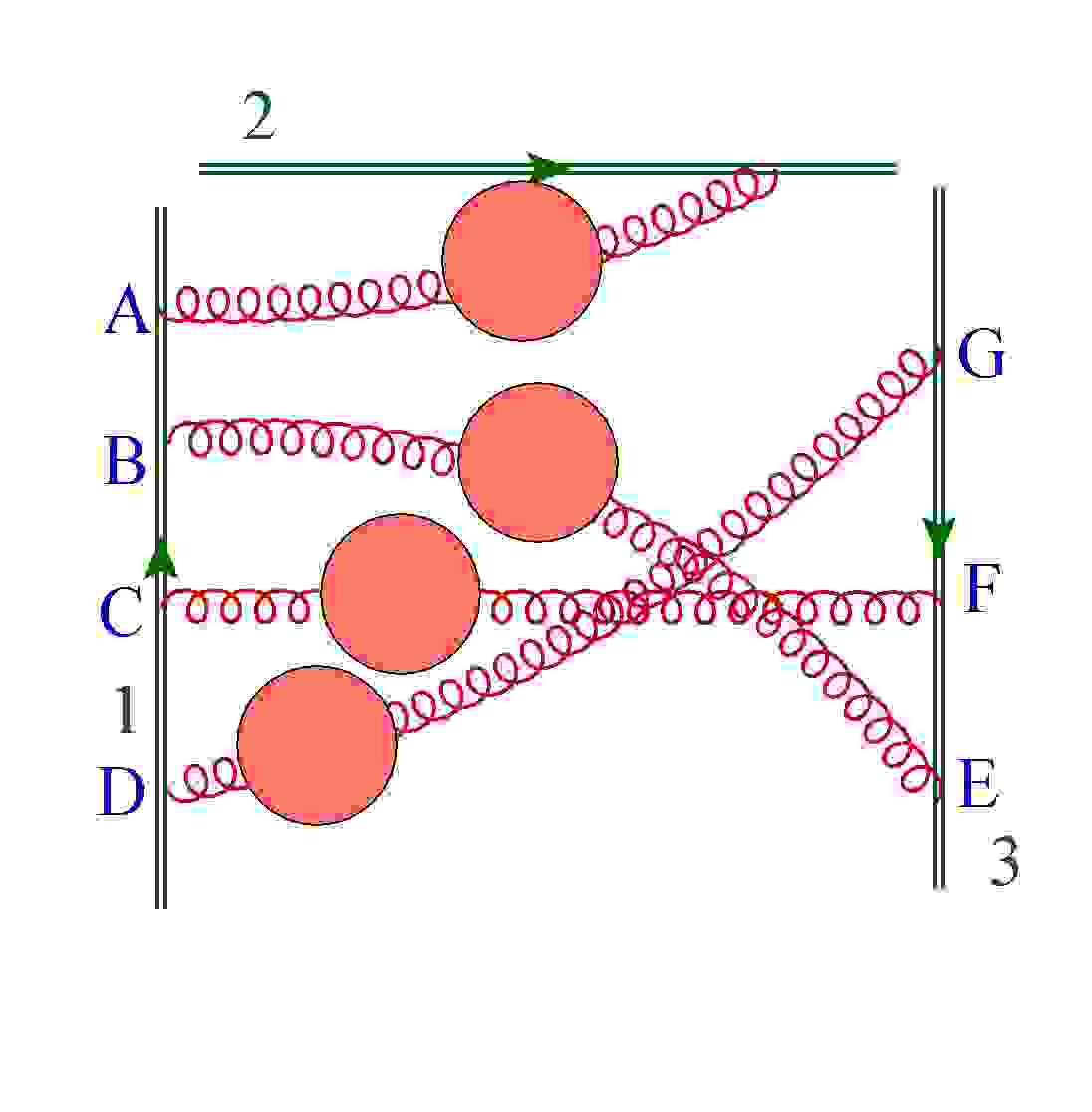}
		\end{figure}
	\end{minipage} 
	\hspace{-2cm}
	\begin{minipage}{0.46\textwidth}
		\vspace{2cm}
		\begin{tabular}{ | c | c | c |}
			\hline
			\textbf{Diagrams} & \textbf{Sequences} & \textbf{s-factors} \\ \hline
			$C_1$ & $\lbrace\lbrace DCBA\rbrace, \lbrace GFE\rbrace \rbrace$ & 1 \\ \hline
			$C_2$ & $\lbrace\lbrace DCBA\rbrace, \lbrace FGE\rbrace \rbrace$ & 0 \\ \hline
			$C_3$ & $\lbrace\lbrace DCBA\rbrace, \lbrace GEF\rbrace \rbrace$ & 0 \\ \hline
			$C_4$ & $\lbrace\lbrace DCBA\rbrace, \lbrace EGF\rbrace \rbrace$ & 0 \\ \hline
			$C_5$ & $\lbrace\lbrace DCBA\rbrace, \lbrace FEG\rbrace \rbrace$ & 0 \\ \hline
			$C_6$ & $\lbrace\lbrace DCBA\rbrace, \lbrace EFG\rbrace \rbrace$ & 0 \\ \hline
			$C_7$ & $\lbrace\lbrace DCAB\rbrace, \lbrace GFE\rbrace \rbrace$ & 1 \\
\hline
		\end{tabular}
		\label{tab:abcd1}	
	\end{minipage}
	
	\noindent The $R$ matrix is given by
	\begin{align}
	\begin{split}
	&
	R=\left(
	\begin{array}{cccccccccccccccccccccccc}
	0 & 0 & 0 & 0 & 0 & 0 & 0 & 0 & 0 & 0 & 0 & 0 & 0 & 0 & 0 & 0 & 0 & 0 & 0 & 0 & 0 & 0 & 0 & 0 \\
	-\frac{1}{6} & \frac{1}{3} & -\frac{1}{6} & 0 & 0 & 0 & \frac{1}{6} & -\frac{1}{6} & 0 & 0 & 0 & 0 & \frac{1}{6} & 0 & -\frac{1}{6} & 0 & 0 & 0 & -\frac{1}{6} & -\frac{1}{6} & \frac{1}{3} & 0 & 0 & 0 \\
	-\frac{1}{6} & -\frac{1}{6} & \frac{1}{3} & 0 & 0 & 0 & \frac{1}{6} & -\frac{1}{6} & 0 & 0 & 0 & 0 & \frac{1}{6} & 0 & -\frac{1}{6} & 0 & 0 & 0 & -\frac{1}{6} & \frac{1}{3} & -\frac{1}{6} & 0 & 0 & 0 \\
	\frac{1}{6} & -\frac{1}{3} & -\frac{1}{3} & \frac{1}{2} & 0 & 0 & \frac{1}{3} & -\frac{1}{3} & 0 & 0 & 0 & 0 & \frac{1}{3} & 0 & -\frac{1}{3} & 0 & 0 & 0 & -\frac{5}{6} & \frac{2}{3} & \frac{2}{3} & -\frac{1}{2} & 0 & 0 \\
	\frac{1}{6} & -\frac{1}{3} & -\frac{1}{3} & 0 & \frac{1}{2} & 0 & \frac{1}{3} & -\frac{1}{3} & 0 & 0 & 0 & 0 & \frac{1}{3} & 0 & -\frac{1}{3} & 0 & 0 & 0 & -\frac{5}{6} & \frac{2}{3} & \frac{2}{3} & 0 & -\frac{1}{2} & 0 \\
	\frac{1}{2} & -\frac{1}{2} & -\frac{1}{2} & 0 & 0 & \frac{1}{2} & \frac{1}{2} & -\frac{1}{2} & 0 & 0 & 0 & 0 & \frac{1}{2} & 0 & -\frac{1}{2} & 0 & 0 & 0 & -\frac{3}{2} & 1 & 1 & 0 & 0 & -\frac{1}{2} \\
	-\frac{1}{6} & 0 & 0 & 0 & 0 & 0 & \frac{1}{2} & 0 & 0 & 0 & 0 & 0 & -\frac{1}{2} & 0 & 0 & 0 & 0 & 0 & \frac{1}{6} & 0 & 0 & 0 & 0 & 0 \\
	\frac{1}{6} & -\frac{1}{6} & -\frac{1}{6} & 0 & 0 & 0 & \frac{1}{6} & \frac{1}{3} & 0 & 0 & 0 & 0 & -\frac{5}{6} & 0 & \frac{1}{3} & 0 & 0 & 0 & \frac{1}{2} & -\frac{1}{6} & -\frac{1}{6} & 0 & 0 & 0 \\
\end{array}
	\right)\,, \qquad \qquad \\&
	D = \D{13} \, 
	\end{split}
	\label{eq:web1app}
	\end{align}
\end{itemize}
%%%%%%%%%%%%%%%%%%%%%%%%%%%%%%%%%%

%%%%%%%%%%%%%%%%%%
\pagebreak
\subsection{Cwebs connecting two Wilson lines}
\vspace{2mm}
%%%%%%%%%%%%%%%%%%
\begin{itemize}
	
	%%%%%%%%%%%%%%%%%%
	
	\item[{\bf 1}.]  $\textbf{W}_{2}^{(1,0,1)}(2,4)$
	
	This Cweb has eight diagrams, one of which is displayed below. The table gives
	the chosen order of the eight shuffles of the gluon attachments, and the corresponding
	$s$ factors.
	\begin{minipage}{0.5\textwidth}
		\begin{figure}[H]
			\vspace{-2mm}
			\includegraphics[height=4cm,width=4cm]{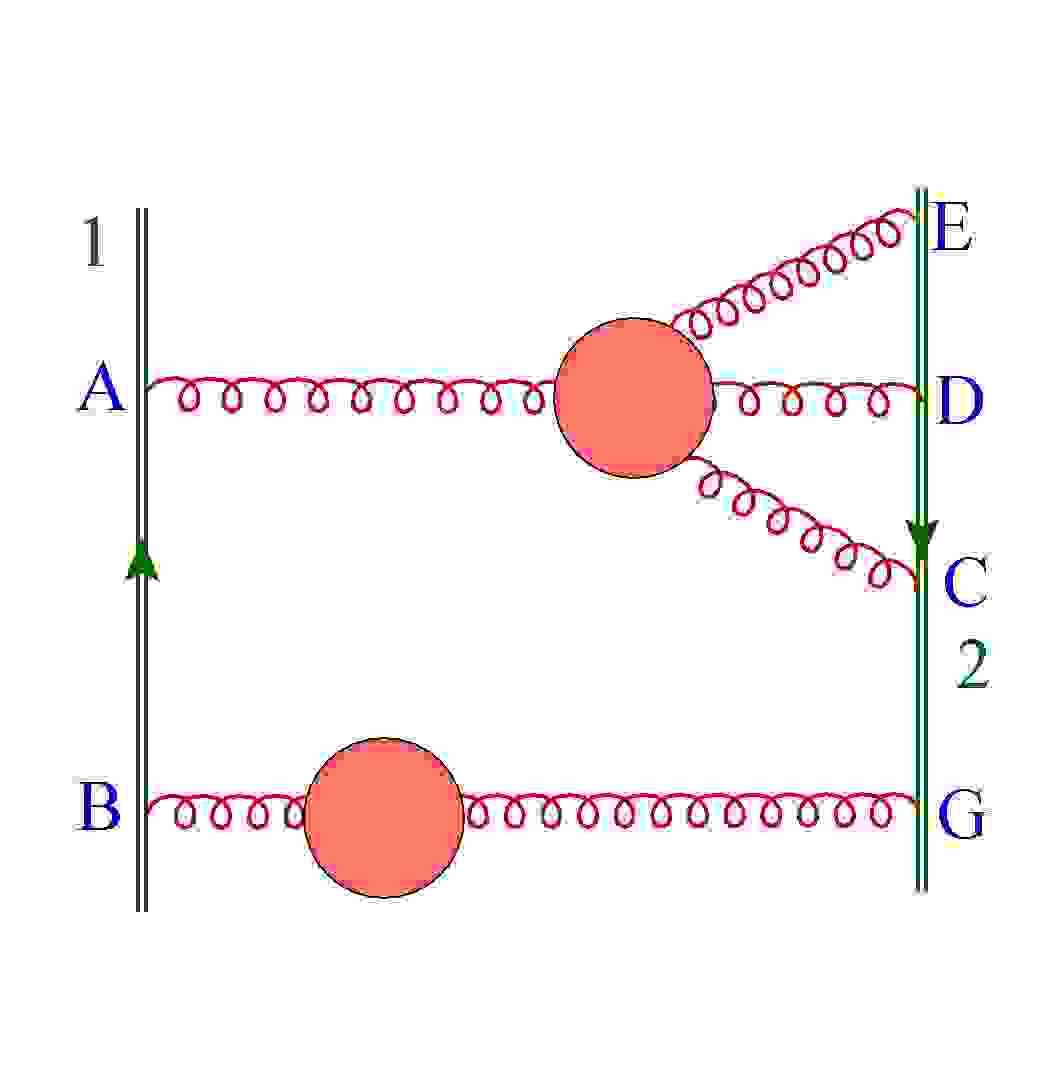}
		\end{figure}
	\end{minipage} 
	\hspace{-2cm}
	\begin{minipage}{0.46\textwidth}
		\vspace{2cm}
\begin{tabular}{ | c | c | c |}
			\hline
			\textbf{Diagrams} & \textbf{Sequences} & \textbf{s-factors} \\ \hline
			$C_1$ & $\lbrace\lbrace BA\rbrace,\lbrace EDCG \rbrace \rbrace$ & 0 \\ \hline
			$C_2$ & $\lbrace\lbrace BA\rbrace,\lbrace EDGC\rbrace \rbrace$ & 0 \\ \hline
			$C_3$ & $\lbrace\lbrace BA\rbrace,\lbrace EGDC \rbrace \rbrace$ & 0 \\ \hline
			$C_4$ & $\lbrace\lbrace BA\rbrace,\lbrace GEDC \rbrace \rbrace$ & 1 \\ \hline
			$C_5$ & $\lbrace\lbrace AB\rbrace,\lbrace EDCG \rbrace \rbrace$ & 1 \\ \hline
			$C_6$ & $\lbrace\lbrace AB\rbrace,\lbrace EDGC\rbrace \rbrace$ & 0 \\ \hline
			$C_7$ & $\lbrace\lbrace AB\rbrace,\lbrace EGDC \rbrace \rbrace$ & 0 \\ \hline
			$C_8$ & $\lbrace\lbrace AB\rbrace,\lbrace GEDC \rbrace \rbrace$ & 0 \\ \hline
		\end{tabular}
		\label{tab:abcd1}	
	\end{minipage}
	
	\noindent The $R$ matrix is given by
	\begin{align}
	\begin{split}
	&
	R=\left(
	\begin{array}{cccccccc}
	1 & 0 & 0 & -\frac{1}{2} & -\frac{1}{2} & 0 & 0 & 0 \\
	0 & 1 & 0 & -\frac{1}{2} & -\frac{1}{2} & 0 & 0 & 0 \\
	0 & 0 & 1 & -\frac{1}{2} & -\frac{1}{2} & 0 & 0 & 0 \\
	0 & 0 & 0 & \frac{1}{2} & -\frac{1}{2} & 0 & 0 & 0 \\
	0 & 0 & 0 & -\frac{1}{2} & \frac{1}{2} & 0 & 0 & 0 \\
	0 & 0 & 0 & -\frac{1}{2} & -\frac{1}{2} & 1 & 0 & 0 \\
	0 & 0 & 0 & -\frac{1}{2} & -\frac{1}{2} & 0 & 1 & 0 \\
	0 & 0 & 0 & -\frac{1}{2} & -\frac{1}{2} & 0 & 0 & 1 \\
	\end{array}
	\right)\,, \qquad \qquad
	D = \D{7} \, .
	\end{split}
	\label{eq:web1app}
	\end{align}
	\item[{\bf 2}.]  $\textbf{W}_{2}^{(1,0,1)}(3,3)$
	
	This Cweb has nine diagrams, one of which is displayed below. The table gives
	the chosen order of the nine shuffles of the gluon attachments, and the corresponding
	$s$ factors.
	\begin{minipage}{0.5\textwidth}
		\begin{figure}[H]
			\vspace{-2mm}
			\includegraphics[height=4cm,width=4cm]{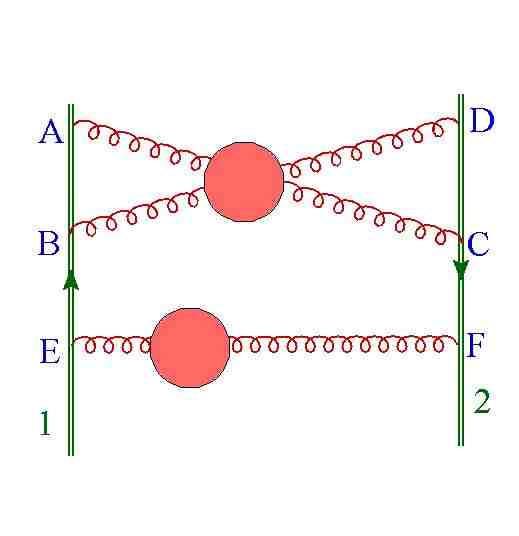}
		\end{figure}
	\end{minipage} 
	\hspace{-2cm}
	\begin{minipage}{0.46\textwidth}
		\vspace{2cm}
		\begin{tabular}{ | c | c | c |}
			\hline
			\textbf{Diagrams} & \textbf{Sequences} & \textbf{s-factors} \\ \hline
			$C_1$ & $\lbrace\lbrace EBA\rbrace,\lbrace DCF \rbrace \rbrace$ & 0 \\ \hline
			$C_2$ & $\lbrace\lbrace EBA\rbrace,\lbrace DFC\rbrace \rbrace$ & 0 \\ \hline
			$C_3$ & $\lbrace\lbrace EBA\rbrace,\lbrace FDC \rbrace \rbrace$ & 1 \\ \hline
			$C_4$ & $\lbrace\lbrace BEA\rbrace,\lbrace DCF \rbrace \rbrace$ & 0 \\ \hline
			$C_5$ & $\lbrace\lbrace BEA\rbrace,\lbrace DFC\rbrace \rbrace$ & 0 \\ \hline
			$C_6$ & $\lbrace\lbrace BEA\rbrace,\lbrace FDC \rbrace \rbrace$ & 0
			\\ \hline
			$C_7$ & $\lbrace\lbrace BAE\rbrace,\lbrace DCF \rbrace \rbrace$ & 1 \\ \hline
			$C_8$ & $\lbrace\lbrace BAE\rbrace,\lbrace DFC\rbrace \rbrace$ & 0 \\ \hline
			$C_9$ & $\lbrace\lbrace BAE\rbrace,\lbrace FDC \rbrace \rbrace$ & 0 \\ \hline
		\end{tabular}
		\label{tab:abcd1}	
	\end{minipage}
	
	\noindent The $R$ matrix is given by
	\begin{align}
	\begin{split}
	&
	R=\left(
	\begin{array}{ccccccccc}
	1 & 0 & -\frac{1}{2} & 0 & 0 & 0 & -\frac{1}{2} & 0 & 0 \\
	0 & 1 & -\frac{1}{2} & 0 & 0 & 0 & -\frac{1}{2} & 0 & 0 \\
	0 & 0 & \frac{1}{2} & 0 & 0 & 0 & -\frac{1}{2} & 0 & 0 \\
	0 & 0 & -\frac{1}{2} & 1 & 0 & 0 & -\frac{1}{2} & 0 & 0 \\
	0 & 0 & -\frac{1}{2} & 0 & 1 & 0 & -\frac{1}{2} & 0 & 0 \\
	0 & 0 & -\frac{1}{2} & 0 & 0 & 1 & -\frac{1}{2} & 0 & 0 \\
	0 & 0 & -\frac{1}{2} & 0 & 0 & 0 & \frac{1}{2} & 0 & 0 \\
	0 & 0 & -\frac{1}{2} & 0 & 0 & 0 & -\frac{1}{2} & 1 & 0 \\
	0 & 0 & -\frac{1}{2} & 0 & 0 & 0 & -\frac{1}{2} & 0 & 1 \\
	\end{array}
	\right)\,, \qquad \qquad
	D = \D{8} \, .
	\end{split}
	\label{eq:web1app}
	\end{align}
	\item[{\bf 3}.]  $\textbf{W}_{2}^{(0,2)}(2,4)$
	
	This Cweb has six diagrams, one of which is displayed below. The table gives
	the chosen order of the six shuffles of the gluon attachments, and the corresponding
	$s$ factors.
	\begin{minipage}{0.5\textwidth}
		\begin{figure}[H]
			\vspace{-2mm}
			\includegraphics[height=4cm,width=4cm]{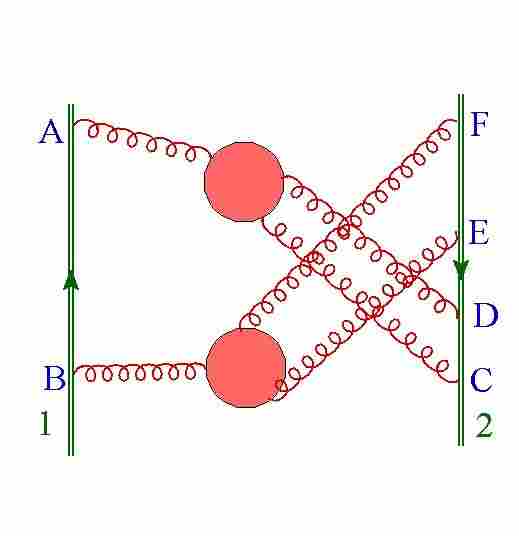}
		\end{figure}
	\end{minipage} 
	\hspace{-2cm}
	\begin{minipage}{0.46\textwidth}
		\vspace{2cm}
		\begin{tabular}{ | c | c | c |}
			\hline
			\textbf{Diagrams} & \textbf{Sequences} & \textbf{s-factors} \\ \hline
			$C_1$ & $\lbrace\lbrace BA\rbrace,\lbrace FEDC \rbrace \rbrace$ & 1 \\ \hline
			$C_2$ & $\lbrace\lbrace BA\rbrace,\lbrace FDEC\rbrace \rbrace$ & 0 \\ \hline
			$C_3$ & $\lbrace\lbrace BA\rbrace,\lbrace DFEC\rbrace \rbrace$ & 0\\ \hline
			$C_4$ & $\lbrace\lbrace BA\rbrace,\lbrace FDCE \rbrace \rbrace$ & 0 \\ \hline
			$C_5$ & $\lbrace\lbrace BA\rbrace,\lbrace DFCE\rbrace \rbrace$ & 0 \\ \hline
			$C_6$ & $\lbrace\lbrace BA\rbrace,\lbrace DCFE\rbrace \rbrace$ & 0
			\\ \hline
		\end{tabular}
		\label{tab:abcd1}	
	\end{minipage}
	
	\noindent The $R$ matrix is given by
	\begin{align}
	\begin{split}
	&
	R=\left(
	\begin{array}{cccccc}
	0 & 0 & 0 & 0 & 0 & 0 \\
	-1 & 1 & 0 & 0 & 0 & 0 \\
	-1 & 0 & 1 & 0 & 0 & 0 \\
	-1 & 0 & 0 & 1 & 0 & 0 \\
	-1 & 0 & 0 & 0 & 1 & 0 \\
	-1 & 0 & 0 & 0 & 0 & 1 \\
	\end{array}
	\right)\,, \qquad \qquad
	D = \D{5} \, .
	\end{split}
	\label{eq:web1app}
	\end{align}
	\item[{\bf 4}.]  $\textbf{W}_{2}^{(0,2)}(3,3)$
	
	This Cweb has nine diagrams, one of which is displayed below. The table gives
	the chosen order of the nine shuffles of the gluon attachments, and the corresponding
	$s$ factors.
	\begin{minipage}{0.5\textwidth}
		\begin{figure}[H]
			\vspace{-2mm}
			\includegraphics[height=4cm,width=4cm]{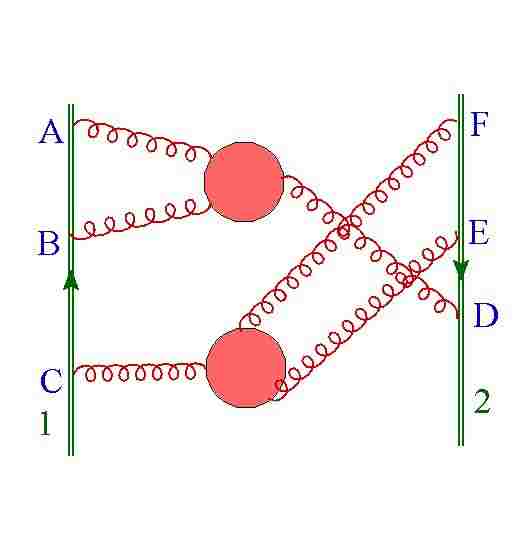}
		\end{figure}
	\end{minipage} 
	\hspace{-2cm}
	\begin{minipage}{0.46\textwidth}
		\vspace{2cm}
		\begin{tabular}{ | c | c | c |}
			\hline
			\textbf{Diagrams} & \textbf{Sequences} & \textbf{s-factors} \\ \hline
			$C_1$ & $\lbrace\lbrace CBA\rbrace,\lbrace FED \rbrace \rbrace$ & 1 \\ \hline
			$C_2$ & $\lbrace\lbrace CBA\rbrace,\lbrace FDE\rbrace \rbrace$ & 0 \\ \hline
			$C_3$ & $\lbrace\lbrace CBA\rbrace,\lbrace DFE \rbrace \rbrace$ & 0 \\ \hline
			$C_4$ & $\lbrace\lbrace BCA\rbrace,\lbrace FED \rbrace \rbrace$ & 0 \\ \hline
			$C_5$ & $\lbrace\lbrace BCA\rbrace,\lbrace FDE\rbrace \rbrace$ & 0 \\ \hline
			$C_6$ & $\lbrace\lbrace BCA\rbrace,\lbrace DFE \rbrace \rbrace$ & 0 \\ \hline
			$C_7$ & $\lbrace\lbrace BAC\rbrace,\lbrace FED \rbrace \rbrace$ & 0 \\ \hline
			$C_8$ & $\lbrace\lbrace BAC\rbrace,\lbrace FDE\rbrace \rbrace$ & 0 \\ \hline
			$C_9$ & $\lbrace\lbrace BAC\rbrace,\lbrace DFE \rbrace \rbrace$ & 1 \\ \hline
		\end{tabular}
		\label{tab:abcd1}	
	\end{minipage}
	
	\noindent The $R$ matrix is given by
	\begin{align}
	\begin{split}
	&
	R=\left(
	\begin{array}{ccccccccc}
	\frac{1}{2} & 0 & 0 & 0 & 0 & 0 & 0 & 0 & -\frac{1}{2} \\
	-\frac{1}{2} & 1 & 0 & 0 & 0 & 0 & 0 & 0 & -\frac{1}{2} \\
	-\frac{1}{2} & 0 & 1 & 0 & 0 & 0 & 0 & 0 & -\frac{1}{2} \\
	-\frac{1}{2} & 0 & 0 & 1 & 0 & 0 & 0 & 0 & -\frac{1}{2} \\
	-\frac{1}{2} & 0 & 0 & 0 & 1 & 0 & 0 & 0 & -\frac{1}{2} \\
	-\frac{1}{2} & 0 & 0 & 0 & 0 & 1 & 0 & 0 & -\frac{1}{2} \\
	-\frac{1}{2} & 0 & 0 & 0 & 0 & 0 & 1 & 0 & -\frac{1}{2} \\
	-\frac{1}{2} & 0 & 0 & 0 & 0 & 0 & 0 & 1 & -\frac{1}{2} \\
	-\frac{1}{2} & 0 & 0 & 0 & 0 & 0 & 0 & 0 & \frac{1}{2} \\
	\end{array}
	\right)\,, \qquad \qquad
	D = \D{8} \, .
	\end{split}
	\label{eq:web1app}
	\end{align}
	\item[{\bf 5}.]  $\textbf{W}_{2}^{(2,1)}(3,4)$
	
	This Cweb has thirty six diagrams, one of which is displayed below. The table gives
	the chosen order of the 36 shuffles of the gluon attachments, and the corresponding
	$s$ factors.
	\begin{minipage}{0.5\textwidth}
		\begin{figure}[H]
			\vspace{-2mm}
			\includegraphics[height=4cm,width=4cm]{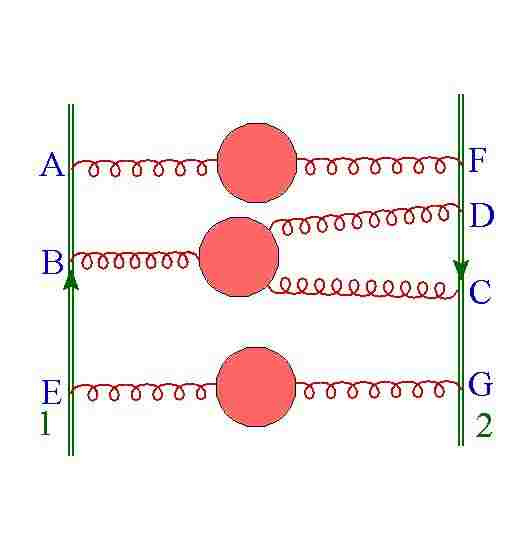}
		\end{figure}
	\end{minipage} 
	\hspace{-2cm}
	\begin{minipage}{0.46\textwidth}
		\vspace{2cm}
		\begin{tabular}{ | c | c | c |}
			\hline
			\textbf{Diagrams} & \textbf{Sequences} & \textbf{s-factors} \\ \hline
			$C_1$ & $\lbrace\lbrace EBA\rbrace,\lbrace FDCG \rbrace \rbrace$ & 0 \\ \hline
			$C_2$ & $\lbrace\lbrace EBA\rbrace,\lbrace DFCG \rbrace \rbrace$ & 0 \\ \hline
			$C_3$ & $\lbrace\lbrace EBA\rbrace,\lbrace DCFG \rbrace \rbrace$ & 0 \\ \hline
			$C_4$ & $\lbrace\lbrace EBA\rbrace,\lbrace FDGC \rbrace \rbrace$ & 0 \\ \hline
			$C_5$ & $\lbrace\lbrace EBA\rbrace,\lbrace DFGC \rbrace \rbrace$ & 0 \\ \hline
			$C_6$ & $\lbrace\lbrace EBA\rbrace,\lbrace FGDC \rbrace \rbrace$ & 0 \\ \hline
			$C_7$ & $\lbrace\lbrace EBA\rbrace,\lbrace GFDC \rbrace \rbrace$ & 0 \\
\hline
		\end{tabular}
		\label{tab:abcd1}	
	\end{minipage}
	
	\noindent The $R$ matrix is given by
	
	\resizebox{.8\linewidth}{!}{
		\begin{minipage}{\linewidth}
			\begin{align}
			\begin{split}
			&
			R=\left(
			\begin{array}{cccccccccccccccccccccccccccccccccccc}
			1 & 0 & 0 & 0 & 0 & 0 & -\frac{1}{2} & 0 & 0 & -\frac{1}{2} & 0 & \frac{2}{3} & 0 & 0 & -\frac{1}{2} & 0 & 0 & 0 & 0 & 0 & 0 & \frac{2}{3} & 0 & -\frac{1}{2} & 0 & 0 & 0 & 0 & 0 & -\frac{1}{2} & \frac{2}{3} & 0 & 0 & 0 & 0 & -\frac{1}{2} \\
			0 & 1 & 0 & 0 & 0 & 0 & 0 & 0 & -\frac{1}{2} & -\frac{1}{2} & 0 & \frac{2}{3} & 0 & 0 & -\frac{1}{2} & 0 & 0 & 0 & 0 & 0 & 0 & \frac{2}{3} & -\frac{1}{2} & 0 & 0 & 0 & 0 & 0 & 0 & -\frac{1}{2} & \frac{2}{3} & 0 & 0 & 0 & 0 & -\frac{1}{2} \\
			0 & 0 & 1 & 0 & 0 & 0 & 0 & 0 & 0 & -\frac{1}{2} & 0 & \frac{1}{6} & 0 & 0 & -\frac{1}{2} & 0 & 0 & 0 & 0 & 0 & 0 & \frac{1}{6} & 0 & 0 & 0 & 0 & 0 & 0 & 0 & -\frac{1}{2} & \frac{2}{3} & 0 & 0 & 0 & 0 & -\frac{1}{2} \\
			0 & 0 & 0 & 1 & 0 & 0 & -\frac{1}{2} & 0 & 0 & 0 & -\frac{1}{2} & \frac{2}{3} & 0 & 0 & -\frac{1}{2} & 0 & 0 & 0 & 0 & 0 & 0 & \frac{2}{3} & 0 & -\frac{1}{2} & 0 & 0 & 0 & 0 & 0 & -\frac{1}{2} & \frac{2}{3} & 0 & -\frac{1}{2} & 0 & 0 & 0 \\
			0 & 0 & 0 & 0 & 1 & 0 & 0 & 0 & -\frac{1}{2} & 0 & -\frac{1}{2} & \frac{2}{3} & 0 & 0 & -\frac{1}{2} & 0 & 0 & 0 & 0 & 0 & 0 & \frac{2}{3} & -\frac{1}{2} & 0 & 0 & 0 & 0 & 0 & 0 & -\frac{1}{2} & \frac{2}{3} & 0 & -\frac{1}{2} & 0 & 0 & 0 \\
			0 & 0 & 0 & 0 & 0 & 1 & -\frac{1}{2} & 0 & 0 & 0 & 0 & \frac{1}{6} & 0 & 0 & -\frac{1}{2} & 0 & 0 & 0 & 0 & 0 & 0 & \frac{2}{3} & 0 & -\frac{1}{2} & 0 & 0 & 0 & 0 & 0 & -\frac{1}{2} & \frac{1}{6} & 0 & 0 & 0 & 0 & 0 \\
			0 & 0 & 0 & 0 & 0 & 0 & \frac{1}{2} & 0 & 0 & 0 & 0 & \frac{1}{6} & 0 & 0 & 0 & 0 & 0 & 0 & 0 & 0 & 0 & \frac{1}{6} & 0 & -\frac{1}{2} & 0 & 0 & 0 & 0 & 0 & 0 & -\frac{1}{3} & 0 & 0 & 0 & 0 & 0 \\
			0 & 0 & 0 & 0 & 0 & 0 & 0 & 1 & -\frac{1}{2} & 0 & -\frac{1}{2} & \frac{2}{3} & 0 & 0 & 0 & 0 & 0 & 0 & 0 & 0 & 0 & \frac{1}{6} & -\frac{1}{2} & 0 & 0 & 0 & 0 & 0 & 0 & 0 & \frac{1}{6} & 0 & -\frac{1}{2} & 0 & 0 & 0 \\
\end{array}
			\right)\,,
			\end{split}
			\label{eq:web1app}
			\end{align}
		\end{minipage}
	}
	\\
	\hspace{-2cm}
	\begin{align}
	D = \D{29}
	\end{align} 
	\item[{\bf 6}.]  $\textbf{W}_{2}^{(4)}(4,4)$
	
	This Cweb has 24 diagrams, one of which is displayed below. The table gives
	the chosen order of the 24 shuffles of the gluon attachments, and the corresponding
	$s$ factors.
	\begin{minipage}{0.5\textwidth}
		\begin{figure}[H]
			\vspace{-2mm}
			\includegraphics[height=4cm,width=4cm]{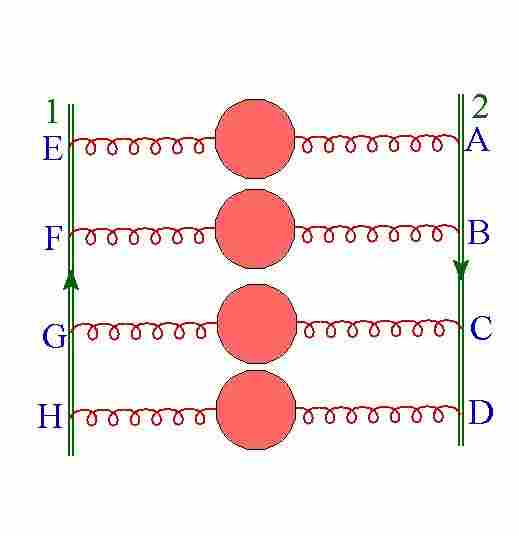}
		\end{figure}
	\end{minipage} 
	\hspace{-2cm}
	\begin{minipage}{0.46\textwidth}
		\vspace{2cm}
		\begin{tabular}{ | c | c | c |}
			\hline
			\textbf{Diagrams} & \textbf{Sequences} & \textbf{s-factors} \\ \hline
			$C_1$ & $\lbrace\lbrace EFGH\rbrace,\lbrace DCBA \rbrace \rbrace$ & 1 \\ \hline
			$C_2$ & $\lbrace\lbrace EFGH\rbrace,\lbrace CDBA \rbrace \rbrace$ & 0 \\ \hline
			$C_3$ & $\lbrace\lbrace EFGH\rbrace,\lbrace DBCA \rbrace \rbrace$ & 0 \\ \hline
			$C_4$ & $\lbrace\lbrace EFGH\rbrace,\lbrace BDCA \rbrace \rbrace$ & 0 \\ \hline
			$C_5$ & $\lbrace\lbrace EFGH\rbrace,\lbrace CBDA \rbrace \rbrace$ & 0 \\ \hline
			$C_6$ & $\lbrace\lbrace EFGH\rbrace,\lbrace BCDA \rbrace \rbrace$ & 0 \\ \hline
			$C_7$ & $\lbrace\lbrace EFGH\rbrace,\lbrace DCAB \rbrace \rbrace$ & 0 \\
\hline
			
		\end{tabular}
		\label{tab:abcd1}	
	\end{minipage}
	
	\noindent The $R$ matrix is given by
	\begin{align}
	\begin{split}
	&
	R=\left(
	\begin{array}{cccccccccccccccccccccccc}
	0 & 0 & 0 & 0 & 0 & 0 & 0 & 0 & 0 & 0 & 0 & 0 & 0 & 0 & 0 & 0 & 0 & 0 & 0 & 0 & 0 & 0 & 0 & 0 \\
	0 & \frac{1}{6} & -\frac{1}{3} & 0 & 0 & 0 & \frac{1}{6} & 0 & 0 & 0 & 0 & 0 & 0 & 0 & 0 & 0 & 0 & 0 & 0 & 0 & 0 & 0 & 0 & 0 \\
	0 & -\frac{1}{3} & \frac{2}{3} & 0 & 0 & 0 & -\frac{1}{3} & 0 & 0 & 0 & 0 & 0 & 0 & 0 & 0 & 0 & 0 & 0 & 0 & 0 & 0 & 0 & 0 & 0 \\
	0 & -\frac{2}{3} & \frac{1}{3} & \frac{1}{2} & 0 & 0 & \frac{1}{3} & 0 & -\frac{1}{2} & 0 & 0 & 0 & 0 & 0 & 0 & 0 & 0 & 0 & 0 & 0 & 0 & 0 & 0 & 0 \\
	0 & -\frac{2}{3} & \frac{1}{3} & 0 & \frac{1}{2} & 0 & \frac{1}{3} & 0 & 0 & 0 & 0 & 0 & -\frac{1}{2} & 0 & 0 & 0 & 0 & 0 & 0 & 0 & 0 & 0 & 0 & 0 \\
	0 & -1 & \frac{1}{2} & 0 & 0 & \frac{1}{2} & \frac{1}{2} & 0 & 0 & 0 & 0 & 0 & 0 & 0 & -\frac{1}{2} & 0 & 0 & 0 & 0 & 0 & 0 & 0 & 0 & 0 \\
	0 & \frac{1}{6} & -\frac{1}{3} & 0 & 0 & 0 & \frac{1}{6} & 0 & 0 & 0 & 0 & 0 & 0 & 0 & 0 & 0 & 0 & 0 & 0 & 0 & 0 & 0 & 0 & 0 \\
	0 & \frac{1}{3} & -\frac{2}{3} & 0 & 0 & 0 & \frac{1}{3} & 0 & 0 & 0 & 0 & 0 & 0 & 0 & 0 & 0 & 0 & 0 & 0 & 0 & 0 & 0 & 0 & 0 \\
\end{array}
	\right)\,, \qquad \qquad \\&
	D = \D{17} \,  \\&
	\end{split}
	\end{align}
	\end{itemize}

\noindent This completes our listing of all Cwebs with a perturbative  expansion 
starting at  ${\cal O} (g^8)$, and connecting two and three Wilson lines.

\bibliographystyle{JHEP}
\bibliography{mybib}

%%%%%%%%%%%%%%%%%%%%%%%%%%%%%%%%%%

\end{document}